\documentclass[12pt,oneside]{scrbook}
\usepackage{scrhack}
\usepackage{hyperref,hypcap,ae,tikz}
\usepackage{graphicx}
\usepackage{bm}
\usepackage{amsmath}
\usepackage[T1]{fontenc}
\usepackage{xcolor}
\usepackage{lmodern}
\usepackage[caption=false]{subfig}
\usepackage{placeins}
\usepackage{amssymb,hyperref,hypcap,
ae,tikz,color}
\usepackage{latexsym,textcomp,slashed,epsfig}
\usepackage{verbatim}
\usepackage{dcolumn}
\usepackage{listings} 
\usepackage{color}
\usepackage{mathtools}

\usepackage{textcomp}
\hypersetup{
	colorlinks=true,
	citecolor=blue,
    linkcolor=magenta,
	}
\usepackage{appendix}

\begin{document}

\begin{titlepage}
   \begin{center}
       \vspace*{1cm}

       \Large{\textbf{ON THE THERMODYNAMIC ASPECTS OF GRAVITY}}

       \vspace{0.5cm}
       \large{ Thesis submitted for the degree of \\
        Doctor of Philosophy (Science) \\
        in Physics}
            
       \vspace{1.5cm}
Submitted by\\
       \textbf{SAMARJIT CHAKRABORTY}

      \vfill
        
       \includegraphics[width=0.20\textwidth]{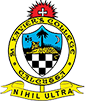} \\
        \vspace{0.8cm}    
       Post Graduate and Research Department
       of Physics\\    
       St. Xavier's College (Autonomous), Kolkata\\
       \textbf{University of Calcutta}\\
       INDIA\\
       2023
           
   \end{center}
\end{titlepage}



            


           

\newpage

\vfill
\begin{center}
\vspace{2cm}
To \\
\vspace{1cm}
My late grand parents\\
\vspace{0.25cm}
MAYA CHAKRABORTY \& PARITOSH CHAKRABORTY\\
\vspace{0.25cm}
SAFALI NAG \& BIMALENDU NAG \\
\vspace{1cm}
My parents\\
\vspace{0.25cm}
SUPRITI CHAKRABORTY \& PRABIR CHAKRABORTY\\
\vspace{1cm}
My respected supervisor\\
\vspace{0.25cm}
SARBARI GUHA\\
\&\\
all my respected teachers\\
\vspace{1cm}
and \\
\vspace{1cm}
My lovely cousin\\
\vspace{0.25cm}
SAMARPITA CHAKRABORTY.

\end{center}

\newpage
\section*{Acknowledgements}

I would like to begin by offering my heartfelt thanks and sincere gratitude to my teacher and supervisor Dr. Sarbari Guha for her generous help, constant care, support, guidance and encouragement during my entire research tenure. I consider myself very fortunate to have her as my teacher, philosopher and guide. I owe her whatever knowledge I have in General Relativity and Cosmology. I want to especially thank Dr. Sarbari Guha and her family for their warm hospitality. Here I must specially thank Dr. Sarbari Guha again for providing me CSIR scholarship and personally monitoring all the  paper work meticulously and helping me understand the official works. I am deeply indebted to my supervisor without whom this work would not be possible.
\\
I must mention Prof. Parthasarathi Majumdar, from whom I learnt a lot during my daily visits to SINP as an early honours student. He used to give us his valuable time from his busy schedule and help us learn relativity.\\
I want to thank Prof. Subenoy Chakraborty, for his valuable comments and suggestions. Also I would like to thank the Relativity and Cosmology Research Centre, Jadavpur
University for their assistance and valuable suggestions several times during my research period. Also I must thank my research advisory committee members i.e., Prof. Amit Ghosh and Prof. Rajesh K. Nayak for their crucial comments and suggestions. I also express my gratitude to Dr. Dibyendu Panigrahi for his collaboration and encouragement.\\
I am grateful to the Indian Association for General Relativity and Gravitation (IAGRG) for providing me multiple opportunities to attend their national and international conferences. It is really an honour and delight to meet all the eminent members of IAGRG helping us learn new things. IAGRG has helped my growth immensely in research and understanding. \\
I would also like to thank Sucheta Dutta and Uttaran Ghosh and other members of the Gravity group of St. Xavier's College (Autonomous), Kolkata for their support and help.
\\
 Simultaneously I would like to acknowledge all the staff and the amenities provided by the Department of Physics, St. Xavier's College (Autonomous), Kolkata. I want to specially mention all the names of my teachers Dr. Indranath Chaudhuri, Dr. Shibaji Banerjee, Dr. Tapati Dutta, Dr. Saunak Palit, Dr. Soma Ghosh, Dr. Suparna Roy Chowdhury, Dr. Tanaya Bhattacharyya, Dr. Sudipto Roy, and Dr. Subhankar Ghosh for their constant encouragement and well wishes. I am deeply thankful to the Inter-University Centre for Astronomy and Astrophysics (IUCAA), Pune for their research facilities and warm hospitality as a significant part of my work was done during visits there. Here I would like to express my gratitude to Prof. Sukanta Bose, Prof. Sanjit Mitra and  Prof. Debarati Chatterjee for their encouraging words. Also I want to thank Dr. Apratim Ganguly for helping and providing me materials regarding my future work prospects and giving me opportunity to present my work at GW group in IUCAA.
\\
 At this moment I should specially thank Prof. Rituparno Goswami for his kind and gracious collaboration. He has helped me immensely in my research and gave valuable suggestions and comments. Here I would also like to mention Prof. Sunil Maharaj for his valuable suggestions and comments regarding our work.
I am also thankful to the Astrophysics and Cosmology Research Unit, School of Mathematics, Statistics and Computer Sciences, UKZN, South Africa, for supporting Dr. Guha's short visit and allowing our work to be presented there. Here I would specially like to thank the National Research Foundation, South Africa, for research support to some of our works.\\
I would like to express my deepest respect to all of my physics teachers from my school days to my bachelor's degree in St. Xavier's college, Kolkata up till my masters in IIT Kanpur, who unveiled the world of Physics to me. They were wonderful and I am greatly indebted to all of them.\\
Above all, I express my gratitude and indebtedness to my parents for their blessings. I can't forget their sacrifices, well wishes, encouragement to make my journey smooth. I must thank my family for the support and care. A special thanks goes to Basabi Bagchi, whose mental support helped me throughout and inspired me to do more. Thanks are also due to my friends for providing several ideas and helping me in life. Last but not the least, I thank the staff of Department of physics, Principals office, PhD office especially Dr. Samrat Roy and the treasury office of St. Xavier's College (Autonomous), Kolkata for their support and cooperation all through, without which it would be very much difficult for me to get along with the entire official work.\\
Last but not the least, I want to thank CSIR, Government of India for giving me the opportunity to work as a JRF and SRF in the major research project No. 03(1446)/18/EMR-II. This entire research work is done under this scholarship.

\vspace{3cm}
\begin{flushright}
    (SAMARJIT CHAKRABORTY)\\
    \vspace{2cm}
    Department of Physics,\\
    St. Xavier's College (Autonomous), Kolkata;\\
    Kolkata 700016,\\
    India.
  \end{flushright}

\linespread{1.5}

\newpage
\section*{Abstract}
This thesis deals with mainly two issues. Firstly we have studied gravitational entropy on different astrophysical and cosmological systems. Secondly the thesis focuses on the generalized thermodynamic laws in gravitational physics. As a result the entire thesis has four parts, where in the first part, titled "\textbf{Prologue}", we have introduced the subject matter in brief. Subsequently in the second part, titled as "\textbf{Gravitational Entropy of Astrophysical and Cosmological systems}", describes our works regarding gravitational entropy. The third part deals with the generalized laws of thermodynamics in presence of gravity titled as "\textbf{Universal Thermodynamics}" and finally the conclusion of this thesis is described in the last part titled as "\textbf{Epilogue}".\\
The very first part of the thesis titled as "\textbf{Prologue}", is the introductory part of the thesis where we have discussed the background of the entire investigation carried out by us. Here we have briefly mentioned the entire history of the thermodynamic studies in gravitational physics, especially emphasising on the thermodynamics of Black holes and horizons in general. In relation to that we have introduced different types of cosmological horizons and their thermodynamics also. Then we briefly introduced the generalized  second law of thermodynamics and summarized our works on this subject matter. Subsequently we introduced the concept of Gravitational entropy and discussed different proposals of it. Finally we ended this part with a summary of our works regarding the Gravitational entropy.\\
In the second part of the thesis, titled as "\textbf{Gravitational Entropy of Astrophysical and Cosmological systems}", the works on gravitational entropy is described in three chapters.\\
In the \textbf{first chapter of the second part} we have examined the validity of a  gravitational entropy proposal in the context of accelerating black hole solutions of the Einstein field equations. We have adopted a phenomenological approach proposed by Rudjord et al,\cite{entropy1} in which the Weyl curvature hypothesis is examined using the proposal for the gravitational entropy. We have considered the $C$-metric to represent the accelerating black holes. We then evaluated the corresponding gravitational entropy and the gravitational entropy density for different types of accelerating black holes. We discussed the merits and demerits of such an analysis and commented on the possible resolutions of the issue.
\\
In the \textbf{second chapter}  we have investigated the entropy of the free gravitational field for some well known isotropic and anisotropic cosmologies. We have utilized the definition of gravitational entropy proposed by Clifton, Ellis and Tavakol,\cite{CET} where the 2-index square root of the 4-index Bel-Robinson tensor is taken as the energy momentum tensor for the free gravitational field. We have then examined whether in the vicinity of the initial cosmic singularity, the ratio of the energy density of free gravity to that of matter density goes to zero or not, examining the Penrose's conjecture on Weyl curvature. We showed that whenever this is true, the gravitational entropy increases monotonically with the structure formation of the universe. Then we discussed the conditions for which the Weyl curvature hypothesis is valid or otherwise.\\
The \textbf{third chapter} of this part of the thesis deals with the validity of two different proposals of gravitational entropy (GE) in the context of traversable wormhole solutions of the Einstein field equations. The first one is the phenomenological approach proposed by Rudjord et al \cite{entropy1,entropy2}, which is a purely geometric method of measuring gravitational entropy. The latter one is the Clifton-Ellis-Tavakol (CET) proposal \cite{CET} for the gravitational entropy which arises in relativistic thermodynamics, and is based on the Bel-Robinson tensor, that represents the effective super-energy-momentum tensor of free gravitational fields.   The application of the CET proposal can provide unique gravitational entropies for spacetimes of Petrov type D and N only, whereas the geometric method can be applied to almost every kind of spacetime, although it has little relation with thermodynamics. We have argued that for any traversable wormhole to be physically realistic, it should have a viable GE. We have found that the GE proposals do provide us a consistent measure of GE in several of wormholes solutions.\\

Subsequently we have the \textbf{third part} of the thesis, titled as "\textbf{Universal Thermodynamics}", where there are two chapters.\\
In the \textbf{first chapter} of this part we have examined the validity of the generalized second law of thermodynamics (GSLT) in an expanding Friedmann Robertson Walker (FRW) universe, which is filled with different variants of Chaplygin gases. We have assumed that the universe is a closed system bound by the cosmological horizon.  Then we presented the general prescription of the entire analysis and in the subsequent part we have analysed the validity of the GSLT on the cosmological horizons for different Chaplygin gas models. We found that for the cosmological apparent horizon, some of these models always obey the GSLT. Whereas the validity of GSLT on the cosmological event horizon depends on the free parameters of the respective models.
\\
Finally, in the \textbf{second chapter} of the third part we have dealt with the evolution of the FRW universe. Here the universe is filled with a variant of the Chaplygin gas model, namely the variable modified Chaplygin gas (VMCG). We have begun with a thermodynamical treatment of the equation of state of the VMCG  and obtained its temperature as a function of redshift $z$. We have shown that the results are consistent with similar works on other types of Chaplygin gas models. In addition to the derivation of the exact expression of temperature we also used observational data to determine the redshift at the epoch of transition from the decelerated to the accelerated phase of the universe. Then the values of other relevant parameters like the Hubble parameter, the equation-of-state parameter and the speed of sound are obtained in terms of the redshift. Subsequently these are compared with the results obtained from previous works on other Chaplygin gas models. We took the redshift of photon decoupling as $ z\simeq 1100 $, and used this value to calculate the temperature of decoupling.\\
Finally in the last part titled as "\textbf{Epilogue}" we have concluded our thesis and discussed our thoughts on it.
\newpage
\section*{Dissemination}

This thesis is based on the following papers listed below:

\begin{itemize}

\item S. Guha and S. Chakraborty,``On the gravitational entropy of accelerating black holes''\\
\textbf{Int.J.Mod.Phys.D 29, 2050034(2020)}\\
\textbf{DOI}: 10.1142/S0218271820500340

\item S. Chakraborty, S. Guha and R. Goswami, ``An investigation on gravitational entropy of cosmological models''\\
\textbf{Int.J.Mod.Phys.D 30, 2150051(2021)}\\
\textbf{DOI}: 10.1142/S0218271821500516

\item S. Chakraborty, S. Guha and R. Goswami, ``How appropriate are the gravitational entropy proposals for traversable wormholes?''\\
\textbf{Gen.Relativ.Gravit.54,47(2022)}\\
\textbf{DOI}: 10.1007/s10714-022-02934-3

\item S. Chakraborty and S. Guha, ``Thermodynamics of FRW universe with Chaplygin gas models''\\
\textbf{Gen.Relativ.Gravit. 51, 158(2019)}\\
\textbf{DOI}: 10.1007/s10714-019-2645-8

\item S. Chakraborty, S. Guha and D. Panigrahi, ``Evolution of FRW universe in variable modified Chaplygin gas model''\\
\textbf{arXiv:1906.12185v1[gr-qc](2019)}

\end{itemize}

\section*{Papers not included in this thesis}
\begin{itemize}
\item S. Chakraborty and S. Guha, ``Density Perturbation and Cosmological Evolution in the Presence of Magnetic Field in $f(R) $ Gravity Models''\\
\textbf{Advances in High Energy Physics
Volume 2022, Article ID 5195251}
\textbf{DOI}:10.1155/2022/5195251

\item S. Datta, S. Guha and S. Chakraborty, ``On the propagation of gravitational waves in matter-filled Bianchi I universe'' \\ \textbf{arXiv:2205.05733 [gr-qc] }
\end{itemize}

\newpage
\section*{Declaration by the author}
All the figures and calculations present in this thesis have been produced by the author using Maple and Mathematica software. The manuscript of this thesis has been checked several times with extreme care to make it free from all discrepancies and typos. Even then the vigilant reader may find mistakes, and some portions of this thesis may appear unwarranted. The author takes the sole responsibility for these unwarranted errors which may have resulted from his limited knowledge of this vast subject or incomplete understanding of the relevant concepts or because some topics escaped his notice.

\newpage

\vspace*{\fill}
\begingroup
\centering
\begin{quote}
  ``It being an established fact that the object and the subject,' that are fit to be the contents of the concepts "you" and "we" (respectively), and are by nature as contradictory as light and darkness, cannot logically have any identity, it follows that their attributes can have it still less. Accordingly, the superimposition of the object, referable through the concept "you", and its attributes on the subject that is conscious by nature and is referable through the concept "we" (should be impossible), and contrariwise the superimposition of the subject and its attributes on the object should be impossible. Nevertheless, owing to an absence of discrimination between these attributes, as also between substances, which are absolutely disparate, there continues a natural human behaviour based on self-identification in the form of "I am this" or "This is mine". This behaviour has for its material cause an unreal nescience and man resorts to it by mixing up reality with unreality as a result of superimposing the things themselves or their attributes on each other.''
  \begin{flushright}
    \small{--- Adi Shankaracharya, Brahmasutra Adhyasa-bhashya}
  \end{flushright}
\end{quote}

\endgroup
\vspace*{\fill}

{\hypersetup{linkcolor=black}
\tableofcontents
}

\part{Prologue}


Gravity is one of the fundamental forces of nature which has always been the most familiar one to human beings like us, right from our childhood. Yet, it is through the studies of numerous scientists and philosophers that the understanding of gravity has developed, and led humanity to the outer space through numerous scientific missions. Beginning with the Newtonian model of gravitation to Einstein's theory of gravity, and further beyond, the real nature of gravity still seems to be a bit of a mystery. \\
Gravity is so fundamental that the concept of spacetime is integral to it, and since all other forces interact among themselves within the same spacetime, it means that in every interaction, however small it maybe, gravity is omnipresent in the background. Unfortunately, despite the fact that a unified theory exists for other fundamental forces, unification with gravity seems illusive and difficult. Nonetheless, we know that gravity plays a crucial role from the very beginning of the universe in creating gas clouds, stars, galaxies and planets, and finally culminating in the creation of the ``supermassive'' black holes. \\
In numerous recent observations starting from the previous century, the general theory of relativity, in which gravity is viewed as a consequence of the geometry of spacetime due to massive objects, has proved to be accurate, and it is the basic fundamental theory for all our astronomical and astrophysical studies. Despite the elegance of this theory or perhaps because of it, although scientists have tried to unify this theory with the standard model of particle physics to create a single theory of quantum gravity to represent all the four fundamental forces of nature, but success have not yet been achieved. The modern theory of Loop Quantum Gravity and String theory are the results of such investigations, which provided further insights into a better theoretical understanding of the problem. This line of investigation is still lying within the realms of theoretical formulations without any solid experimental evidence, but it has shown us that gravitational field can be described statistically in terms of micro states. \\~~~\\
In these circumstances, it is very natural for one to seek for other alternative explanations of the phenomenon of gravity, among which the thermodynamic description seems to be very effective.

\section{Thermodynamics in presence of gravity}
The work on relativistic thermodynamics in flat spacetime  was initiated  way back in 1926 by Lenz in Zeitschrift fuer Physik \cite{lenz}, where he investigated the equilibrium between the radiation and matter in an Einstein's closed universe. Lenz had based his work on the principles of ordinary thermodynamics in flat spacetime, which was inadequate for the physics of curved spacetimes that encompasses Einstein's theory of gravity (GR). This work of Lenz was mentioned by R.C. Tolman in his 1928 paper \cite{Tolm1}, where he extended the principles of thermodynamics so as to hold in curved spacetime, where the methods of general relativity (GR) are applicable \cite{Tolm2,Tolm3,Tolm4}. \\
Tolman considered two principles expressed in the form of equations (true for all sets of coordinates) which would serve as the analogues of the ordinary first and second laws of thermodynamics in GR.
 The First Law was the analogue of the law of the conservation of energy in GR (Vanishing divergence of energy-momemtum tensor, converted into tensor densities and integrated over 4-volume).
 For the Second Law he considered two guiding principles: \\
(1) The postulate must be expressible in a form valid in all coordinate systems.
(2) It must be equivalent to the ordinary second law of thermodynamics in flat spacetime.

Consequently Tolman defined the entropy vector in the following way: \\
At any given point in spacetime defined by the equation $$S^{\mu}=\phi_0\frac{dx_{\mu}}{ds},$$
where $dx_{\mu}/ds$ refers to the macroscopic motion of matter (or energy) at the point, and $\phi_0$ is the proper density of entropy as measured by a comoving observer, the divergence of $S^{\mu}$ integrated over 4-volume (which includes the universe as a whole as an isolated system) should be positive or zero, subjected to the validity of the conservation of the energy-momentum tensor. Tolman in his subsequent papers \cite{Tolm5,Tolm6} and in several other works studied this subject in detail and expanded on it.

\section{Black hole as a thermodynamic system}
The thermodynamic studies of gravity gathered momentum in a major way with the semi classical study of black holes (BHs) which tells us that these objects emit thermal radiation (unlike cold classical BHs) and behave like black bodies, which indicated that black holes behave like thermodynamical systems. To save the sacred second law of thermodynamics (SLT), the Hawking-Bekenstein (HB) entropy was introduced \cite{Bekenstein} on the event horizon of a blackhole, where the temperature of the BH is proportional to the surface gravity and the entropy is proportional to the area of the horizon. Thus, the foundations of the laws of black hole thermodynamics in physics were established when it was found that the BH temperature, mass and entropy satisfied the first law of thermodynamics \cite{SWH1,BCH}.\\
Black hole physics provides us with at least two examples in which the second law of thermodynamics may be violated or transcended. In the first example, provided by J.D. Bekenstein \cite{bekenstein1}, we imagine that someone drops a cup of coffee into a black hole (BH). Then the outside observer would see that the environment entropy went down because the information of that cup will never leave the BH to the outside observer. This scenario depicts a perfect situation where BH mechanics violates the second law of thermodynamics. The second method was proposed by R. Geroch \cite{geroch} where we lower a mass tied by a string towards a BH and let the radiation from the rest mass energy get sucked into the BH, and then again haul the body up. In this process the BH remains unchanged and some amount of energy from the suspended mass gets completely converted into the work done. This again violates the second law of thermodynamics. This posed a problem for the whole framework of physics. In the year 1972 Jacob D. Bekenstein argued that the only solution to this problem is to endow the black hole with some entropy so that the sum of the BH entropy plus the environment entropy always increases \cite{bekenstein1}. In the next paper \cite{Bekenstein}, Bekenstein argued that the BH area and entropy have similarities in the sense that both of them always increase in any physical transformation. Floyd and Penrose also suggested that the area increase maybe a general feature of BH transformations \cite{penrose1,penrose2}. Independently Christodoulou had also shown that no process whose ultimate outcome is the capture of a particle by a Kerr BH can result in the decrease of irreducible mass of BH. He showed that the irreducible mass is proportional to the area of the BH. Christodoulou's results are consistent with that of Penrose and Floyd, i.e. in most of the processes, BH area increases \cite{chris1,chris2,chris3}. Hawking also illustrated a general proof that the BH surface area cannot decrease in any physical process. For a system of BHs, Hawking's theorem says that the area of the individual BHs cannot decrease and for the merger of two BHs, the resulting BH area cannot be smaller than the sum of the initial area of the individual BHs \cite{hawking1}.
From the works of all the above authors and many others, it became clear that BH area behaves like entropy, and that a BH has some kind of mechanism which resembles the second law of thermodynamics in action, which is: for any closed thermodynamic system, the changes take place in the direction of increasing entropy. For the BHs, therefore, its area represents the measure of its entropy.
\\
In 1973, the four laws of BH mechanics were proposed by J.M. Bardeen, B. Carter and S. W. Hawking where they argued that the area of the BH and its surface gravity in their derived relations resembles entropy and temperature, respectively, in the ordinary laws of thermodynamics. Based on this, they proposed the famous four laws \cite{bh} as mentioned below.
\begin{itemize}
\item \textbf{The Zeroth Law:} The surface gravity $(\kappa) $ of a stationary BH is constant over the Event Horizon.
\item \textbf{The First Law:} Any two neighbouring stationary axisymmetric solutions containing a perfect fluid with circular flow and a central black hole are related by
\begin{equation}\label{first_law}
\partial M=\dfrac{\kappa}{8\pi} \partial A + \Omega_{H} \partial J_{H} + \Phi \partial Q .
\end{equation}
It is very interesting to note that the quantity $\dfrac{\kappa}{2\pi} $ is analogous to temperature. Similarly
the area of the event horizon $A/4$ is analogous to entropy. Here $T$ is known as the Hawking temperature and $S$ is the HB entropy.
\item \textbf {The Second Law:} The event horizon area ($A$) of individual black hole does not decrease with time, i.e.
\begin{equation}
\partial A=0.
\end{equation}
In other words, if two black holes gets merged, the area of the final event horizon of the resulting BH is greater than the sum of the areas of the individual initial horizons, i.e.
\begin{equation}
A_{f}>A_{i1}+A_{i2}
\end{equation}
\item \textbf{The Third Law:} No matter how idealized the procedure is, it is impossible to reduce the surface gravity of a stationary BH ($\kappa $) to zero by a finite sequence of operations.
\end{itemize}
In 1974, Hawking showed by applying quantum mechanics that BHs are not cold but are hot bodies and its temperature is proportional to the horizon surface gravity, and that BHs are not eternal objects but they will evaporate one day via the Hawking radiation \cite{hawking2}. This established the surface gravity as a measure of the physical temperature of BH. This result became the final cornerstone of black hole thermodynamics.\\
Consequently physicists searched for a possible connection between BH thermodynamics and the gravitational field equations. Ted Jacobson \cite{Jacobson} was the first to derive the Einstein field equations from the proportionality of the black hole entropy and the horizon area together with the fundamental relation $\delta Q = TdS$. He showed that this relation is valid for all the local Rindler causal horizons through each space time point,  with $\delta Q$ and $T$ as the energy flux and Unruh temperature with respect to (wrt) an accelerated observer just inside the horizon. Subsequently, Hayward \cite{Hayward} proposed a unified first law of black hole dynamics and relativistic thermodynamics for spherically symmetric GR.

\section{Thermodynamics on horizons}
Within a few years after Hayward's work, Thanu Padmanabhan \cite{Paddy} used the Einstein equations to formulate the first law of thermodynamics on ``any'' horizon for a general static spherically symmetric space time.
Therefore the Einstein's equation for the gravitational field has a deep connection with the thermodynamics on horizons. Further to this, the uncanny similarity between the laws of BH mechanics and the laws of thermodynamics are also well known. In addition, if we are to believe that gravity has some microscopic degrees of freedom, then a strong connection between quantum physics and gravity must be there.\\
Clearly, the role of thermodynamics is not limited solely to BH mechanics / BH horizon as it has important cosmological implications which cannot be avoided if we assume the laws of thermodynamics to be true.

On this subject, Verlinde did a detailed study \cite{Ver} and observed that for the radiation dominated FLRW universe, the Friedmann equation can be expressed in the form of the Cardy-Verlinde formula, which is an entropy formula for a conformal field theory. Here radiation is described by a conformal field theory.
Consequently, the entropy formula describing the thermodynamics of radiation in the universe has the same mathematical form of the Friedmann equation, which describes the dynamics of spacetime.
Verlinde's studies indicates some hidden relation between thermodynamics and Einstein equations.
In a four dimensional de Sitter spacetime with radius $l$, there exists a cosmological event horizon. This horizon is like a black hole
horizon and is associated with thermodynamic properties, like the Hawking temperature $T$ and entropy $S$ ~\cite{GH},
\begin{equation}
 \label{1eq1}
 T = \frac{1}{2\pi l}, \ \ \ S = \frac{A}{4G},
\end{equation}
where $A= 4\pi l^2$ is the cosmological horizon area and $G$ is
the Newton constant. For an asymptotic de Sitter space such as a
Schwarzschild-de Sitter space, there still exists the cosmological
horizon, for which the area law of the entropy holds $S=A/4G$,
where $A$ denotes the cosmological horizon area, and whose Hawking
temperature is given by $T=\kappa/2\pi$, where $\kappa$ is the
surface gravity of the cosmological horizon. Suppose that some
matter with energy $dE$ passes through the cosmological horizon,
one then has
 \begin{equation}
 \label{1eq2}
 -dE=TdS.
 \end{equation}
Interestingly the cosmological horizon in the Schwarzschild-de Sitter space satisfies the relation (\ref{1eq2}).

Therefore, on the cosmological scale, the SLT can be applied by assuming that the universe is a closed system bounded by horizon, preferably the cosmological apparent horizon.\\
Cai and Kim \cite{caikim} applied the first law of thermodynamics to the apparent horizon of a FRW universe and considered the Bekenstein entropy on the apparent horizon, leading to the Friedmann equations for a universe with any spatial curvature. They applied the formulae of entropy for the static spherically symmetric BH horizons in Gauss-Bonnet gravity and in Lovelock gravity, to obtain the Friedmann equations in these theories. Another important work is where Paranjpe et al \cite{PSP} demonstrated that the gravitational field equations for the Lanczos-Lovelock action in a spherically symmetric spacetime can also be expressed in the form of the first law of thermodynamics. Then Akbar and Cai \cite{AC1} extended the work of Cai and Kim to the cases of scalar–tensor gravity and $f(R)$ gravity, and subsequently showed that \cite{AC2} the Friedmann equation of a FRW universe can be rewritten as the first law of thermodynamics on the apparent horizon of the universe. They also extended their procedure to the Gauss-Bonnet and Lovelock gravity. Again Cai and Cao \cite{CC} showed that the unified first law proposed by Hayward for the outer trapping horizon of a dynamical black hole, can be applied to the apparent horizon of the FLRW universe for the Einstein's field theory, Lovelock gravity, and the scalar-tensor theories of gravity.
\\

\subsection{Cosmological Horizons}
Before discussing the thermodynamics on cosmological horizons, we briefly mention below the relevant types of cosmological horizons for the FLRW universe.\\
The FLRW spacetimes are spherically symmetric about every spatial point. This makes it trivially spherically symmetric.  Nevertheless FLRW spacetimes are very important as cosmological models. Though they are simpler than the BH spacetimes but they still have horizons. Usually, FLRW spacetimes possess time-dependent apparent horizons and are very interesting as well as convenient from our point of view. As event horizons (EHs) are relevant both in the study of BHs and cosmology, we will begin by describing such cosmological EHs.

\subsubsection{Event Horizon}
An event horizon (EH) is defined as the proper distance to the most distant event wrt a comoving observer that he will ever see. Therefore if we want to define the EH we have to know the entire future of the universe from time $t$ to infinity. This also means that the EH is defined globally, not locally. The EH can  therefore be said to be the complement of the particle horizon (which we will discuss a bit later). In GR, the FLRW event horizon exists only for accelerated universes with $ P<-\rho/3 $ (for perfect fluid). So the event horizon does not exist for every FLRW spacetime. Also the cosmological event horizon turns out to be a null surface \cite{Faraoni} which evolves according to the equation
\begin{equation}
\dot{R}_{EH}=HR_{EH}-1,
\end{equation}
and the acceleration of the event horizon is governed by the following relation:
\begin{equation}
\ddot{R}_{EH}=(\dot{H}+H^2)R_{EH}-H.
\end{equation}
Thus the EH is more of a mathematical concept rather than a real viable horizon, the best example of such a mathematical entity being in the case of the Schwarzschild BH.\\
A more relevant and astrophysicaly viable horizon is the Apparent horizon.

\subsubsection{Apparent Horizon}
The cosmological apparent horizon depends on the observer, much like horizons in flat space, acting as a sphere which surrounds the observer and hides information. The FLRW apparent horizon for a comoving observer is a sphere with the proper radius
\begin{equation}
R_{AH}=\dfrac{1}{\sqrt{H^2 + k/a^2}},
\end{equation}
where $ k $ is the curvature of universe. The apparent horizon (AH), unlike the EH, is not a null surface in general. Consequently the AH evolves according to an equation different from the EH evolution equation. The AH evolution equation is given by the following:
\begin{equation}
\dot{R}_{AH}=4\pi H R^3_{AH}(P+\rho).
\end{equation}
The AH is null iff the cosmic fluid resembles the cosmological constant, i.e. $ P=-\rho $. If the EoS parameter is less than $ -1 $, the AH becomes timelike, but if the fluid is phantom-type, the AH becomes spacelike. Unlike the black hole dynamical horizons which are usually  spacelike, cosmological horizons on the other hand are timelike in the presence of non-exotic matter \cite{Faraoni}.

\subsubsection{Trapping Horizon}

The FLRW apparent horizon comes under the general class of trapping horizons when $ \mathcal{L}_{\ell}\theta_{n}=\mathcal{R}/3>0 $ just as in the BH case. Here $\mathcal{R} $ is the Ricci scalar of FLRW spacetime. Consequently, in the theory of GR, for a perfect fluid the condition of a trapping horizon becoming an AH becomes,  $\mathcal{L}_{\ell}\theta_{n}= \dfrac{8\pi}{3} (\rho-3P) $ \cite{ellis,Faraoni}.

\subsubsection{Particle Horizon}

As discussed before, the particle horizon (PH) at time $ t $ complements the EH, i.e., it is a sphere centered on and around the comoving observer at $ r=0 $. From the definition itself  we can understand that the particle horizon encapsulates all the particle signals that have reached the observer from the time of the Big Bang ($ t=0 $) to the time $ t $. From the definition it is evident that the PH is observer dependent. The cosmological particle horizon is also a null surface similar to EH. The cosmological PH evolves according to the equation
\begin{equation}
\dot{R}_{PH}=HR_{PH}+1,
\end{equation}
and the acceleration of the particle horizon is given by
\begin{equation}
\ddot{R}_{PH}=-\dfrac{4\pi}{3}(\rho+3P)R_{PH}+H.
\end{equation}
Here, $R_{PH}(t)$ must diverge as $ t $ approaches it's maximum value, possibly undefined, otherwise there will always be some region in spacetime which will be inaccessible to the comoving observers.

\subsubsection{Hubble Horizon}
Lastly we will mention the Hubble horizon, which is basically a conceptual horizon indicating the boundary between particles that are moving slower and faster than the speed of light wrt an observer at a given time. The radius of the Hubble horizon is given by:
\begin{equation}
R_{H}=\dfrac{1}{H}.
\end{equation}
This expression also gives us an estimate of the radius of curvature of a FLRW space. This is also used as an estimate of the radius of the event horizon during the slow-roll inflation, where the universe was close to being a de Sitter space. Another interesting feature is that the Hubble horizon coincides with the AH for spatially flat universes and also with the horizon of a de Sitter space. Unfortunately, this horizon does not have much physical significance from a thermodynamic view point \cite{Faraoni}.


\subsection{Thermodynamics of Cosmological Horizons}
Gary Gibbons and Stephen Hawking \cite{GH} have demonstrated that the BH thermodynamics is much more general than just the black holes. They also showed that the cosmological event horizons, like BH horizons, also possesses an entropy and temperature. Later t'Hooft and Susskind utilized the laws of BH thermodynamics to argue for a general holographic principle, which states that any consistent theory of gravity and quantum mechanics must be lower-dimensional.
\\
The thermodynamics is not well defined for the event horizon of FLRW spacetimes. Therefore we will discuss the AH. The AH is often considered as a causal horizon (horizon which separates the spacetime into two parts causally between observable events and non-observable events) associated with a gravitational temperature, entropy and surface gravity in dynamical spacetimes. If these arguments are true, then the cosmological horizon should also have the same characteristics. A review of the thermodynamical properties of the FLRW AH, as well as the derivation of the Kodama vector, Kodama-Hayward surface gravity, and the Hawking temperature in various coordinate systems, are available in Ref. \cite{appa}. The Kodama-Hayward temperature of the FLRW apparent horizon is given by the following expression:
\begin{equation}
k_{B}T=\left(\dfrac{\hbar G}{c}\right) \dfrac{R_{AH}}{3}(3P-\rho),
\end{equation}
 and the Kodama-Hayward surface gravity ($\kappa_{Kodama}$) is given by
 \begin{equation}
 \kappa_{Kodama}=\dfrac{R_{AH}}{12}\mathcal{R}.
 \end{equation}
\\
The entropy on the AH in FLRW spacetime would be
\begin{equation}
S_{AH}=\left(\dfrac{k_{B}c^3}{\hbar G}\right)\dfrac{\pi}{H^2 + k/a^2}=\left(\dfrac{k_{B}c^3}{\hbar G}\right)\dfrac{A_{AH}}{4}.
\end{equation}
The Hamiltonian constraint gives $ S_{AH}=\dfrac{3}{8\rho} $ and $ \dot{S}_{AH}=\dfrac{9H}{8\rho^2}(P+\rho) $.
\\
All the four generalized thermodynamic laws can be written for the cosmological AH. Though the zeroth law is obvious, the rest of the laws are still debatable \cite{Faraoni,poisson}. The thermodynamics of AH is very interesting. The AH thermodynamics is formulated for the horizons that are changing in an arbitrary manner. However, it is pertinent to state that the equilibrium thermodynamics could only be applied for physical systems that are in equilibrium or in near-equilibrium states. Therefore the appropriate application should be for slowly varying apparent horizons. Apparent horizons are important also for that fact that they seem to have implications for the BH information loss paradox, and are considered as an alternative to firewalls, but this viewpoint needs much more development.
\\
The thermodynamics of cosmological horizon is a vast subject right now and each and every argument has many pros and cons, which is beyond the scope of this brief introduction. But one must proceed very carefully before defining any quantity on cosmic horizons. The details have been omitted intentionally for the sake of brevity.

\section{Universal thermodynamics}
Now we will introduce our works related to gravitational thermodynamics in the subsequent portion of this report. As mentioned before, the introduction of the Hawking-Bekenstein entropy on the black hole event horizon paved the way for the complete development of the laws of black hole thermodynamics. The concept of BH entropy was necessary so that the second law of thermodynamics (SLT) could be preserved. Another interesting thing to note from the section of BH thermodynamics which appeared earlier in this thesis, is that, the BH temperature and the entropy are proportional to the surface gravity on the horizon and the area of the horizon, respectively. Moreover, the BH temperature, entropy and the mass of the BH were found to satisfy a relation very similar to the first law of thermodynamics \eqref{first_law}. This means that the BH thermodynamical  parameters rather the BH thermodynamics itself is deeply connected to the geometry of the BH horizon. All these visible similarities prompted physicists to find a possible connection between black hole thermodynamics and the gravitational field equations.
As mentioned before, the works of Ted Jacobson becomes very important
because the Einstein field equations are obtained from the proportionality of the black hole entropy to the horizon area. Interestingly this is valid for all local Rindler causal horizons, with an energy flux and the Unruh temperature seen by an accelerated observer just inside the horizon \cite{Jacobson}. The contribution of T. Padmanabhan in this field is also very significant as he formulated the first law of thermodynamics on ``any'' horizon for a general static spherically symmetric spacetime, using the Einstein field equations \cite{Paddy}.
Thus, the similarity between the laws of thermodynamics and the laws of black hole mechanics on one hand and the equivalence between the first law and the Einstein equations on the other side, reveals a strong connection between gravity and thermodynamics.

\subsection{Generalized Second Law of Thermodynamics}
As in the case of ordinary thermodynamics, in the same way, we can apply the SLT on the cosmological scale by assuming that the universe is a closed system bounded by a horizon, presumably the cosmological apparent horizon. This extension is known as the generalized second law of thermodynamics (GSLT).\\
Although the apparent horizon is physically much more relevant in a dynamical situation, but the event horizon is also significant. As we know that in a dynamically evolving universe or a black hole, both of these horizons appear, so it is justified to check for the validity of the GSLT on these horizons for a universe filled with various types of matter and/or energy, which in our case of study is assumed to be a fluid modelled as the Chaplygin gas. Chaplygin gas models are very versatile and useful cosmological models suitable for representing the different phases of evolution of the universe. In fact, the necessity of a model which can explain the evolutionary history of the universe successfully, led to the birth of the Chaplygin gas cosmology. Since the Chaplygin gas models can describe the accelerating expansion of the universe in the current epoch, hence they provide us a satisfactory model to take into account the role of the mysterious Dark Energy (DE).

\subsection{Our Works with Chaplygin gas models}
With the above utility in mind, we have compared the different Chaplygin gas models from a thermodynamic point of view, to identify the viability of the different models in this group, and have commented on their merits \cite{CG}. For this purpose, we have examined the validity of the GSLT both on the cosmological apparent horizon and the cosmological event horizon for the different Chaplygin gas models. As each model in this group is distinct in its own way, we obtained different cosmological consequences for the validity of the GSLT on both the horizons in these models. \\

For the analysis of the cosmological apparent horizon, we have considered the Kodama-Hayward temperature because the Kodama-Hayward surface gravity is more relevant for the description of dynamical horizons. In the case of the Variable modified Chaplygin gas (VMCG), we have already determined the temperature of the FLRW universe in another paper (described in the next part of this report). This temperature is the bulk temperature. In this paper, after calculating the Kodama-Hayward temperature of the VMCG dominated FLRW universe for the apparent horizon, we have compared these two types of temperatures to see how their behavior affects the thermodynamics of the universe. Here we want to mention that our approach is much more generalized compared to other works as we did not assume any specific definition of surface gravity (i.e., temperature) for our analysis in the case of the cosmological event horizon. \\

The analysis of generalized thermodynamics of FLRW universe for models like the Variable modified Chaplygin gas (VMCG), New Variable modified Chaplygin gas (NVMCG), Generalized cosmic Chaplygin gas (GCCG), and Modified cosmic Chaplygin gas (MCCG) on both the cosmological horizons is also a completely new study. For each case we have given a detailed study and determined the conditions of validity of GSLT on the cosmological horizons. This will help further analysis on such models in future from the thermodynamic point of view.\\

In the last work we have specifically focused on the Variable modified Chaplygin gas (VMCG) and its evolution in the FLRW universe.  In this paper \cite{CGP} we have studied the evolution of the FLRW universe filled with variable modified Chaplygin gas (VMCG). We used the thermodynamical treatment on the equation of state of the VMCG, and obtain its temperature as a function of redshift. We showed that the results are consistent with similar works on other types of Chaplygin gas models.  This temperature is used in the previously described thermodynamic analysis of Chaplygin gas models. In addition to deriving the exact expression of temperature of the fluid in terms of the boundary conditions and redshift, we also used observational data to determine the redshift at the epoch of transition from the decelerated to the accelerated phase of expansion of the universe. The values of other relevant parameters like the Hubble parameter, the equation-of-state parameter and the speed of sound are obtained in terms of the redshift parameter, and these values are compared with the results obtained from previous works on Modified Chaplygin gas (MCG) and other Chaplygin gas models for the various values of $n$ permitted by thermodynamic stability (where $n$ is the index of VMCG). We also assumed the present value of temperature of the microwave background radiation to be given by $2.7$K, and the parameter $A$ in the equation of state is taken as $1/3$, since it corresponds to the radiation-dominated phase of the universe. As it is known that the redshift of photon decoupling is $1100$, we used this value to calculate the temperature of decoupling.\\

\section{Gravitational Entropy}

\subsection{ Initial entropy problem leads to violation of second law of thermodynamics in cosmology}

As mentioned before, the SLT is one of the most fundamental laws of physics, according to which an ensemble of ideal gas molecules confined to a closed chamber will always spread to fill the entire space once the chamber is opened, thereby achieving a state of maximum entropy.  However, this is not true when we take the matter content of the entire universe as a fluid. The universe started from a homogeneous state. Then density fluctuations appeared due to the  effects of gravity, which in course of evolution led to structure formation in the universe. This is in contrast to the SLT, according to which the matter should spread out rather than clump together. There is another issue in the evolution of the universe as it was much hotter in the beginning and there was thermal equilibrium between matter and radiation (hence the perfect blackbody spectrum in the early universe) making it a state of maximum entropy. Apparently this makes the evolution of the universe impossible because a universe which is already in a state of maximum entropy at the beginning is not expected to follow the SLT during its evolution. This poses a fundamental problem in cosmology, that the universe violates the second law of thermodynamics in its evolution driven by gravity if we only consider the usual thermodynamic entropy. \\
\subsection{What is Gravitational entropy and why is it necessary}
In order to resolve this problem, Sir Roger Penrose proposed the concept of ``Gravitational entropy'' (GE) which represents the entropy carried by the free gravitational field. This GE is also contributing to the total entropy of the universe along with the usual thermodynamic entropy (TE). The main idea is that - in the initial epoch of the universe the total entropy (which includes both the GE and the TE) was very small and as time passed, with structure formation in the universe, the GE contribution in the total entropy increases and approaches a maximum value at the end state of the universe.

\subsection{Weyl Curvature hypothesis}
To measure this GE, Penrose proposed the famous “Weyl Curvature hypothesis” (WCH) according to which the GE can be determined by a function of the Weyl tensor, because the Weyl tensor is a measure of inhomogeneity/anisotropy of the universe, and the inhomogeneity increases in course of time as gravitational condensation leads to structure formation in the universe.\\
To put it in a proper perspective, we know that in Einstein's general theory of relativity the gravitational field can be split into two parts i.e., the Ricci and the Weyl parts. The former i.e., the Ricci tensor is related to the energy-momentum tensor (via the gravitational field equations), where standard definitions for the thermodynamic entropy of matter fields hold. Therefore counting the entropy of spacetime in the Ricci curvature would mean counting the entropy in the matter fields twice. As the objective is to characterise the gravitational entropy of free gravitational fields, we therefore must concentrate on the Weyl part of the curvature tensor. This gives us a tensorial description of the free part of the gravitational field, which is present even in the absence of matter fields.
Moreover, after the introduction of GE \cite{Penrose1,Penrose2}, scientists found that when applied to the Schwarzschild black hole, it reproduces the HB entropy, indicating that the BH entropy is also linked to the free gravitational field.  In fact, for any proposal of GE it is necessary to check whether the proposal can reproduce the HB entropy for Schwarzschild black hole or one can use this to normalize the free parameters in the GE proposal. Taking the WCH as the guide, many proposals were introduced to offer a measure of GE. We must emphasize here that the study of GE is a fairly new and open field. Consequently, a universal definition of GE (if it exists at all) which is applicable to every spacetime, is yet to be formalized.  \\

\subsection{Different proposals for Gravitational entropy}
Several definitions have been proposed for GE using the WCH as a guiding principle. One can simply take the Weyl curvature tensor and construct scalars out of it to represent the GE. Similarly, we can have the Weyl curvature scalar and take its ratio to the Ricci scalar. In this way, several other definitions can be used to represent the GE, among which we have used the ratio of the Weyl curvature scalar to the Kretschmann scalar as one of the measure of GE as described and elaborated in \cite{entropy1,entropy2}.  Using this ratio as the measure of GE, studies were conducted to show that this definition indeed produces the HB entropy for the stationary BHs. In fact, it can also be used to study the evolution of GE in worm hole (WH) spacetimes. From the definition it becomes clear that it is a purely geometric description and one can use it to compute the GE for the BH horizon and see whether it is producing the desired HB entropy for any spacetime.


\subsection{Weyl scalar proposal}

In this proposal \cite{entropy1} we consider a surface integral to determine the gravitational entropy of a black hole which can be described as the following:
\begin{equation}
S_{\sigma}=k_{s}\int_{\sigma}\mathbf{\Psi}.\mathbf{d\sigma},
\end{equation}
where $ \sigma $ represents the surface of the horizon of a black hole and the vector field $\mathbf{\Psi}$ is given by
\begin{equation}
\mathbf{\Psi}=P \mathbf{e_{r}},
\end{equation}
with $ \mathbf{e_{r}} $ as a unit radial vector. The scalar $ P $ is defined in terms of the Weyl scalar ($ W $) and
the Krestchmann scalar ($ K $) in the form
\begin{equation}\label{P_sq}
P^2=\dfrac{W}{K}=\dfrac{C_{abcd}C^{abcd}}{R_{abcd}R^{abcd}}.
\end{equation}
In order to compute the gravitational entropy, we have to consider the 3-space. Therefore, we consider the spatial metric which is defined as
\begin{equation}\label{sm}
h_{ij}=g_{ij}-\dfrac{g_{i0}g_{j0}}{g_{00}},
\end{equation}
where $ g_{\mu\nu} $ is the concerned 4-dimensional space-time metric and the indices denote spatial components, $i, j = 1, 2, 3$. Therefore the infinitesimal surface element is expresses as:
\begin{equation}
d\sigma=\dfrac{\sqrt{h}}{\sqrt{h_{rr}}}d\theta d\phi.
\end{equation}
Utilizing the Gauss's divergence theorem, we can now find out the entropy density \cite{entropy1} as
\begin{equation}
s=k_{s}|\mathbf{\nabla}.\mathbf{\Psi}|.
\end{equation}

There are other conditions to be satisfied by the GE in order to be plausible, as for example, the GE should be non-negative and it should vanish whenever the Weyl tensor vanishes (which follows from the WCH). Moreover the GE must measure the local anisotropy in the free gravitational field and it must increase monotonically with the structure formation in the universe. Finally, it is expected that the GE and the entropy of matter fields are to be additive, to render the total entropy to be an extrinsic quantity of the fields.

\subsection{CET proposal}

All of the scalars and the ratio of scalars considered earlier suffer from different limitations but they were a natural progression and a necessity for the study of the evolution of GE in a gravitating system. Subsequently it was discovered that the Bel-Robinson (BR) tensor is quite suitable for describing the energy density of the free gravitational field \cite{bel,bel2,robinson,bonsen}. So attempts were made to propose a measure of GE involving the Bel-Robinson tensor \cite{PL,PC}. As the BR tensor is constructed from the Weyl tensor and its dual tensor, therefore proposals using the BR tensor also follows the WCH. Such a definition was proposed by Clifton, Ellis and Tavakol (CET) \cite{CET} which is based on the square root of the BR tensor. The advantage of this proposal is that it was constructed using the considerations of relativistic thermodynamics and they showed that the CET scheme is observer dependent, i.e., the GE will change according to the observer we choose.  It is important to mention that unlike the purely geometric measures of scalar ratios, the CET proposal is useful in algebraically special spacetimes of Petrov type D and N. This is because the CET proposal gives us a unique GE only for such spacetimes. Finally, in the context of the CET proposal we want to mention that although the geometric measures are applicable to every kind of spacetime, the CET proposal gives us a more nuanced GE as it was constructed from relativistic thermodynamics. Another interesting feature of the CET proposal is that one can choose the gravitational temperature independently, as the proposal only defines the gravitational energy density from the BR tensor. It is important to mention that a number of recent studies have been done by different authors using the CET proposal on different gravitational systems \cite{PR,cet1,cet2}, which analyzed the scope of the proposal in detail. \\
In the subsequent portion of this thesis our contribution in this field is being discussed in detail.

\section{Our works on gravitational entropy}
In one of our studies we have considered a special class of BHs that are accelerating \cite{GC1}. We know that BHs in the universe are not static. In fact, they are indeed in acceleration due to the effects of gravity, however small it maybe. This becomes more important when BH binary mergers are happening, where BHs are rapidly accelerating. Also, there are cases in which after the collision of two galaxies, some BHs leave the galactic system and travels solo. In such a case also, acceleration becomes very important. We have showed that for the accelerating BHs, the chosen geometric definition of GE (which we call the ‘Weyl scalar proposal’) produces plausible results and reproduces HB entropy on the horizon and it follows the WCH. In this work we considered the C-metric and its variations, namely four different types of accelerating BHs: non-rotating accelerating BH, non-rotating charged accelerating BH, uncharged rotating BH and charged rotating BH.  We also evaluated the GE density and showed that in most of the cases we get a viable GE density. The definition of GE density can be found from the definition of GE using the Gauss’s divergence theorem.  Accelerating BHs are mathematically special as it contains a conical singularity due to acceleration, and as the acceleration tends to zero, the deficiency factor of this conical singularity reduces to unity rendering the spacetime metric to one of stationary BH. In the four different cases, the deficiency factors are different and contain the BH parameters like mass, charge and spin along with acceleration. This deficiency factor prevents the usual polar coordinate $\phi$ to run from zero to $2\pi$. Instead it runs from zero to $2\pi C$ where $C$ is the deficiency factor. We have analyzed each case in detail and have shown that for accelerating BHs also we obtain HB-like entropy, establishing that GE reduces to BH entropy even for accelerating BHs.

In another work \cite{CGG} we studied several cosmological models representing the different epochs of evolution of the universe and showed that the measure of GE is viable and is consistently increasing with time with the evolution of the universe. In this study we validated the WCH in the cosmological context using the previously described CET proposal.  We also investigated the ratio of free gravitational energy density to matter energy density in the vicinity of the initial singularity to check whether it is decreasing to zero following the WCH, and found it to be exactly so.  In this work, first we used the 1+3 covariant decomposition of the spacetime and discussed the relevant Ricci identities and Bianchi identities to show the two-way relationship between the shear and the electric part of Weyl tensor. The electric part of the Weyl tensor together with the shear drives the evolution of shear and the matter density, which then drives the evolution of the electric Weyl. Consequently, we clearly identified and showed the physical processes behind the generation of gravitational entropy, i.e., it is indeed generated from the anisotropies of the universe. Here we have considered various kinds of spacetimes to show that the ‘CET-proposal GE’ is evolving consistently in each of them. In the very beginning we considered the FLRW spacetime and showed that the GE is zero where the gravitational temperature is related to the cosmological constant in the dark energy dominated era. Next, moving from the previously discussed isotropic case to the anisotropic case, we considered the LRS Bianchi I model, where two of the spatial directions have the same scale factor.  Here we analyzed the conditions for the validity of WCH in detail and computed the conditions for a monotonous GE function. Next, we examined a spacetime model proposed by Liang representing the early phase of the evolution of the universe, i.e., the radiation dominated epoch.  In the analysis of the expansion anisotropy, we found that the spacetime begins from an isotropic singularity and the anisotropy increases with time as the universe expands. The GE found in this case completely corresponds with the evolution of other parameters and we showed that the GE indeed increases with the structure formation in the universe. Subsequently we studied the spatially inhomogeneous models with irrotational dust as source, i.e., the class II Szekeres solution of the Einstein’s field equations, which is a Petrov type D spacetime, as all the previous spacetimes. (This is because the CET proposal gives us unique GEs only for Petrov type D and type N spacetimes). Here also we found a consistently behaving GE with the gravitational temperature decreasing with time i.e., near the initial singularity the gravitational temperature blows up whereas with time it consistently decreases to lower values. Further, we found that the gravitational energy density blows up in the vicinity of the initial singularity and with time decreases with the evolution of the universe.  Finally, we considered a spacetime which fits a general class of solutions of the Einstein’s field equations but simple enough to study a perturbed kind of flat spacetime (like the perturbed FLRW spacetime) i.e., Bianchi VI$_{h}$ model. We showed that the deviation from conformal flatness and isotropy lead us to an inhomogeneous spacetime where gravitational entropy is generated. Without any exception, in this case too, the GE goes to zero near the initial singularity and with increasing time it increases with the increase in structure formation. In conclusion, we found that the CET proposal of the GE conforms to the WCH and provides us a robust tool to analyze the evolution of GE in such Petrov type D cosmological spacetime models.

In our next study \cite{CGG2} we have taken the previously discussed geometric proposal of the GE where the different combinations of the Weyl scalar to other curvature scalars are used with the thermodynamically derived CET proposal and compared these two approaches in traversable WH systems.  Here we wanted to study the GE proposals in the more exotic gravitational systems and to see whether they can give us a consistent measure of GE.  Considering some of the Lorentzian traversable wormholes along with the Brill solution for NUT wormholes and the AdS wormholes, we have evaluated the gravitational entropy for these systems. As more and more studies are revealing that these systems may exist as possible astrophysical objects, the study of traversable WHs becomes important in the context of GE. If traversable WHs do exist, then the different proposals of GE must be tested on them and for any traversable wormhole to be physically realistic, it should have a viable GE. We found that the GE proposals do give us a consistent measure of GE in several of them. This also means that the existence of a viable GE strictly depends on its definition. We also wanted to see whether the CET proposal can give us a satisfactory measure of GE when applied to WHs, because such studies were not available in the literature. For this purpose, we examined various traversable wormholes from the simplest to the more involved ones. The Ellis wormhole is the simplest case we have analysed. It is a zero-mass traversable WH which connects two asymptotically flat regions at its throat. There have been many proposals for the energy source of such WHs making it not only an ideal toy model to study, but also having rich physical content.  In the Weyl scalar proposal, we found a viable measure of GE but for the CET proposal, though the gravitational energy density is positive, the gravitational temperature turned out to be zero, making the computation of GE impossible. To get a more suitable observer we chose the class of observers who crosses the WH from one side to another in the equatorial plane and still found a null gravitational temperature. For this reason, in the Ellis WH case, we considered another temperature function according to the Gibbs one form and computed the GE and found that the GE obtained in the two different proposals (Weyl scalar proposal and CET proposal) behave a bit differently although both of them are giving viable results. In the same spirit we have considered the exponential metric WH, which is a much more general spherically symmetric spacetime, and is traversable at the throat. Here also we computed both the Weyl scalar proposal GE and the GE using the CET proposal and compared them. In both the cases the WH have a viable GE to work with. There are WHs which can also mimic BHs, and the simplest case of such a WH is the Darmour-Solodukhin (DS) WH. This is of great physical interest as it is not only traversable, but also because it mimics the Schwarzschild BH to an outside observer, for all practical purposes. In the spherically symmetric static cases, both the exponential metric WH and the DS WH possess rich mathematical and physical structure. For the DS WH also we found that the GEs in both proposals are viable and compared them. In order to examine the consequence of charge present in the WH system and to study its effect on the GE of that system, we considered the Maldacena ansatz, which connects two oppositely charged BHs. Here the CET proposal despite having a nonzero gravitational temperature gives us a null GE.  As for the stationary cases, we considered the Brill NUT WH (where both the magnetic and electric charges are present), and is an extension of the Reissner-Nordström solution with a Newman–Unti–Tamburino (NUT) parameter. This enabled us to study the behavior of GE in Einstein-Maxwell systems, where the NUT parameter controls the WH neck. We have utilized the geometric definition of the GE (Weyl scalar proposal) to check the validity of GE as this spacetime is not strictly a Petrov type D and we found that the GE is behaving consistently. Extending this to cosmological settings, we considered an AdS NUT WH in the presence of a negative cosmological constant, so as to study the effect of the cosmological constant on the GE of NUT WH. Here again we have used two geometric proposals of GE, specifically the ratio of Weyl scalar to Kretschmann scalar and simply the Weyl scalar. In both the cases we found desirable behaviors of the GE. We have also discussed briefly the Tolman law in WH spacetimes and concluded that the gravitational temperature and the Tolman temperature are related. Thus, we have considered these widely different traversable WHs, in order to study the behavior of GE explicitly in such scenarios, and to determine whether the GE proposals considered by us are physically viable or not.\\
It becomes clear from our extensive studies that the concept of GE can be applied to various types of spacetimes and it can provide us a viable measure of GE. Our comparative study also showed that GE of a system strictly depends on the definition used, despite following the WCH. Here we must clarify once again that the GE is different from the usual thermodynamic entropy (TE).  The utility of GE cannot be understated as it not only explains the entropy problem in the early universe but also gives us a function which can keep track of the structure formation with the evolution of the universe. Further, the GE can reproduce entropy of black holes and other cosmological objects. \\

Hence the entire thesis can be divided into two parts. \\
The initial part of this thesis is dedicated to the study of the purpose and utility of gravitational entropy (GE). There we briefly discuss different types of proposals of GE and then we have discussed in detail our three works on GE using two of these proposals. First, we described our work on the accelerating BHs and then the cosmological models. Lastly, we presented our work on the traversable WHs and comparison of different GE proposals in this context. \\
The later part of the thesis is dedicated to the thermodynamic study of gravitational systems, especially to the generalized laws of thermodynamics. In this part we first described the motivation and utility of such a theory and then described our other two works on this study. Firstly, we present a very detailed thermodynamic study of various kinds of Chaplygin gas models on cosmological horizons with their validity conditions.
Lastly, we end our study with the description of the thermodynamic study of the variable modified Chaplygin gas. \\

Now that I have introduced our work briefly, I will describe all the studies mentioned above in detail with proper citations in the subsequent chapters of this thesis.

\part{ Gravitational Entropy of Astrophysical and Cosmological systems}

\chapter{ Gravitational entropy of accelerating black holes}

The contents of this chapter have been published in a journal, details of which are given below:\\

\textbf{JOURNAL REFERENCE:} International Journal of Modern Physics D, Vol. 29, No. 5 (2020) 2050034 (24 pages)

\textbf{ARTICLE NAME:} On the gravitational entropy of accelerating black holes \\
DOI: 10.1142/S0218271820500340 \\~~~\\

The paper is quoted below:\\

``

\section{Introduction}
The $C$-metric was independently discovered by Levi-Civita \cite{Levi} and Weyl \cite{Weyl} in 1917. Ehlers and Kundt \cite{EK} while working on the classification of the degenerated static vacuum fields, constructed a table in which this metric was placed in the slot ``$C$'', leading to the name `$C$-metric'. Kinnersley and Walker \cite{KW} pointed out that this metric is an exact solution of Einstein's equations which describes the combined electromagnetic and gravitational field of a uniformly accelerating object having mass $m$ and charge $e$, and is an example of ``almost everything''. It is for this reason that the $C$-metric is the focus of our attention in this paper.

Dray and Walker \cite{DW} showed that this spacetime represents the gravitational field of a pair of uniformly accelerating black holes. Letelier and Oliveira \cite{LO}, studied the static and stationary $C$-metric and sought its interpretation in details, in particular those cases charaterized by two event horizons, one for the black hole and another for the acceleration. For spacetimes with vanishing or positive cosmological constant, the $C$-metric represents two accelerated black holes in asymptotically flat or de Sitter (dS) spacetime, and for a negative $\Lambda$ term, depending on the magnitude of acceleration \cite{DL}, it may represent a single accelerated black hole or a pair of causally separated black holes which accelerate away from each other \cite{Krtous}. The acceleration $A$ is due to forces represented by conical singularities arising out of a strut between the two black holes or because of two semi-infinite strings connecting them to infinity \cite{Podolsky,GKP}.

The second law of thermodynamics is one of the most fundamental laws of physics. We know that for an ensemble of ideal gas molecules confined to a closed chamber, the gas spreads out to fill the entire space once the chamber is opened, thereby reaching a state of maximum entropy. However, in the case of the universe with its matter content modelled as a fluid (or gas), this is not exactly true. The universe was born from a very homogeneous state and later on, small density fluctuations appeared due to the effect of gravity, that ultimately led to the formation of structures in the universe. This evolution is contrary to our expectations from the thermodynamic point of view, since the ``gas'' condenses into clumps of matter, instead of spreading out. Moreover in the past, the universe was much hotter and at some point of time, matter and radiation were in thermal equilibrium, and the entropy was maximum. So, how can the entropy increase if it was maximum in the past? It appears that if the evolution of the universe is dominated solely by gravity, then we may encounter a violation of the second law of thermodynamics, if we are considering the contribution of the thermodynamic entropy only.

To resolve this problem and to provide a proper sequence to the occurrence of gravitational processes, Penrose \cite{Penrose1} proposed that we must assign an entropy function to the gravitational field itself. He suggested that the Weyl curvature tensor could be used as a measure of the gravitational entropy. The Weyl tensor $C_{\alpha\beta\gamma\delta}$ in $ n $ dimensions is expressed as \cite{Chandra}
\begin{eqnarray}\label{decom}
\nonumber
C_{\alpha\beta\gamma\delta}= R_{\alpha\beta\gamma\delta} - \dfrac{1}{(n-2)}(g_{\alpha\gamma}R_{\beta\delta} + g_{\beta\delta}R_{\alpha\gamma} - g_{\beta\gamma}R_{\alpha\delta} - g_{\alpha\delta}R_{\beta\gamma}) + \\
 \dfrac{1}{(n-1)(n-2)}R(g_{\alpha\gamma}g_{\beta\delta}-g_{\alpha\delta}g_{\beta\gamma}),
\end{eqnarray}
where $R_{\alpha\beta\gamma\delta}$ is the covariant Riemann tensor, $R_{\alpha\beta}$ is the Ricci tensor and $R$ is the Ricciscalar.

According to Penrose, initially after the `big bang', when the universe started evolving, the Weyl tensor component was much smaller than the Ricci tensor component of the spacetime curvature. This hypothesis sounds credible because the Weyl tensor is independent of the local energy–momentum tensor. Moreover, the universe was in a nearly homogeneous state before structure formation began, and the FRW models successfully describe this homogeneous phase of the evolution. Further, the Weyl curvature is zero in the FRW models. However, the Weyl is large in the Schwarzschild spacetime. Thus we need a description of gravitational entropy, which should increase throughout the history of the universe on account of formation of more and more structures leading to the growth of inhomogeneity \cite{Penrose2,Bolejko}, and thus preserve the second law of thermodynamics. But there is still doubt regarding the definition of gravitational entropy in a way analogous to the thermodynamic entropy, which would be applicable to all gravitational systems \cite{CET}. The definition of gravitational entropy as the ratio of the Weyl curvature and the Ricci curvature faces problems with radiation \cite{Bonnor}. Once Senovilla showed that the Bel-Robinson tensor is suitable for constructing a measure of the ``energy'' of the gravitational field \cite{Senovilla}, several attempts were made to define the gravitational entropy based on the Bel-Robinson tensor and also in terms of the Riemann tensor and its covariant derivatives \cite{PL,PC}.

Many efforts has been made to explain the entropy of black holes using the quantized theories of gravity, such as the
string theory and loop quantum gravity. However, in this paper we will handle the problem from a phenomenological approach proposed in \cite{entropy1} and expanded in \cite{entropy2}, in which the Weyl curvature hypothesis is tested against the expressions for the entropy of cosmological models and black holes. They considered a measure of gravitational entropy in terms of a scalar derived from the contraction of the Weyl tensor and the Riemann tensor, and matched it with the Bekenstein-Hawking entropy \cite{SWH1,Bekenstein}. In our current work we will consider the accelerating black holes only, which represent more realistic black holes for several reasons. For instance, collision of galaxies is a rather common phenomenon occurring in the universe, and it inevitably leads to black hole mergers with the associated production of gravitational waves \cite{POK}. In such situations, we may imagine that the black holes at the centre of these galaxies are accelerating towards each other, although we can always think of any black hole as accelerating since no black hole is gravitationally isolated from the neighboring massive systems. Moreover, a static black hole may be considered as the limiting case of an accelerating black hole. Thus the study of accelerating black holes is very important. Here we will investigate whether the calculations for gravitational entropy proposed in \cite{entropy1} and \cite{entropy2} can be applied in this context. The organization of our paper is as follows: Sec. II deals with the definition of gravitational entropy and Sec. III enlists the metrics of accelerating black holes considered by us. Sec. IV provides the main analysis of our paper where we evaluate the gravitational entropy and the corresponding entropy density for these black holes. We discuss our results in Sec. V and present the conclusions in Sec. VI.

\section{Gravitational Entropy}
The entropy of a black hole can be described by the surface integral \cite{entropy1}
\begin{equation}
S_{\sigma}=k_{s}\int_{\sigma}\mathbf{\Psi}.\mathbf{d\sigma},
\end{equation}
where $ \sigma $ is the surface of the horizon of the black hole and the vector field $\mathbf{\Psi}$ is given by
\begin{equation}
\mathbf{\Psi}=P \mathbf{e_{r}},
\end{equation}
with $ \mathbf{e_{r}} $ as a unit radial vector. The scalar $ P $ is defined in terms of the Weyl scalar ($ W $) and
the Krestchmann scalar ($ K $) in the form
\begin{equation}\label{P_sq}
P^2=\dfrac{W}{K}=\dfrac{C_{abcd}C^{abcd}}{R_{abcd}R^{abcd}}.
\end{equation}
In order to find the gravitational entropy, we need to do our computations in a 3-space. Therefore, we consider the spatial metric which is defined as
\begin{equation}\label{sm}
h_{ij}=g_{ij}-\dfrac{g_{i0}g_{j0}}{g_{00}},
\end{equation}
where $ g_{\mu\nu} $ is the concerned 4-dimensional space-time metric and the Latin indices denote spatial components, $i, j = 1, 2, 3$. So the infinitesimal surface element is given by
\begin{equation}
d\sigma=\dfrac{\sqrt{h}}{\sqrt{h_{rr}}}d\theta d\phi.
\end{equation}
Using Gauss's divergence theorem, we can easily find out the entropy density \cite{entropy1} as
\begin{equation}
s=k_{s}|\mathbf{\nabla}.\mathbf{\Psi}|.
\end{equation}

\section{Accelerating Black holes}

\subsection{Non-rotating black hole}
The $C$-metric in spherical type coordinates is given by
\begin{equation}\label{cmetric}
ds^2=\dfrac{1}{(1-\alpha r cos\theta)^2}\left(-Qdt^2+\dfrac{dr^2}{Q}+\dfrac{r^2d\theta^2}{P}+Pr^2sin^2\theta d\phi^2\right),
\end{equation}
where $ P=(1-2\alpha m cos\theta)$, and $ Q=\left(1-\dfrac{2m}{r}\right)(1-\alpha^2r^2) $. This metric represents an accelerating massive black hole, which has two coordinate singularities, one is at $ r_{a}=\dfrac{1}{\alpha} $ and the other is at $ r_{h}=2m $. The $ r_{h}=2m $ singularity stands for the familiar \emph{event horizon}, but the $ r_{a}=\dfrac{1}{\alpha} $ singularity is the \emph{acceleration horizon} formed due to the acceleration of the black hole \cite{GP}. Here $m$ is the mass of the black hole and $ \alpha $ is the acceleration parameter. The important feature of this metric is that  $ \phi\in [0,2\pi C) $ unlike the full $ 2\pi $ range for stationary black holes because of the conical singularity arising due to acceleration. Here $ C=\dfrac{1}{(1+2\alpha m)} $ is the deficiency factor in the range of $ \phi $. If the acceleration of the black hole vanishes, i.e., $ \alpha=0 $, then the deficiency factor $ C $ becomes unity and $ \phi $ reduces to the conventional polar coordinate running from $ 0 $ to $ 2\pi $. All the accelerated black hole metrics discussed below also have this property.

\subsection{Non-rotating charged black hole}
We now consider the metric representing charged accelerating black holes. The charged $C$-metric in spherical type coordinate is \cite{GP}
\begin{equation}\label{cmetric1}
ds^2=\dfrac{1}{(1-\alpha r cos\theta)^2}\left(-Qdt^2+\dfrac{dr^2}{Q}+\dfrac{r^2d\theta^2}{P}+Pr^2sin^2\theta d\phi^2\right),
\end{equation}
where $ P=(1-2\alpha m cos\theta+\alpha^2e^2cos^2\theta)$, and $ Q=\left(1-\dfrac{2m}{r}+\dfrac{e^2}{r^2}\right)(1-\alpha^2r^2) $.
This is just the charged version of the previous metric with the parameter $ e $ representing the charge of the black hole. We can also think of it as an accelerated Reissner–Nordstrom (RN) black hole, because as $ \alpha\rightarrow 0 $, the metric reduces to the familiar RN metric. In the case of this metric also, we have $ r=\dfrac{1}{\alpha} $ as the acceleration horizon, and because of the introduction of charge, we have the outer and inner horizons at $ r_{\pm}=m\pm \sqrt{m^{2}-e^{2}}.$ Here the corresponding deficiency factor is given by $ C=\dfrac{1}{(1+2\alpha m+\alpha^2 e^2)} $.

\subsection{Rotating black hole}
The general line element for \textbf{an accelerating} rotating black hole is given by
\begin{equation}\label{htn}
ds^2=\dfrac{1}{\Omega^2}\left(-\dfrac{Q}{R}(dt-asin^2\theta d\phi)^2 + \dfrac{R}{Q}dr^2+\dfrac{R}{P}d\theta^2+\dfrac{P}{R}sin^2\theta[adt-(r^2+a^2)d\phi]^2\right),
\end{equation}
where $ \Omega=1-\alpha rcos\theta $, $ R=r^2+a^2cos^2\theta $, $ P=(1-2\alpha m cos\theta +\alpha^2a^2cos^2\theta) $, and $ Q=(a^2-2mr+r^2)(1-\alpha^2r^2) $. This metric represents the rotating version of the $ C$-metric, and contains three coordinate singularities, namely $ r_{\pm}=m\pm \sqrt{m^2-a^2} ,$ representing the outer and inner horizons, and $ r=\dfrac{1}{\alpha} $ representing the acceleration horizon \cite{GP} with the deficiency factor $C=\dfrac{1}{(1+2\alpha m+\alpha^2 a^2)}$ .

\subsection{Rotating charged black hole}
This is the charged version of the previous accelerating rotating metric. It may be regarded as the most general case among all the black holes considered by us, and is given by
\begin{equation}\label{ht}
ds^2=\dfrac{1}{\Omega^2}\left(-\dfrac{Q}{R}(dt-asin^2\theta d\phi)^2 + \dfrac{R}{Q}dr^2+\dfrac{R}{P}d\theta^2+\dfrac{P}{R}sin^2\theta[adt-(r^2+a^2)d\phi]^2\right),
\end{equation}
where $ \Omega=1-\alpha rcos\theta $, $ R=r^2+a^2cos^2\theta $, $ P=(1-2\alpha m cos\theta +\alpha^2(a^2+e^2)cos^2\theta) $, and $ Q=(a^2+e^2-2mr+r^2)(1-\alpha^2r^2) $. In this case the deficiency factor is given by $ C=\dfrac{1}{(1+2\alpha m+\alpha^2 (a^2+e^2))}. $ As in the previous case, the acceleration horizon is at $ r=\dfrac{1}{\alpha} $, however, the outer (or inner) horizons are located at $ r_{\pm}=m\pm \sqrt{m^2-a^2-e^2}. $

\section{Analysis}

\subsection{Non-rotating accelerating black hole}

We know that the Kretschmann scalar for a given spacetime geometry is defined by the relation
\begin{equation}
K=R_{abcd}R^{abcd},
\end{equation}
where $R_{abcd} $ is the covariant Riemann curvature tensor. For the $C$-metric (\ref{cmetric}), the Kretschmann scalar turns out to be
\begin{equation}\label{non-rot_Kscalar}
K_{c}=\dfrac{48m^2(\alpha r cos\theta-1)^6}{r^6}.
\end{equation}
The Weyl scalar is defined by
\begin{equation}
W=C_{abcd}C^{abcd},
\end{equation}
where the $C_{abcd} $ is the  Weyl curvature tensor. For the $C$-metric (non-rotating black hole) the Weyl scalar is evaluated as
\begin{equation}\label{non-rot_Wscalar}
W_{c}=\dfrac{48m^2(\alpha r cos\theta-1)^6}{r^6}.
\end{equation}
This result is expected since the Ricci tensor for this metric turns out to be zero. As the Riemann tensor can be decomposed into the Ricci and the Weyl parts according to equation (\ref{decom}), the vanishing Ricci component renders the Riemann and Weyl tensors identical as evident from equations (\ref{non-rot_Kscalar}) and (\ref{non-rot_Wscalar}).
The scalar function $P$ is defined by the relation (\ref{P_sq}) as
\begin{equation}
P^2=\dfrac{C_{abcd}C^{abcd}}{R_{abcd}R^{abcd}}.
\end{equation}
For this $C$-metric, we get $ P^2=1 $. Therefore we assume that $ P=+1 $ for our entropy calculations, since the entropy must be non-negative.

Now the \emph{spatial section} corresponding to this metric is
\begin{eqnarray}
\nonumber
 h_{ij}=diag\Big[\dfrac{1}{( 1 - \alpha r cos\theta )^2 (1-2m/r) (1-\alpha^2 r^2)},\\
\dfrac{r^2}{(( 1 - \alpha r cos\theta )^2 ( 1 - 2\alpha m cos\theta ))},
\dfrac{r^2 sin^2 \theta ( 1 - 2 \alpha m cos\theta )}{(1 - \alpha r cos\theta )^2}\Big],
\end{eqnarray}
with the determinant given by
\begin{equation}
h=\dfrac{sin^2(\theta)r^5}{(\alpha^2 r^2-1)(-r+2m)(\alpha r cos(\theta)-1)^6}.
\end{equation}
Therefore, the infinitesimal surface element has the form
\begin{equation}
d\sigma=\dfrac{\sqrt{h}}{\sqrt{h_{rr}}}d\theta d\phi=\dfrac{r^2 sin\theta}{(\alpha r cos\theta-1)^2} d\theta d\phi.
\end{equation}

\begin{figure}
\centering
\includegraphics[width=0.34\textwidth]{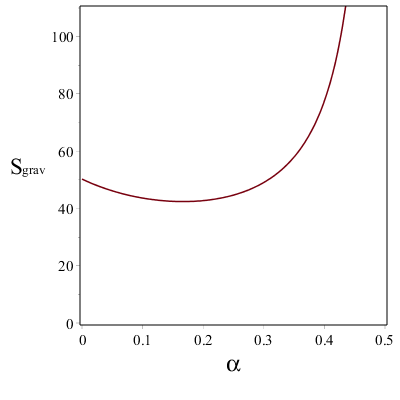}
\caption{Plot showing the variation of the total gravitational entropy for the accelerating non-rotating BH with respect to the acceleration parameter $ \alpha $, where we have taken $m=1 \: \textrm{and} \: k_{s}=1$.}\label{plot3}
\end{figure}

We are now in a position to calculate the magnitude of the gravitational entropy on the event horizon $H_{0}$ at the location $ r_{h}=2m $ for this metric, which is
\begin{equation}\label{s_grav_nonrot}
S_{grav}=k_{s}r_{h}^2\int_{\theta=0}^{\pi}\dfrac{sin\theta}{(\alpha r_{h} cos\theta-1)^2}d\theta \int_{\phi=0}^{2\pi C} d\phi=k_{s}\dfrac{4 \pi C r_{h}^2}{(1-r_{h}^2\alpha^2)}=k_{s}\dfrac{4 \pi r_{h}^2}{(1-r_{h}^2\alpha^2)(1+2\alpha m)}.
\end{equation}
From equation (\ref{s_grav_nonrot}) it is evident that the gravitational entropy is proportional to the area of the event horizon of the black hole, as in the case of the Bekenstein-Hawking entropy \cite{SWH1,Bekenstein}. Here $ C=\dfrac{1}{(1+2\alpha m)} $ is the deficiency factor in the limit of $ \phi $ as it runs from $ 0\rightarrow 2\pi C $ (as mentioned earlier). In FIG. \ref{plot3}, we have shown the variation of the total entropy on the horizon with the acceleration parameter $ \alpha $.

Similarly we can compute the entropy density as
\begin{equation}\label{s_nonrot}
s=k_{s}\frac{1}{\sqrt{h}}\frac{\partial}{\partial r}\left(\sqrt{h}\dfrac{P}{\sqrt{h_{rr}}} \right)=\dfrac{2k_{s}}{r}\sqrt{\left(1-\alpha^{2}r^{2}\right)\left(1-\frac{r_{h}}{r}\right)}.
\end{equation}
In the above equation (\ref{s_nonrot}), inserting $ \alpha=0 $, we get the entropy density for the Schwarzschild black hole. In FIG. \ref{fig2}, the dependence of the gravitational entropy density corresponding to this metric on other relevant parameters have been indicated. From equation (\ref{s_nonrot}) we can see that the zeroes of the gravitational entropy density function are located at the acceleration horizon $ r=\dfrac{1}{\alpha} $, and at the event horizon $ r=2m $, which is clearly evident from FIG. \ref{fig2}. Specifically, FIG. \ref{fig2}(a) shows that for $ \alpha=0 $, the acceleration horizon goes to infinity where the entropy density reduces to zero, and at the event horizon $r=2$, the entropy density becomes zero. Similarly for $ \alpha=0.5 $, the acceleration horizon and the event horizon coincide at $ r=2 $, where the entropy density becomes zero. FIG. \ref{fig2}(b) indicates that for $\alpha=0.25$, the acceleration horizon lies at $r=4$ and the event horizon is at $ r=2 $, the entropy density going to zero at both these places, and diverges at the singularity $ r=0 $ which is in agreement with equation (\ref{s_nonrot}).

\begin{figure}[ht]
    \centering
    \subfloat[Subfigure 1 list of figures text][]
        {
        \includegraphics[width=0.4\textwidth]{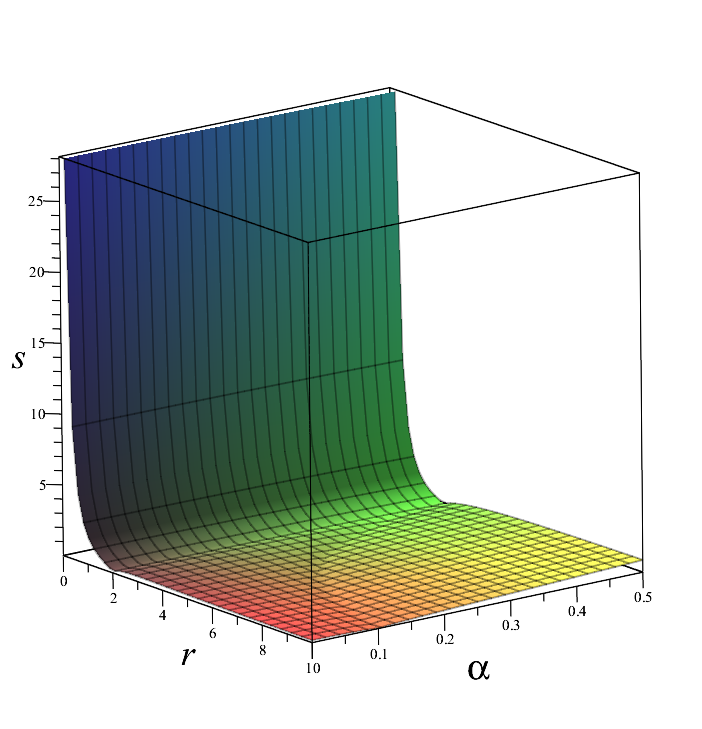}
        \label{fig:subfig1}
        }
    \hspace{0.5cm}
    \subfloat[Subfigure 2 list of figures text][]
        {
        \includegraphics[width=0.34\textwidth]{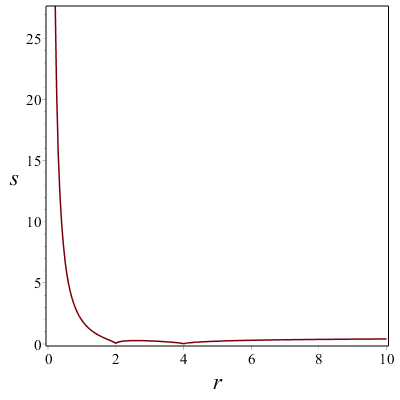}
        \label{fig:subfig2}
        }
    \caption{(a) Plot showing the variation of the gravitational entropy density for an accelerating non-rotating BH with respect to the acceleration parameter $ \alpha $ and the radial coordinate $ r $, for $ m=1$ and $ k_{s}=1 $. (b) Plot showing the variation of the gravitational entropy density  for the accelerating non-rotating BH with respect to the radial coordinate $ r $, where $ \alpha=0.25$, $ m=1$, and $ k_{s}=1 $.}
    \label{fig2}
\end{figure}

\subsection{Non-rotating charged accelerating black hole}

The Kretschmann scalar for the non-rotating charged black hole given by the metric (\ref{cmetric1}) is evaluated to be
\begin{equation}
K=\dfrac{56\left(\alpha rcos\theta-1\right)^6 \left(cos^2\theta \alpha^2 e^4 r^2+\dfrac{10}{7} \left(e^2-\dfrac{6}{5}mr\right)re^2 \alpha cos\theta+e^4-\dfrac{12}{7} e^2mr+\dfrac{6}{7}m^2r^2 \right)}{r^8},
\end{equation}
and the corresponding Weyl scalar is
\begin{equation}
W=\dfrac{4}{3}\dfrac{\left(\alpha r cos\theta -1\right)^4 \left(5cos^2\theta \alpha^2 e^2 r^2-sin^2\theta\alpha^2 e^2 r^2+\alpha^2 e^2 r^2-6 m \alpha cos\theta r^2-6e^2+6mr\right)^2}{r^8}.
\end{equation}

Therefore the quantity $P$ is given by the expression
\begin{equation}
P^2=\dfrac{6(e^2\alpha cos\theta r+e^2-mr)^2}{(7 cos^2\theta\alpha^2e^4r^2+10\alpha r cos\theta e^4-12\alpha r^2 cos\theta e^2 m+7e^4-12e^2 mr+6 m^2 r^2)}.
\end{equation}

The spatial metric for this case is
\begin{eqnarray}
  h_{ij} &=& \dfrac{1}{(1-\alpha r cos\theta)^2} diag\left[\dfrac{1}{\left(1-\dfrac{2m}{r}+\dfrac{e^2}{r^2}\right)\left(-\alpha^2 r^2+1 \right)},
   \dfrac{r^2}{\beta}, \beta r^2 sin^2\theta \right],
\end{eqnarray}
where
\begin{equation}
\beta=\left(1-2\alpha m cos\theta +\alpha^2 e^2 cos^2\theta\right). \nonumber
\end{equation}
Consequently the determinant of the spatial $ h_{ij} $ metric is given by
\begin{equation}
h=-\dfrac{sin^2\theta r^6}{(\alpha^2 r^2-1)(e^2-2mr+r^2)(\alpha r cos\theta-1)^6},
\end{equation}
and the infinitesimal surface element is
\begin{equation}
d\sigma=\dfrac{\sqrt{h}}{\sqrt{h_{rr}}}d\theta d\phi=\dfrac{r^2 sin\theta}{(\alpha r cos\theta-1)^2} d\theta d\phi.
 \end{equation}

Next we calculate the gravitational entropy on the horizon $H_{0}$ at $ r_{h}=r_{\pm}=m \pm \sqrt{m^2-e^2} ,$ which turns out to be
\begin{equation}\label{s_grav_nonrot_chrg}
S_{grav}=k_{s}r_{h}^2\int_{\theta=0}^{\pi}\dfrac{P(r_{h},\theta) sin\theta}{(\alpha r_{h} cos\theta-1)^2}d\theta \int_{\phi=0}^{2\pi C} d\phi=\dfrac{k_{s}(4\pi r_{\pm}^2)}{(1+2\alpha m+\alpha^2 e^2)}\int_{\theta}\dfrac{P_{\pm}(\theta)sin\theta d\theta}{2(\alpha r_{\pm} cos\theta-1)^2},
\end{equation}
where the quantity $P_{\pm}$ corresponds to the value calculated for $r_{\pm}$.

From equation (\ref{s_grav_nonrot_chrg}) we find that the gravitational entropy is proportional to the area of the event horizon of the black hole, just as in the case of the Bekenstein-Hawking entropy. We can further check the validity of our result by setting $\alpha=0$ in (\ref{s_grav_nonrot_chrg}), to see whether it leads us to the desired expression for the entropy of the Reissner–Nordstrom (RN) black hole. This exercise yields the result
\begin{equation}
S_{grav}^{RN}=k_{s}(4\pi r_{\pm}^{2})\int_{\theta}P^{RN}_{\pm}(\theta)\dfrac{\sin\theta}{2} d\theta.
\end{equation}
We can easily see that $P^{RN}_{\pm}(\theta)=P_{\pm}(\alpha=0)=\dfrac{6e^{4}-12e^{2}mr+6m^{2}r^{2}}{7e^{4}-12e^{2}mr+6m^{2}r^{2}} , $ and
therefore the gravitational entropy for the RN black hole is
\begin{equation}
S_{grav}^{RN}=k_{s}(4\pi r_{\pm}^{2})\sqrt{\dfrac{6e^{4}-12e^{2}mr+6m^{2}r^{2}}{7e^{4}-12e^{2}mr+6m^{2}r^{2}}}\int_{\theta}\dfrac{\sin\theta}{2}d\theta=k_{s}(4\pi r_{\pm}^{2})\sqrt{\dfrac{6e^{4}-12e^{2}mr+6m^{2}r^{2}}{7e^{4}-12e^{2}mr+6m^{2}r^{2}}}.
\end{equation}
This result matches with the expression of gravitational entropy for the RN black hole derived in \cite{entropy2} by Romero et al.
The entropy density for the non-rotating charged black hole is obtained as
\begin{align} \label{s_nonrot_chrg}
\left.s\right. & = \frac{16\sqrt {6}k_{s}\sqrt { \left( -{\alpha}^{2}{r}^{2}+1 \right)
 \left( {e}^{2}-2mr+{r}^{2} \right) }}{{{r}^{2} \left( 7{e}^{4}{\alpha}^{2} \cos^{2}\theta{r}^{2} + 10 \left( {e}^{2}-\dfrac{6mr}{5} \right) r\alpha{e}^{2}\cos\theta + 7{e}^{4}-12{e}^{2}mr+6{m}^{2}{r}^{2} \right) ^{3/2}}}   \nonumber \\
& \times \left[ \cos^{3}\theta{\alpha}^{3}{e}^{6}{r}^{3} + {\frac {15{e}^{4}{\alpha}^{2} \cos^{2}\theta{r}^{2}}{8} \left( {e}^{2} - {\frac{13mr}{10}} \right) } + \dfrac{9 r\alpha{e}^{2} \cos\theta }{4} \left( {e}^{4}-{\frac {11{e}^{2}mr}{6}}+{m}^{2}{r}^{2} \right) \right. \nonumber \\
 & \qquad\qquad\qquad + \left. {\frac {7{e}^{6}}{8}} - \frac{3mr}{4} \left({ \frac {13{e}^{4}}{4}} - 3{e}^{2}mr + {m}^{2}{r}^{2} \right) \right] .
 \end{align}
If in this expression we substitute $ e=0 $, and consider the absolute value of this quantity, then we get back the expression (\ref{s_nonrot}) for the entropy density of the accelerating black hole. In FIG. \ref{fig3} we have shown the dependence of the gravitational entropy density of the non-rotating charged black hole on different parameters appearing in (\ref{s_nonrot_chrg}). From FIG. \ref{fig3}(a) we can determine the zeroes of the gravitational entropy function (\ref{s_nonrot_chrg}), e.g., for $ \alpha=0, \theta=\frac{\pi}{2} $, the acceleration horizon goes to infinity and by solving the entropy density function, we obtain the zeroes at $ r=0.13, 1.87 $, and also at $ r=0.18 $, where $ r=0.13, 1.87 $ are the horizons. Again from (\ref{s_nonrot_chrg}), using $ \alpha=0.45, \theta=\frac{\pi}{2} $, we find that the zeroes of the entropy density function are located at the acceleration horizon $ r=\frac{1}{\alpha}=2.22 $, and at the event horizon $ r=m\pm\sqrt{m^2 - e^2}=1 $, which is evident from FIG. \ref{fig3}(b). The additional zero can be found by solving for the roots of the second factor in (\ref{s_nonrot_chrg}) which gives us the only real root at $ r=0.74 $.  We have also analyzed the case for $ \alpha=0.25 $ (shown in FIG. \ref{plot4a}) from which we can identify the zeroes clearly, i.e., at the acceleration horizon $ r=\frac{1}{\alpha}=4 $, and at the horizons $ r=0.13, 1.87 $. Further, another zero arises from the second term of the entropy density function at $ r=0.19 $. The overall behavior is also as we expected, that is, the entropy density diverges near the $r=0$ singularity and it increases inside the horizon, encountering some zeroes in between.

\begin{figure}[ht]
    \centering
    \subfloat[Subfigure 1 list of figures text][]
        {
        \includegraphics[width=0.36\textwidth, height=0.3\textheight]{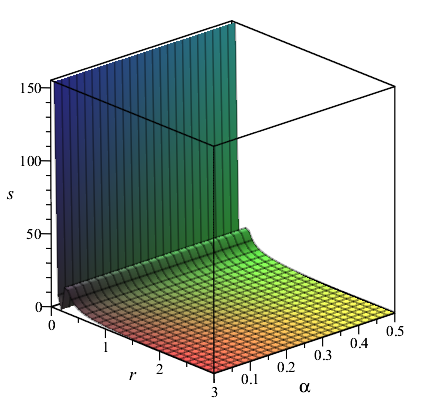}
        \label{fig:subfig3}
        }
        \hspace{1.0cm}
    \subfloat[Subfigure 2 list of figures text][]
        {
        \includegraphics[width=0.36\textwidth, height=0.32\textheight]{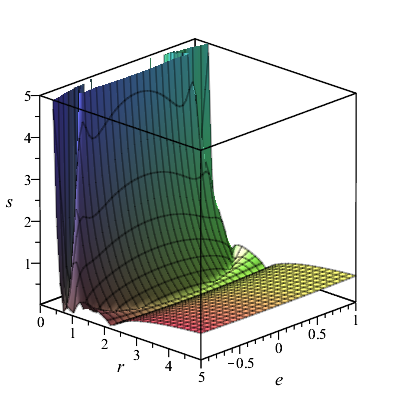}
        \label{fig:subfig4}
        }
    \caption{(a) Plot showing the variation of the gravitational entropy density for the accelerating non-rotating charged BH with respect to the acceleration parameter $ \alpha $ and the radial coordinate $ r $, where $ m=1, \; k_{s}=1, \; e=0.5, \; \textrm{and} \; \theta=\dfrac{\pi}{2} $. (b) Plot showing the variation of the gravitational entropy density for the accelerating non-rotating charged BH with respect to the radial coordinate $ r $ and the charge $ e $, where $ \alpha=0.45, \: m=1, \: k_{s}=1, \: \textrm{and} \: \theta=\dfrac{\pi}{2} $.}
    \label{fig3}
\end{figure}

\subsection{Rotating accelerating black hole}
Using the metric for the rotating accelerating black hole we have calculated the Ricci tensor, which turns out to be zero. Thus the Kretschmann scalar $K$ and the Weyl scalar $W$ are identical. Thus we have
\begin{eqnarray}
\nonumber
\small K= W = 48{m}^{2} \left( \alpha r\cos \left( \theta \right) -1 \right) ^{6} \big(  \left({a}^{4}\alpha+{a}^{3} \right) \cos^{3}\theta
+ 3{a}^{2}r \left( a \alpha-1 \right) \cos^{2}\theta \\ \nonumber
- 3a {r}^{2} \left( a\alpha+1 \right) \cos \theta -{r}^{3}  \left( a\alpha-1 \right)  \big)  \\
\times \dfrac{ \left(  \left( {a}^{4}\alpha - {a}^{3} \right) \cos^{3}\theta - 3{a}^{2}r \left( a \alpha+1 \right)  \cos^{2}\theta
- 3a {r}^{2}\left(a\alpha-1 \right) \cos \theta + {r}^{3}  \left( a\alpha+1 \right)  \right) } { \left( {r}^{2} + {a}^{2} \cos^{2}\theta \right) ^{6} }.
\end{eqnarray}
Therefore $ P^{2}=\dfrac{W}{K}=1 $, i.e. $ P=+1 $. Hence the total gravitational entropy in this case is given by
\begin{equation}\label{s_grav_rot}
S_{grav}=k_{s}\int_{\sigma}\mathbf{\Psi}.\mathbf{d\sigma}=k_{s}\int_{\sigma}d\sigma=k_{s}\int_{\theta=0}^{\pi}\int_{\phi=0}^{2\pi C}\sqrt{g_{\theta\theta}g_{\phi\phi}}d\theta d\phi.
\end{equation}
The entropy evaluated at $r_{\pm}$ is obtained as
\begin{equation}\label{s_grav_pm}
S_{grav_{\pm}}=k_{s}\dfrac{4\pi C(r^{2}_{\pm}+a^{2})}{(1-\alpha^{2}r_{\pm}^{2})}=k_{s}\dfrac{4\pi (r^{2}_{\pm}+a^{2})}{(1-\alpha^{2}r_{\pm}^{2})(1+2\alpha m+\alpha^2a^2)}.
\end{equation}
If we substitute $ a=0 $ in (\ref{s_grav_pm}), then we get back the expression (\ref{s_grav_nonrot}) for the entropy of the non-rotating accelerating black holes. We see that as the acceleration parameter vanishes, i.e., $ \alpha\rightarrow 0 $, the equation (\ref{s_grav_pm}) reduces to the expression of gravitational entropy for Kerr black holes  derived in \cite{entropy2}. However for this axisymmetric metric, it is not possible to evaluate the spatial metric using equation (\ref{sm}) because the object is rotating, and so there is a nonzero contribution from the component of $ g_{t\phi} $, which changes the spatial positions of events in course of time. Therefore the entropy density is calculated by using the full four-dimensional metric determinant $ g $ in the expression involving the covariant derivative \cite{entropy2}, and we get
\begin{equation}\label{enden1}
s=k_{s}|\mathbf{\nabla}.\mathbf{\Psi}|=\dfrac{k_{s}}{\sqrt{-g}}\left(\dfrac{\partial}{\partial r}\sqrt{-g}P\right)=2k_{s}\dfrac{(2\cos^{3}\theta a^{2}\alpha+\cos\theta \alpha r^{2}+r)}{(1-\alpha r \cos\theta)(r^{2}+a^{2}cos^{2}\theta)},
\end{equation}
where $ g=-\sin^{2}\theta \dfrac{(a^{2}\cos^{2}\theta+r^{2})^{2}}{(\alpha r \cos\theta -1)^{8}} $.

From equation (\ref{enden1}) we see that the entropy density diverges at the ring singularity and at $ r=\dfrac{1}{\alpha\cos\theta}, $ which is the conformal infinity in this spherical type coordinate system, as is evident from the metric (\ref{htn}). This can also be further verified from the expressions of the Kretschmann scalar and the Weyl scalar in this case, since they vanish at the conformal infinity but diverge at the ring singularity. To compute the zeroes of the entropy density function we only need to find the roots of the numerator in (\ref{enden1}), which is a quadratic function in $ r $.

Substituting $ \alpha=0 $ in the above expression of entropy density, we get the entropy density for the Kerr black hole:
\begin{equation}
s_{kerr}=\dfrac{2k_{s}r}{(r^{2}+a^{2}\cos^{2}\theta)}.
\end{equation}

\begin{figure}
\centering
\includegraphics[width=0.55\textwidth]{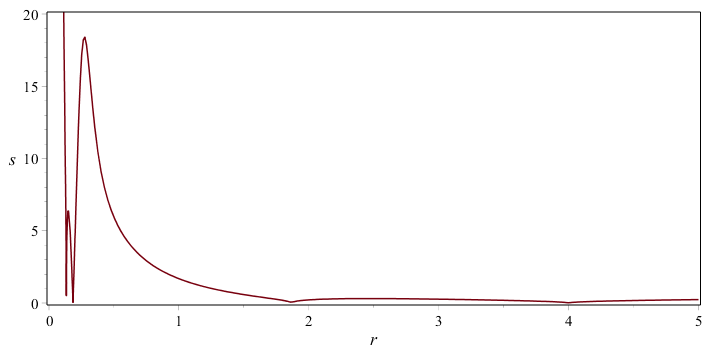}
\caption{Plot showing the variation of the gravitational entropy density for the accelerating non-rotating charged BH with respect to the  radial coordinate $ r $, where $ m=1, \; k_{s}=1, \; \alpha=0.25, \;  e=0.5, \; \textrm{and} \; \theta=\dfrac{\pi}{2} $.}\label{plot4a}
\end{figure}

\begin{figure}
\centering
\includegraphics[width=0.40\textwidth]{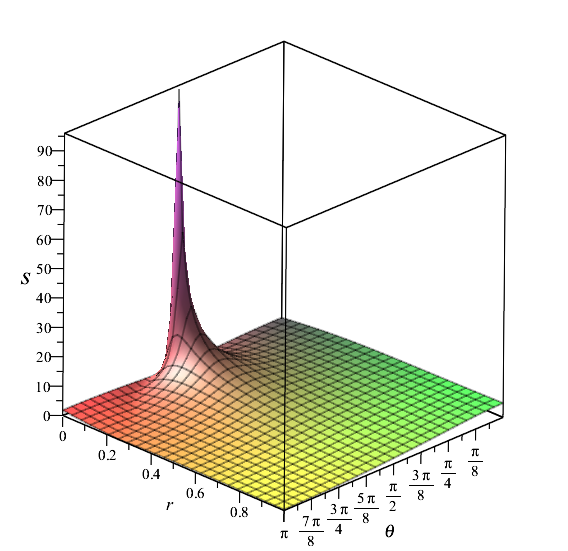}
\caption{Plot showing the variation of the gravitational entropy density for the accelerating rotating BH with respect to the radial coordinate $ r $ and the angular coordinate $ \theta $, where $ \alpha=0.45, \: m=1, \: {\color{black}{a=0.5,}} \: \textrm{and} \: k_{s}=1 $. This figure clearly indicates that at the ring singularity $\left( r=0, \: \theta=\dfrac{\pi}{2} \right) $ the gravitational entropy density diverges.}\label{plot4b}
\end{figure}

\begin{figure}[ht]
    \centering
    \subfloat[Subfigure 1 list of figures text][]
        {
        \includegraphics[width=0.42\textwidth]{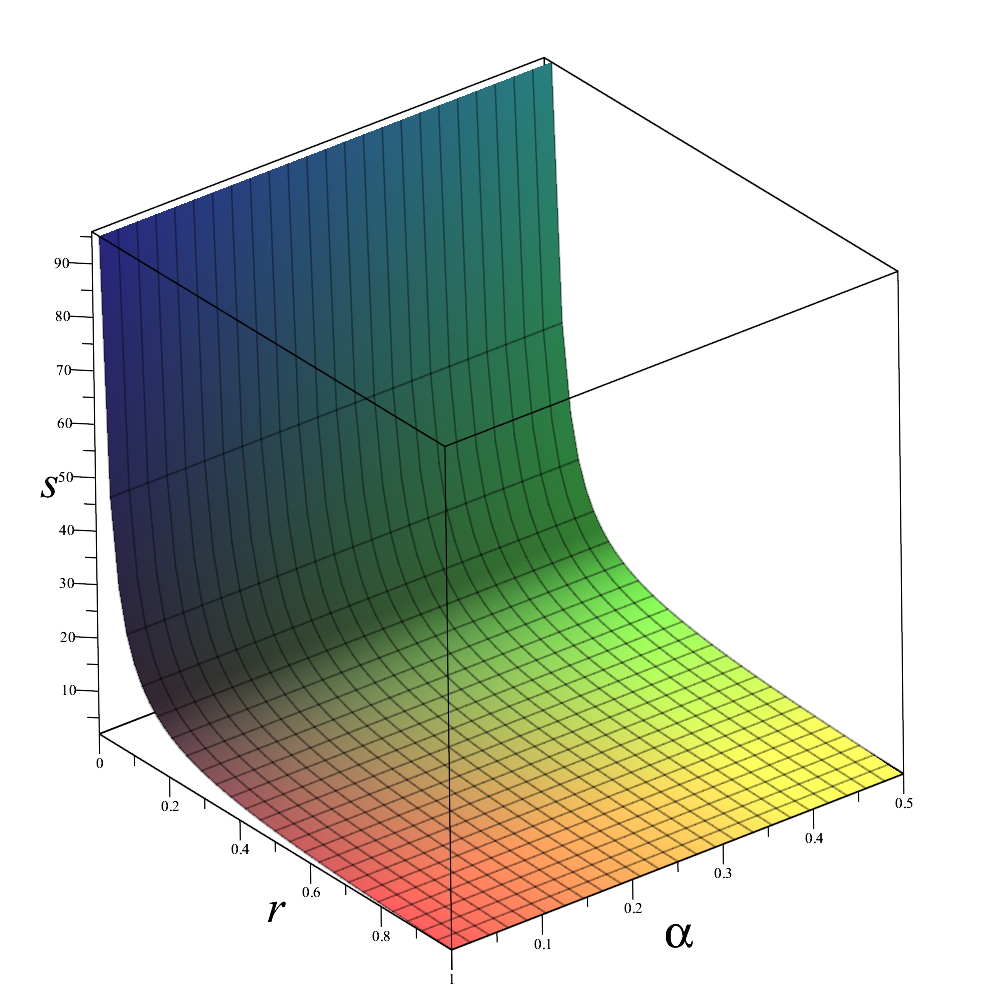}
        \label{fig:subfig7}
        }
        \hspace{1.0cm}
    \subfloat[Subfigure 2 list of figures text][]
        {
        \includegraphics[width=0.37\textwidth]{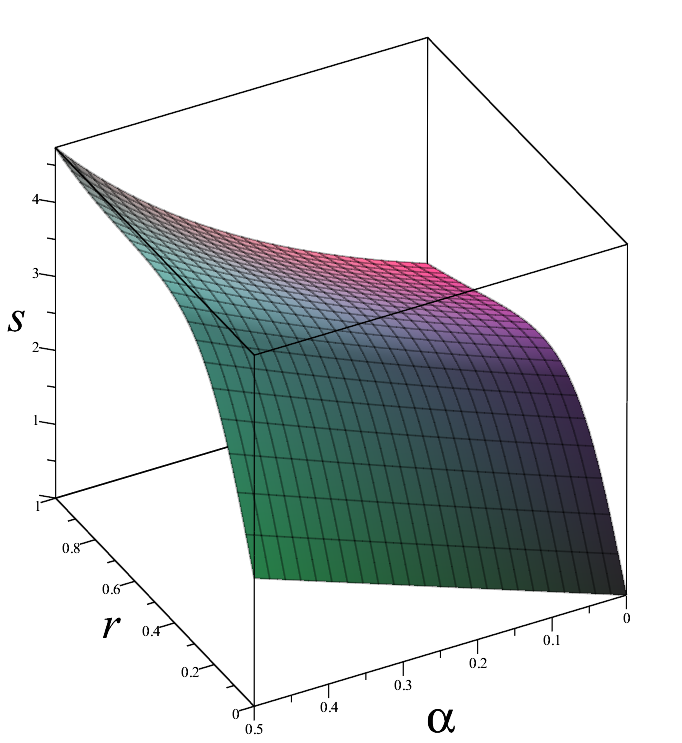}
        \label{fig:subfig8}
        }
    \caption{(a) Plot showing the variation of the gravitational entropy density for the accelerating rotating BH with respect to the radial coordinate $ r $ and the acceleration parameter $ \alpha $, where $ a=0.5, \: m=1, \: k_{s}=1, \: \textrm{and} \: \theta=\dfrac{\pi}{2} $. This figure clearly shows that at the ring singularity $ \left(r=0, \: \theta=\dfrac{\pi}{2}\right) $ the gravitational entropy density diverges. (b) Plot showing the variation of the gravitational entropy density  for the accelerating rotating BH with respect to the radial coordinate $ r $ and the acceleration parameter $ \alpha $, where $ a=0.5, \: m=1, \: k_{s}=1, \: \textrm{and} \: \theta=\dfrac{\pi}{6} $. This figure shows that at $ r=0 $ and $ \theta=\dfrac{\pi}{6} $, the gravitational entropy density is finite.}
    \label{fig4}
\end{figure}

\begin{figure}[ht]
    \centering
    \subfloat[Subfigure 1 list of figures text][]
        {
        \includegraphics[width=0.42\textwidth]{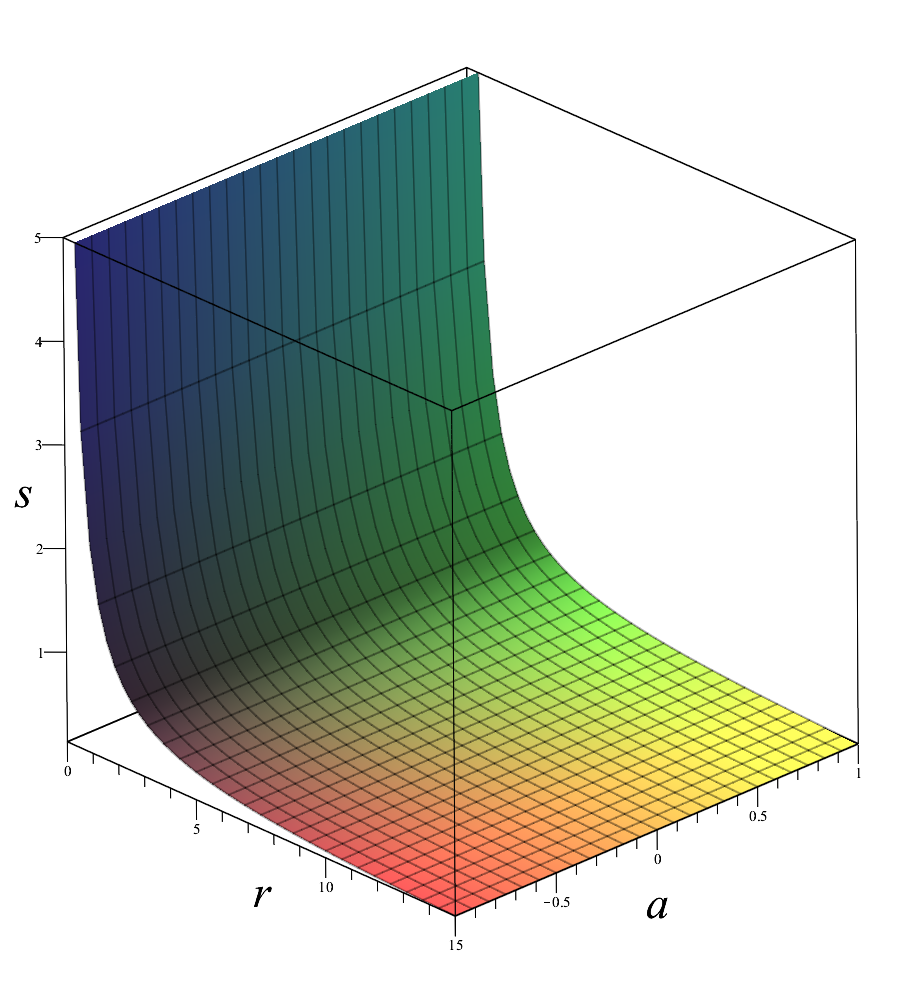}
        \label{fig:subfig9}
        }
    \subfloat[Subfigure 2 list of figures text][]
        {
        \includegraphics[width=0.47\textwidth]{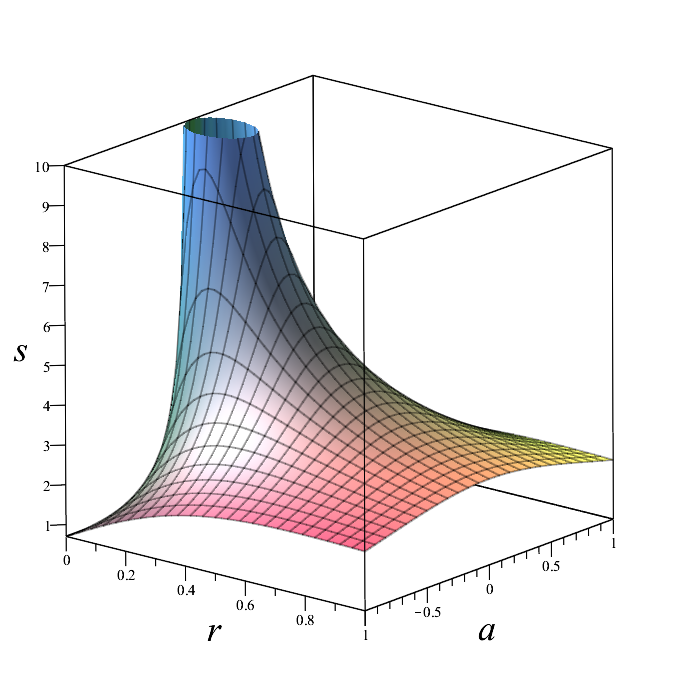}
        \label{fig:subfig10}
        }
    \caption{(a) Plot showing the variation of the gravitational entropy density for the accelerating rotating BH with respect to the radial coordinate $ r $ and the rotation parameter $ a $, where $ \alpha=0.25, \: m=1, \: k_{s}=1, \: \textrm{and} \: \theta=\dfrac{\pi}{2} $. (b) Plot showing the variation of the gravitational entropy density for the accelerating rotating BH with respect to the radial coordinate $ r $ and the rotation parameter $ a $, where $ \alpha=0.25, \: m=1, \: k_{s}=1, \: \textrm{and} \: \theta=\dfrac{\pi}{4} $.}
    \label{fig5}
\end{figure}

\begin{figure}
\centering
\includegraphics[width=0.6\textwidth]{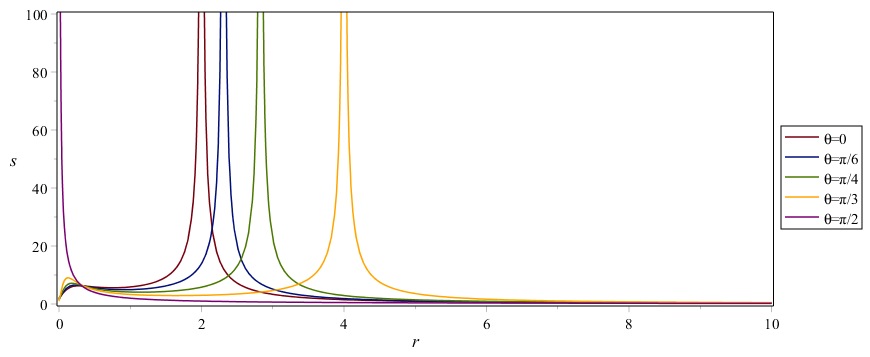}
\caption{Plot showing the variation of the  gravitational entropy density for the accelerating rotating BH with respect to the radial coordinate $ r $ for different values of $ \theta $, where we have taken $m=1 \: a=0.25, \: \alpha=0.5, \textrm{and} \: k_{s}=1 $.}\label{plottn}
\end{figure}

FIG. \ref{plot4b} clearly shows that the measure of entropy density is well behaved everywhere except at the ring singularity. In figures FIG. \ref{fig4}(a) and FIG. \ref{fig4}(b), we have shown that for different values of $\theta$ we can have a diverging or finite entropy density at $r=0$. When $ \theta=\frac{\pi}{2} $, the expression of entropy density in (\ref{enden1}) simply becomes $ \dfrac{2k_{s}}{r} $, which can be seen clearly in FIG. \ref{fig4}(a) and it also diverges at the ring singularity at $ r=0 $ for this case. Whereas in FIG. \ref{fig4}(b), we can see that for $ \theta=\frac{\pi}{6} $ the entropy density is finite at $ r=0 $ for nonzero values of acceleration parameter whereas for  $ \alpha=0 $ the entropy density becomes zero at the central singularity at $ r=0 $. FIG. \ref{fig5}(a) shows us that the entropy density simply behaves like inverse squared in $ r $ when $ \theta=\frac{\pi}{2} $, for different values of $ a $, as it becomes independent of the rotation parameter and the acceleration parameter, which can be easily seen from the expression (\ref{enden1}). In FIG. \ref{fig5}(b), the entropy density diverges for the condition $ r=0, \: \textrm{and} \: a=0 $, because it corresponds to the central singularity of an accelerating non-rotating BH. These behaviors of the entropy density are in conformity with our expectations and so we can say that this definition of entropy density is quite suitable for these kinds of black holes.

In FIG. \ref{plottn}, the nature of the gravitational entropy density for accelerating rotating black hole is studied for different values of $ \theta $, where we have fixed the values of the black hole parameters as the following: $m=1, \: a=0.25, \: \alpha=0.5, \textrm{and} \: k_{s}=1 $. The conformal infinity lies at $ r=\dfrac{1}{\alpha\cos\theta} $. From (\ref{enden1}) it is clear that as the value of $ \theta $ goes from $ 0 $ to $ \dfrac{\pi}{2} $, the conformal infinity shifts towards infinity, giving rise to the $ \sim \dfrac{1}{r^2} $ behavior which diverges only at the ring singularity at $ r=0, \theta=\dfrac{\pi}{2} $. Moreover, we observe that except for $ \theta=\dfrac{\pi}{2} $, the gravitational entropy density remains finite at $ r=0 $ for all other values of $\theta$.

\subsection{Rotating charged accelerating black hole}
For the rotating charged black hole, the Weyl scalar $W$ is
\begin{align}
\left.W\right. &=48\frac { \left( \alpha r\cos \left( \theta \right) -1 \right) ^{6}}
{ \left( {r}^{2}+{a}^{2} \left( \cos \left(\theta \right)  \right) ^{2} \right) ^{6}} \times\nonumber\\
 &(  \left( {e}^{2}r\alpha+am \left( a\alpha+1 \right)  \right) {
a}^{2}\cos^{3}\theta+ \left( 2a{e}
^{2}{r}^{2}\alpha+3{a}^{2}m \left( a\alpha-1 \right) r+{a}^{2}{e}^{2
} \right)   \cos^{2}\theta \nonumber\\
&+ \left( -{e
}^{2}\alpha\,{r}^{3}-3am \left( a\alpha+1 \right) {r}^{2}+2 a{e}^{2
}r \right) \cos \left( \theta \right) -{r}^{2} \left( m \left( a\alpha
-1 \right) r+{e}^{2} \right)  ) \nonumber\\
&  (  \left( {e}^{2}r\alpha+a
m \left( a\alpha-1 \right)  \right) {a}^{2} \left( \cos \left( \theta
 \right)  \right) ^{3}+ \left( -2a{e}^{2}{r}^{2}\alpha-3{a}^{2}m
 \left( a\alpha+1 \right) r+{a}^{2}{e}^{2} \right)  \left( \cos
 \left( \theta \right)  \right) ^{2} \nonumber\\
 &+ \left( -{e}^{2}\alpha{r}^{3}-3
am \left( a\alpha-1 \right) {r}^{2}-2a{e}^{2}r \right) \cos
 \left( \theta \right) +{r}^{2} \left( m \left( a\alpha+1 \right) r-{e
}^{2} \right)  ),
\end{align}
and the Kretschmann scalar $K$ is
\begin{align}
\small \left.K\right. &=  48\frac {\left( \alpha r \cos \left( \theta \right)-1 \right)^{6}}{\left({r}^{2}+{a}^{2}\left(\cos\left(\theta \right)\right)^{2}\right)^{6}}({a}^{4}\left({a}^{4}{\alpha}^{2}{m}^{2}+2{a}^{2}{\alpha}^{2}{e}^{2}mr+7/6{\alpha}^{2}{e}^{4}{r}^{2}
-{a}^{2}{m}^{2} \right)  \left( \cos \left( \theta \right)  \right) ^{6}  \nonumber\\
&+2 \left( {a}^{2}m+5/6{e}^{2}r \right)  \left( {e}^{2}-6mr\right) {a}^{4}\alpha \left(\cos\left(\theta \right)\right)^{5} +\big(15{a}^{4}{m}^{2}{r}^{2}-20{a}^{4}{\alpha}^{2}{e}^{2}m{r}^{3}
\nonumber\\
& -15{a}^{6}{\alpha}^{2}{m}^{2}{r}^{2}-{\frac {17{a}^{2}{\alpha}^{2}{e}^{4}{r}^{4}}{3}}-10{a}^{4}{e}^{2}mr+7/6{a}^{4}{e}^{4}\big)  \times\left( \cos \left( \theta \right)  \right) ^{4} -20 \big(-{e}^{2}m{r}^{2}+ \nonumber\\
&  \left( -2{a}^{2}{m}^{2}+{\frac {19{e}^{4}}{30}}\right) r+{a}^{2}{e}^{2}m \big) {r}^{2}{a}^{2}\alpha \left( \cos\left( \theta \right)  \right) ^{3}+ \big( 7/6{\alpha}^{2}{e}^{4}{r}^{6}-{\frac {17{a}^{2}{e}^{4}{r}^{2}}{3}}-15{a}^{2}{m}^{2}{r}^{4
}\nonumber\\
 & +10{a}^{2}{\alpha}^{2}{e}^{2}m{r}^{5}+15{a}^{4}{\alpha}^{2}{m}^{2}{r}^{4}+20{a}^{2}{e}^{2}m{r}^{3} \big)  \left( \cos \left( \theta\right)  \right) ^{2} + 10 \left( {e}^{2}-6/5mr \right) \nonumber\\
 & {r}^{4}\alpha \left( {a}^{2}m+1/6{e}^{2}r \right) \cos \left( \theta
 \right)+ \left( -{a}^{2}{\alpha}^{2}{m}^{2}+{m}^{2} \right) {r}^{6}-2{e}^{2}m{r}^{5}+7/6{e}^{4}{r}^{4} ).
\end{align}

From the above scalars we can calculate the ratio $ P=\sqrt{\dfrac{W}{K}} $, defined in \cite{entropy1}, which serves as the measure of gravitational entropy, $S_{grav}$ of black holes. The four-dimensional determinant of the metric is
\begin{equation}
g=-{\frac { \left( \sin \left( \theta \right)\right) ^{2} \left( {r}^{2}+{a}^{2} \left( \cos \left( \theta
 \right)  \right) ^{2} \right) ^{2}}{ \left( \alpha\,r\cos \left( \theta \right) -1 \right) ^{8}}}.
\end{equation}
Here again the axisymmetric metric denies us the calculation of the spatial metric due to the nonzero metric component $ g_{t\phi} $.  Therefore as in the previous calculation for rotating black holes, the entropy density is calculated by using the metric determinant $ g $ in the covariant derivative. We thus have
\begin{equation}\label{enden2}
s=k_{s}|\mathbf{\nabla}.\mathbf{\Psi}|=\dfrac{k_{s}}{\sqrt{-g}}\left(\dfrac{\partial}{\partial r}\sqrt{-g}P\right).
\end{equation}
Here we have intentionally avoided writing the exact expression of entropy density as it is lengthy and too much complicated, but we can easily check the validity of the result. We have checked that if we substitute $ e=0 $ in these calculations, then we get back the result for the accelerating rotating black hole.

\begin{figure}[ht]
    \centering
    \subfloat[Subfigure 1 list of figures text][]
        {
        \includegraphics[width=0.38\textwidth]{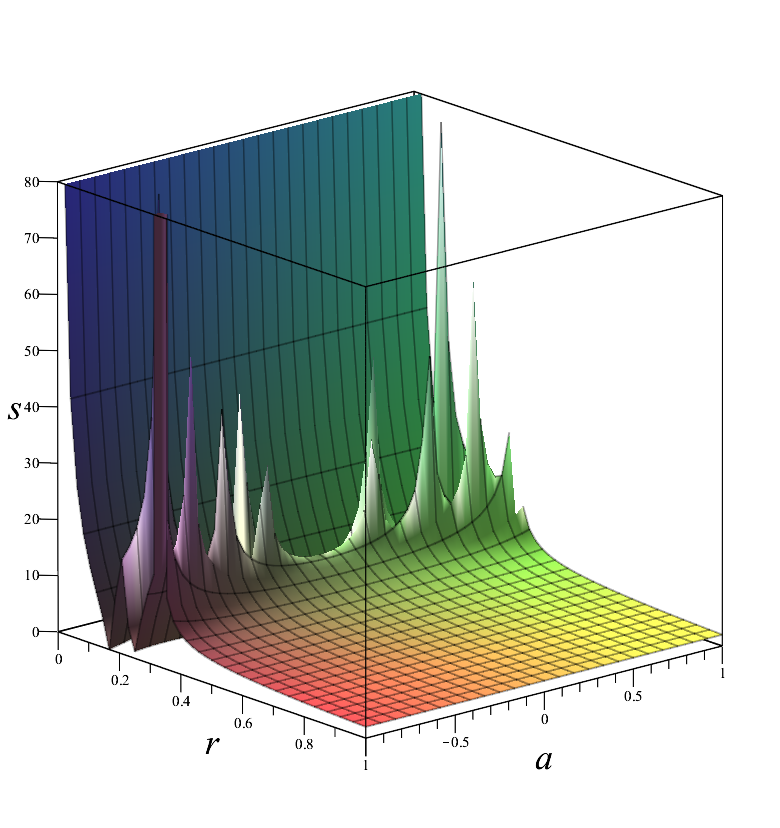}
        \label{fig:subfig11}
        }
        \hspace{1.0cm}
    \subfloat[Subfigure 2 list of figures text][]
        {
        \includegraphics[width=0.4\textwidth]{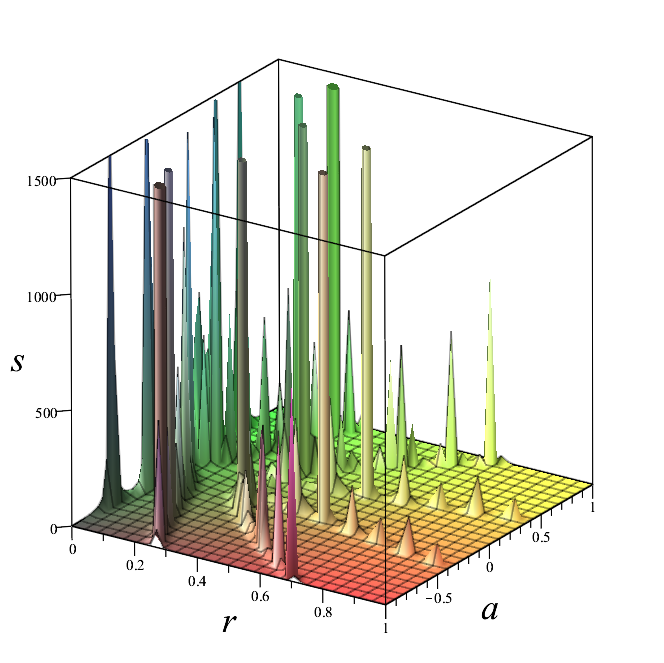}
        \label{fig:subfig12}
        }
    \caption{(a) Plot showing the variation of the gravitational entropy density for the accelerating rotating charged BH with respect to the radial coordinate $ r $ and the rotation parameter $ a $, where $ e=0.6, \: \alpha=0.45, \: m=1, \: k_{s}=1, \: \textrm{and} \: \theta=\dfrac{\pi}{2} $. (b) Plot showing the variation of the gravitational entropy density for the accelerating rotating charged BH with respect to the radial coordinate $ r $ and the rotation parameter $ a $, for $e=0.6, \: \alpha=0.45, \: m=1, \: k_{s}=1, \: \textrm{and} \: \theta=\dfrac{\pi}{4} $.}
    \label{fig6}
\end{figure}
In FIG. \ref{fig6} we find that the gravitational entropy density is not smooth, but contains several singularities. The above analysis clearly shows that the measure of the gravitational entropy used above is not adequate to explain the case of the accelerating rotating charged black holes. Therefore we have to use the measure proposed in \cite{entropy2} for the expression of $ P $, which is
\begin{equation}\label{mod_P}
P=C_{abcd}C^{abcd}.
\end{equation}
Using the definition (\ref{mod_P}) of $ P $, we have calculated the gravitational entropy density, which is given in equation (\ref{new3}):
\begin{align}\label{new3}
\left.s\right.&=\dfrac{k_{s}}{(a^2\cos^2(\theta)+r^2)^7}\bigg(96\Big(a^6\alpha(a^4 \alpha^2 m^2+3a^2\alpha^2 e^2 mr+2\alpha^2 e^4 r^2 -a^2m^2)\cos^9(\theta)\nonumber\\
&+(-20e^2mr^2+(-18a^2m^2+2e^4)r+a^2e^2m)\alpha^2a^6\cos^8(\theta)- 34\alpha a^4(21/34e^4r^4\alpha^2+\nonumber\\
&57/34a^2e^2mr^3\alpha^2+(a^4\alpha^2m^2-a^2m^2)r^2+5/34a^2e^2mr-3/17a^4m^2)\cos^7(\theta)+\nonumber\\
&(90a^4e^2mr^4\alpha^2+(142a^6\alpha^2m^2-21a^4\alpha^2e^4)r^3-9a^6e^2m\alpha^2r^2+(20a^8\alpha^2 m^2-20a^6m^2)r\nonumber\\
&+5a^6e^2m)\cos^6(\theta)+30\alpha a^2(8/15e^4r^5\alpha^2+17/6a^2e^2 m r^4\alpha^2+(3a^4\alpha^2 m^2-3a^2m^2)r^3\nonumber\\
&+1/6a^2e^2mr^2+(-19/5a^4m^2+11/30e^4a^2)r+a^4e^2m)r\cos^5(\theta)+(-48a^2e^2m\alpha^2r^6+ \nonumber\\
&(-142a^4\alpha^2m^2+16a^2\alpha^2e^4)r^5-5a^4e^2mr^4\alpha^2+(-90a^6\alpha^2m^2+90a^4m^2)r^3-75a^4e^2mr^2\nonumber\\
&+11a^4e^4r)\cos^4(\theta)-100(1/100e^4r^5\alpha^2+3/20a^2e^2mr^4\alpha^2+(17/50a^4\alpha^2m^2- \nonumber\\
&17/50a^2m^2)r^3-9/100a^2e^2mr^2+(-17/10a^4m^2+13/50e^4a^2)r+a^4e^2m)\alpha r^3\cos^3(\theta) \nonumber\\
&+(2e^2m\alpha^2r^8+(18a^2\alpha^2 m^2-\alpha^2e^4)r^7+5a^2e^2m\alpha^2 r^6+(48a^4\alpha^2m^2-48a^2m^2)r^5\nonumber\\
&+75a^2e^2mr^4-26e^4a^2r^3)\cos^2(\theta)+30((1/30a^2\alpha^2m^2-1/30m^2)r^3-1/30e^2mr^2+ \nonumber\\
&(-a^2m^2+1/10e^4)r+a^2e^2m)\alpha r^5\cos(\theta)+(-2a^2\alpha^2m^2+2m^2)r^7-5e^2 mr^6+3e^4r^5\Big) \nonumber\\
&(r\alpha  \cos(\theta)-1)^5\bigg).
\end{align}

\begin{figure}[ht]
    \centering
    \subfloat[Subfigure 1 list of figures text][]
        {
        \includegraphics[width=0.46\textwidth]{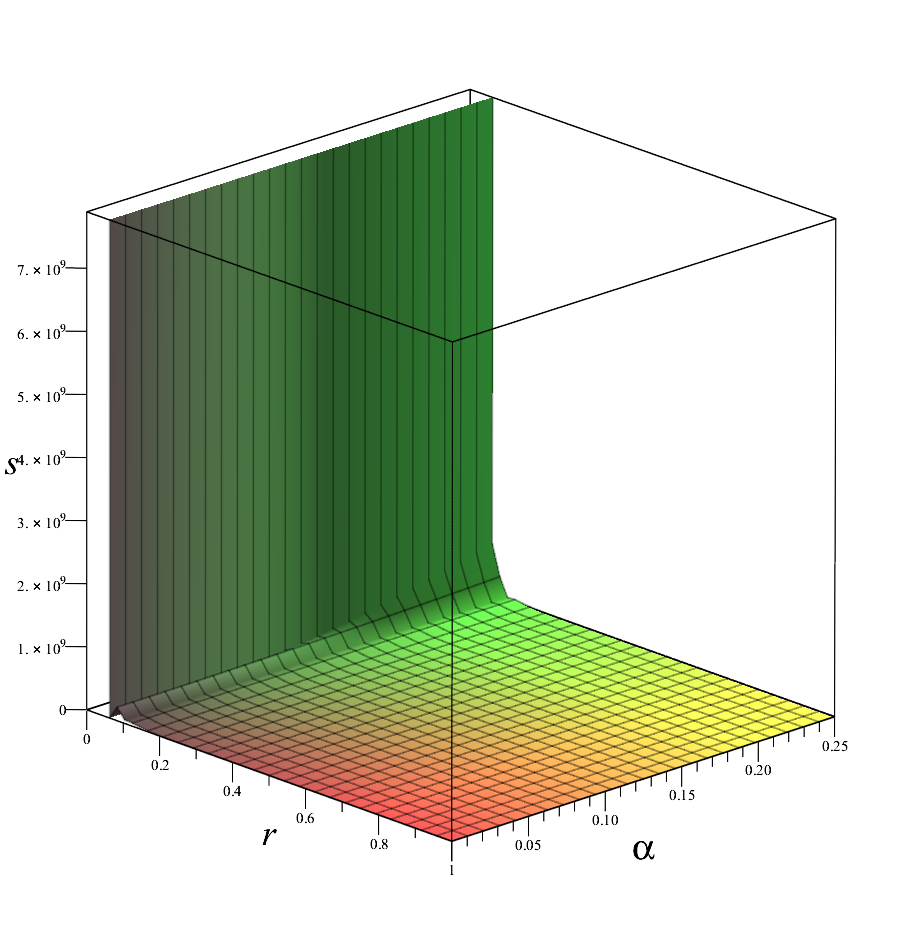}
        \label{fig:subfig13}
        }
    \subfloat[Subfigure 2 list of figures text][]
        {
        \includegraphics[width=0.45\textwidth]{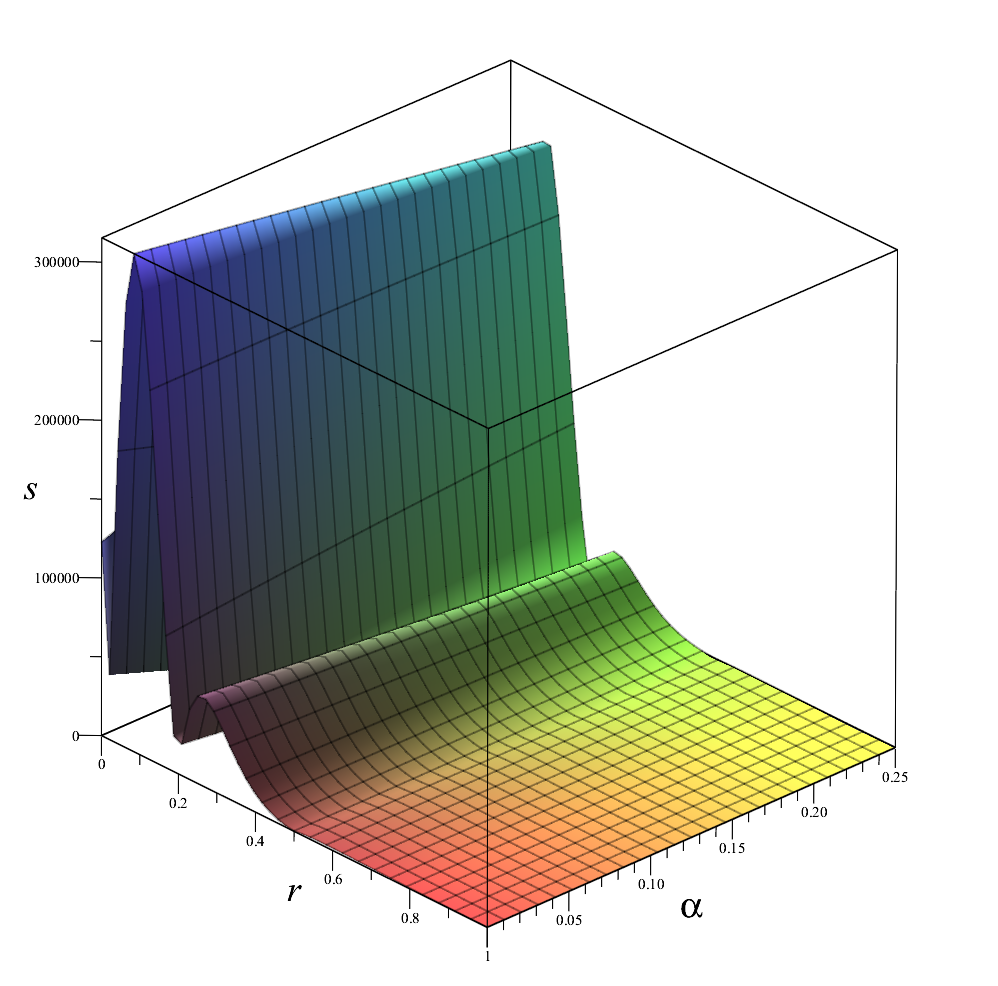}
        \label{fig:subfig14}
        }
    \caption{(a) Plot showing the variation of the gravitational entropy density for the accelerating rotating charged BH with respect to the radial coordinate $ r $ and the acceleration parameter $ \alpha $, for $ e=0.25, \: a=0.5, \: m=1, \: k_{s}=1, \: \textrm{and} \: \theta=\dfrac{\pi}{2} $. (b) Plot showing the variation of the gravitational entropy density for the accelerating rotating charged BH with respect to the radial coordinate $ r $ and the acceleration parameter $ \alpha $, where $e=0.25, \: a=0.5, \: m=1, \: k_{s}=1, \: \textrm{and} \: \theta=\dfrac{\pi}{4} $.}
    \label{fig7}
\end{figure}

In FIG. \ref{fig7} we have shown the variation of the gravitational entropy density with the radial distance and the acceleration parameter using this new definition (\ref{mod_P}) of the scalar $ P $. The entropy density function is now well-behaved and all the singularities vanish, except the ring singularity, on account of the introduction of this new definition. In FIG. \ref{fig7}(a), the entropy density function diverges at $ r=0 $ and $ \theta=\frac{\pi}{2} $, as it encounters the ring singularity, whereas in FIG. \ref{fig7}(b) the entropy density stays finite at $ r=0 $ and $ \theta=\frac{\pi}{4} $. Although the entropy density function (\ref{new3}) vanishes at the conformal infinity $ r=\dfrac{1}{\alpha\cos(\theta)} $, we cannot simply associate the zeroes of the entropy density function with the horizons, because according to this modified definition, the expression (\ref{new3}) does not have such factors, and so we have to solve the function explicitly in order to determine the zeroes of the entropy density.

\section{Discussions}
We now discuss the possibility of having an angular component in the vector field $ \mathbf{\Psi} $ for axisymmetric spacetimes as proposed in \cite{entropy2}. Using this modified definition of $ \mathbf{\Psi} $, and the modified expression (\ref{mod_P}), we now calculate the gravitational entropy density for axisymmetric space-times, using the following expression:
\begin{equation}\label{news}
s=k_{s}|\mathbf{\nabla}.\mathbf{\Psi}|=\dfrac{k_{s}}{\sqrt{-g}}\left|\left(\dfrac{\partial}{\partial r}(\sqrt{-g}P)+\dfrac{\partial}{\partial \theta}(\sqrt{-g}P) \right)\right|.
\end{equation}

\begin{figure}[ht]
    \centering
    \subfloat[Subfigure 1 list of figures text][]
        {
        \includegraphics[width=0.45\textwidth]{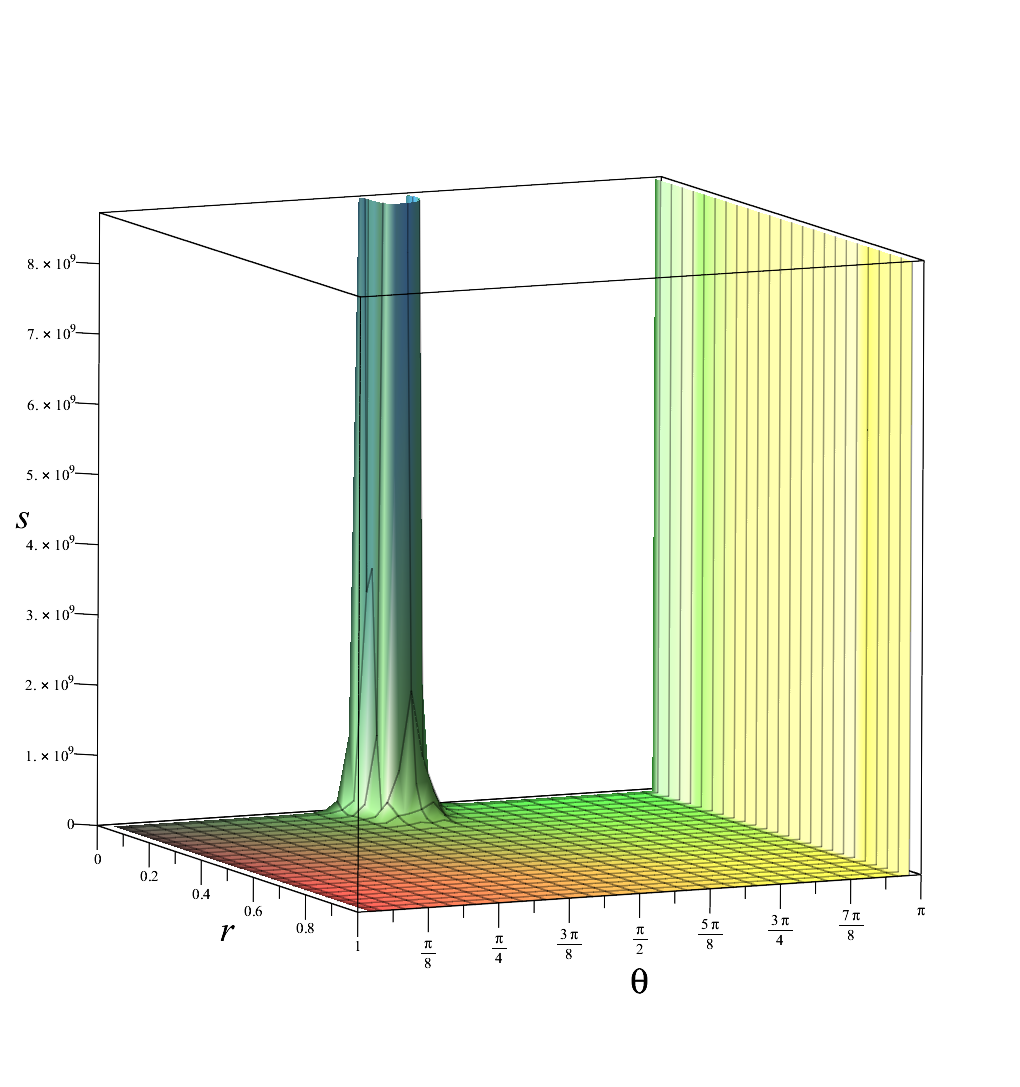}
        \label{fig:subfig15}
        }
    \subfloat[Subfigure 2 list of figures text][]
        {
        \includegraphics[width=0.48\textwidth]{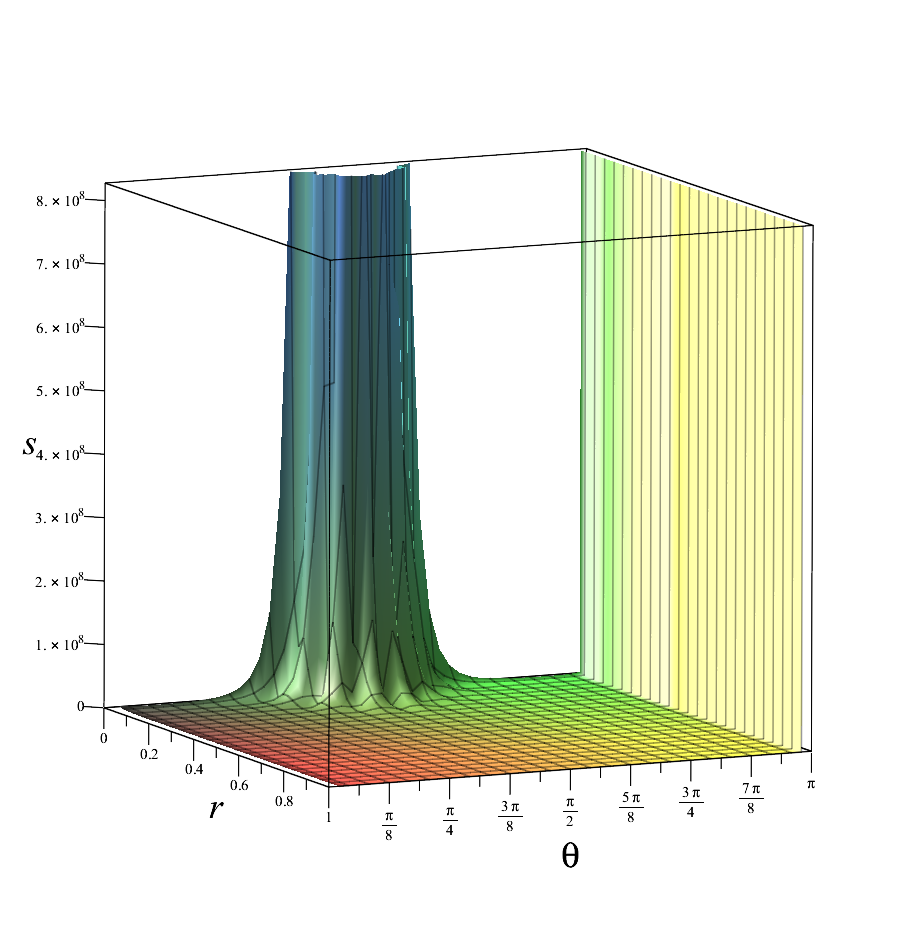}
        \label{fig:subfig16}
        }
    \caption{(a) Plot showing the variation of the gravitational entropy density for the accelerating rotating BH with respect to the radial coordinate $ r $ and the angular coordinate $ \theta $, using the modified expression given in \ref{new1}, where $\alpha=0.25, \: e=0, \: a=0.95, \: m=1, \: \textrm{and} \: k_{s}=1 $. (b) Plot showing the variation of the gravitational entropy density for the accelerating rotating charged BH with respect to the radial coordinate $ r $ and the angular coordinate $ \theta $, using the modified expression given in \ref{new2}, where $\alpha=0.25, \: e=0.2, \: a=0.45, \: m=1, \: \textrm{and} \: k_{s}=1 $.}
    \label{fig8}
\end{figure}

The gravitational entropy density for the uncharged rotating accelerating black hole is given by
{\footnotesize
\begin{align}\label{new1}
\left.s\right. &= \dfrac{k_{s}}{\sqrt{\dfrac{\sin^2(\theta)(a^2\cos^2(\theta)+r^2)^2}{(r\alpha\cos(\theta)-1)^8}}}\nonumber\\
&\Bigg(\Bigg|\dfrac{48}{\sqrt{\dfrac{\sin^2(\theta)(a^2\cos^2(\theta)+r^2)^2}{(r\alpha\cos(\theta)-1)^8}}(a^2\cos^2(\theta)+r^2)^5(r\alpha \cos(\theta)-1)^3} \nonumber\\
&\bigg(\sin(\theta)\Big(((2a^{10}\alpha^{3} m^{2} r-2a^{8}\alpha m^{2}r)\cos^{8}(\theta)-4(a^{4}m^{2}\alpha^{2}+(9\alpha^{2}r^{2}-1)m^{2}a^{2})a^{6}\cos^{7}(\theta)  \nonumber\\
&-10(m^{2}(\frac{34}{5}r^{3}\alpha^{2}-6r)a^{4}-\frac{34}{5}r^{3}a^{2}m^{2})\alpha a^{4}\cos^{6}(\theta)
+96 a^{4}(a^{4}m^{2}r^{2}\alpha^{2}+(\frac{71}{24}\alpha^{2}r^{4}-r^{2}) \nonumber\\
&m^{2}a^{2})\cos^{5}(\theta)+150((\frac{6}{5}r^{3}\alpha^{2}-\frac{34}{15}r)m^{2}a^{4}-\frac{6}{5}r^{3}a^{2}m^{2}) \alpha a^{2}r^{2}\cos^{4}(\theta)-180a^{2}(a^{4}m^{2}r^{2}\alpha^{2} \nonumber\\
&+(\frac{71}{45}\alpha^{2}r^{4}-r^2)m^2a^2)r^2\cos^{3}(\theta)-150\alpha((\frac{34}{75} r^3\alpha^{2}-\frac{38}{25} r)m^2a^4-\frac{34}{75} r^3 a^2 m^2)r^4\cos^{2}(\theta) \nonumber\\
&+40(a^4 m^2 r^2 \alpha^2+(\frac{9}{10} \alpha^2 r^4-r^2)m^2a^2)r^4\cos(\theta)+10\alpha((\frac{1}{5}r^3\alpha^2-\frac{6}{5}r)m a^2-\frac{1}{5}r^3 m)mr^6)\nonumber\\
& \sin^2(\theta)+(2\alpha a^6 (a^4\alpha^2m^2-a^2m^2)\cos^{9}(\theta)-36\cos^8(\theta) a^8 \alpha^2 m^2 r-68 \alpha a^4(m^2(\alpha^2 r^2-\frac{3}{17}) \nonumber\\
&a^4 -a^2 m^2 r^2)\cos^7(\theta)+40 a^4(a^4 m^2 r\alpha^2-\frac{9}{20}(-\frac{142}{9} r^3\alpha^2+\frac{20}{9}r)m^2a^2)\cos^6(\theta)+60\alpha a^2  \nonumber\\
&((3 r^3 \alpha^2-\frac{19}{5}r)m^2 a^4-3 r^3 a^2 m^2)r\cos^5(\theta)-180 a^2(a^4 m^2 r^2\alpha^2+(\frac{71}{45} \alpha^2r^4-r^2)m^2a^2)\nonumber\\
&r\cos^{4}(\theta)-200((\frac{17}{50}r^3\alpha^2-\frac{17}{10}r)m^2a^4-\frac{17}{50}r^3 a^2 m^2)\alpha r^3 \cos^3(\theta)+96 r^3(a^4 m^2 r^2\alpha^2+ \nonumber\\
& (\frac{3}{8}\alpha^2 r^4-r^2)m^2a^2)\cos^2(\theta)+60((\frac{1}{30} r^3\alpha^2-r)m^2 a^2-\frac{1}{30}r^3 m^2)\alpha r^5\cos(\theta)-4a^2\alpha^2 m^2 r^7 \nonumber\\
&+4 m^2 r^7)\sin(\theta)+(a^2\cos^{2}(\theta)+r^2)(a^2(a^2\alpha m -a m)\cos^3(\theta)+(-3 a^3\alpha m r-3 a^2 m r) \nonumber\\
&\cos^2(\theta)+ (-3 a^2 \alpha m r^2+3 a m r^2)\cos(\theta)+r^2(a\alpha m r+m r))(r\alpha\cos(\theta)-1)(a^2(a^2\alpha m \nonumber\\
&+a m)\cos^3(\theta)+(3a^3\alpha m r-3a^2mr)\cos^{2}(\theta)+(-3 a^2\alpha m r^2-3 a m r^2)\cos(\theta) \nonumber\\
&-r^3 a m\alpha+r^3 m)\cos(\theta)\Big)\bigg)\Bigg|\Bigg)
\end{align}
}%

In FIG. \ref{fig8}(a), we indicate the variation of the gravitational entropy density with radial distance and the angular coordinate, using the modified expression in (\ref{new1}).

Similarly the gravitational entropy density for the charged rotating accelerating black hole is given by equation (\ref{new2}), which is quoted below. We can always check that if we substitute $ e=0 $ in (\ref{new2}), we get back the expression in (\ref{new1}). In FIG. \ref{fig8}(b), we have shown the corresponding variation of the gravitational entropy density with radial distance and angular coordinate. The expressions of gravitational entropy density in both (\ref{new1}) and (\ref{new2}) diverges at the ring singularity i.e. at $ (r^2 + a^2 \cos^2(\theta))=0 $ and vanishes at the conformal infinity i.e. at $ r=\dfrac{1}{\alpha \cos(\theta)} $.

{\footnotesize
\begin{align}\label{new2}
 \left.s\right. &= \dfrac{k_{s}}{\sqrt{\dfrac{\sin^2(\theta)(a^2\cos^2(\theta)+r^2)^2}{(r\alpha\cos(\theta)-1)^8}}} \nonumber\\
&\Bigg(\Bigg|\dfrac{48}{\sqrt{\dfrac{\sin^2(\theta)(a^2\cos^2(\theta)+r^2)^2}{(r\alpha\cos(\theta)-1)^8}}(a^2\cos^2(\theta)+r^2)^5(r\alpha \cos(\theta)-1)^3} \nonumber\\
&\bigg(\sin(\theta)\Big(((2a^{10} m^2r\alpha^3+(4\alpha^3e^2m r^2-2\alpha m^2 r)a^8+2a^6e^4r^3\alpha^3)\cos^8(\theta)- 4(a^4m^2\alpha^2+ \nonumber\\
&((9\alpha^2r^2-1)m^2+1/2e^2mr\alpha^2)a^2 -1/2e^2r^2\alpha^2(e^2-15mr))a^6\cos^7(\theta)-10(m((34/5r^3\alpha^2- \nonumber\\
& 6r)m+e^2)a^4+3/5(-34/3m^2r^2+16e^2r(\alpha^2r^2-5/48)m+e^4)ra^2 +16/5e^4r^5\alpha^2)\alpha a^4\cos^6(\theta)  \nonumber\\
&+96a^4(a^4m^2r^2\alpha^2+((71/24\alpha^2r^4-r^2)m^2+3/16e^2(\alpha^2r^2+10/3)rm-1/16e^4)a^2-1/3e^2\alpha^2(e^2 \nonumber\\
&-85/16mr)r^4)\cos^5(\theta)+150(((6/5r^3\alpha^2-34/15r)m+e^2)ma^4+26/75(-45/13m^2r^2+45/13(\alpha^2r^2-  \nonumber\\
&\frac{1}{18})e^2rm+e^4)ra^2  +\frac{7}{25}e^4r^5\alpha^2)\alpha a^2r^2\cos^4(\theta)-180a^2(a^4m^2r^2\alpha^2+((\frac{71}{45}\alpha^2r^4-r^2)m^2  \nonumber\\
&-\frac{1}{18}e^2r(\alpha^2r^2-20)m-\frac{13}{45}e^4)a^2-\frac{7}{30}e^2\alpha^2r^4(e^2-\frac{19}{7}mr))r^2\cos^3(\theta)-150\alpha(((\frac{34}{75}r^3\alpha^2  \nonumber\\
&-\frac{38}{25}r)m+e^2)ma^4+\frac{11}{75}r(-\frac{34}{11}m^2r^2+\frac{20}{11}e^2(\alpha^2r^2+\frac{9}{20})rm +e^4)a^2+\frac{2}{75}e^4r^5\alpha^2)r^4 \nonumber\\
& \cos^2(\theta)+40(a^4m^2r^2\alpha^2+((\frac{9}{10}\alpha^2r^4-r^2)m^2-\frac{1}{4}e^2r(\alpha^2r^2-6)m-\frac{11}{20}e^4)a^2-\frac{1}{10}(e^2  \nonumber\\
& -\frac{3}{2}mr)e^2\alpha^2r^4)r^4\cos(\theta)+10\alpha(((\frac{1}{5}r^3\alpha^2-\frac{6}{5}r)m+e^2)a^2+\frac{1}{5}r^2(e^2-mr))mr^6)\sin^2(\theta) \nonumber\\
& +(2\alpha a^6(a^4m^2\alpha^2+  (3\alpha^2e^2mr-m^2)a^2+2\alpha^2e^4r^2)\cos^9(\theta)+2\alpha^2a^6((e^2m-18m^2r)a^2 \nonumber\\
&+2re^4-20e^2mr^2)\cos^8(\theta)-68\alpha a^4(m^2(\alpha^2r^2 -\frac{3}{17})a^4+(-m^2r^2 +\frac{57}{34}e^2r(\alpha^2r^2+\frac{5}{57})   \nonumber\\
& m)a^2+\frac{21}{34}e^4r^4\alpha^2)\cos^7(\theta)+40a^4(a^4m^2r\alpha^2-\frac{9}{20}((-\frac{142}{9}r^3\alpha^2+\frac{20}{9}r)m +e^2(\alpha^2r^2-\frac{5}{9})) \nonumber\\
&ma^2-\frac{21}{20}e^2\alpha^2(e^2-\frac{30}{7}mr)r^3)\cos^6(\theta)+60\alpha a^2(((3r^3\alpha^2-\frac{19}{5}r)m+e^2)ma^4+(-3r^3m^2 \nonumber\\
&+\frac{17}{6}e^2  (\alpha^2r^2+\frac{1}{17})r^2m+\frac{11}{30}re^4)a^2+\frac{8}{15}e^4r^5\alpha^2)r\cos^5(\theta)-180a^2(a^4m^2r^2\alpha^2+((\frac{71}{45}   \nonumber\\
&\alpha^2r^4-r^2)m^2+\frac{1}{18}e^2r(\alpha^2r^2+15)m-\frac{11}{90}e^4)a^2 -\frac{8}{45}e^2r^4\alpha^2(e^2-3mr))r\cos^4(\theta) \nonumber\\
&-200(((17/50r^3\alpha^2-17/10r)m+e^2)ma^4+13/50(-17/13m^2r^2+15/26e^2(\alpha^2r^2 -3/5) rm+e^4)ra^2   \nonumber\\
& +1/100 e^4r^5\alpha^2)\alpha r^3\cos^3(\theta)+96r^3(a^4m^2r^2\alpha^2+((3/8\alpha^2r^4-r^2)m^2+5/48e^2r(\alpha^2r^2+15)m   \nonumber\\
&-13/24e^4)a^2 -1/48e^2r^4\alpha^2(e^2-2mr))\cos^2(\theta)+60(((1/30r^3\alpha^2-r)m+e^2)ma^2+ \nonumber\\
&1/10r(e^4-1/3 e^2mr-1/3 m^2r^2)) \times\alpha r^5\cos(\theta) -4a^2\alpha^2m^2r^7+6e^4r^5-10e^2mr^6+4m^2r^7) \nonumber\\
&\sin(\theta)+(a^2\cos^2(\theta)+r^2)(a^2(a^2\alpha m+\alpha e^2r-am)\cos^3(\theta)+ (-3a^3mr\alpha+(e^2-3mr) \nonumber\\
&a^2-2ae^2\alpha r^2)\cos^2(\theta)+(-3a^2mr^2\alpha+(-2e^2r+3mr^2)a-e^2\alpha r^3)\cos(\theta)+r^2(a\alpha mr- \nonumber\\
&e^2+mr))(r\alpha \cos(\theta)-1)(a^2(a^2\alpha m+\alpha e^2 r+am)\cos^3(\theta)+(3a^3mr\alpha+(e^2-3mr)a^2+ \nonumber \\
& 2ae^2\alpha r^2)\cos^2(\theta)+ (-3a^2mr^2\alpha+(2e^2r-3mr^2)a-e^2\alpha r^3)\cos(\theta)-r^3am\alpha-e^2r^2+r^3m)\cos(\theta)\Big)\bigg)\Bigg|\Bigg)
\end{align}
}%

Therefore from FIG. \ref{fig8}, we find that the entropy density measure diverges not only at the ring singularity but also at $ \theta=\pi $, which renders this measure inappropriate for determining the gravitational entropy in these cases. There is another singularity at $ \theta=0 $, though not visible in FIG. \ref{fig8}, but can be inferred from the mathematical analysis. This is in agreement with the observations in \cite{entropy2} for non-accelerating axisymmetric black holes. This is a disturbing feature of this method of analysis. For a possible resolution of this problem we want to point out that in the case of rotating black holes, the existence of stationary observer is not well defined because of the effect of frame dragging. Nevertheless, we have worked with the chosen definition of gravitational entropy density to get an overall idea of the way things work out. For such cases of axisymmetric space-times, it is not possible to determine the spatial metric $h_{ij} $ because of the presence of the metric coefficient $g_{t\phi}$ in the metric (\ref{ht}) and in metric (\ref{htn}). This is because the object is rotating and the spatial position of each event in the space-time depends on time. Therefore the covariant divergence is calculated from the determinant of the full metric and is given in equations (\ref{enden1}) and (\ref{enden2}).

\section{Conclusions}

In this paper we have adopted a phenomenological approach of determining the gravitational entropy of accelerating black holes as done in \cite{entropy1} and \cite{entropy2}. We find that the gravitational entropy proposal \cite{entropy1} for the accelerating black holes and charged accelerating black holes works pretty well, except for the rotating charged metric where we faced difficulties in this regard. We then considered the alternative definition of $ P $ given in \cite{entropy2} to compute the entropy density and showed that the gravitational entropy is well defined in this case. In the end we considered the vector $ \mathbf{\Psi} $ to have additional angular components for axisymmetric spacetimes, as proposed in \cite{entropy2}, to compute the entropy density for accelerating rotating and accelerating charged rotating black holes. From our calculations and the corresponding plots, we can conclude that for the rotating black holes the entropy density will be well-defined if we change our definition of the vector field $ \mathbf{\Psi} $, be it in the magnitude ($ P $) of it, or in the vector directions (having additional angular components).  ''



\chapter{ Gravitational entropy of cosmological models}



The contents of this chapter have been published in a journal, details of which are given below:\\

\textbf{JOURNAL REFERENCE:} International Journal of Modern Physics D, Vol. 30, No. 7 (2021) 2150051 (28 pages)

\textbf{ARTICLE NAME:} An investigation on gravitational entropy
of cosmological models \\
DOI: 10.1142/S0218271821500516 \\~~~\\

The paper is quoted below:\\

``

\section{Introduction}

The proposal of gravitational entropy attempts to provide a sense of sequence to gravitational processes, and remains as one of the open problems in General Relativity (GR) till today. Although a suitable definition of gravitational entropy in the case of stationary black holes was available in literature for quite some time \cite{bh}, but a universally agreeable analogue in the case of cosmology has been under the process of formulation till late.

It is well-known that GR is plagued by the problem of spacetime singularities. However, Roger Penrose \cite{Penrose} put forward his belief that this problem was a consequence of the limitations of the ``very notion of spacetime geometry'' and the corresponding physical laws. According to him, the problem of spacetime singularities held the key to the ``origin of the \emph{arrow of time}''. Several researchers \cite{arrow,PL} based their studies on the notion of this arrow of time.

Originally Penrose proposed the \textit{Weyl curvature hypothesis} (WCH), which merely required that the Weyl tensor should be zero at the big bang singularity \cite{Rothman}, and the subsequent evolution of the universe must be close to the homogeneous and isotropic Friedman-Lemaitre-Robertson-Walker (FLRW) model. Assuming the existence of ``gravitational entropy'', Penrose argued that \emph{the principle of increase
of entropy} implies that the big bang singularity should be of low entropy, which is related to the ``absence of clumping of matter'', and hence to the absence of Weyl tensor, so that the big bang singularity must have been a regular one. Thus the universe could evolve to its observed FLRW form and be consistent with the second law of thermodynamics \cite{PL,PC}. If matter was approximately under thermodynamic equilibrium during the big bang, then it requires a corresponding low entropy of the gravitational field. Although the big bang would normally be considered as a state of maximum entropy (assuming thermal equilibrium), but in reality the entropy of the universe is increasing. This apparent paradox may be due to the omission of gravitational degrees of freedom, and the gravitational entropy of the big bang was actually low. Penrose \cite{Penrose} argued that gravitational entropy must be defined from the free gravitational field. An increase in gravitational entropy would imply an increase in the local anisotropy (thereby facilitating structure formation), which can be quantified by the shear tensor. By analyzing the trace free Bianchi identities we can also suggest that this shear tensor affects the evolution of the Weyl tensor \cite{MB}, thereby establishing a physical relationship between gravitational entropy and the Weyl tensor. Although a conformally flat perfect fluid spacetime has vanishing shear and vorticity, and the metric is of FLRW type \cite{ST}, but conformally flat spacetimes with diagonal trace-free anisotropic pressure and zero cosmological constant is found to possess a simple equation of state where the energy per particle density depends only on the shear scalar \cite{BR}. This provides us with one more reason to connect local anisotropy i.e. gravitational entropy to the Weyl tensor.

Penrose proposed that the gravitational entropy should be related to a suitable measure of the Weyl curvature, and the condition of low entropy should enforce constraints on the Weyl curvature. All these implied that some suitable dimensionless scalar must be asymptotically zero. Therefore, the determination of the gravitational entropy function requires the construction of this scalar function.

In 1982, Goode and Wainwright \cite{GW1982} presented a new formulation of the two classes of Szekeres solutions of the Einstein field equations, and provided a general analysis of the scalar polynomial curvature singularities of these solutions, and of their time-evolution. They identified the solutions
which are close to an FLRW model near the initial singularity, or in the later stages of evolution.

Wainwright and Anderson  discussed the evolution of a class of exact spatially homogeneous cosmological models of Bianchi type $VI_{h}$ \cite{wainander}. It is known that solutions of type $VI_{h}$ cannot approach isotropy asymptotically at large times. They infact  become asymptotic to an anisotropic vacuum plane wave solution. Nevertheless, for these solutions the initial anisotropy decays and  leads to a stage of finite duration in which the model is close to isotropy. Depending on the choice of parameters in the solution, this quasi-isotropic stage can commence at the initial singularity, in which case the singularity is of the type known as ``isotropic'' or ``Friedmann-like''. The existence of this quasi-isotropic stage implies that these models can be compatible in principle with the observed universe. Inspired by the WCH, Goode and Wainwright \cite{GW} gave the geometric definition of the concept of `isotropic singularity' and showed that the Weyl tensor is dominated by the Ricci tensor at this scalar polynomial curvature singularity. Husain \cite{Husain} examined the WCH in the context of the Gowdy cosmology. He calculated the expectation values of the square of the Weyl curvature in states of clumped and unclumped gravitons and found that the curvature contains information about the gravitational entropy.

Senovilla \cite{Senovilla} showed that the Bel-Robinson tensor is quite suitable for providing a measure of the energy of gravitational fields. Following his work, there were several attempts to define the gravitational entropy on the basis of the Bel-Robinson tensor on one hand, and also in terms of the Riemann tensor and its covariant derivatives \cite{PL,PC}. Lake and Pelavas \cite{PL} introduced a class of ``gravitational epoch'' functions which were dimensionless scalars, one of which was built from the Riemann tensor and its covariant derivatives only, denoted by $P$. Other alternative functions involving the Bel-Robinson tensor were also suggested by them. They analyzed whether such functions could be regarded as gravitational entropy function or not. Other dimensionless scalars have also been considered, for example by \cite{GCW}, which are constructed from the Riemann tensor and its covariant derivatives.

In spite of all these efforts, there was still doubt regarding the definition of gravitational entropy in a way analogous to the thermodynamic entropy, which would be applicable to all gravitational systems. Attempts were also made to explain the gravitational entropy of black holes. Among them one interesting approach is to handle the problem from a phenomenological point of view as proposed in \cite{entropy1} and expanded in \cite{entropy2}, for the purpose of testing the WCH against the expressions for the entropy of cosmological models and black holes. They considered a measure of gravitational entropy in terms of a scalar derived from the contraction of the Weyl tensor and the Riemann tensor, and matched it with the Bekenstein-Hawking entropy \cite{SWH1,Bekenstein}. Recently, Guha and Chakraborty \cite{GC1} investigated whether the prescriptions for calculating gravitational entropy as proposed in \cite{entropy1} and \cite{entropy2} could be applied to the case of the accelerating black holes. They found that such a definition of gravitational entropy works pretty well for the accelerating black holes and charged accelerating black holes, except for the rotating charged accelerating metric.

An important proposal was offered by Clifton et al. \cite{CET}, who provided a measure of gravitational entropy based on the square root of the Bel–Robinson tensor which was motivated by thermodynamic considerations, and has a natural interpretation as the effective super-energy-momentum tensor of free gravitational fields. They applied this construction to several cases, including cosmological ones, and found that the specific form of this measure depended on the nature of the gravitational field, namely, whether it was Coulomb-like or wave-like. However, this definition of gravitational entropy is only valid for General Relativity, where the Bel-Robinson tensor can be defined in this way. In the subsequent text, we refer to this formulation as the ``CET proposal''.

Bolejko \cite{cet5} showed that both the notion of gravitational entropy of the universe (associated with inhomogeneity) and the cosmic
no-hair conjecture (that a universe dominated by dark energy should asymptotically approach a homogeneous and isotropic de Sitter state) are simultaneously valid and are not contradictory. It was found that a universe with a positive cosmological constant and nonpositive spatial curvature in fact approaces the de Sitter state, but at the same time keeps generating the gravitational entropy.

In the paper \cite{CET}, the authors considered scalar perturbations of a FLRW geometry. They found that the gravitational entropy function behaved like the Hubble weighted anisotropy of the gravitational field, which therefore increases with structure formation. On the other hand, the FLRW metric is conformally flat \cite{Ellis}, and so the Bel–Robinson tensor has vanishing components. The gravitational epoch function $W$ generated by the Bel–Robinson tensor, is also vanishing, resulting in a vanishing gravitational entropy. In fact, gravitational entropy is identified with the presence of inhomogeneity, which requires both anisotropy and a non-zero $W$.

The intermediate homogenization of inhomogeneous cosmological models was studied in \cite{cet6} along with the problem of gravitational entropy. All definitions of entropy examined in this paper yielded decreasing gravitational entropy during the homogenization process, which implies that the gravitational entropy may actually decrease in some cases.

Gr{\o}n and Hervik \cite{gron} investigated the evolution of different measures of `gravitational entropy' in Bianchi type I and Lema\^{\i}tre–Tolman–Bondi (LTB) universe models. They found that the WCH remains secure if one considers the non-local version of the conjecture.
In their paper, Mishra and Singh \cite{cet7} investigated whether the inhomogeneous cosmological models could be motivated on the basis of thermodynamic grounds and a particular minimal void LTB inhomogeneous model was chosen for the analysis. They examined several definitions of gravitational entropy and found that the Weyl curvature entropy exhibits satisfactory thermodynamic behavior in the case of inhomogenous cosmologies.

Sussman \cite{cet0} introduced a weighed scalar average formalism (the `q-average' formalism) for the study of the theoretical properties and the dynamics of spherically symmetric LTB dust models and explored the application of this formalism to a definition of a gravitational entropy functional proposed by Hosoya et al (HB proposal) \cite{HB}. Subsequently, Sussman and Larena \cite{cet1} considered the generic LTB dust models to probe the CET proposal and the HB proposal, along with a variant of the HB proposal, suggesting that the notion of gravitational entropy is a theoretically robust concept which can also be applied to other general spacetimes.

The evolution of the CET gravitational entropy was also studied by the same authors in \cite{cet2} for the local expanding cosmic CDM voids using a non-perturbative approach. Marozzi et al \cite{cet4} calculated the gravitational entropy of the large scale structures of the universe in the linear regime, where it can be described by the perturbed FLRW spacetime. This entropy arises from the averaging made over an extended region and explains the formation of large-scale structure in the Universe. The results obtained in \cite{cet2} for the gravitational entropy agreed well with their results, when the LTB evolution is in its linear regime, thus providing us with a connection between the local physics and the large scale linear regime.

In \cite{cet3} it was pointed out that the formation of numerous astronomical objects like the supermassive black holes (quasars), super-novae and dust, and the occurrence of several phenomenona like gamma ray bursts in the early universe are contradictory to the conventional mechanisms of its possible origin. The $\Lambda$CDM cosmology fails to explain several observations like the absence of central cusps with $\rho \sim r^{-1}$ in the dark matter distribution \cite{Primack}, presence of too many bright satellite galaxies at high $z$ \cite{list1,list2,list3}, or the larger value of observed angular momentum of galaxies. Although the standard cosmological model successfully describes the gross properties of the universe, yet fails in terms of several smaller details, both in the early universe at redshifts $z \sim 10$ and in the present time. The early universe is abundantly populated by quasars (\cite{quasar} and references therein), but it is practically impossible to create so many quasars in the young universe assuming the standard mechanism of BH formation by the process of matter accretion. These discrepancies may be removed if there exists a dark matter particle having life-time greater than the age of the universe at the time of recombination \cite{BDT}. Such observations suggest the existence of New physics beyond the standard theory, which therefore requires suitable modification (see also the references in \cite{cet3}).

A possible solution may lie in a different cosmological expansion law as indicated in \cite{DHT} and \cite{MM}. The supersymmetric Grand Unified models \cite{cet3,AD} consider the action of a scalar field, $\chi$, with non-zero baryonic number, $B$, due to which bubbles may be generated. Initially (after inflation) $\chi$ was away from the origin and when inflation is over it starts to evolve down to the equilibrium point, $\chi = 0$. Because of the inflationary expansion, the bubbles could become astrophysically large. Immediately after the formation of bubbles with large value of $\chi$, inhomogeneities developed in the energy density due to different equations of state in the regions inside and outside these bubbles. The big bang nucleosynthesis (BBN) inside or in the vicinity of the high-$B$ bubbles creates heavy elements more efficiently than that predicted in the standard model. This may lead to the observed distribution of the celestial bodies and a lot of dust at $z \sim 10$ \cite{cet3}. Also recently astronomers have shown a very strong possibility of an anisotropic universe unlike the standard assumption of a homogeneous and isotropic universe \cite{ani1,ani2}.

These informations provide a very strong motivation for us to explore the inhomogeneous and anisotropic cosmologies. Such models have also been studied recently \cite{ani3} for the resolution of the discrepancy between the Hubble parameter measured locally as opposed to its value derived from the cosmic microwave background radiation (CMB).

In this paper we have examined the validity of the CET definition of gravitational entropy in the context of some exact cosmological solutions of the Einstein's field equations. In the next section we will first discuss the formalism of calculating gravitational energy density and temperature, which then provides us with the formalism of defining the gravitational entropy of a physical system. We begin Section III by describing the covariant 1 + 3 splitting of spacetime, which we have used to analyse the behaviour of some cosmological models with regard to the Weyl curvature. We have calculated all the relevant functions like the normalized epoch function, gravitational energy density, gravitational temperature and the gravitational entropy for these models. The discussions and conclusions are presented in Section IV.

\section{Gravitational Entropy}

In the following proposals, the measure of gravitational entropy is based on the Bel-Robinson tensor, which is defined in terms of the Weyl tensor in the form \cite{bel,robinson}

\begin{equation}
T_{abcd}\equiv \frac{1}{4}\left(C_{eabf}C^{e\; \; \; \; f}_{\; \; c d \; \;}+C^{*}_{\; e a b f}C^{* \; e \; \; \; \; f}_{\; \; \; \; \; c d \; \;}\right),
\end{equation}
where $ C^{*}_{abcd}=\frac{1}{2}\eta_{abef}C^{ef}_{\phantom{ef}cd} $ is the dual of the Weyl tensor.

The important property of this tensor is that it is overall symmetric, tracefree, and is covariantly conserved in vacuum or in presence of the cosmological constant. The factor of $1/4$ gives a natural interpretation of the Bel-Robinson tensor in terms of the Weyl spinor \cite{PR}. A measure of gravitational entropy was constructed by Pelavas and Lake \cite{PL} and Pelavas and Coley \cite{PC}, which had the form
\begin{equation}
S=\int W d\tau,
\end{equation}
where the epoch function $ W $ was defined using the Bel-Robinson tensor $T_{abcd}$ and the observer four velocity. The epoch function so constructed was therefore observer dependent and non-negative.

Exploring the correspondence between electromagnetism and general relativity, Maartens and Basset \cite{MB} considered a $1 + 3$ covariant, nonperturbative approach, where the free gravitational field was covariantly characterized by the Weyl gravito-electric and gravito-magnetic spatial tensor fields. They demonstrated the covariant analogy between the tensor Bianchi equations and the vector Maxwell equations, and presented the important result that the Bel-Robinson (BR) tensor is a ``unique Maxwellian tensor'' which could be constructed from the Weyl tensor, which behaves as the ``super energy-momentum'' tensor for the gravitational fields. The only problem was that the dimension of the BR tensor is $L^{-4}$ and not $L^{-2}$ (where $L$ is the unit of length), which is the expected dimension for the energy momentum tensor. Based on this work, Clifton, Ellis and Tavakol \cite{CET} proposed that the symmetric $2$-index square root $t_{ab}$, of the BR tensor should act as the effective energy momentum tensor for free gravitational field. Subsequently Goswami and Ellis \cite{GE} constructed a tensor describing the interaction between free gravity and matter, which is taken to be the symmetric two index square root of the BR tensor. Since we intend to examine the CET proposal of gravitational entropy as applied to a few interesting cosmological models, so we will now present a brief review of the CET proposal.

\subsection{The CET proposal}
In the CET proposal of gravitational entropy \cite{CET}, rather than constructing the entropy measure as an integral along a timelike curve, they employed integrals over spacelike hypersurfaces. In section 2 of their paper they laid down a list of requirements for a viable measure of gravitational entropy, $S_{grav}$. They used the gravito-electromagnetic properties of the Weyl tensor and the 1+3 decomposition of the equations, to express the epoch function as follows:
\begin{equation}
W=T_{abcd}u^{a}u^{b}u^{c}u^{d}=\dfrac{1}{4}\left(E_{a}^{b}E_{b}^{a}+ H_{a}^{b}H_{b}^{a}\right).
\end{equation}

Here $ W $ is the ``Super energy density'' and $ E_{ab},H_{ab} $ are the electric and magnetic parts of the Weyl tensor respectively. The inhomogeneous distribution would require that either $E_{ab}$ or $H_{ab}$ is non-zero, so that $W > 0$, which means that inhomogeneity requires both
anisotropy and a non-zero $W$.

From this symmetric and tracefree four-index tensor $T_{abcd}$, one can define a symmetric two-index “square-root”, $t_{ab}$, which is a solution of the equation
\begin{equation}\label{belrob}
T_{abcd}=t_{(ab}t_{cd)}-\frac{1}{2} t_{e (a} t_{b}^{\phantom{b} e} g_{c d)} - \frac{1}{4} t_{e}^{\phantom{e} e} t_{(a b} g_{cd)} +\frac{1}{24} \left( t_{ef} t^{ef} +\frac{1}{2} (t_{e}^{\phantom{e}e})^2 \right)g_{(ab}g_{cd)}.
\end{equation}
The right hand side of this equation constitutes the only totally symmetric and tracefree four index
tensor, the quadratic $ t_{ab}$, that may be constructed. For any solution, $ t_{ab}$, there exists another solution $\epsilon t_{ab} + f g_{ab}$, where $ \epsilon=\pm 1 $, and $ f $ is an arbitrary function.
In spacetimes of Petrov type D or N, although the solution to the above equation is unique for a tracefree $t_{ab}$, but that does not necessarily lead to a quantity that is conserved in vacuum. Therefore the square-root of the Bel-Robinson may be chosen to inherit its tracefree property, or for its conservation in vacuum, but not necessarily both at the same time. For Petrov type D spacetimes, with two double principal null directions, and a Coulomb-like gravitational field, the tracefree square-root can be written in the form
\begin{equation}\label{tracel}
t_{ab}=3\epsilon|\Psi_{2}|(m_{(a}\bar{m}_{b)}+l_{(a}k_{b)}),
\end{equation}
where $\Psi_2= C_{abcd} k^a m^b \bar{m}^c l^d$ is the only non-zero Weyl scalar. The complex null tetrad is defined as
\begin{equation}
\label{nullt}
m^a = \frac{1}{\sqrt{2}} \left( x^a - i y^a \right), \quad
l^a = \frac{1}{\sqrt{2}} \left( u^a - z^a \right), \quad {\mathrm{and}} \quad
k^a = \frac{1}{\sqrt{2}} \left( u^a + z^a \right),
\end{equation}
where $x^a$, $y^a$ and $z^a$ are spacelike unit vectors, which form an orthonormal basis together with $u^a$, $g_{ab} =2 m_{(a} \bar{m}_{b)} - 2 k_{(a} l_{b)}$, with $l^a$ and $k^a$ being aligned with the principal null directions.
The effective energy-momentum, $ \tau_{ab} $, of the Coulomb-like gravitational fields present in a Petrov type D spacetime, was assumed to be given by the solution to \eqref{belrob}, with a traceless part described by \eqref{tracel}, so that
\begin{align}\label{diff}
8\pi\tau_{ab}&=\alpha[3\epsilon\vert\Psi_{2}\vert(m_{(a}\bar{m}_{b)}+l_{(a}k_{b)})+fg_{ab}] \nonumber\\
 &=\alpha \left[ \left( \frac{3}{2} \epsilon \vert \Psi_2 \vert +f \right) \left( x_a x_b +y_a y_b\right) - \left( \frac{3}{2} \epsilon \vert \Psi_2 \vert - f \right)\left(  z_a z_b -u^a u^b \right) \right].
\end{align}
Here $ \alpha $ is an unknown constant which must be determined. By contracting this effective energy-momentum tensor with the timelike unit vector, $ u^{a} $, and the projection tensor, $ h_{ab} $, one obtains the effective energy density, pressure and momentum density, which are given below:
\begin{align}\label{var}
& 8\pi \rho_{\mathrm{grav}} = \alpha \left( \frac{3}{2} \epsilon \vert \Psi_2 \vert -f \right), \quad
8\pi p_{\mathrm{grav}} = \alpha\left( \frac{1}{2} \epsilon \vert \Psi_2 \vert +f \right), \quad  \nonumber\\
& 8\pi \pi^{\mathrm{grav}}_{ab} = \frac{\alpha}{2} \epsilon \vert \Psi_2 \vert  \left( x_a x_b +y_a y_b -z_a z_b + u^a u^b \right),
\end{align}
and $q^{\mathrm{grav}}_a = 0$. These quantities seem to obey the equations that are closely analogous to those of matter fields, and therefore these equations were used to construct a definition of gravitational entropy. The free function $ f $ can be removed by imposing the energy conservation condition in vacuum for the effective energy momentum tensor, and simultaneously (\ref{belrob}) is to be imposed to get the functional form of $ f $. To do so, equation \eqref{diff} is to be differentiated, and at the same time one has to use the relation
\begin{equation}\label{psi2}
|\Psi_{2}|=\sqrt{\dfrac{2W}{3}},
\end{equation}
and equation $ (45) $ of \cite{MB}, which represents the covariant non-perturbative generalization of Bel's linearized conservation equation in \cite{bel}. Finally one can derive the functional form of $ f $ as follows:
\begin{equation}
f=-\frac{1}{2} \epsilon \vert \Psi_2 \vert +\lambda_1,
\end{equation}
where $ \lambda_1 $ is an arbitrary constant which may be set to zero, as it does not affect the relevant thermodynamic quantities. Consequently the effective energy-momentum tensor is obtained in the form
\begin{equation}
8 \pi \tau_{ab} =  \epsilon \alpha \sqrt{\frac{2 W}{3}} \left( x_a x_b +y_a y_b\right) - 2\epsilon \alpha \sqrt{\frac{2 W}{3}}  \left(  z_a z_b -u^a u^b \right),
\end{equation}
and the effective energy density and the pressure in the free gravitational field are obtained as
\begin{equation}
8\pi\rho_{grav}=2\alpha\sqrt{\dfrac{2W}{3}},  \qquad  {\mathrm{and}} \qquad
p_{\mathrm{grav}} =0,
\end{equation}
where $\epsilon =+1$, so that $\rho_{\mathrm{grav}} \geq 0$. The anisotropic pressure and momentum density remain unaffected according to equation (\ref{var}) and the relation $q^{\mathrm{grav}}_a=0$, because the function $f$ occurs only in the trace of $\tau_{ab}$. These relations constitute the ``unique expressions'' for the effective energy density and pressure of free gravitational field in Petrov type D spacetimes \cite{CET}, which are determined from the square-root of the Bel-Robinson tensor $ T_{abcd} $ by imposing the condition of energy conservation in vacuum, i.e., $u_a \tau^{ab}_{\phantom{ab};b}=0$, and the positivity of energy density. Finally the  relation (\ref{psi2}) implies that many properties of the Bel-Robinson tensor are inherited by the effective energy-momentum tensor $ \tau_{ab} $.
In order to calculate the entropy according to thermodynamic prescriptions, one must know the {``temperature'', $T_{\mathrm{grav}}$}, of the free gravitational fields. For that purpose, one must have possess some knowledge about the underlying microscopic theory. Naturally, the CET proposal assumed that a thermodynamic treatment of the free gravitational field is very much similar to that of standard thermodynamics. This was the motivation for looking into the results of black hole thermodynamics, and quantum field theory in curved spacetimes. Therefore they required the definition of temperature to be local (instead of being defined for horizons only), which reproduced the expected results from semi-classical calculations in Schwarzschild and de Sitter spacetimes. The temperature at any point in spacetime was given by the following expression:
\begin{equation}
T_{grav}=\dfrac{|u_{a;b}l^{a}k^{b}|}{\pi}=\dfrac{|\dot{u_{a}}z^{a}+H+\sigma_{ab}z^{a}z^{b}|}{2\pi},
\end{equation}
where $z^{a}$ is a spacelike unit vector aligned with the Weyl principal tetrad, and  $ H=\dfrac{\Theta}{3} $ is the isotropic Hubble rate.

\section{Gravitational entropy of some cosmological models}

From earlier studies \cite{CET} it is known that the CET proposal of gravitational entropy is applicable only to Petrov type D spacetimes in four dimensional GR. Therefore, in the following sequel we will explore some Petrov type D spacetimes representing the various phases of evolution of the universe filled with ideal irrotational fluids.

We will use the covariant $1+3$ splitting of spacetime \cite{Elst1,Elst2,Elst3} with the timelike vector field $ u^{a} $ and the projection tensor $ h_{ab} $ satisfying the following relations
\begin{equation}
  U^{a}\,_{b} \coloneqq u^{a}u_{b},\;\;\;\;\;   g_{ab} \coloneqq -u_{a}u_{b}+h_{ab},
\end{equation}
\begin{equation}
 U^{a}\,_{c} U^{c}\,_{b}=  U^{a}\,_{b},\;\;\;\;\;  U^{a}\,_{b}u^{b}= u^{a},\;\;\;\;\;  U^{a}\,_{a} =-1,
\end{equation}
\begin{equation}
 h^{a}\,_{c} h^{c}\,_{b}= h^{a}\,_{b},\;\;\;\;\;  h^{a}\,_{b}u^{b}=0,\;\;\;\;\; h^{a}\,_{a}=3.
\end{equation}
The covariant time derivative (represented by a dot) and the projected spatial derivative (represented by $\mathbf{D})$ using the projection tensor $ h_{ab} $ are given by:
\begin{equation}
\dot{A}^{a...}_{b...}=u^{c}\bigtriangledown_{c}A^{a...}_{b...}, \;\;\;\;\;    \mathbf{D}_{a}A^{b...}_{c...} \equiv h_{a}^{p}h^{b}_{q}...h_{c}^{r} \mathbf{\bigtriangledown}_{p}A^{q...}_{r...},
\end{equation}

\begin{equation}
\dot{U}^{<ab>}=\dot{h}^{<ab>}=0,\;\;\;\;\; \mathbf{D}_{a}U_{bc} = \mathbf{D}_{a}h_{bc}=0\,.
\end{equation}
Here the angular brackets `$ <> $' denote the symmetric and trace-free part of a tensor.
The $ 3-$volume element is defined as:
\begin{equation}
\epsilon \coloneqq -\epsilon_{defg}h^{d}\,_{a}h^{e}\,_{b}h^{f}\,_{c}u^{g}=u^{g}\epsilon_{gdef}h^{d}\,_{a}h^{e}\,_{b}h^{f}\,_{c}\,.
\end{equation}

The kinematical variables are obtained from the covariant derivative of $ \mathbf{u}$, which is given by
\begin{equation}
\nabla_{a}u_{b}=-u_{a}\dot{u}_{b}+\mathbf{D}_{a}u_{b}\coloneqq -u_{a}\dot{u}_{b}+\frac{1}{3}\Theta h_{ab}+\sigma_{ab}+\epsilon_{abc}\omega^{c},
\end{equation}
where the kinematical variables are defined by
\begin{equation}
\dot{u}^{a} \coloneqq u^{b}\nabla_{b}u^{a},\;\;\;  \Theta \coloneqq \mathbf{D}_{a}u^{a},\;\;\;  \sigma_{ab} \coloneqq \mathbf{D}_{<a}u_{b>},\;\; \;  \omega^{a}\coloneqq \epsilon^{abc} \mathbf{D}_{b}u_{c}\; .
\end{equation}
The matter variables are defined as the following:
\begin{equation}
\mu \coloneqq T_{ab}u^{a}u^{b},\;\;\;  q^{a}  \coloneqq -T_{cb}h^{ca}u^{b},\;\;\;  p \coloneqq \dfrac{1}{3} T_{ab}h^{ab},\;\;\;  \pi_{ab} \coloneqq T_{cd} h^{c}\,_{<a}h^{d}\,_{b>}\;.
\end{equation}

The electric and magnetic parts of the Weyl curvature tensor are given by
\begin{equation}
E_{ab} \coloneqq C_{cdef}h^{c}\,_{a}u^{d}h^{e}\,_{b}u^{f},\;\;\;\;\;  H_{ab} \coloneqq (-\frac{1}{2}\epsilon_{cdgh}C^{gh}\,_{ef})h^{c}\,_{a}u^{d}h^{e}\,_{b}u^{f}\;.
\end{equation}

In order to drive in to the point that the gravitational entropy measured in terms of the Weyl tensor indeed fulfills the physical requirement that it reflects the inherent anisotropy of the spacetime, we will take into consideration two equations.

One of the Ricci identities is given by \cite{Elst1}:
\begin{equation}\label{d1}
\dot{\sigma}^{<ab>}-\mathbf{D}^{<a}\dot{u}^{b>}=-\dfrac{2}{3}\Theta\sigma^{ab}+\dot{u}^{<a}\dot{u}^{b>}-\sigma^{<a}\,_{c}\,\sigma^{b>c}-\omega^{<a} \omega^{b>}-\left(E^{ab}-\frac{1}{2}\pi^{ab}\right),
\end{equation}
and also one of the contracted Bianchi identities \cite{Elst1} is
\begin{align}\label{d2}
\left(\dot{E}^{<ab>}+\frac{1}{2}\dot{\pi}^{<ab>}\right)-\epsilon^{cd<a}\mathbf{D}_{c}H^{b>}\,_{d}+\frac{1}{2}\mathbf{D}^{<a}q^{b>}
=-\frac{1}{2}(\mu+p)\sigma^{ab}-\Theta\left(E^{ab}+\dfrac{1}{6}\pi^{ab}\right) \nonumber\\
+3\sigma^{<a}\,_{c}\left(E^{b>c}-\frac{1}{6}\pi^{b>c}\right)-\dot{u}^{<a}q^{b>}
\nonumber\\
+\epsilon^{cd<a}\left[2\dot{u}_{c}H^{b>}\,_{d}+\omega_{c}\left(E^{b>}\,_{d}+\frac{1}{2}\pi^{b>}\,_{d}\right)\right]
\end{align}
A very important point is to be noted in this context, i.e., equations (\ref{d1}) and (\ref{d2}) together define a two way relationship between the shear and the electric part of Weyl tensor. The electric Weyl drives the evolution of shear and matter density, together with the shear which drives the evolution of the electric Weyl. From these two equations we can clearly identify the physical processes behind the generation of gravitational entropy, i.e. it is indeed generated from the anisotropies of the universe.

\subsection{FLRW model}
Let us begin with the most common and universally accepted cosmological model, the FLRW metric. Since it represents a homogeneous and isotropic universe, the gravitational entropy function is expected to vanish in this case. However, that does not violate the WCH, since gravitational entropy is associated with inhomogeneity.

Here we are only considering the zero spatial curvature case ($k=0$) but the results are valid for any of the three cases: $k=0,\ \pm1$. The generality of the property of conformal flatness for any spatial curvature in FLRW models has been shown explicitly in \cite{frw1}. The flat FLRW metric is given by
\begin{equation}
ds^2=-dt^2 + A^2(t)(dx^2+dy^2+dz^2),
\end{equation}
where $ A(t) $ is the scale factor. This is a conformally flat algebraically special metric, i.e., the Weyl curvature is zero in this case, thereby making it a Petrov type O spacetime, which is a subclass of the algebraically more general Petrov type D spacetime. Consequently all the Weyl curvature components are zero: $ \Psi_{0}=..=\Psi_{4}=0 $. For the sake of completeness let us consider the following four vectors in conformity with the Weyl principal tetrad:
\begin{equation}
u^{a}=\left(1,0,0,0\right),
\end{equation}
\begin{equation}
z^{a}=\left(0,\frac{1}{A},0,0\right),
\end{equation}
where $ u^{a} $ and $ z^{a} $ forms the null cone, and the $ (m,\bar{m}) $ plane is covered by following two four vectors:
\begin{equation}
x^{a}=\left(0,0,\frac{1}{A},0\right),
\end{equation}
\begin{equation}
y^{a}=\left(0,0,0,\frac{1}{A}\right).
\end{equation}
Using our chosen $ u^{a} $, we can determine the expansion scalar $ \Theta $ along with other quantities like the acceleration, shear tensor and rotation tensor. The expression for the expansion is obvious and matches with the isotropic three dimensional volume expansion:
\begin{equation}\label{expfrw}
\Theta=\frac{3\dot{A}}{A}.
\end{equation}
Subsequently all the components of the shear tensor, $ \sigma_{ab} $, are zero along with vanishing acceleration, and rotation tensor. We thus obtain
\begin{equation}
\rho_{grav}=\frac{\alpha}{4\pi}\vert\Psi_{2}\vert=0,
\end{equation}
\begin{equation}\label{Tfrw}
T_{grav}=\left\vert\dfrac{\dot{u_{a}}z^{a}+H+\sigma_{ab}z^{a}z^{b}}{2\pi}\right\vert=\frac{1}{2\pi}\left\vert\frac{\dot{A}}{A}\right\vert,
\end{equation}
and
\begin{equation}
S_{grav}=\int_{V}\dfrac{\rho_{grav}v}{T_{grav}}=0,
\end{equation}
which should be obvious because the metric is conformally flat, and the measure of free gravitational energy density depends on the Bel–Robinson tensor which is constructed out of the Weyl tensor and its dual \cite{CET}. The purpose of this rather simple exercise is to show that the gravitational temperature is nonzero and finite in spite of the fact that its free gravitational energy density vanishes. From the expression of the gravitational temperature \eqref{Tfrw}, it is clear that this quantity depends on the expansion of the FLRW spacetime, as given in \eqref{expfrw}. To put this argument in the proper context, let us consider different epochs of evolution of the universe. In the radiation dominated era, we know that the scale factor $A(t)$ varies as $ \sqrt{t} $, yielding the gravitational temperature as $ T_{grav}=\frac{1}{2\pi}\left(\frac{1}{2t} \right)$. Similarly for the matter dominated era we can take the scale factor to be $ A(t)\varpropto t^{2/3} $, so that $ T_{grav}=\frac{1}{3\pi}\left(\frac{1}{t} \right)$. Note that in both the radiation and matter dominated eras, the gravitational temperature varies inversely with time. Lastly, for the dark energy dominated era, $ A(t)\varpropto \exp(H_{0}t) $, where $H_{0} $ is the Hubble constant. Consequently, the gravitational temperature for the dark energy dominated era is $ T_{grav}=\frac{H_{0}}{2\pi}=\frac{1}{2\pi}\sqrt{\frac{\Lambda}{3}} $, which turns out to be directly related to the cosmological constant, $ \Lambda $. Analyzing the expression of gravitational temperature from the radiation dominated era through the matter dominated era into the dark energy dominated era, we can say that the gravitational temperature decreases in course of time and ultimately tends to a value determined by the cosmological constant. Although the gravitational entropy carried by the free gravitational field is zero in these cases, which is in agreement with our understanding of \cite{CET}, but the difference in the gravitational temperature among different cosmological eras hints towards a hidden physics behind it. It is a matter of independent investigation, which we leave aside for the moment.

\subsection{LRS Bianchi I model}
If we break the isotropy, then the next step would be to consider the Bianchi class of spacetimes, namely the most general one: Bianchi type I. This spacetime is algebraically general and doesn't help us to apply the CET proposal. Therefore a more physically interesting case would be to consider the Locally Rotationally Symmetric (LRS) Bianchi type I spacetimes \cite{lrsbi1}, where two of the spatial directions have the same directional scale factor, whereas the third one has a different scale factor. The FLRW metric can be considered to be a special case of this spatially homogeneous and anisotropic model. A general form of the LRS Bianchi I metric is the following \cite{lrsbi2}:
\begin{equation}
ds^2=-dt^2 + A^2(t)dx^2+B^2(t)(dy^2+dz^2).
\end{equation}
Imposing the LRS restriction makes the Bianchi I spacetime algebraically special, with Petrov D classification. Therefore we will have one nonzero weyl component $ \Psi_{2} $ which will give us the free gravitational energy density. Choosing our vectors in accordance with the Weyl principal tetrad, we can calculate the necessary variables as given below. The expression for the expansion scalar and the components of the shear tensor are the following:
\begin{equation}
\Theta=\left(\dfrac{\dot{A}}{A}+\dfrac{2\dot{B}}{B}\right),
\end{equation}
\begin{equation}
\sigma_{xx}=-\left(\dfrac{2A(\dot{B}A-\dot{A}B)}{3B}\right), \qquad \sigma_{yy}=\dfrac{B(\dot{B}A-\dot{A}B)}{3A}, \qquad \sigma_{zz}=\dfrac{B(\dot{B}A-\dot{A}B)}{3A}.
\end{equation}
Using these components, the shear scalar $\sigma^2 $ can be evaluated as follows:
\begin{equation}
\sigma^2=\frac{1}{3}\left(\frac{\dot{B}}{B}-\frac{\dot{A}}{A}\right)^2.
\end{equation}
Consequently the expansion anisotropy is
\begin{equation}
\frac{\sigma^2}{\Theta^2}=\frac{1}{3}\left(\dfrac{\dot{B}A-\dot{A}B}{2A\dot{B}+B\dot{A}}\right)^2.
\end{equation}
The gravitational epoch function can be obtained using the Bel-Robinson tensor, which is given by
\begin{equation}
W=T_{abcd}u^{a}u^{b}u^{c}u^{d}=\dfrac{(B\ddot{B}A-\dot{B}^2A+\dot{A}B\dot{B}-\ddot{A}B^2)^2}{24A^2B^4}.
\end{equation}
For our calculations we have used the following four-vectors:
\begin{equation}
u^{a}=\left(1,0,0,0\right),
\end{equation}
\begin{equation}
z^{a}=\left(0,\frac{1}{A},0,0\right).
\end{equation}
Here $u^{a} \, \textrm{and} \, z^{a}  $ form the null vectors $ l^{a} $ and $ k^{a} $ generating the null cone. The vectors
\begin{equation}
x^{a}=\left(0,0,\frac{1}{A},0\right),
\end{equation}
and
\begin{equation}
y^{a}=\left(0,0,0,\frac{1}{A}\right),
\end{equation}
form the $ (m,\bar{m}) $ plane. Using these values, we calculate the free gravitational energy density $\rho_{grav}  $, which is found to be
\begin{equation}
\rho_{grav}=\frac{\alpha}{4\pi} \left\vert\dfrac{B\ddot{B}A-\dot{B}^2A+\dot{A}B\dot{B}-\ddot{A}B^2}{6B^2A}\right\vert.
\end{equation}
We can use our null tetrad to determine the Weyl-NP scalars. As it is of Petrov class D, the only nonzero component is $\Psi_{2}  $. The values of $ \Psi_{i} $ are the following: $ \Psi_{0}=0, \Psi_{1}=0, \Psi_{2}=\dfrac{(B\ddot{B}A-\dot{B}^2A+\dot{A}B\dot{B}-\ddot{A}B^2)}{6B^2A}, \Psi_{3}=0 $ and $ \Psi_{4}=0 $. Therefore the relation $ |\Psi_{2}|=\sqrt{\dfrac{2W}{3}} $ is satisfied here, indicating that the choice of this nonholonomic tetrad is in accordance with the local light cone structure. Subsequently the gravitational temperature is given by
\begin{equation}
T_{grav}=\frac{1}{2\pi}\left\vert\frac{\dot{A}}{A}\right\vert .
\end{equation}
Surprisingly, the contribution from the shear and the expansion oppose each other, and the only contributing component remains when the isotropy is broken. The anisotropy in the geometry gives rise to this opposing feature between the expansion and the shear, giving us a net contribution to the temperature, meaning that the gravitational temperature is powered by the anisotropy of a homogeneous system. In order to have a physically viable universe, the directional Hubble parameter $\left\vert\frac{\dot{A}}{A}\right\vert  $ should decrease with time $ t $. Finally the gravitational entropy is given by the expression:
\begin{equation}\label{enbi0}
S_{grav}=\int_{V}\dfrac{\rho_{grav}v}{T_{grav}}=\frac{\alpha}{12}\left\vert\dfrac{A(B\ddot{B}A-\dot{B}^2A+\dot{A}B\dot{B} -\ddot{A}B^2)}{\dot{A}}\right\vert\int_{V}dxdydz .
\end{equation}
In order to analyze the above expression of gravitational entropy, we notice that
\begin{equation}\label{wbi}
C_{abcd}C^{abcd}=\dfrac{4(\dot{B}^2A-\dot{A}B\dot{B}-B\ddot{B}A+\ddot{A}B^2)^2}{3A^2B^4} .
\end{equation}
Now let us impose a strong condition like the one which Penrose originally proposed, i.e., the Weyl scalar should increase with time as the universe expands. In the denominator of the Weyl curvature scalar we have the volume $ V (=AB^2) $ squared, and in order that the Weyl curvature scalar may increase, the numerator must dominate and it should increase faster than the denominator. Therefore $ \left\vert\dfrac{A(B\ddot{B}A-\dot{B}^2A+\dot{A}B\dot{B}-\ddot{A}B^2)}{\dot{A}}\right\vert $ is increasing monotonically with time as $\left\vert\frac{A}{\dot{A}}\right\vert$ is also increasing with time along with the term $ \left\vert(B\ddot{B}A-\dot{B}^2A+\dot{A}B\dot{B}-\ddot{A}B^2)\right\vert $. Therefore it is clear that the expression (\ref{enbi0}) for the gravitational entropy is non negative, and increases monotonically with time.
The above equation (\ref{enbi0}) can be rearranged as follows:
\begin{equation}\label{enbi}
S_{grav}=\dfrac{\alpha}{12}\left\vert \frac{V}{\dot{A}}\left(\sqrt{3}\dot{\sigma}A-\Theta+3\dot{A}\left(\frac{\dot{B}}{B}\right)\right) \right\vert V ,
\end{equation}
where we have taken $ V=AB^2 $. From the equation (\ref{enbi}) it is clear that the gravitational entropy depends on the shear $ \sigma $, and as the rate of shear increases, it contributes positively to the gravitational entropy. Therefore in order to have a physically relevant universe, the gravitational entropy should increase monotonically with time, i.e.
\begin{equation}\label{condition_0}
\dot{S}_{grav}>0,
\end{equation}
which means that
\begin{equation}\label{condition}
\dfrac{d}{dt}\left[\dfrac{A}{\dot{A}}(B\ddot{B}A-\dot{B}^2A+\dot{A}B\dot{B}-\ddot{A}B^2)AB^2\right]>0.
\end{equation}
Now if we choose the condition $ -\beta>0 $, where $ \beta=(\dot{B}^2A-\dot{A}B\dot{B}-B\ddot{B}A+\ddot{A}B^2) $, the above condition \eqref{condition} reduces to the following identity:
\begin{equation}\label{betacond_bi}
\dfrac{\dot{\beta}}{\beta}>\dfrac{\ddot{A}}{\dot{A}}-\left(\dfrac{\dot{A}}{A}+\Theta\right).
\end{equation}
To put the entire analysis in the proper perspective, let us observe the rate of change of Weyl scalar:
\begin{equation}\label{wtbi}
\dfrac{d}{dt}(C_{abcd}C^{abcd})=\dfrac{8}{3A^2B^4}(\beta\dot{\beta}-\beta^2 \Theta).
\end{equation}
From the above condition \eqref{wtbi} the condition of monotonicitally increasing Weyl curvature scalar can be obtained as the following:
\begin{equation}
\dfrac{\dot{\beta}}{\beta}>\Theta, \,\, \textrm{if}  \,\, \beta>0; \qquad \textrm{and} \qquad
\,\,\,\dfrac{\dot{\beta}}{\beta}<\Theta, \,\, \textrm{if}  \,\, \beta<0.
\end{equation}

Let us fix our condition as $ \beta<0 $ for the following analysis.
Note that to obtain a monotonically increasing $ C_{abcd}C^{abcd} $ we need to satisfy the condition $\dfrac{\dot{\beta}}{\beta}<\Theta$, which is not always true when we want to have monotonically increasing gravitational entropy. Now to determine the restriction on $ \dfrac{d}{dt}(C_{abcd}C^{abcd}) $ for monotonically increasing gravitational entropy, i.e., $ \dot{S}_{grav}>0 $, once again we consider the identity \eqref{betacond_bi}.
We know that
\begin{equation}\label{monC}
\dfrac{d}{dt}(C_{abcd}C^{abcd})=\dfrac{8}{3A^2B^4}(\beta\dot{\beta}-\beta^2 \Theta)=\dfrac{8}{3A^2B^4}\beta^2\left(\dfrac{\dot{\beta}}{\beta}-\Theta\right).
\end{equation}
Imposing the condition of monotonically increasing gravitational entropy, i.e., \eqref{betacond_bi} on \eqref{monC} we get the following condition :
\begin{equation}
\dfrac{d}{dt}(C_{abcd}C^{abcd})>\dfrac{8}{3A^2B^4}\beta^2\left(\dfrac{\ddot{A}}{\dot{A}}-\dfrac{\dot{A}}{A}-2\Theta\right).
\end{equation}
Now if $\left(\dfrac{\ddot{A}}{\dot{A}}-\dfrac{\dot{A}}{A}-2\Theta\right)>0 , $ then $d/dt({C_{abcd}C^{abcd}})$ is always positive, i.e., the Weyl curvature scalar is monotonically increasing at all times. But if $\left(\dfrac{\ddot{A}}{\dot{A}}-\dfrac{\dot{A}}{A}-2\Theta\right)<0 ,$ then $d/dt(C_{abcd}C^{abcd})$ can be negative, implying that the Weyl curvature may decrease while the gravitational entropy is increasing, which is similar to the situation illustrated by the authors in \cite{greg}. Thus we can conclude that the LRS Bianchi I spacetime with different kinds of matter as their source, must satisfy the above condition \eqref{condition} for the monotonic increase of gravitational entropy. We are keeping this analysis general as it is clear that the validity of condition \eqref{condition} depends on the nature of the source. In short, in order to have a monotonically increasing gravitational entropy, the LRS Bianchi I spacetimes with various matter sources must satisfy the condition \eqref{condition}, or in other words, it must have an increasing Weyl curvature scalar for the condition $\left( \dfrac{\ddot{A}}{\dot{A}} - \dfrac{\dot{A}}{A} - 2\Theta \right)>0 $. The same analysis can be done by assuming $ \beta $ to be positive. Summing up, first we have shown that, if the Weyl curvature is diverging at the initial singularity or is decreasing with increasing time, then the LRS Bianchi I spacetime can have decreasing gravitational entropy thereby violating the Weyl curvature hypothesis, but in the later part of our analysis we also demonstrated that if we need a monotonically increasing gravitational entropy, then depending on the conditions we may either have an increasing or a decreasing Weyl scalar.

\subsection{Liang model}
Let us now consider a spacetime representing the early phase of evolution of the universe to see whether the CET gravitational entropy proposal holds good in this era.
An example of an exact solution of the Einstein's field equation with an irrotational fluid source with the equation of state $p=1/3\mu$, energy density $ \mu=T_{ab}u^{a}u^{b}=3/(4t^2A^2)$, and fluid velocity $u=A^{-1/2}\dfrac{\partial}{\partial t}$, representing the radiation dominated universe, and whose initial singularity is `Friedmann-like' as considered by Liang (1972) \cite{Liang}, is given by the metric
\begin{equation}
ds^2=-Adt^2 + t[A^{-1}dx^2+A^2b^{-2}(dy^2+f^2dz^2)],
\end{equation}
where $ A=1-(4\tilde{\epsilon} b^2 t)/9 $, $b\equiv$ constant, $ f(y)= \sin y , \tilde{\epsilon}=+1$ and $f(y)= \sinh y ,\tilde{\epsilon}=-1$.
In our subsequent calculations we will assume $ \tilde{\epsilon}=1 $.

The expansion scalar obtained in this model is
\begin{equation}
\Theta=\dfrac{9}{2t}\dfrac{(9-8b^{2}t)}{(9-4b^{2}t)^{3/2}}.
\end{equation}
Apparently this model expands in the interval $ 0<t<\dfrac{9}{8b^{2}} $, since $ \Theta>0 $ in this range, and then shrinks in the interval $\dfrac{9}{8b^{2}}<t<\dfrac{9}{4b^{2}}$, for which $ \Theta<0 $ . Therefore we will consider the range of $ t $ as $ 0<t<\dfrac{9}{8b^{2}} $, since it represents an expanding universe. The acceleration vector and the vorticity tensor turns out to be zero in this case.

The corresponding components of the shear tensor are
\begin{equation}
\sigma_{xx}=\dfrac{108b^{2}t}{(-4b^{2}t+9)^{5/2}}, \qquad \sigma_{yy}=\dfrac{-2\sqrt{-4b^{2}t+9}t}{27}, \qquad \sigma_{zz}=\dfrac{-2sin^{2}y\sqrt{-4b^{2}t+9}t}{27}.
\end{equation}

Therefore we can evaluate the shear scalar and the expression is given by the following:
\begin{equation}
\sigma^2=\dfrac{108 b^4}{(9-4b^{2}t)^3}.
\end{equation}
From the above expression it is evident that as time $ t $ increases, $\sigma^2 $ i.e. the shear scalar also increases. An important parameter in these models is the ratio $\sigma^2 / \Theta^2$, which is found to be given by
\begin{equation}
\dfrac{\sigma^2}{\Theta^2}=\dfrac{24b^4t}{(9-8b^2t)}.
\end{equation}
It is already known that the ratio of the shear scalar to the expansion scalar (known as \emph{expansion anisotropy}) is a good measure of anisotropy \cite{ans1,ans2}, and we can easily check that the ratio in this case is increasing in the allowed range of time. Thus the universe begins from an isotropic singularity (as the ratio vanishes at $ t=0 $) and then the anisotropy increases with time as the universe expands, thereby fulfilling the requirement of inhomogeneity \cite{CET}.

We now compute the velocity dependent gravitational epoch function for this metric using the Bel-Robinson tensor:
\begin{equation}
W=T_{abcd}u^{a}u^{b}u^{c}u^{d}=\dfrac{2b^{4}}{27A^{6}t^{2}}.
\end{equation}
As we have both the anisotropy and the nonzero $ W $ (corresponding to tidal forces as the magnetic part of the Weyl tensor is zero for Petrov type D spacetimes), it is eligible for the calculation of gravitational entropy, as per the criterion set in \cite{CET}.
The normalized dimensionless scalar constructed from this quantity has the form
\begin{equation}
\tilde{P}=\dfrac{W}{\mu^{2}}=\left(\dfrac{32}{243}\right)\dfrac{b^{4}t^{2}}{A^{2}}.
\end{equation}

As $t\rightarrow 0^{+}$, the normalized Bel-Robinson epoch function vanishes, i.e. $ \tilde{P}\rightarrow 0 $, and $ \tilde{P} $ increases monotonically  as one moves away from the isotropic singularity.

For the sake of computation, we will use the following timelike and spacelike unit vectors in accordance with the Weyl principal tetrad:
\begin{equation}
u^{a}=\left(\dfrac{3}{\sqrt{-4b^{2}t+9}},0,0,0\right),
\end{equation}
and
\begin{equation}
z^{a}=\left(0,\dfrac{1}{3}\sqrt{\dfrac{-4b^2 t+9}{t}},0,0\right).
\end{equation}

The null cone is defined by the vectors $ k^{a}$ and $ l^{a}$ (which therefore lie in the $t,x$ plane). The $ (m,\bar{m}) $ plane is defined by $ m^{a}$, where the spacelike vectors are defined as $ x^{a}=\left(0,0,\dfrac{9b}{\sqrt{t(9-4b^{2}t)^{2}}},0\right) $ and $y^{a}=\left(0,0,0,\dfrac{9b}{\sqrt{t(9-4b^{2}t)^{2}\sin^{2}y}}\right)  $.

The gravitational energy density for this Petrov type D spacetime, obtained from the epoch function $W$, is given by
\begin{equation}
\rho_{grav}=\dfrac{\alpha}{18\pi}\dfrac{b^{2}}{A^{3}t}.
\end{equation}
This spacetime is of Petrov D type and the Weyl scalars are $ \Psi_{0}=0, \Psi_{1}=0, \Psi_{2}=-\dfrac{2b^{2}}{9tA^{3}}, \Psi_{3}=0, \textrm{and} \Psi_{4}=0$. Therefore the relation $ |\Psi_{2}|=\sqrt{\dfrac{2W}{3}} $ is satisfied in this case (as given in eqn.(\ref{psi2})).

The gravitational temperature is given by the expression
\begin{equation}\label{t1}
T_{grav}=\dfrac{1}{8\pi t A^{3/2}}.
\end{equation}
We can see that in order to have a non-negative gravitational energy density and temperature, we require the condition $ A>0 $, which also implies that $0<t<\dfrac{9}{4b^{2}} $. Finally, from the definition of gravitational entropy we have
\begin{equation}\label{s1}
S_{grav}=\int_{V} \dfrac{\rho_{grav}v}{T_{grav}}=\dfrac{4\alpha t^{3/2}}{9}\int_{V} dx \textrm{sin}y dy dz=\dfrac{4\alpha t^{3/2}}{9}\int_{0}^{x} dx \int_{0}^{y}\textrm{sin}y dy \int_{0}^{z}dz.
\end{equation}
Thus
\begin{equation}\label{s1n}
S_{grav}= \dfrac{4\alpha t^{3/2}}{9}x(1-cosy)z .
\end{equation}
Here we can identify $(x,y,z)$ as $(r,\theta,\phi)$, where $x$ acts as the ``radial'' co-ordinate, $\theta \in [0, \pi]$, $\phi \in [0, 2\pi]$ and the $(y,z)$ plane is the $(m,\bar{m})$ plane. Therefore the resulting expression for gravitational entropy will be obtained as

\begin{equation}\label{s1n1}
S_{grav}= \dfrac{16\pi\alpha t^{3/2}r}{9} .
\end{equation}

From the above equations (\ref{s1n}) and (\ref{s1n1}), we can see that the gravitational entropy is non-negative and monotonically increasing, leading to structure formation in the universe \cite{CET}. Further, the increase of shear tensor with time corresponds to the evolution of anisotropy in the universe, which leads to an increase in the above mentioned gravitational entropy.
The above analysis clearly shows us that as $ t\rightarrow 0 $, $ A\rightarrow 1 $, so that both the gravitational energy density $\rho_{grav} $ and the temperature $T_{grav} $ blow up. Consequently, in the limit $ t\rightarrow 0 $, the gravitational entropy $ S_{grav}\rightarrow 0 $, which is in agreement with the Weyl curvature hypothesis.

\subsection{Szekeres model}
We now turn to the spatially inhomogeneous models with irrotational dust as source, i.e., the class $\textbf{II}  $  Szekeres solution of the Einstein's field equations, which is known to be a Petrov type D spacetime. The metric under our consideration is the following \cite{GW1982}:
\begin{equation}
ds^{2}=t^{4}[-dt^{2}+dx^{2}+dy^{2}+(A-\beta_{+}t^{2})^{2}dz^{2}],
\end{equation}

where the function $ A $ is defined as
\begin{equation}
A=a(z)+b(z)x+c(z)y-5\beta_{+}(z)(x^{2}+y^{2}).
\end{equation}

In the class $\textbf{II}  $ Szekeres models, the parameters $ a(z), b(z), c(z), \textrm{and}  \beta_{+} $ are arbitrary smooth functions of $ z $, which gives us the freedom of choosing coordinates. For $ \beta_{+}=0 $, the class $\textbf{II}$ Szekeres solution reduces to FLRW metric. We also observe that if we assume $ a=1,b=0,c=0 $ further, i.e, $A=1$, we get the Cartesian form of FLRW metric directly.
The fluid four velocity is defined as $ u=t^{-2}\dfrac{\partial}{\partial t} $ and the energy density is given by
\begin{equation}
\mu=\dfrac{12}{t^{6}\left(1-\left(\dfrac{\beta_{+}}{A}\right)t^{2}\right)}.
\end{equation}
If the energy density is non-negative, then we need the following conditions to be satisfied: $A>0, (A-\beta_{+}t^{2})>0$. This imposes a bound on $ t $ because $ 0<t<\sqrt{\dfrac{A}{\beta_{+}}}$.  The expansion scalar of the universe is obtained as
\begin{equation}
\Theta=\dfrac{2(3A-4\beta_{+}t^{2})}{t^{3}(A-\beta_{+}t^{2})}.
\end{equation}
Thus this model is expanding throughout the cosmic time since $ \Theta>0 $ in the allowed range of $ t $ due to the fact that $ 0<t<\dfrac{\sqrt{3}}{2}\sqrt{\dfrac{A}{\beta_{+}}} $.

The shear tensor in this case is given by
\begin{equation}
\sigma_{xx}=\dfrac{2\beta_{+}t^{3}}{3(A-\beta_{+}t^{2})}, \qquad \sigma_{yy}=\dfrac{2\beta_{+}t^{3}}{3(A-\beta_{+}t^{2})}, \qquad
\sigma_{zz}=\dfrac{4\beta_{+}t^{3}}{3}(\beta_{+}t^{2}-A).
\end{equation}

The shear scalar is given by
\begin{equation}
\sigma^2=\dfrac{8\beta_{+}^2}{9t^2}.
\end{equation}
The expansion anisotropy in this universe is therefore given by
\begin{equation}
\dfrac{\sigma^2}{\Theta^2}=\dfrac{2\beta_{+}^2t^4(A-\beta_{+}t^2)^2}{9(3A-4\beta_{+}t^2)^2}.
\end{equation}
The above ratio vanishes at $ t=0 $, representing an isotropic initial singularity, and subsequently increases with time. Therefore the anisotropy increases with the evolution and expansion of the universe giving rise to structure formation.

Once again using the fluid $ 4- $velocity, we construct the positive scalar from the Bel-Robinson tensor:
\begin{equation}
W=T_{abcd}u^{a}u^{b}u^{c}u^{d}=\dfrac{6\beta_{+}^{2}}{t^{8}(\beta_{+}t^{2}-A)^{2}}.
\end{equation}
Therefore we get the normalized dimensionless scalar in the form
\begin{equation}
\tilde{P}=\dfrac{W}{\mu^{2}}= \dfrac{t^{4}\beta_{+}^{2}}{24A^{2}}.
\end{equation}
Thus, as $ t \rightarrow 0^{+} $, the normalized Bel-Robinson epoch function vanishes $( \tilde{P}\rightarrow 0 )$.
Let us construct the timelike and spacelike unit vectors in accordance with the Weyl principal tetrad so that $ u_{a}u^{a}=-1 $, $ z_{a}z^{a}=1 $ and $ u_{a}z^{a}=0 $, to get

\begin{equation}
u^{a}=\left(\dfrac{1}{t^{2}},0,0,0\right),
\end{equation}
and
\begin{equation}
z^{a}=\left(0,0,0,\dfrac{1}{t^{2}(A-\beta_{+}t^{2})}\right).
\end{equation}
The $ (m,\bar{m}) $ plane is defined by $ m^{a}$ which is defined in Section II, where the spacelike vectors are now defined as $ x^{a}=\left(0,\dfrac{1}{t^{2}},0,0\right) $ and $y^{a}=\left(0,0,\dfrac{1}{t^{2}},0\right)  $.

From the definition of the gravitational energy density of Petrov type D spacetimes, we get
\begin{equation}
\rho_{grav}=\dfrac{\alpha\beta_{+}}{2\pi t^{4}(A-\beta_{+}t^{2})},
\end{equation}
The above expression of gravitational energy density clearly indicates that for the non-negativity of the gravitational energy density, the following conditions must be fulfilled: $ \beta_{+}>0, A> \beta_{+}t^{2}.$ Therefore $ A $ must be a positive quantity.
The Weyl scalars are: $ \Psi_{0}=0, \Psi_{1}=0, \Psi_{2}=-\dfrac{2\beta_{+}}{t^{4}(\beta_{+}t^{2}-A)}, \Psi_{3}=0, \textrm{and} \Psi_{4}=0$. So the relation $ |\Psi_{2}|=\sqrt{\dfrac{2W}{3}} $ is now satisfied.

The gravitational temperature is given by
\begin{equation}\label{t2_1}
T_{grav}=\dfrac{(A-2\beta_{+}t^{2})}{\pi t^{3}(A-\beta_{+}t^{2})}.
\end{equation}
From the above equation (\ref{t2_1}) it is clear that we require an additional constraint in the form $(A-2\beta_{+}t^{2})>0  $ in order to ensure the non-negativity of the temperature. Thus the allowed range of cosmic time should be $ 0<t<\sqrt{\dfrac{A}{2\beta_{+}}} $.

As before, using the relevant definition, we obtain the expression of gravitational entropy as follows
\begin{equation}\label{s2_1}
S_{grav}=\dfrac{\alpha t^{5}}{2} \int_{0}^{x}\int_{0}^{y}\int_{0}^{z}\left[1+\dfrac{\beta_{+}t^{2}}{(A(x,y,z)-2\beta_{+}(z)t^{2})}\right]\beta_{+}dx dy dz =\dfrac{\alpha t^{5}}{2}T(t),
\end{equation}
where $$ T(t)=\int_{0}^{x}\int_{0}^{y}\int_{0}^{z}\left[1+\dfrac{\beta_{+}t^{2}}{(A(x,y,z)-2\beta_{+}(z)t^{2})}\right]\beta_{+}dx dy dz .$$ The term $ T(t) $ is not directly integrable because of the presence of unknown functions, but the term in the parenthesis is increasing monotonically with $ t $ as the denominator of the second term is decreasing with time and the numerator is directly proportional to $ t^2 $.
Therefore although it is not possible to integrate this equation further, the expression (\ref{s2_1}) of the gravitational entropy is not only non-negative but is also monotonically increasing, thereby satisfying the conditions of structure formation as laid down in \cite{CET}. Moreover, as $ t\rightarrow 0^{+} $, both the gravitational energy density $ \rho_{grav} $ and the temperature $ T_{grav} $ diverge, and as a result the gravitational entropy vanishes, i.e., $ S_{grav}\rightarrow 0 $. Further we know that $ \beta_{+}=0 $ gives us the FLRW metric and indeed the expression of gravitational entropy (\ref{s2_1}) reduces to zero in that case.

\subsection{Bianchi VI$_{h}$ model}
Finally we consider a spacetime which fits a general class of solutions of the Einstein's field equations but simple enough to study a perturbed kind of flat spacetime like the perturbed FLRW spacetime. We will show that the deviation from conformal flatness and isotropy leads us to an inhomogeneous spacetime where gravitational entropy is generated.

Wainwright and Anderson \cite{wainander}, showed that in the Bianchi VI$_h$ class of models, a suitable choice of parameters may help to represent the quasi-isotropic stage beginning at the initial singularity, leading to an isotropic singularity for these spacetimes. By assuming the parameter $\alpha_{c}  $ to be small in that model, one can consider deviations about this flat FLRW model. In the line element in \cite{wainander}, we set $\alpha_{s}=0$ and  $\alpha_{m}=1$, so that the line-element in conformal time coordinates is obtained as follows:
\begin{equation}
ds^{2}=\tau^{4/(3\gamma-2)}(-A^{2(\gamma-1)}d\tau^2+A^{2q_{1}}dx^{2}+A^{2q_{2}}e^{2r[s+(3\gamma-2)]x}dy^{2}+A^{2q_{3}}e^{2r[s-(3\gamma-2)]x}dz^{2}),
\end{equation}
where $ A^{2-\gamma}=1+\alpha_{c}\tau^{2}, \quad q_{1}=\dfrac{\gamma}{2}, \quad q_{2}=\dfrac{2-\gamma+s}{4}, \quad q_{3}=\dfrac{2-\gamma-s}{4}, \quad s^{2}=(3\gamma+2)(2-\gamma) , \quad \textrm{and} \quad r^{2}=\dfrac{(3\gamma+2)\alpha_{c}}{4(2-\gamma)(3\gamma-2)^{2}}. $
The parameter denoted by $ \alpha_{c} $ determines the curvature of the spacelike hypersurfaces orthogonal to $ u=A^{1-\gamma} \tau^{-2/(3\gamma-2)}\dfrac{\partial}{\partial \tau}.$
For $ \alpha_{c}=0 $, we obtain the flat FLRW solution.

In full generality, this metric is of Petrov type I, but the CET gravitational entropy measure only works on the Petrov types D and N. Therefore we will only consider the case for $ \gamma=4/3 $ which reduces the spacetime to Petrov type D.
The resulting Petrov type D metric is given by
\begin{equation}
ds^{2}=\tau^{2}\left(-(\alpha_{c} \tau^{2}+1)d\tau^{2}+(\alpha_{c} \tau^{2}+1)^{2}dx^{2}+(\alpha_{c} \tau^{2}+1)^{2}e^{6\sqrt{\alpha_{c}}x}dy^{2}+\dfrac{1}{(\alpha_{c} \tau^{2}+1)}dz^{2}\right).
\end{equation}

The expansion scalar is given by the following expression
\begin{equation} \label{expn_sclr_bian}
\Theta=\dfrac{3(2\alpha_{c}\tau^{2}+1)}{\tau^{2}(\alpha_{c}\tau^{2}+1)^{3/2}},
\end{equation}
and the shear tensor is
\begin{equation}
\sigma_{xx}=\alpha_{c}\tau^{2}\sqrt{\alpha_{c}\tau^{2}+1}, \qquad \sigma_{yy}=\alpha_{c}\tau^{2}e^{6\sqrt{\alpha_{c}}x}\sqrt{\alpha_{c}\tau^{2}+1}, \qquad
\sigma_{zz}=\dfrac{-2\alpha_{c}\tau^{2}}{(\alpha_{c}\tau^{2}+1)^{5/2}}.
\end{equation}
This model is also expanding with time as the expansion $ \Theta>0 $ for all $ \tau $.
In this case, the shear scalar is given by
\begin{equation}
\sigma^2=\dfrac{3\alpha_{c}^2}{(1+\alpha_{c}\tau^2)^3}.
\end{equation}

Once again, as a measure of expansion anisotropy we compute $ \sigma/\Theta $ which is given by the following expression:
\begin{equation}
\dfrac{\sigma^2}{\Theta^2}=\dfrac{\alpha_{c}^2\tau^4}{3(2\alpha_{c}\tau^2+1)^2}.
\end{equation}
Thus the ratio vanishes at $ \tau=0 $ indicating an initial isotropic singularity as in the previous models. It then increases with time and becomes constant as time tends to infinity. Therefore the time evolution of this universe is such that as time increases, the expansion anisotropy increases from the isotropic initial singularity, and gradually the rate of increase of this ratio decreases and finally it becomes more or less constant in the distant future. It may be mentioned here that exact spatially homogeneous cosmologies in which this ratio is constant, were studied in \cite{CGW}.

We can now easily compute the energy density as
\begin{equation}
\mu=\dfrac{3}{\tau^{4}(\alpha_{c}\tau^{2}+1)^{2}}.
\end{equation}
Next we construct the velocity dependent gravitational epoch function from the Bel-Robinson tensor, which yields
\begin{equation}
W=T_{abcd}u^{a}u^{b}u^{c}u^{d}=\dfrac{6\alpha_{c}^{2}}{\tau^{4}(\alpha_{c} \tau^{2}+1)^{6}}.
\end{equation}
In order to construct a dimensionless scalar from this quantity, we normalize the standard epoch function with the square of $ \mu=T_{ab}u^{a}u^{b} $ to get
\begin{equation}
\tilde{P}=\dfrac{W}{\mu^{2}}= \dfrac{2\alpha_{c}^{2}\tau^{4}}{3(\alpha_{c}\tau^{2}+1)^{2}}.
\end{equation}
As $ \tau \rightarrow 0^{+} $, the normalized Bel-Robinson epoch function vanishes: $ \tilde{P}\rightarrow 0 $. Therefore $ \tilde{P} $ behaves appropriately as the isotropic singularity is approached.

Now, for the analysis of the CET gravitational entropy for this spacetime, we will use the following unit vectors, where $ u^{a} $ is a timelike and $ z^{a} $ is a spacelike unit vector. Here we choose our vectors such that they specify a Weyl principal tetrad:

\begin{equation}
u^{a}=\left(\dfrac{1}{\tau\sqrt{\alpha_{c} \tau^{2}+1}},0,0,0\right),
\end{equation}
and
\begin{equation}
z^{a}=\left(0,0,0,\dfrac{\sqrt{\alpha_{c}\tau^{2}+1}}{\tau}\right).
\end{equation}
We note that this choice of tetrads is also supported by the work of Pelavas and Coley in \cite{PC}. In this case, the $ (m,\bar{m}) $ plane is defined by the spacelike vectors $ x^{a}=\left(0,\dfrac{1}{\tau(\alpha_{c}\tau^{2}+1)},0,0\right) $ and $ y^{a}=\left(0,0,\dfrac{1}{\tau(\alpha_{c}\tau^{2}+1)e^{3\sqrt{\alpha_{c}}x}},0\right) $, with the null cone defined by $l^a$ and $k^a$, as mentioned in Section II, along with the definition of $ m^{a}$.

From the definition of gravitational energy density we now obtain
\begin{equation}\label{r3}
\rho_{grav}=\dfrac{\alpha \alpha_{c}}{2\pi \tau^{2}(\alpha_{c}\tau^{2}+1)^{3}}.
\end{equation}
As this spacetime is a Petrov D spacetime, the Weyl scalars are obtained as $ \Psi_{0}=0, \Psi_{1}=0, \Psi_{2}=\dfrac{2\alpha_{c}}{\tau^{2}(1+\alpha_{c}\tau^{2})^{3}}, \Psi_{3}=0, \textrm{and}  \Psi_{4}=0$. Therefore the relation (\ref{psi2}) is satisfied in this case. Similarly the gravitational temperature can be calculated as
\begin{equation}\label{t3}
T_{grav}=\dfrac{1}{2\pi \tau^{2}(\alpha_{c}\tau^{2}+1)^{3/2}}.
\end{equation}
From the above two expressions of the gravitational energy density (\ref{r3}) and the temperature (\ref{t3}), we can clearly observe that as $ \tau \rightarrow 0^{+} $, both $ \rho_{grav} $ and $ T_{grav} $ diverge near the isotropic singularity.

Now integrating over a volume $ V $ on a hypersurface of constant $ \tau $, we get the expression of gravitational entropy
\begin{align}\label{t4}
S_{grav}&=\alpha \alpha_{c}\tau^{3}\int_{V}e^{3\sqrt{\alpha_{c}}x} dx dy dz=\alpha \alpha_{c}\tau^{3}\int_{0}^{x}e^{3\sqrt{\alpha_{c}}x} dx \int_{0}^{y}dy\int_{0}^{z} dz \nonumber\\
&=\dfrac{\alpha \sqrt{\alpha_{c}}}{3}\tau^{3}(e^{3\sqrt{\alpha_{c}}x}-1)yz = \dfrac{\alpha \sqrt{\alpha_{c}}}{3}V\tau^{3},
\end{align}
where $ V =(e^{3\sqrt{\alpha_{c}}x}-1)yz$ is a term which depends on the spatial volume, and will monotonically increase with increasing spatial dimensions.
We note that it is not possible to determine the bound of $ \tau $ in (\ref{t4}). However, the final expression indicates that not only the gravitational entropy increases with time $ \tau $, but for any increasing volume (i.e., larger values of $ x,y,z $), the gravitational entropy is increasing monotonically. Therefore, once again we find that $ S_{grav}$ is non-negative and monotonic in nature, and as $ \tau \rightarrow 0^{+} $, the gravitational entropy vanishes, i.e. $S_{grav}\rightarrow 0 $, in accordance with Penrose's Weyl curvature hypothesis. It is also evident that for $ \alpha_{c}\simeq0 $, the spacetime becomes FLRW-like, with vanishing gravitational entropy.

\section{Discussions and Conclusions}

In the very first case for the homogeneous and isotropic FLRW universe, we have shown that the gravitational entropy is zero because the space-time is conformally flat, thereby supporting the Weyl curvature hypothesis, as the free gravitational field in this case does not carry any gravitational energy density while maintaining a finite gravitational temperature. In the LRS Bianchi I case, we have shown explicitly that the gravitational entropy is monotonically increasing with time if its Weyl curvature increases with time, but if there are matter sources in the spacetime which cause the Weyl curvature to decrease in course of time, then the LRS Bianchi I spacetime will have a gravitational entropy which decreases with time, thereby violating the Weyl curvature hypothesis. That is, in order to have a non negative monotonically increasing gravitational entropy, the LRS Bianchi I spacetime must have monotonically increasing Weyl curvature. We are neglecting the cases where $\left\vert \frac{\dot{A}}{A} \right\vert$ is increasing with time as it will give rise to unrealistic situations where the gravitational temperature will increase with time and at the initial singularity it is zero or finite. We have also shown explicitly that if we want to have a non negative monotonically increasing gravitational entropy, depending on the conditions imposed by us, the Weyl curvature scalar can either increase or decrease. In other words, under certain conditions the time derivative of gravitational entropy is positive, i.e., the CET entropy is increasing monotonically.

The above analysis indicates that in the Liang, Szekeres and Bianchi $VI_{h}$ models considered by us, the gravitational entropy goes to zero as we approach the initial singularity and increases monotonically with non-negative value in course of time, thereby fulfilling the necessary requirements to be satisfied by the gravitational entropy in these space-times, in order to ensure structure formation in the universe \cite{CET}. We found that in each of these cases, representing different phases of evolution of the universe, the expansion anisotropy increases as times elapses after the initial isotropic singularity, and it appears that there is a correlation between the expansion anisotropy and the gravitational entropy. We also showed in a general formalism, how the shear tensor is related to the Weyl tensor. Therefore the CET formalism in these cases clearly gives us a well behaved entropy measure which increases as the structure formation progresses, resulting in an increase in anisotropy of these universes. These features make these models physically more realistic for describing the actual evolution of the universe. In all the cases we found that the gravitational entropy vanishes at the initial isotropic singularity. Moreover in each of these cases, the gravitational energy density and the temperature are well-behaved throughout the evolution of the conformal time associated with the metric.

In a recent paper \cite{greg} the authors have illustrated a counterexample involving a class of inhomogeneous universes that are supported by a chameleon massless scalar field and exhibit anisotropic spacetime shearing effects. We will now present a careful review of this work and compare it with our present work.

In \cite{greg}, the shear scalar depends on the Hubble parameter and it is increasing with time. Also the gravitational temperature which they are getting, is independent of time, which can always happen, as in the case of dark energy dominated FLRW universe. In their model, the CET entropy is increasing with time, whereas their gravitational energy density is decreasing with time, which is being compensated by gravitational anisotropic pressure. Finally the authors have shown that the time derivative of gravitational entropy is always positive. The interesting point that the authors in \cite{greg} are making is that the Weyl curvature is decreasing with time despite having increasing CET gravitational entropy, claiming that this violates the Weyl curvature hypothesis.

Here, in this paper, we have considered a variety of different classes of cosmological models and have showed explicitly that for each of them the CET gravitational entropy is increasing. The authors in \cite{greg} stated that the difference between our studies is that their calculation yields time independent gravitational temperature whereas we have found the gravitational temperature to be time dependent. We want to clarify that this is a model dependent phenomenon, and we also have such a case for the dark energy dominated FLRW model where the gravitational temperature is time independent. The important thing to note is that although for the LRS Bianchi I case the Weyl curvature must be increasing in time for a monotonically increasing gravitational entropy, it is not so for the other models considered by us. For the Liang, Szekeres and also for the Bianchi VI$_{h}$ model, although the CET entropy is clearly increasing, the magnitude of $\Psi_{2}$ is decreasing with time, similar to that in \cite{greg}. We want to emphasize that this does not the violate the Weyl curvature hypothesis, as the CET proposal was arrived at after various authors worked with different measures of gravitational entropy, especially the initial case, i.e., where the simple Weyl curvature scalar gives the measure of gravitational entropy. This is precisely the motivation for CET and other measures, as there are plenty of spacetimes where the Weyl curvature is decreasing, giving us a decreasing gravitational entropy if we take the simple Weyl curvature entropy as our measure.

We also want to draw attention to the requirements listed in the original CET paper \cite{CET} for a viable gravitational entropy in the form of E1...E5 at the end of Section 2 in their paper. There is no such requirement where the gravitational entropy should increase with the Weyl curvature, although this was the original proposal of Penrose, but this condition was too strong to be satisfied and subsequently new measures of gravitational entropy were proposed in the form of a suitable function of the Weyl curvature. Therefore we must be very clear when we say that it is satisfying the Weyl curvature hypothesis. What we really mean is that the spacetime has a viable non negative gravitational entropy which is monotonically increasing with time, which is also the case for \cite{greg}. Therefore our conclusions are in accordance with each other, i.e., both these papers are getting non negative and monotonically increasing CET gravitational entropy. The surprising result in \cite{greg} is that, throughout the entire evolution, the matter curvature dominates over the Weyl curvature unlike our cases, but in spite of that the CET gravitational entropy is non negative and monotonically increasing, indicating the robustness of the CET definition of gravitational entropy. The authors in \cite{greg} mention that the shear also plays a very important role, which is true as it is not only affecting the dynamics of the electric part of the Weyl tensor resulting in a change in gravitational energy density, but it is also contained in the gravitational temperature. It is to be noted that \cite{greg} showed that increasing Weyl is NOT necessary but is a sufficient condition for an increasing entropy, and throughout this paper we have worked with this sufficient condition.

Further, we must remember that the CET proposal is independent of any specific definition of gravitational temperature. Therefore, theoretically speaking, different measures of gravitational temperature can be employed. Regarding the role of the definition of gravitational temperature, one can say that it will change the exact results for the gravitational entropy but whether it will affect the monotonicity of the gravitational entropy is a matter of separate investigation. In the definition proposed in \cite{CET}, it depends on the acceleration, expansion, shear tensor and the rotation tensor (if we generalize the definition), capturing all the necessary variables. The time independent or dependent cases may arise from the internal dynamics of these variables, i.e., it is model dependent.

In conclusion, we can clearly state that the definitions of gravitational entropy proposed by Pelavas et al. \cite{PC} and Clifton et al. \cite{CET}, i.e.,  $ \tilde{P} $ and  $ S_{grav} $, are in conformity with the Weyl curvature hypothesis in the case of the models considered by us, and provides a very good description of the gravitational entropy on a local scale. It is to be noted that for a large scale description, one needs to employ the method of averaging, similar to that considered in \cite{cet4}. ''



\chapter{ Gravitational entropy proposals for traversable wormholes}




The contents of this chapter have been published in a journal, details of which are given below:\\

\textbf{JOURNAL REFERENCE:} General Relativity and Gravitation, 54:47 (2022)

\textbf{ARTICLE NAME:} How appropriate are the gravitational entropy proposals for traversable wormholes? \\
DOI: 10.1007/s10714-022-02934-3 \\~~~\\

The paper is quoted below:\\

``

\section{Introduction}
Gravitational entropy is the entropy measure reflecting the degrees of freedom associated with the free gravitational field. In any physical process involving gravity, the clumping of matter (structure formation) or the intensity of gravitational field in a local region of spacetime can be measured in terms of this quantity. Historically, this idea was proposed to justify the low entropy state of the initial universe, i.e., entropy is associated with the free gravitational field even during the time of big bang, so that a gravity dominated evolution of the universe does not violate the second law of thermodynamics. Another importance of gravitational entropy is that it puts the black hole entropy in the proper context, making it the special case of the gravitational entropy of free gravitational fields. Black hole (BH) entropy has a special position in physics because it is the only entropy which is proportional to the area of the gravitating object, unlike other thermodynamic entropies which are proportional to the volume.

Gravitational entropy proposals (GE proposals) can be studied in both local and global contexts. Locally, BH entropy (or the entropy of any astrophysical object) represents the immense concentration of entropy in a given region of spacetime. Thus gravity condenses matter leading to the increase in entropy of the universe, and BH entropy is the ultimate result of that process. Global studies on the evolution of the universe also indicate that the GE is monotonically increasing and well-behaved near the initial singularity. The study of GE is also important in thermodynamics. In gravitational theories, geometry and energy are interrelated, and both kinds of measures (geometric and thermodynamic) are available for GE. It not only gives us an idea of the nature of some specific geometry but also encapsulates the overall energetics of that region. The study of GE also tells us how matter and free gravitational fields behave in a particular region or in an overall fashion. Thus the concept of GE is still developing, as pointed out in \cite{CET}.

Roger Penrose’s Weyl curvature hypothesis (WCH) \cite{Penrose1} was the first effort to understand GE in a formal way. He argued that entropy can be assigned to the free gravitational field, and the Weyl curvature serves as a measure of it. It was assumed that the universe began from a singular state where the Weyl component was much smaller compared to the Ricci component. To this end, the FLRW models provide us an approximate description of the  early phase of the universe. The Weyl curvature was zero at early times for the FLRW case, but is large in the Schwarzschild-like spacetimes, which represents the geometry of a spherically symmetric star or a black hole formed during the later phases of evolution, thereby indicating the validity of the Weyl curvature hypothesis, as free gravitational entropy is larger in strongly gravitating systems than in flat spacetimes. This is the kind of behavior that we expect from a description of GE, i.e., it should increase throughout the history of the universe, in agreement with the second law of thermodynamics \cite{Penrose2,Bolejko}. Even then, there is still doubt about the definition of gravitational entropy in a way similar to thermodynamic entropy, which may be applied to all gravitating systems \cite{CET}.

In $ 2008 $, Rudjord, Grøn and Hervik \cite{entropy1} pointed out that the Weyl scalar is not a good measure of GE. The same is true for the ratio of the Weyl scalar squared to the squared Ricci tensor. They explicitly showed that their proposal of GE, i.e., the ratio of the Weyl scalar to the Kretschmann scalar, serves as a good measure and reproduces the Hawking-Bekenstein entropy for black holes \cite{SWH1,Bekenstein}. Later in $ 2012 $, Romero, Thomas and Pérez \cite{entropy2} applied it to other systems of black holes and wormholes, validating and extending the proposal of Rudjord et al. In $2013$, Clifton et al. \cite{CET} provided a measure of gravitational entropy based on the square root of the Bel-Robinson tensor, an idea motivated by thermodynamic considerations, which has a natural interpretation as the effective super-energy-momentum tensor of free gravitational fields. This is the so-called Clifton-Ellis-Tavakol (CET) proposal of gravitational entropy. But this definition is only valid for General Relativity (GR), where the Bel-Robinson tensor can be defined in such a way. Among the above two measures of gravitational entropy, the one due to Rudjord et al. and Romero et al. represents a geometrical measure, whereas the CET proposal gives us a thermodynamic measure of GE.

Sussman \cite{suss1} introduced a weighed scalar average formalism (the ``q-average'' formalism) for the study  of spherically symmetric LTB dust models and considered the application of this formalism to the definition of gravitational entropy functional proposed by Hosoya et al. (HB proposal) \cite{HB}. Subsequently, Sussman and Larena \cite{suss2} analyzed the generic LTB dust models to probe the CET proposal and the HB proposal (also a variant of the HB proposal), suggesting that the notion of gravitational entropy is theoretically robust, and can also be applied to many other generic spacetimes. The same authors studied the evolution of the CET gravitational entropy for the local expanding cosmic CDM voids using a nonperturbative approach \cite{suss3}.

Gravitational entropy can also be computed for objects like black holes and wormholes. Wormholes are shortcuts between two spacetime points. In most cases, these holes are unstable or causally disconnected, i.e., observers encounter closed timelike geodesics while travelling through them. Here we are interested in the traversable wormholes as they are physically more interesting because observers do not violate any causality while crossing them, there are no horizons, nor any curvature singularities giving rise to exotic physical scenarios \cite{visser1}. The study of gravitational entropy in the case of the traversable wormholes is important because it gives us a thermodynamic perspective of the physical reality of these solutions. If the gravitational entropy is well behaved, then these solutions are thermodynamically robust. Otherwise it may be thermodynamically unstable, and thus traversability may not be guaranteed. We now present a brief review of the important solutions of traversable wormholes.

The idea of traversable wormhole was first proposed by H. G. Ellis in 1973 \cite{ellis}, who showed that a coupling of geometry with a scalar field $ \phi $ produces a static, spherically symmetric, geodesically complete, horizonless space-time manifold with a topological hole (termed as a \emph{drainhole}) in its center. In 2005, Das and Kar \cite{ellis2} showed that the Ellis wormhole can also be obtained using tachyon matter as the source with a positive cosmological constant in $3+1$ dimensions. The Ellis wormhole has been studied by several authors \cite{ellis3,ellis4,ellis5} mainly in the field of deflection of light by massive objects, gravitational microlensing and wormhole shadows \cite{ellis1}. In \cite{ellis6} it was shown that when the throat is set into rotation, the static wormhole evolves into a rotating 4-dimensional Ellis wormhole supported by phantom scalar field. Recently, the Ellis wormhole without a phantom scalar field has been demonstrated \cite{ellis7} in $ 3+1 $ dimensional Einstein-scalar-Gauss-Bonnet theory (EsGB) in electrovacuum. By nonminimally coupling the phantom scalar field with the Maxwell field in \cite{ellis8}, the authors obtained charged Ellis wormhole and black hole solutions in the Einstein-Maxwell-scalar theory.

Independently, Bronnikov in 1973 proposed the same idea in scalar tensor theories on vacuum and electrovacuum spherically symmetric static solutions \cite{bro}. Subsequently Morris and Thorne also discussed the traversable wormholes in their 1988 paper \cite{mt}. Matt Visser in 1989 \cite{visser2} discussed some examples of traversable wormholes. In 1995, Cramer et.al. \cite{cramer} proposed another kind of traversable wormhole with negative mass cosmic strings which might have occurred in the early universe. In 1954, Papapetrou proposed the exponential spherically symmetric metric induced by scalar and antiscalar background fields \cite{Papapetrou} which represents a counterpart to the Schwarzschild black hole. Recently \cite{exp2}, it was shown that this metric has its origin within a wide class of scalar and antiscalar solutions of the Einstein equations parameterized by scalar charge. The exact rotational generalization of the antiscalar Papapetrou spacetime as a viable alternative to the Kerr black hole has been studied in \cite{exp4}. The Darmour-Solodukhin wormhole was proposed in $2007$ by T. Darmour and S. N. Solodukhin \cite{DS1}. Recently Matyjasek \cite{DS2} have demonstrated that for scalar fields there is a parameter space which allows this wormhole to be a traversable one.

The solution of the Einstein-Maxwell system of equations was found by Brill \cite{brill} in $1964$. Recently, the thermodynamics of the Taub-NUT solution has been studied in the Euclidean sector by imposing the condition for the absence of Misner strings \cite{NUT1}. More work have been done recently by several authors \cite{NUT2,NUT3,NUT4,NUT6,NUT7} on the various thermodynamic issues. Clement et.al. \cite{NUT5} investigated the electrically charged particle motions for such a metric. More recently \cite{NUT8}, nonlinear extensions of gravitating dyon-like NUT wormholes have been studied. A recent work by J. Podolský on accelerating NUT blackholes \cite{pod} shows that accelerating NUT blackholes act as a throat of maximal curvature connecting our universe $(r > 0)$ with the second  parallel universe in the region $(r < 0)$.

Richarte and Simeone constructed thin-shell Lorentzian wormholes with spherical symmetry in 5-dimensional Einstein-Gauss-Bonnet theory. For certain values of the parameters, these wormholes could be supported by ordinary matter \cite{WH0}. Clement et. al \cite{WH1} showed that traversable wormholes can exist without exotic matter but with a NUT parameter, and there is no causality violation in such cases. Beato et. al. \cite{WH2} found that the self-gravitating, analytic and globally regular Skyrmion solution of the Einstein–Skyrme system in presence of a cosmological constant has a non-trivial byproduct representing traversable AdS wormholes with NUT parameter, in which the only ``exotic matter'' required for their construction is a negative cosmological constant. Carvente et.al. \cite{lWH} studied traversable $ \ell $-wormholes supported by ghost scalar fields. Lima et. al. \cite{lima} have recently calculated the gravitational entropy for wormholes with exotic matter and in galactic halos.

G. Horowitz et. al. in 2019 discussed the nucleation process of traversable wormholes through a nonperturbative process in quantum gravity \cite{WHn}. In the same year Mattingly et.al. \cite{WHi} determined the curvature invariants of Lorentzian traversable wormholes. In \cite{torsion} the authors have shown that the violation of the null energy condition by matter, required by the traversable wormholes, can be avoided in spacetimes with torsion. Sebastiani et. al. \cite{wkb} proposed a unified classical approach for the studying idealized gravitational compact objects like wormholes (WHs) and horizonless stars by using the characteristic echoes generated in the ringdown phase.
Therefore, the study of traversable WHs becomes important in the context of GE, as more and
more of such studies have revealed that these systems may exist as possible astrophysical objects. If traversable WHs do exist, then the different proposals of GE must be tested on them to see whether they exhibit a viable GE.

There have been many definitions of black hole entropy and wormhole entropy using quantized theories of gravity, such as string theory and loop quantum gravity. However, in this paper we have addressed the problem from two different perspectives: the first one is the phenomenological approach proposed in \cite{entropy1} and expanded in \cite{entropy2}, in which the Weyl curvature hypothesis was tested against the expressions for the entropy of cosmological models and black holes. Comprehensive study of the various proposals of gravitational entropy in the case of various traversable wormholes is not available in literature. This has prompted us to analyze the gravitational entropy of wormholes in terms of the second perspective: the CET proposal, which yields us a pure thermodynamic measure of GE.

In our previous works on the accelerating black holes \cite{GC} and the cosmological models \cite{CGG}, we examined the validity of the Weyl scalar proposal of GE (proposed by Rudjord et al.), and the CET proposal of GE, respectively, for two different types of systems. In this paper we are extending our previous studies on gravitational entropy to the case of traversable wormholes using these two proposals. As the Weyl scalar proposal is based on a purely geometric approach, whereas the CET proposal includes the details of relativistic thermodynamics, it is interesting to study them side by side on the same system to see how these proposals differ from each other when applied to a specific spacetime geometry. A similar study was done by Pérez et al \cite{kerr} in the context of Kerr black holes. One purpose of this current work is also to see how our results for traversable wormholes compare with the results found in \cite{kerr}. These comparisons will provide us a better understanding of these different estimators of GE.

In this paper, we have examined various traversable wormholes from the simplest to the more involved ones. The Ellis wormhole is the simplest case. It is a zero mass traversable WH which connects two asymptotically flat regions at its throat. There have been many proposals for the energy source of such WHs making it not only an ideal toy model to study, but also having rich physical content. In the same spirit we have considered the exponential metric WH, which is a much more general spherically symmetric spacetime, and is traversable at the throat. There are WHs which can also mimic BHs, and the simplest case of such a WH is the Darmour-Solodukhin (DS) WH. This is of great physical interest as it is not only traversable, but also because it mimics the Schwarzschild BH to an outside observer, for all practical purposes. In the spherically symmetric static cases, both the exponential metric WH and the DS WH possess rich mathematical and physical structure. In order to examine the consequence of charge present in the WH system on the GE of that system, we considered the Maldacena ansatz, which connects two oppositely charged BHs. As for the stationary cases, we considered the Brill NUT WH (where both the magnetic and electric charges are present), and is an extension of the Reissner-Nordström solution with a Newman–Unti–Tamburino (NUT) parameter. This would enable us to study the behaviour of GE in Einstein-Maxwell systems, where the NUT parameter controls the WH neck. Extending this to cosmological settings, we consider an AdS NUT WH in the presence of a negative cosmological constant, so as to study the effect of $ \Lambda $ on the gravitational entropy of NUT WH. We have considered these widely different traversable WHs, in order to study the behaviour of GE explicitly in such scenarios, and to determine whether the GE proposals considered by us are physically viable or not.

Our paper is organized as follows: In Section II we have explained the gravitational entropy proposal given by Rudjord et al \cite{entropy1} and expanded by Romero et al \cite{entropy2}, and the CET proposal of gravitational entropy \cite{CET}. In the next section we have analysed the gravitational entropy for these systems. In section IV we have tested the validity of the Tolman law in the case of static spherically symmetric WHs, and subsequently presented the summary of our work and the concluding remarks in sections V and VI. Finally, in the appendix we have also included a brief analysis of the traversable AdS wormhole and followed it up for a general wormhole ansatz recently proposed by Maldacena.

\section{Gravitational Entropy}
In this section we will describe briefly the two proposals of gravitational entropy, namely the Weyl scalar proposal proposed by Rudjord et al. \cite{entropy1} and extended by Romero et al. \cite{entropy2}, and the CET proposal given by Clifton et al. \cite{CET}. As chronologically the Rudjord et al. proposal came first, and then the CET proposal, we will be following this sequence in our description and subsequent analysis.

\subsection{\textbf{{Weyl scalar proposals}}}

Here we provide a brief description of the proposal given in \cite{entropy1} for the determination of gravitational entropy.
The entropy on a surface is described by the surface integral
\begin{equation}
S_{\sigma}=k_{s}\int_{\sigma}\boldsymbol{\Psi}.d\sigma ,
\end{equation}
where $ \sigma $ denotes the surface and the vector field $\boldsymbol{\Psi}$ is given by \cite{entropy1}:
\begin{equation}
\boldsymbol{\Psi}=P_{i} \boldsymbol{e_{r}},
\end{equation}
with $ e_{r} $ as a unit radial vector. Here $P_{i}$ represents either $ P_{1} $ or $ P_{2} $, which are described below. The scalar $ P_{1} $ is defined in terms of the Weyl scalar ($ W $) and the Krestchmann scalar ($ K $) in the form \cite{entropy1}:
\begin{equation}\label{P_sq}
P_{1}^2=\dfrac{W}{K}=\dfrac{C_{abcd}C^{abcd}}{R_{abcd}R^{abcd}},
\end{equation}
where the Weyl tensor in $ n $ dimensions is given by \cite{Chandra}
\begin{equation}
C_{\alpha\beta\gamma\delta}= R_{\alpha\beta\gamma\delta}+\dfrac{2}{(n-2)}(g_{\alpha[\gamma})R_{\delta]\beta}-g_{\beta[\gamma})R_{\delta]\alpha})
+\dfrac{2}{(n-1)(n-2)}Rg_{\alpha[\gamma}g_{\delta]\beta}.
\end{equation}
Equation \eqref{P_sq} is a purely geometric measure of GE, and hence it nicely encompasses the curvature dynamics. Here, the gravitational entropy is evaluated by doing computations in a 3-space. The spatial metric $ h_{ab} $ is defined as:
\begin{equation}
h_{ij}=g_{ij}-\dfrac{g_{i0}g_{j0}}{g_{00}},
\end{equation}
where $ g_{\mu\nu} $ is the corresponding 4-dimensional space-time metric, and Latin indices denote spatial components, $i, j = 1, 2, 3$. So the infinitesimal surface element is given by:
\begin{equation}
d\sigma=\dfrac{\sqrt{h}}{\sqrt{h_{rr}}}d\theta d\phi.
\end{equation}
Since wormholes does not have any horizons, it is preferable to switch into entropy density, $s$. We imagine an enclosed hypersurface, and apply Gauss's divergence theorem to find the entropy density \cite{entropy1} as the following:
\begin{equation}
s=k_{s}|\nabla.\Psi|.
\end{equation}
If we only consider the radial contribution of the vector $ \Psi $ then the gravitational entropy density becomes
\begin{equation}\label{sd1}
s=k_{s}|\nabla.\Psi|=\dfrac{k_{s}}{\sqrt{-g}}\left\vert\dfrac{\partial}{\partial r}(\sqrt{-g}P_{i})\right\vert.
\end{equation}
This is useful in the spacetimes with spherical symmetry.

We have also discussed the possibility of having an angular component in the vector field $\boldsymbol{ \Psi} $ for axisymmetric spacetimes as proposed in \cite{entropy2}. Using this modified definition of $ \boldsymbol{\Psi} $, we can calculate the gravitational entropy density for axisymmetric space-times, which is given by the following expression:
\begin{equation}\label{sd2}
s=k_{s}|\nabla.\Psi|=\dfrac{k_{s}}{\sqrt{-g}}\left\vert\left(\dfrac{\partial}{\partial r}(\sqrt{-g}P_{i})+\dfrac{\partial}{\partial \theta}(\sqrt{-g}P_{i}) \right)\right\vert.
\end{equation}
Therefore the $ P_{1} $, is not the only case that we have considered. We have also used the measure proposed in \cite{entropy2} for the expression of $ P_{i} $ in the case of metrics having nonzero $ g_{t\phi} $ component, which is given below:
\begin{equation}\label{p2}
P_{2}=C_{abcd}C^{abcd}.
\end{equation}
Using this definition of $ P_{2} $ we have calculated the gravitational entropy density for the relevant wormholes.
It may be noted that the pure Weyl square proposal (in Eq. \eqref{p2}) fails at isotropic singularities, and
cannot handle the decaying and growing perturbation modes. As these quantities are purely geometric in nature that incorporate the connection of the Weyl tensor with the free gravitational field, they do not provide a theoretical connection with thermodynamics or Information theory. Further, as the ratio in Eq. \eqref{P_sq} is a dimensionless scalar, it cannot be related to the Hawking-Bekenstein entropy. Moreover, these geometric proposals are frame-independent, and hence have no connection with the worldlines of physical fluids. While $ P_{1} $ can address the above objections, but it does not seem to give the correct sense of time for a radiating source (see \cite{CET} and references therein).

\subsection{\textbf{Clifton-Ellis-Tavakol (CET) proposal}}
For the static spherically symmetric WHs we have also used the CET proposal \cite{CET} to examine the behaviour of GE for these systems. To establish the validity of this proposal, the CET paper has shown that it not only reproduces the Hawking-Bekenstein entropy for BHs, but also its entropy production rate, $ \dot{s}_{grav} $, is always non-negative.
This proposal begins with the construction of the second order symmetric traceless tensor $ t_{ab} $ which is obtained from the algebraic
``square root'' of the fourth order Bel-Robinson tensor $T_{abcd}$, because $T_{abcd}$ is the only totally symmetric traceless tensor that can be constructed out of the conformal Weyl tensor $ C_{abcd} $, and secondly, $T_{abcd}$ is fourth order with dimensions as $ L^{-4} $ making its ``square root'' necessary. This $ t_{ab} $ helps us to derive the ``effective'' or ``super energy–momentum tensor'' $ \mathcal{T}_{ab} $ of the free gravitational field. We can also compute other variables like gravitational energy density $ \rho_{grav} $, gravitational pressure $ p_{grav} $, anisotropic stresses $ \Pi^{ab}_{grav} $, and heat flux $ q^{a}_{grav} $ by contracting with the matter 4-velocity $ u^{a} $ and projector tensor $h_{ab} = u_{a}u_{b} + g_{ab}$. Subsequently, by analogy with the standard laws of fluid thermodynamics applied on the quantities associated with $\mathcal{T}_{ab}$, the notion of gravitational entropy emerges clearly. In \cite{CET}, the authors have considered two types of gravitational fields: the ``Coulomb-like'' (Petrov type D) and the ``wave-like'' (Petrov type N) fields for which $ \mathcal{T}_{ab} $ reduces to expressions involving the Newman–Penrose conformal invariants $ \Psi_{2} $ and $ \Psi_{4} $. We will be restraining our discussions to Coulomb-like fields, and therefore will only discuss the Petrov type D case. For this case, CET derived the following $ \mathcal{T}_{ab} $ and the associated fluxes:\\
\begin{align}\label{effT}
&\frac{{\mathcal{T}}^{ab}}{8\pi}=\alpha|\Psi_2|\left[ x^{a}x^{b}+y^{a}y^{b}-2\left(z^{a}z^{b}-u^{a}u^{b}\right)\right]=\rho_{grav} u^{a}u^{b} + p_{grav} h^{ab}+2q^{(a}_{grav} u^{b)}+\Pi_{grav}^{ab},\nonumber\\
& \left.\right. \\
& 8\pi\rho_{grav}=2\alpha|\Psi_2|,\quad p_{grav}= q_{grav}=0,\quad 8\pi\Pi_{grav}^{ab}=\frac{\alpha|\Psi_2|}{2}(x^{a}x^{b}+y^{a}y^{b}-z^{a}z^{b}+u^{a}u^{b}).\nonumber
\end{align}
Further, the gravito-electromagnetic properties of the Weyl tensor, and the 1+3 decomposition of the equations is used to express the gravitational ``Super energy density'' function $w$ as follows:
\begin{equation}
w=T_{abcd}u^{a}u^{b}u^{c}u^{d}=\dfrac{1}{4}\left(E_{a}^{b}E_{b}^{a}+ H_{a}^{b}H_{b}^{a}\right).
\end{equation}
Here, $ \alpha $ is a positive constant which provides appropriate physical units, and $\left[u^{a}, x^{a}, y^{a}, z^{a}\right]$
is an orthonormal tetrad. The quantities $ E_{ab}$, and $ H_{ab} $ are the electric and magnetic parts of the Weyl tensor $ C_{abcd} $ respectively.
For the mathematical computations, the complex null tetrad is defined as the following:
\begin{equation}
\label{nullt}
m^a = \frac{1}{\sqrt{2}} \left( x^a - i y^a \right), \quad
l^a = \frac{1}{\sqrt{2}} \left( u^a - z^a \right), \quad {\mathrm{and}} \quad
k^a = \frac{1}{\sqrt{2}} \left( u^a + z^a \right),
\end{equation}
where $x^a$, $y^a$ and $z^a$ are spacelike unit vectors, which constitute an orthonormal basis together with $u^a$. Using these, the entire metric can be rewritten in terms of these tetrads: $g_{ab} =2 m_{(a} \bar{m}_{b)} - 2 k_{(a} l_{b)}$, with $l^a$ and $k^a$ being aligned with the principal null directions. Therefore in this scheme of the free gravitational field \cite{CET}, the effective gravitational energy density can be written as:
\begin{equation}\label{Psi2}
8\pi\rho_{grav}=2\alpha\sqrt{\dfrac{2w}{3}}, \quad |\Psi_{2}|=\sqrt{\dfrac{2w}{3}},
\end{equation}
with  $\rho_{\mathrm{grav}} \geq 0$. Here $ \Psi_{2} $ is the nonzero Weyl scalar component for Petrov type D spacetimes.
In the CET paper \cite{CET}, the authors by analogy with the off-equilibrium Gibbs equation, obtained the following expression for the entropy production in the presence of perfect fluid matter field:
\begin{equation}
 T_{grav}\dot{s}_{grav} = (\rho_{grav} v)^{\cdot}=-v\sigma_{ab}\left[\Pi^{ab}_{grav}+\frac{4\pi(\rho+p)}{3\alpha|\Psi_2|}E^{ab}\right]. \label{gone}
\end{equation}
The independently defined local gravitational temperature at any point in spacetime was given by the following expression:
\begin{equation}
T_{grav}=\dfrac{|u_{a;b}l^{a}k^{b}|}{\pi}=\dfrac{|\dot{u_{a}}z^{a}+H+\sigma_{ab}z^{a}z^{b}|}{2\pi},
\end{equation}
where $z^{a}$ is a spacelike unit vector aligned with the Weyl principal tetrad, $ \sigma_{ab} $ is the shear tensor, $\dot{u_{a}}=u^{b}\nabla_{a}u_{b}$ is the $4-$acceleration, $ H=\dfrac{\Theta}{3} $ is the isotropic Hubble rate, and $\Theta\equiv\tilde{\nabla}_{c}u^{c}=h^{b}_{c}\nabla_{b}u^{c} $ is the isotropic expansion scalar.
As the above described variables depend on all the four spacetime variables, naturally the gravitational entropy one-form can be expressed as: $ ds_{grav}=\partial_{0}(s_{grav})dx^{0}+\partial_{i}(s_{grav})dx^{i} $. Consequently the Gibbs one-form becomes a system of four partial differential equations where $ T_{grav} $ acts an integrating factor, and these systems need not be integrable. On careful inspection it is evident that one needs to know about the physics of the microscopic theory of gravity to know the temperature of gravitational fields, and therefore, in \cite{CET} the authors considered the results of BH thermodynamics and quantum field theory in curved spacetime to propose the definition of $T_{grav}$. This definitionis not limited only to horizons, but is local, and can reproduce the Hawking temperature, the Unruh temperature and the temperature of de Sitter spacetime in the appropriate limits. Therefore, although the $ T_{grav} $ provided in \cite{CET} is an extra ingredient appearing along with their main proposal, and it is always possible to define a new temperature, still the concept of $ T_{grav} $ is rather well-motivated.

In the original CET paper \cite{CET}, the authors were interested in the gravitational entropy production, considering only the time derivative $\dot{s}_{grav}=u^{a}\partial_{a}s_{grav}$, along the worldlines of comoving observers. The full integrability of the Gibbs one-form have been discussed explicitly by Sussman and Larena in \cite{suss2,suss3}. The spherically symmetric static WHs that we have considered here, are the equilibrium cases where the condition $ \dot{s}_{grav}=u^{a}\partial_{a}s_{grav}=0 $ holds strictly, and therefore, for the one-form coordinate basis $ [dt,dr,d\theta,d\phi] $, the gravitational entropy one-form becomes
\begin{equation}
ds_{grav}=\partial_{r}(s_{grav})dr,
\end{equation}
and the Gibbs one-form reduces to a single ordinary differential equation:
\begin{equation}
T_{grav}\partial_{r}[s_{grav}]=\partial_{r}[\rho_{grav}],
\end{equation}
which leads to the rate of variation of the local piecewise gravitational entropy along the radial direction as:
\begin{equation}
\partial_{r}s_{grav}(r)=\dfrac{\rho_{grav}(r)v(r)}{T_{grav}(r)}.
\end{equation}
Finally, the variation of the local piecewise gravitational entropy of static spherically symmetric spacetimes $ s_{grav}(r) $ can be determined as:
\begin{equation}\label{totalen}
s_{grav}(r)=\int\dfrac{\rho_{grav}(r)v(r)}{T_{grav}(r)}dr,
\end{equation}
where the volume element is represented by $ v(r)=4\pi\sqrt{h} $, where $ h $ is the determinant of the projector tensor $ h_{ab} $. It may be further noted that the CET proposal is applicable only to Einstein's gravity. If we try to apply this proposal to different kinds of spacetimes, it is found that the CET proposal can provide unique gravitational entropy only for the Petrov type D and type N spacetimes. The algebraic decomposition of the Bel-Robinson tensor into a second order effective energy momentum tensor can be done for other Petrov type spacetimes also, but it is only in the Petrov type D and N spacetimes that the second order effective energy momentum tensor is unique \cite{bonsen}.

\section{Analysis of gravitational entropy}

In this section we will study some important traversable wormholes extending our previous works \cite{GC,CGG}. In each case we will be applying the two separate proposals of gravitational entropy if possible, and discuss their implications and limitations. We will also compare the corresponding results wherever possible.
We want to emphasize here that the energy momentum tensor for the  matter source supporting the WHs has no effect on the effective energy momentum tensor, $\mathcal{T}^{ab}$, for the free gravitational field given by CET and described in \eqref{effT}. It is clear from this equation that the geometric fluid for $\mathcal{T}^{ab}$ is not associated to the  matter source ($T^{ab}$) of WH.

\subsection{\textbf{Ellis wormhole}}
The Ellis WH obtained in 1973 \cite{ellis} is the first example of a non-singular wormhole solution. This is an exact solution of the Einstein-phantom scalar system with a scalar field having negative kinetic energy. Ellis used the negative kinetic energy term (‘phantom’) in the scalar field action to achieve the violation of energy condition, which is  necessary to support the wormhole. The metric of this wormhole is given by \cite{ellis}:
\begin{equation}\label{ellism}
ds^2=-dt^2+dr^2+(r^2+a^2)(d\theta^2+sin^2\theta d\phi^2),
\end{equation}
where $r=a $ is the throat radius of the wormhole.  The radial coordinate $ r $ runs from $ -\infty $ to $ +\infty $ to cover the entire wormhole geometry, where $ r=0 $ corresponds to the throat of the wormhole. This metric has no singularity, and the throat connects the two separate regions $ r\rightarrow +\infty $ and $ r\rightarrow -\infty $. The stress-energy tensor for this WH is: $-T^{tt}=-T^{rr}=T^{\theta\theta}=T^{\phi\phi}=\dfrac{a^2}{(a^2+r^2)^2} $, where the energy density is negative, which is only possible with exotic phantom matter.

\begin{enumerate}
\item \textbf{Weyl scalar proposal:} The spatial section of the Ellis wormhole is given by the following:
\begin{equation}
h_{ij}=diag\big[1,(r^2+a^2),(r^2+a^2)sin^2\theta \big].
\end{equation}\\
The determinant of the above mentioned matrix is given by $ h $:
$ h=(r^2+a^2)^2 sin^2\theta .$\\
The Weyl curvature scalar $ W $ and the Kretschmann curvature scalar $ K $ are obtained in the form:
$$ W= \frac{16}{3}\dfrac{a^4}{(a^2+r^2)^4}, \qquad \qquad \qquad K= \dfrac{12a^4}{(a^2+r^2)^4}.$$
Therefore the ratio of the two curvature scalars is given by:
$ P_1^2=\frac{W}{K}=\frac{4}{9} .$

\begin{figure}[ht]
    \centering
    \subfloat[Subfigure 1 list of figures text][]
        {
        \includegraphics[width=0.45\textwidth]{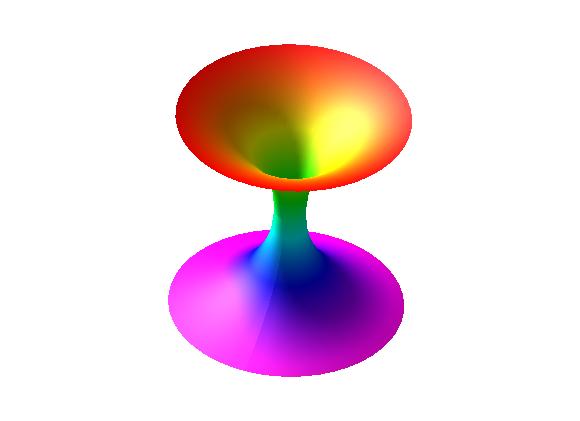}
        \label{fig:subfig1}
        }
    \subfloat[Subfigure 2 list of figures text][]
        {
        \includegraphics[width=0.45\textwidth]{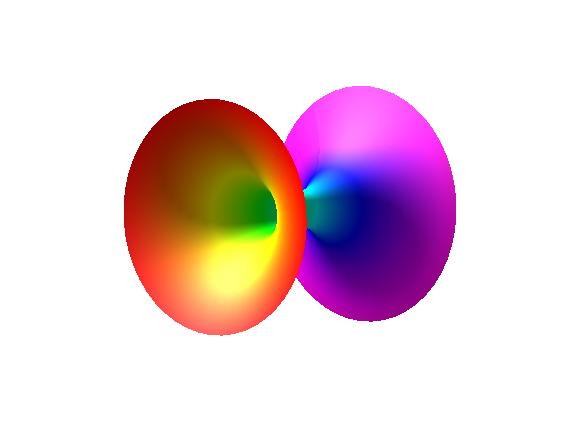}
        \label{fig:subfig2}
        }
    \caption{Embedding diagram of the Ellis wormhole where we have taken $ a=2 $.}
    \label{ellisem}
\end{figure}
The gravitational entropy density is obtained as
\begin{equation}\label{el}
s=k_{s}\vert\nabla.\Psi \vert= k_{s}\left\vert\frac{1}{\sqrt{h}}\dfrac{\partial}{\partial r}(\sqrt{h}\dfrac{P}{\sqrt{h_{rr}}}) \right\vert=\dfrac{4k_{s}}{3}\left\vert\dfrac{r}{(r^2+a^2)}\right\vert.
\end{equation}
Fig.\ref{ellisem} illustrates the embedding diagram of the Ellis wormhole for the values of parameters as mentioned.

\begin{figure}[ht]
    \centering
    \subfloat[Subfigure 1 list of figures text][]
    {\begin{minipage}[t][9cm][b]{.5\textwidth}
\centering
        {
        \includegraphics[width=0.97\textwidth]{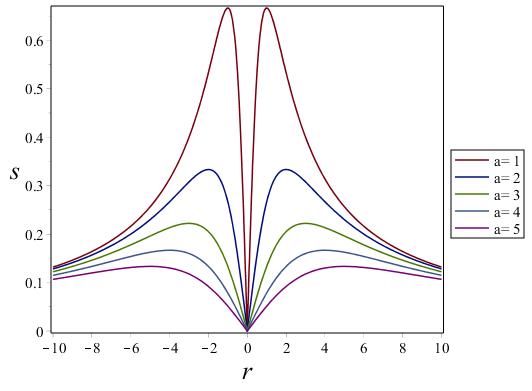}
        \label{fig:subfig1a}
        }
        \end{minipage}}
\subfloat[]{\begin{minipage}[t][9cm][b]{.5\textwidth}
\centering
\includegraphics[width=0.97\textwidth, height=0.2\textheight]{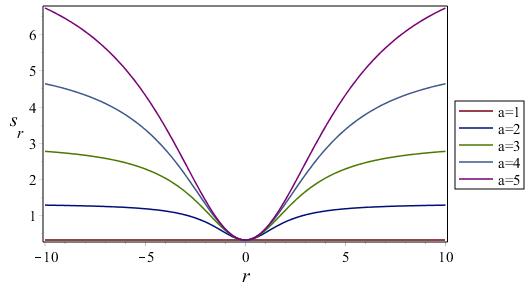}
\includegraphics[width=0.97\textwidth,height=0.2\textheight]{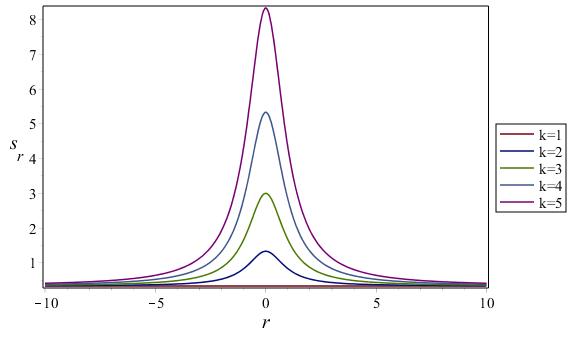}
\end{minipage}}
    \caption{(a) Variation of GE density $ s $ of Ellis wormhole with $ r $ for various throat lengths $ a $. (b) Rate of variation of CET gravitational entropy along the radial direction $\partial_{r}s_{grav}^{IF}\equiv s_{r} $ for different values of $ a $, where $ k=1 $, and for different values of $ k  $ where $ a=1 $, respectively.}
    \label{ellisentrop}
\end{figure}

In Fig.\ref{ellisentrop}(a) the gravitational entropy density derived in the equation \eqref{el} is depicted for different throat lengths. The maxima of gravitational entropy density lies at the wormhole throat radius, and the maxima of the entropy density is shifting to a lower value with the increasing throat radius. Further, the gravitational entropy density is always zero at the throat region of the wormhole as there is no central singularity in the metric of the Ellis wormhole. The finiteness of the gravitational entropy density also conforms to the traversability of this wormhole. It maybe noted that Romero et al. \cite{entropy2} evaluated the gravitational entropy density of another kind of traversable wormhole with exotic matter using the Weyl scalar proposal. They have also found that the gravitational entropy density is always zero at the throat of the wormhole irrespective of the value of the throat radius. The discussion of Fig.\ref{ellisentrop}(b) is given later.
\item \textbf{CET proposal:} Next we examine the CET proposal of gravitational entropy for the traversable Ellis wormhole. The gravitational epoch function $ w $ in this case is
\begin{equation}
w=T_{tttt}u^{t}u^{t}u^{t}u^{t}=\dfrac{a^4}{6(a^2+r^2)^4}.
\end{equation}
We have calculated all the Weyl scalars and we find that the only nonzero component is $\Psi_{2}  $ given by:
\begin{equation}
\Psi_{2}=-\frac{1}{3}\dfrac{a^2}{(a^2+r^2)^2}.
\end{equation}
Therefore the identity $ \vert \Psi_{2} \vert=\sqrt{\dfrac{2w}{3}} $ is satisfied in this Petrov type D spacetime. The energy density of the gravitational field is
\begin{equation}
\rho_{grav}=\frac{\alpha}{4\pi}\vert\Psi_{2}\vert=\frac{\alpha}{12\pi} \left\vert \dfrac{a^2}{(a^2+r^2)^2} \right\vert.
\end{equation}
As $ \dot u_{a}=0, \Theta=0, \sigma_{ab}=0  $ and $ \omega_{ab}=0 $, the gravitational temperature provided by CET in \cite{CET} emerges as
\begin{equation}
T_{grav}=0.
\end{equation}
The vanishing gravitational temperature is problematic because the GE will blow up in that case. As the CET GE is frame dependent, and we have considered observers only along a static frame, it would be very interesting to check for observers whose $ 4- $velocity is tangential to the worldlines of observers traversing the WH, to see whether it solves the problem of vanishing $ T_{grav} $. Let us consider the radial timelike geodesics on the  equatorial plane, and with the condition $ a^2>0 $, these geodesics can traverse the WH from one side to the other. Consequently the $ 4- $velocity in such a frame turns out to be $ u^{a}=(\sqrt{2},-1,0,0) $. For the sake of clarity, we state our orthonormal basis $ [u^{a},r^{a},\theta^{a},\phi^{a}] $ in the expressions below:
\begin{align}
 &u^{a}=(\sqrt{2},-1,0,0), &  &\theta^{a}=\left(0,0,\frac{1}{\sqrt{2}}\dfrac{1}{\sqrt{r^2+a^2}},\frac{1}{\sqrt{2}sin\theta}\dfrac{1}{\sqrt{r^2+a^2}}\right),\nonumber\\
 &r^{a}=(1,-\sqrt{2},0,0), &   &\phi^{a}=\left(0,0,\frac{1}{\sqrt{2}}\dfrac{1}{\sqrt{r^2+a^2}},-\frac{1}{\sqrt{2}sin\theta}\dfrac{1}{\sqrt{r^2+a^2}}\right).
\end{align}
Here $ u^{a} $ is a timelike unit vector, and the rest are spacelike unit vectors orthogonal to $ u^{a} $ and to each other. Consequently, we can form the complex null tetrad according to \eqref{nullt}. We obtain the gravitational energy density as $ \rho_{grav}=\dfrac{\alpha}{12\pi}\dfrac{a^2}{(a^2+r^2)^2} $.

Using the above mentioned $ 4- $velocity we found the congruence properties, which are: the $ 4- $acceleration $ \dot{u}_{a}=0 $, the expansion scalar $ \Theta(r)=\dfrac{-2r}{a^2+r^2} $, and the stress tensor $ \sigma_{ab} $ is given below:
\begin{align}
&\sigma_{00}=\dfrac{2r}{3(a^2+r^2)},\, \sigma_{01}=\sigma_{10}=\dfrac{2r\sqrt{2}}{3(a^2+r^2)},\,  \nonumber\\
&\sigma_{11}=\dfrac{4r}{3(a^2+r^2)},\,\sigma_{22}=-\frac{r}{3},\,\sigma_{33}=-\dfrac{r\sin^2\theta}{3}.
\end{align}
Therefore, the quantities required for the determination of $ T_{grav} $ becomes: $ \sigma_{ab}r^{a}r^{b}=\dfrac{2r}{3(a^2+r^2)} $,
and the isotropic Hubble rate is $ H=\frac{\Theta}{3}=-\dfrac{2r}{3(a^2+r^2)} $. Once again this yields $ T_{grav}=0 $.
This means that the ratio blows up, i.e., $\frac{\rho_{grav}}{T_{grav}}\rightarrow \infty $, if we consider the $ T_{grav} $ proposed by CET. Here the free gravitational energy density is finite locally but the gravitational temperature according to the CET proposal for the free gravitational field is vanishing, indicating the divergent local piecewise gravitational entropy. However, physically this is impossible. Although the CET proposal is independent of the definition of $T_{grav}$, at most we can say that the definition of temperature given in the CET paper is not suitable in this case. Since the CET definition of $T_{grav}$ is completely \emph{ad hoc}, and as $T_{grav}$ is simply an integrating factor of the Gibbs one-form, we can choose an ad hoc expression of $ T_{grav} $ in order to integrate the expression for GE. Now, if we want a finite entropy density at the throat and a non negative function of $ r $, then let us define the simplest integrating factor as $ T^{IF}_{grav}\sim\dfrac{1}{k^2+r^2}$, where $ k $ is a constant. It is to be noted that this temperature is for pure mathematical convenience and has no physical standing, whereas the temperature $T_{grav}$ given by CET was very robust since it could reproduce the Hawking-Bekenstein temperature for BH. This \textbf{\emph{temperature function}} $ T^{IF}_{grav}$ actually replicates the CET definition far away from the wormhole, at large values of ``r'', whereas near the wormhole it gives a finite non zero value. We will restrict our calculations to the equatorial plane ($ \sin\theta=1 $). The function $ T^{IF}_{grav}$ gives us the rate of variation of GE along the radial direction as
\begin{equation}
\partial_{r}s_{grav}^{IF}=\dfrac{\rho_{grav}v}{T^{IF}_{grav}}\sim\bigg\vert\dfrac{\alpha}{3}\dfrac{a^2 (k^2+r^2)}{(a^2+r^2)}\bigg\vert .
\end{equation}
In Fig.\ref{ellisentrop}(b) the rate of radial variation of CET GE, $\partial_{r}s_{grav}^{IF} $, is shown as a function of $ r $. Unlike the Weyl scalar proposal, here the entropy density is not always zero at the throat. The local piecewise CET gravitational entropy can be obtained by integrating over the radial coordinate as the following:
\begin{equation}
s_{grav}^{IF}(r)=\int_{0}^{r} \dfrac{\rho_{grav}v}{T^{IF}_{grav}}dr\sim\bigg\vert\dfrac{\alpha a}{3}\left((a^2-k^2)\arctan\left(\frac{r}{a}\right)-ar\right)\bigg\vert.
\end{equation}
Both the radial rate of variation of CET gravitational entropy, $ \partial_{r}s_{grav}^{IF} $, and the local piecewise GE, $ S_{grav}^{IF}(r) $, vanish at the throat for $ k=0 $, and for $ k\neq 0 $ the rate of variation of GE along the radial direction at the throat depends on the values of $ a $ and $ k $.
\end{enumerate}

\subsection{\textbf{Darmour-Solodukhin wormhole}}
The Darmour-Solodukhin wormhole \cite{DS1} is a good candidate for cosmological observations, the ``black hole foils'', i.e. these wormholes are objects that mimic some aspects of black holes, while lacking some of their defining features, such as the event horizon. This is a modification of the well known Schwarzschild metric in order to make it horizonless. The stress energy tensor and the information for the source required to maintain this WH are discussed in \cite{DS2}, where the authors used the Schwinger-DeWitt expansion to derive an approximated stress-energy tensor of the quantized massive scalar, spinor and vector field for the DS WH. They found that for the scalar field there is a region in the parameter space for which the stress-energy tensor has the desired properties. The stress energy of the massive scalar field with a general curvature coupling $ \xi $ is:
\begin{equation}
T^{b}_{\,a}=\dfrac{1}{96\pi^{2}\mu^{2}m^{6}}\left(1-\frac{2}{x}-\lambda^2\right)^{-6} \sum_{k=0}^{7} \beta^{(k)b}_{a}\frac{1}{x^{k+8}}.
\end{equation}
Here $ \mu $ is the mass of the field with $ x=r/m $. The coefficients $ \beta^{(k)b}_{a} $ depend parametrically on $ \lambda $ and $ \xi $.
Considering this stress-energy tensor, they found that there is a region in the $(\lambda,\xi)-$plane in which the stress-energy tensor has the form required to support the wormhole. The exact form of the stress-energy tensor of quantized massive fields are complicated and can be found in the supplementary of the same paper.

The Darmour-Solodukhin (DS) wormhole metric is
\begin{equation}
ds^2=-(f(r)+\lambda^2)dt^2+\dfrac{dr^2}{f(r)}+r^2(d\theta^2+\sin^2\theta d\phi^2),
\end{equation}
where $ f(r)=(1-\frac{2m}{r})$, and $ \lambda $ is a dimensionless parameter. Here, if $ \lambda=0 $, then the Darmour-Solodukhin wormhole metric reduces to the usual Schwarzschild black hole metric having an event horizon at $ r=2m $. But for $ \lambda \neq 0 $, however small, there is no event horizon. Instead this Lorentzian wormhole have a throat at $ r=2m $ connecting two isometric and asymtotically flat regions $ 2m\leq r \leq \infty $.
\begin{enumerate}
\item \textbf{Weyl scalar proposal:} In order to begin our analysis we first need to identify the spatial section of the Darmour-Solodukhin wormhole metric, which is given by the following expression:
$$ h_{ij}= \text{diag}\left(\dfrac{1}{(1-\frac{2m}{r})},r^2,r^2\sin^2\theta\right) .$$
We also need to compute the ratio $(P_{1}^2)$, which is given by the expression \eqref{DS1}:
\begin{align}\label{DS1}
\left.P_{1}^2\right. & =\dfrac{C_{abcd}C^{abcd}}{R_{abcd}R^{abcd}}\nonumber\\ & =\frac{1}{3} \left( 3\,{\lambda}^{4}{r}^{2}-19\,{\lambda}^{2}mr+9\,{
\lambda}^{2}{r}^{2}+24\,{m}^{2}-24\,mr+6\,{r}^{2} \right) ^{2} \times \nonumber\\
& \big(6\,{\lambda}^{8}{r}^{4}-48\,{\lambda}^{6}m{r}^{3}+24\,{\lambda}^{6}{r}^{4}
+177\,{\lambda}^{4}{m}^{2}{r}^{2}-172\,{\lambda}^{4}m{r}^{3}+42\,{\lambda}^{4}{r}^{4}- 304\,{\lambda}^{2}{m}^{3}r \nonumber\\
&  +448\,{\lambda}^{2}{m}^{2}{r}^{2}-220\,{\lambda}^{2}m{r}^{3}+36\,{\lambda}^{2}{r}^{4}+192\,{m}
^{4}-384\,{m}^{3}r+288\,{m}^{2}{r}^{2}-96\,m{r}^{3} \nonumber\\
&+12\,{r}^{4}\big)^{-1}.
\end{align}
\begin{figure}[ht]
    \centering
\includegraphics[width=1.0\textwidth, height=0.32\textheight]{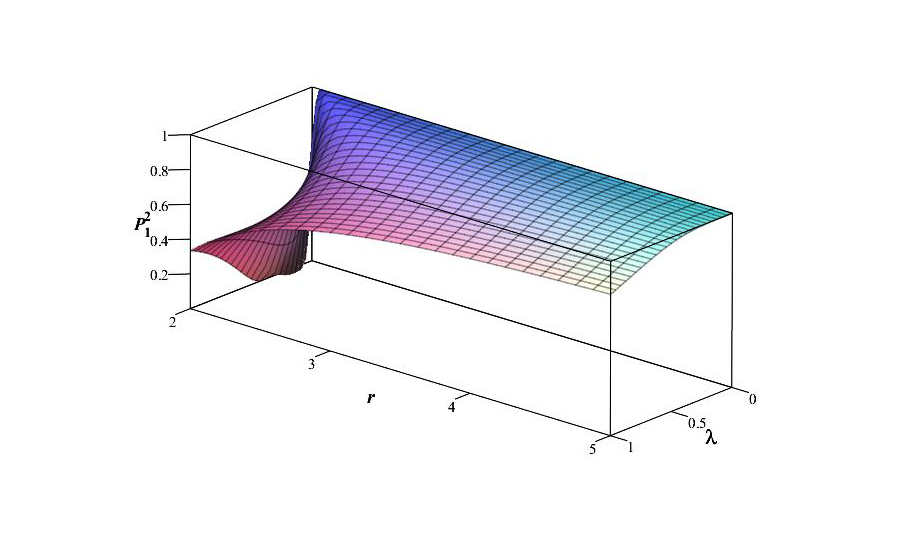}
\caption{Variation of the ratio of curvature scalars $P_{1}^2 $ as a function of $ r,\lambda $ for Darmour-Solodukhin WH.}
\label{DSW}
\end{figure}
The ratio of curvature scalars computed in \eqref{DS1} is depicted in Fig.\ref{DSW}. We find that as we approach the throat region at $ r=2 $, the contribution of the matter dominates over the Weyl, resulting in a dip of the value near that region, which is expected as the mass is located in that region. Using the proposal of Rudjord et al., we arrive at the expression of GE density in \eqref{DS2} which is depicted in Fig.\ref{DSentrop}(a). We note that the gravitational entropy density is peaking around the throat of the wormhole. Further, as the value of $ \lambda $ increases, the peak of the gravitational entropy density decreases, with the highest value in the case of the black hole itself.

\begin{align}\label{DS2}
s=& k_{s}\vert\nabla.\Psi\vert  =  3072k_{s}\,\sqrt {3} \Bigg| \frac {\sqrt {-r+2\,m}}{{r}^{3/2}} \Bigg( {\frac {{\lambda}^{12}{r}^{6}}{256}}-{\frac {79\,{r}^{5}{\lambda}^{10}}{1536} \left( m-{\frac {42\,r}{79}} \right) }+  \nonumber\\
 & \left( {\frac {329\,{m}^{2}{r}^{4}}{1024}}-{\frac {165\,m{r}^{5}}{512}}+{\frac {21\,{r}^{6}}{256}} \right) {\lambda}^{8}-\frac {583\,{r}^{3}{\lambda}^{6}}{512}\big( {m}^{3}-{\frac {5147\,{m}^{2}r}{3498}}+{\frac {423\,m{r}^{2}}{583}}\nonumber\\
 & -{\frac {70\,{r}^{3}}{583}} \big) +{\frac {41\,{r}^{2} \left(
-r/2+m \right) {\lambda}^{4}}{18} \left( {m}^{3}-{\frac {3837\,{m}^{2}
r}{2624}}+{\frac {3747\,m{r}^{2}}{5248}}-{\frac {153\,{r}^{3}}{1312}}
 \right) }-  \nonumber\\
 & {\frac {19\,r \left( -r/2+m \right) ^{4}{\lambda}^{2}}{8}
 \left( m-{\frac {9\,r}{19}} \right) }+\left( -r/2+m \right) ^{6}
 \Bigg)\Bigg( 6\,{\lambda}^{8}{r}^{4}+ \big( -48\,m{r}^{3} \nonumber\\
&+24\,{r}^{4} \big) {\lambda}^{6}+\left( 177\,{m}^{2}{r}^{2}-172\,m{r}^{3}+
42\,{r}^{4} \right) {\lambda}^{4}-
 304\,r \left( -\frac{r}{2}+m \right) ^{2} \nonumber\\
& \left( m-{\frac {9\,r}{19}} \right) {\lambda}^{2}+192\, \left( -\frac{r}{2}+m
 \right) ^{4} \Bigg) ^{-\frac{3}{2}} \Bigg|.
\end{align}
If we put $ \lambda\rightarrow 0 $ in \eqref{DS2}, then the entire expression for $ s $ reduces to that of the Schwarzschild black hole, i.e. $ s=\dfrac{2k_{s}}{r} \Big|\sqrt{1-\frac{2m}{r}}\Big|$, which matches with the result in \cite{entropy1}. This is depicted in Fig.\ref{DSentrop}(a), validating the expression we have obtained here. We note that the GE density for the Schwarzschild black hole has a maxima at $ r=\frac{3}{2}\times(2m)$ and goes to zero at the event horizon at $ r=2m $. We have taken $ m=1 $ for our plots.
\item \textbf{CET proposal:} Let us now consider the CET proposal of gravitational entropy. We have chosen the four vectors to be aligned with the principal null tetrads, and computed the Weyl curvature scalar $ \Psi_{2} $ as follows:
\begin{equation}
\Psi_{2}= -4\,{\frac {m}{ \left( -{\lambda}^{2}r+2\,m-r \right) ^{2}{r}^{3}}
 \left(  \left( \frac{{\lambda}^{4}}{8}\,+\frac{3 {\lambda}^{2}}{8}\,+\frac{1}{4}\right) {r}^
{2}+m \left( -{\frac {19\,{\lambda}^{2}}{24}}-1 \right) r+{m}^{2}
 \right) }.
\end{equation}
Now using the definition of gravitational energy density $ \rho_{grav} $ and the expression of $ \Psi_{2} $, we have obtained the following expression \eqref{DS3}:
\begin{equation}\label{DS3}
\rho_{grav}=\dfrac{\alpha}{4\pi}\vert\Psi_{2}\vert={\frac {\alpha}{24\pi} \left| {\frac {m \left( 3\,{\lambda}^{4}{r}
^{2}-19\,{\lambda}^{2}mr+9\,{\lambda}^{2}{r}^{2}+24\,{m}^{2}-24\,mr+6
\,{r}^{2} \right) }{ \left( {\lambda}^{2}r-2\,m+r \right) ^{2}{r}^{3}}
} \right| }.
\end{equation}
Similarly using the definition of gravitational temperature $ T_{grav} $ we obtain \eqref{DS4}:
\begin{equation}\label{DS4}
T_{grav}=\frac{1}{2\pi}\left\vert\dfrac{m\sqrt{r-2m}}{(r-2m+\lambda^2 r)r^{3/2}}\right\vert.
\end{equation}
Here we chose the equatorial plane for our calculations. Finally, we obtain the expression for the rate of variation of CET gravitational entropy along the radial coordinate given in \eqref{DS5} (depicted in Fig.\ref{DSentrop}(b)):
\begin{equation}\label{DS5}
\partial_{r}s_{grav}=\frac{\alpha \pi r}{3}\, \left| {\frac {3\,{\lambda}^{4}{r}^{2}-19\,{\lambda}^{2
}mr+9\,{\lambda}^{2}{r}^{2}+24\,{m}^{2}-24\,mr+6\,{r}^{2}}{(r-2
\,m) \left( {\lambda}^{2}r-2\,m+r \right) }} \right|.
\end{equation}
The variation of the local piecewise gravitational entropy of the DS wormhole is obtained as:
\begin{align}
 s_{grav}(r)= & \dfrac{4\pi}{\left( {\lambda}^{2}+1 \right) ^{2}}\times{\mathit {signum}}
 \left( {\frac { \left( 3\,{\lambda}^{4}+9\,{\lambda}^{2}+6 \right) {r
}^{2}+ \left( -19\,{\lambda}^{2}-24 \right) mr+24\,{m}^{2}}{ \left( -r
+2\,m \right)  \left( -{\lambda}^{2}r+2\,m-r \right) }} \right) \nonumber \\
& \bigg( \frac{{m}^{2}}{6}\ln  \left( {\lambda}^{2}r-2m+r
 \right) + \Big( {m}^{2} \left( {\lambda}^{2}+1 \right)  \left( {
\lambda}^{2}-\frac{1}{6} \right) \ln  \left( r-2m \right)
\nonumber \\
&+\frac{r}{2} \Big(
 \left( m+\frac{r}{4} \right) {\lambda}^{4}+ \left( \frac{5}{6}m+\frac{3}{4}r \right) {
\lambda}^{2}+\frac{r}{2} \Big)  \Big)
 \left( {\lambda}^{2}+1 \right)
 \bigg)\bigg\vert_{2m+\epsilon}^{r},
\end{align}
where $\epsilon>0$ is a very small quantity as the above expression is not valid for $ r=2m $.
\end{enumerate}
\begin{figure}[ht]
    \centering
    \subfloat[Subfigure 1 list of figures text][]
        {
        \includegraphics[width=0.48\textwidth]{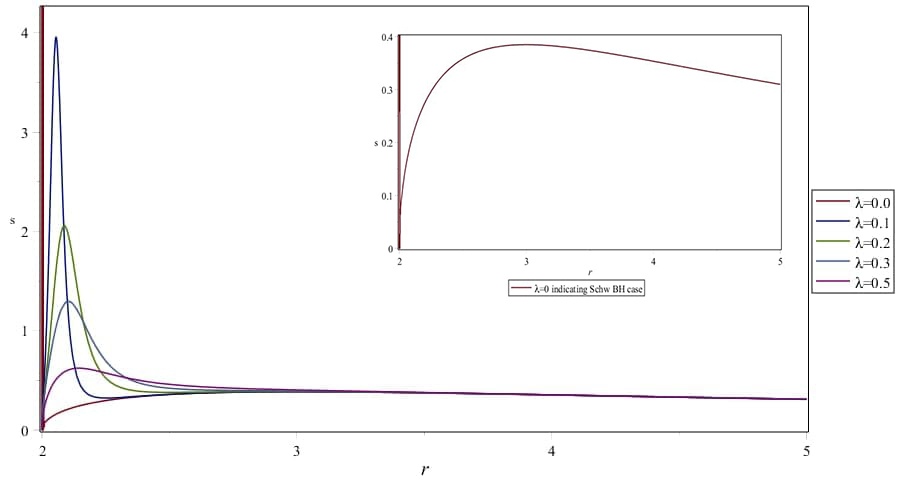}
        \label{fig:subfig3}
        }
    \subfloat[Subfigure 2 list of figures text][]
        {
        \includegraphics[width=0.48\textwidth]{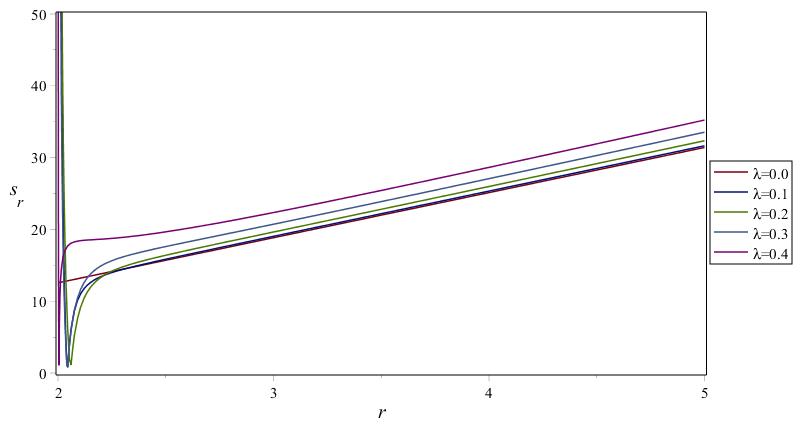}
        \label{fig:subfig4}
        }
    \caption{(a) Variation of GE density $ s $ of DS wormhole with $ r $ for various parameter strength $ \lambda $ using the Rudjord proposal. (b) Rate of variation of CET gravitational entropy  $ \partial_{r}s_{grav}\equiv s_{r} $ of DS wormhole with $ r $ for various parameter strength $ \lambda $ using the CET proposal.}
   \label{DSentrop}
\end{figure}
We find that in this case, the rate of variation of gravitational entropy along the radial direction, $\partial_{r}s_{grav}$, decreases as the throat region is approached. Over here, as the value of $ \lambda $ increases, the magnitude goes towards a higher value. We also observe that the rate of variation of CET GE goes towards zero near the throat region for nonzero $ \lambda $.

\subsection{\textbf{Exponential metric wormhole}}
Next we will analyze the exponential metric wormhole. The exponential or Papapetrou metric represents the counterpart to the Schwarzschild black hole with antiscalar background fields, and can be considered as a special case of the Fisher–(Newman–Janis–Winicour)
–Wyman–Ellis–Bronnikov solutions for a massless scalar field coupled to gravity. It is related to the Morris–Thorne traversable wormhole. A detailed description of the antiscalar source field can be found in \cite{exp2,exp4}, where the authors analysed the origin of the metric within a wide class of scalar and antiscalar solutions of the Einstein equations parameterized by scalar charge. The exponential metric wormhole in isotropic coordinates is given by :
\begin{equation}
ds^2=-e^{\frac{-2m}{r}}dt^2+e^{\frac{2m}{r}}[dr^2+r^2(d\theta^2+\sin^2\theta d\phi^2)],
\end{equation}
with the Einstein tensor $ G^{a}_{\,b}=\dfrac{m^{2}e^{\dfrac{-2m}{r}}}{r^4}\text{diag}\left\lbrace 1,-1,1,1 \right\rbrace^{a}_{\,b}  $, and the antiscalar source field described by: $ T^{SF}_{ab}=\frac{1}{4\pi}\left(   \phi_{a}\phi_{b}-\frac{1}{2}g_{ab}\phi_{c}\phi^{c}\right) $. The energy momentum tensor is quadratic in $ \phi_{a}=\nabla_{a}\phi=\phi_{,a} $, with the field equation $ G_{ab}=-8\pi T^{SF}_{ab} $. The Lagrangian for the antiscalar field is given by $ L=\frac{1}{16\pi}\left(R+2\phi_{a}\phi^{a}\right) $. Recently, Boonserm et.al. \cite{exp1} demonstrated that this metric represents a traversable wormhole. Here $ m $ is the mass of the wormhole. The throat of the WH is located at $ r=m $. This WH does not have any horizon because $ g_{tt} $ is nonzero for all non negative values of $ r\in(0,+\infty) $. The region $ r<m $ represents the other universe, and the curvature invariants are nonzero at the throat \cite{WHi} while they go to zero asymptotically as $ r $ goes to infinity (as it should), in order to connect two asymptotically flat regions. In \cite{exp3} the authors have studied the geodesics using the Jacobi metric approach.
\begin{enumerate}
\item \textbf{Weyl scalar proposal:} For the exponential metric wormhole we find that the spatial section is the following:
\begin{equation}
h_{ij}= e^{\frac{2m}{r}} \text{diag}(1,r^2,r^2 \sin^2\theta),
\end{equation}
where the determinant of $ h_{ij} $ is given by: $ h= e^{\frac{6m}{r}}r^4 \sin^2\theta$. The ratio of the curvature scalars is given in \eqref{expw}:
\begin{equation}\label{expw}
P_1^2=\frac{4}{3}\dfrac{(2m-3r)^2}{(7m^2-16mr+12r^2)}.
\end{equation}
From this expression it is clear this ratio becomes zero at $ r=\frac{2m}{3} $. For the sake of clarity, the variation of $ P_{1}^2 $ is shown in FIG.\ref{EXPP} for different values of mass parameter $ m $. From the figure it is clear that $ P_{1}^2 $ becomes zero at specific points as mentioned above, and there is also a dip in the values near the throat of the wormhole where the mass is located, implying a more dominant contribution from the Riemann tensor than the Weyl component.
\begin{figure}[ht]
    \centering
\includegraphics[width=0.65\textwidth]{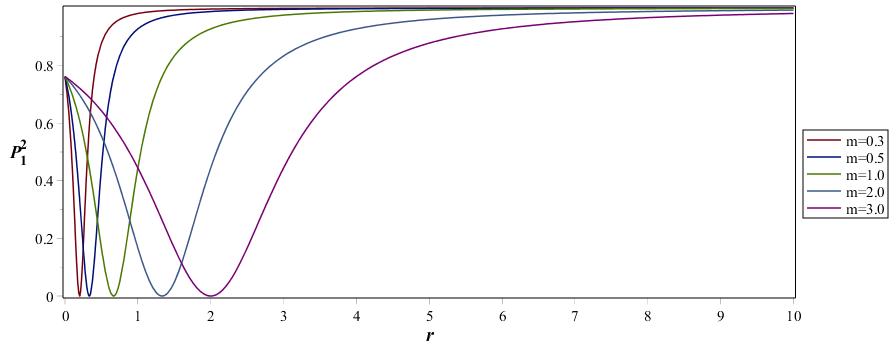}
\caption{ Variation of the ratio of curvature scalars $P_{1}^2 $ as a function of $ r $ for different $ m $ in the case of exponential metric wormhole.}
\label{EXPP}
\end{figure}
In this case we will use the following definition of gravitational entropy density:
\begin{equation}\label{e2}
s=k_{s}\vert\nabla.\Psi \vert= k_{s}\left\vert\frac{1}{\sqrt{h}}\dfrac{\partial}{\partial r}(\sqrt{h}\dfrac{P_{1}}{\sqrt{h_{rr}}}) \right\vert.
\end{equation}
\begin{figure}[ht]
    \centering
    \subfloat[Subfigure 1 list of figures text][]
        {
        \includegraphics[width=0.46\textwidth]{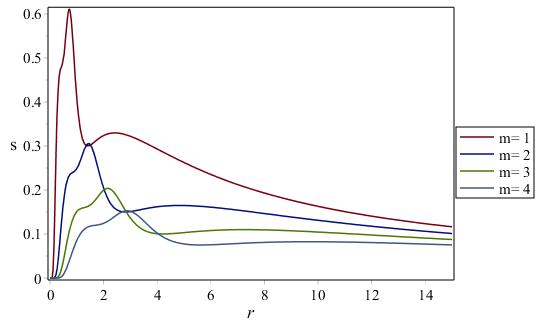}
        \label{fig:subfig3a}
        }
    \subfloat[Subfigure 2 list of figures text][]
        {
        \includegraphics[width=0.46\textwidth]{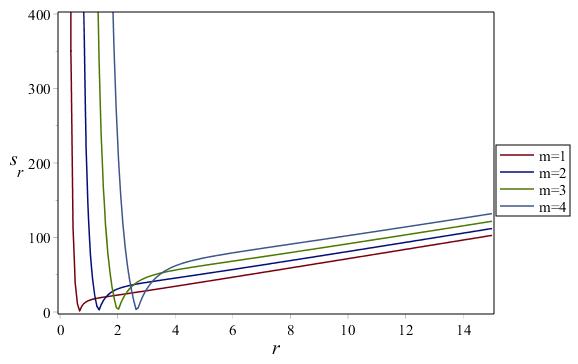}
        \label{fig:subfig4a}
        }
    \caption{(a)Variation of GE density $ s $ of exponential metric wormhole with $ r $ for various throat lengths $ r=m $ using Rudjord proposal. (b) Rate of variation of gravitational entropy along the radial direction $ \partial_{r}s_{grav}\equiv s_{r} $ of exponential metric wormhole with $ r $ for various throat lengths $ r=m $ using CET proposal.}
   \label{expentrop}
\end{figure}
The above definition of gravitational entropy density yields the result in \eqref{exps1}.
\begin{align}\label{exps1}
   s = & \frac{2k_{s}}{3}  \frac {\sqrt {3}}{{r}^{2} \left( 7\,{m}^{2}-16\,mr+12\,{r}^{2}
 \right) ^{2}} \times\nonumber\\
 & \Bigg| {{\frac {(56\,{m}^{5}-352\,{m}^{4}r+912\,{m}^{3}{r}^{2}-
1197\,{m}^{2}{r}^{3}+792\,m{r}^{4}-216\,{r}^{5})}{\sqrt {{
\frac { \left( 2\,m-3\,r \right) ^{2}}{7\,{m}^{2}-16\,mr+12\,{r}^{2}}}
}}}} \Bigg|  \left( {{\mathrm{e}}^{{\frac {m}{r}}}} \right) ^{-1}.
\end{align}
In Fig.\ref{expentrop}(a) we have shown the variation of gravitational entropy density using  \eqref{exps1} with different wormhole throat values. We find that the entropy density increases and becomes maximum near the throat and decays to zero as we move away from the central throat region. As the mass increases, so does the throat radius and the height of the maximum of the gravitational entropy density decreases with it, which is expected. Also the entropy density goes to zero on the other side of the throat as it approaches $ r=0. $
\item \textbf{CET proposal:} Next we consider the CET gravitational entropy proposal. At first the gravitational epoch function $ w $ is calculated:
\begin{equation}
w=T_{tttt}u^{t}u^{t}u^{t}u^{t}=\dfrac{m^2 e^{\frac{-4m}{r}}(2m-3r)^2}{6r^8}.
\end{equation}
Once again all the Weyl scalars are calculated independently and $ \Psi_{2} $ turns out to be the only nonzero component which is given below:
\begin{equation}
\Psi_{2}=\dfrac{m e^{\frac{-2m}{r}}(2m-3r)}{3r^4}.
\end{equation}
Here also the identity $ \vert \Psi_{2} \vert=\sqrt{\dfrac{2w}{3}} $ is satisfied as this is a Petrov type D spacetime. The gravitational energy density $\rho_{grav}  $ is given by
\begin{equation}
\rho_{grav}=\frac{\alpha}{4\pi}\vert\Psi_{2}\vert=\frac{\alpha}{12\pi} \left\vert\dfrac{m e^{\frac{-2m}{r}}(2m-3r)}{r^4}\right\vert.
\end{equation}
As $ \dot u_{r}=\dfrac{m}{r^2}, \Theta=0, \sigma_{ab}=0  $ and $ \omega_{ab}=0 $, we can calculate the gravitational temperature as
\begin{equation}
T_{grav}={\dfrac{1}{2\pi}}\left\vert \dfrac{m}{r^2 e^{\frac{m}{r}}}\right\vert.
\end{equation}
Once again we consider the equatorial plane, and compute the rate of variation of GE along the radial direction to get the following expression:
\begin{equation}\label{sgravcet}
\partial_{r}s_{grav}=\dfrac{\rho_{grav}v}{T_{grav}}=\frac{2\alpha\pi}{3} \left\vert e^{\frac{2m}{r}}(2m-3r)\right\vert.
\end{equation}
Finally, we obtain the piecewise local gravitational entropy of the exponential metric wormhole as the following:\\
\begin{equation}
s_{grav}(r)=\int_{0}^{r}\dfrac{\rho_{grav}v}{T_{grav}}dr=\frac {2\alpha\pi}{3}
\begin{cases}
\left\vert-\frac{3}{2}\,{{\mathrm{e}}^{{\frac {2m}{r}}}}{r}^{2}-{{\mathrm{e}}^{{\frac {2m}{r}}}}rm-2\,{\mathit{Ei}} \left( 1,-{\frac {2m}{r}} \right) {m}^{2}\right\vert; \quad r< \frac{2m}{3},\\
\textit{\text{undefined}}; \quad r=\frac{2m}{3},\\
\left\vert\frac{3}{2}\,{{\mathrm{e}}^{{\frac {2m}{r}}}}{r}^{2}+{{\mathrm{e}}^{{\frac {2m}{r}}}}rm+2\,{\mathit{Ei}} \left( 1,-{\frac {2m}{r}} \right) {m}^{2}\right\vert; \quad \frac{2m}{3}<r.
\end{cases}
\end{equation}
where $ Ei(a,z) $ are exponential integrals. In Fig.\ref{expentrop}(b), the rate of variation of gravitational entropy along the radial direction is plotted using \eqref{sgravcet} for different wormhole throat values. The zeroes are the points where the Weyl curvature becomes zero (as also in \eqref{expw}), i.e. at $ r=\frac{2m}{3} $. Therefore, as $ m $ increases, the throat radius increases, and the zero of the rate of radial variation of GE in the CET convention shifts toward positive infinity. As the distance increases from the central throat, i.e., at $r=m$ to the other side, it rises sharply from zero. This is also evident from equation \eqref{sgravcet}. A similar behaviour can also be found for $ s_{grav} $: it decreases monotonically as we approach the throat, crosses the throat smoothly, and then sharply increases on the other side.
\end{enumerate}

\subsection{\textbf{Traversable NUT wormhole}}
The solution of the Einstein-Maxwell system of equations found by Brill \cite{brill} in $1964$ is given by
\begin{equation}\label{trav_NUT_wmhl}
ds^2=-f(dt-2n(cos\theta +C)d\phi)^2+f^{-1}dr^2+(r^2+n^2)(d\theta^2+sin^2\theta d\phi^2).
\end{equation}
The Brill solution with no horizon, which connects two asymptotically locally flat regions, is the wormhole of our interest \cite{WH1}. The unknown parameters are as follows:
$$ f= \dfrac{(r-m)^2+b^2}{r^2+n^2}, \quad \textrm{and} \quad b^2=q^2+p^2-m^2-n^2 =e_{2}-m^2-n^2, $$
where $ n $ is the NUT parameter, $ m $ is the mass parameter, $ q $ and $ p $ are the electric and magnetic charges respectively. For the sake of simplicity we will combine these two charges and call them as $ e_{2}= q^2+p^2$.
For $b^2<0 $ it has two horizons, just as the Reissner-Nordstr\"{o}m (RN) solution. For $ b^2=0 $, it has, just as the extreme RN solution, a double horizon. However, for $b^2>0$, contrary to the RN solution, it is not singular, but has the (Lorentzian) wormhole topology, the coordinate $ r $ varying along the whole real axis, with two asymptotic regions $r = \pm \infty$. As $r (> 0)$ decreases, the $r =$ constant $2$-spheres shrink until a minimal sphere of area $4 \pi n^2$ (the wormhole neck) is reached for $r=0$, and then expand as $r (< 0)$ continues to decrease.
The determinant of the metric \eqref{trav_NUT_wmhl} for the above mentioned traversable NUT wormhole is given by
$ g=- \left( \sin \left( \theta \right)  \right) ^{2} \left( {n}^{4}+2\,{n}^{2}{r}^{2}+{r}^{4} \right). $\\
Subsequently the Weyl scalar is given by the expression:
\begin{align}
 W= & -\frac{48}{\left( {n}^{2}+{r}^{2}
 \right) ^{6}} \Big( -{n}^{4}+ \left( m-3\,r \right) {n}^{3}+ \left( -
3\,mr+3\,{r}^{2}+e_{{2}} \right) {n}^{2}+ \Big( -3\,m{r}^{2}+{r}^{3}+
\nonumber \\
& 2\,re_{{2}} \Big) n+
{r}^{2} \left( mr-e_{{2}} \right)  \Big)
\Big( {n}^{4}+ \left( m-3\,r \right) {n}^{3}+ \left( 3\,mr-3\,{r}^{2
}-e_{{2}} \right) {n}^{2}+ \Big( -3\,m{r}^{2}+ \nonumber \\
&{r}^{3}+2\,re_{{2}} \Big) n- m{r}^{3}+e_{{2}}{r}^{2} \Big).
\end{align}
\\
\begin{enumerate}
\item \textbf{Weyl scalar proposal:} To evaluate the gravitational entropy density of this spacetime we need the Kretschmann curvature scalar for this metric, which is the following:
\begin{equation}
 K=-\frac{8}{\left( {n}^{2}+{r}^{2} \right) ^{6}} \Sigma .
\end{equation}
The square root of the ratio of the Weyl scalar and the Kretschmann curvature scalar gives us the magnitude of the vector $\boldsymbol{\Psi}$, since $P_{1}^{2}=\dfrac{W}{K} $. Thus $ P_{1} $ is obtained in the form
\begin{align}
\left. P_{1}\right.&=\Big[ -{n}^{4}+ \left( m-3\,r \right) {n}^{3
}+ \left( -3\,mr+3\,{r}^{2}+e_{{2}} \right) {n}^{2}+ \left( -3\,m{r}^{
2}+{r}^{3}+2\,re_{{2}} \right) n \nonumber\\
&+{r}^{2} \left( mr-e_{{2}} \right) \Big]^{1/2} \times
   \sqrt{6}\Big[ {n}^{4}+ \left( m-3\,r \right) {n}^{3}+ \left( 3\,mr-3\,{r}^{2}-e_{{2}} \right) {n}^{2}+ \nonumber\\
&\left( -3\,m{r}^{2}+{r}^{3}+2\,re_
{{2}} \right) n-m{r}^{3}+e_{{2}}{r}^{2} \Big]^{1/2} \times \Sigma^{-\frac{1}{2}} ,
\end{align}
where $ \Sigma $ is given by the following expression:
\begin{align}
\left. \Sigma\right.&=\big[  6\,{m}^{2}({n}^{6}-15\,{n}^{4}{r}^{2}+15\,{n}^{2}{r}^{4}-{r}^{6})-24\,m(3{n}^{6}r-10{n}^{4}{r}^{3}+3
{n}^{2}{r}^{5})- \nonumber\\
& 6\,{n}^{2}({n}^{6}-15\,{n}^{4}{r}^{2}+15\,{n}^{2}{r}^{4}-{r}^{6})
+e_{{2}}(60\,m{n}^{4}r-120\,m{n}^{2}{r}^{3}+12\,m{r}^
{5}+ \nonumber\\
& 12\,{n}^{6}-120\,{n}^{4}{r}^{2}+ 60\,{n}^{2}{r}^{4}-7\,{n}^{4}{e_{{2}}}+34\,{n}^{2}{r}^{2}{e_{{2}}}-7
\,{r}^{4}{e_{{2}}} )\big].
\end{align}
\\
\begin{figure}[ht]
    \centering
    \subfloat[Subfigure 1 list of figures text][]
        {
        \includegraphics[width=0.48\textwidth, height=0.16\textheight]{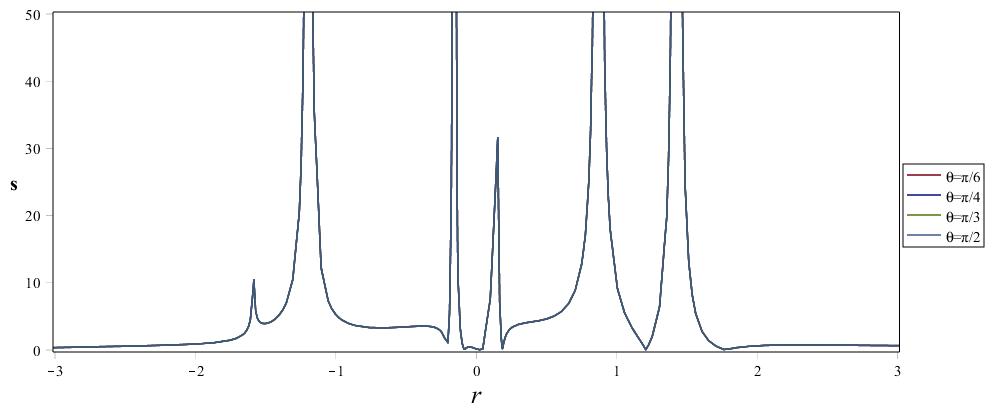}
        \label{fig:subfig3b}
        }
    \subfloat[Subfigure 2 list of figures text][]
        {
        \includegraphics[width=0.48\textwidth]{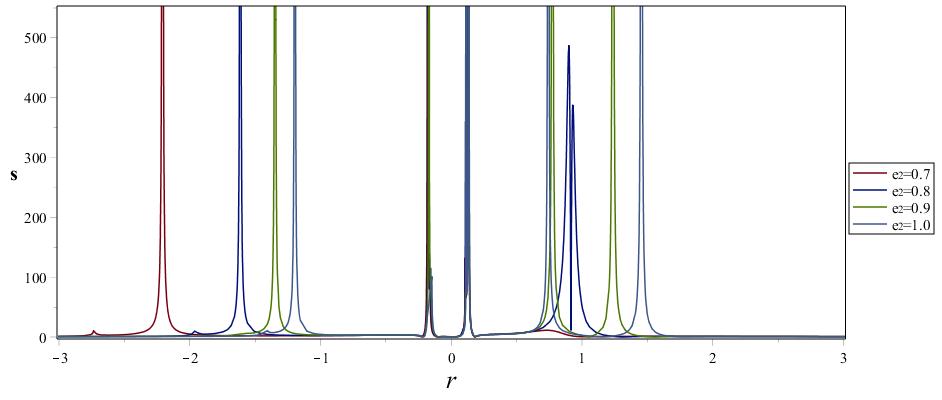}
        \label{fig:subfig4b}
        }
    \caption{Variation of GE density $ s $ of traversable NUT wormhole with $ r $ for different parameters. Here we have only considered the radial contribution as in \eqref{sd1}. In this case the expression of GE density have been calculated using $ P_{1} $.}
   \label{NUT1r}
\end{figure}



\begin{figure}[ht]
    \centering
    \subfloat[Subfigure 1 list of figures text][]
        {
        \includegraphics[width=0.48\textwidth]{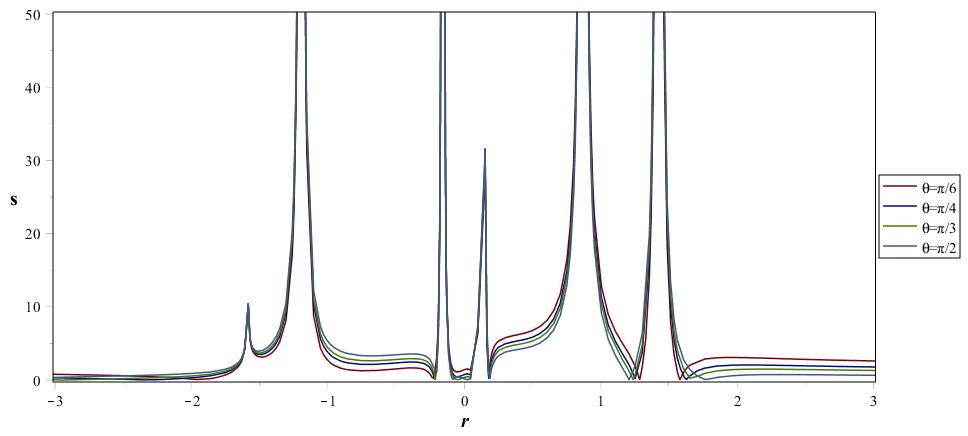}
        \label{fig:subfig3c}
        }
    \subfloat[Subfigure 2 list of figures text][]
        {
        \includegraphics[width=0.48\textwidth, height=0.165\textheight]{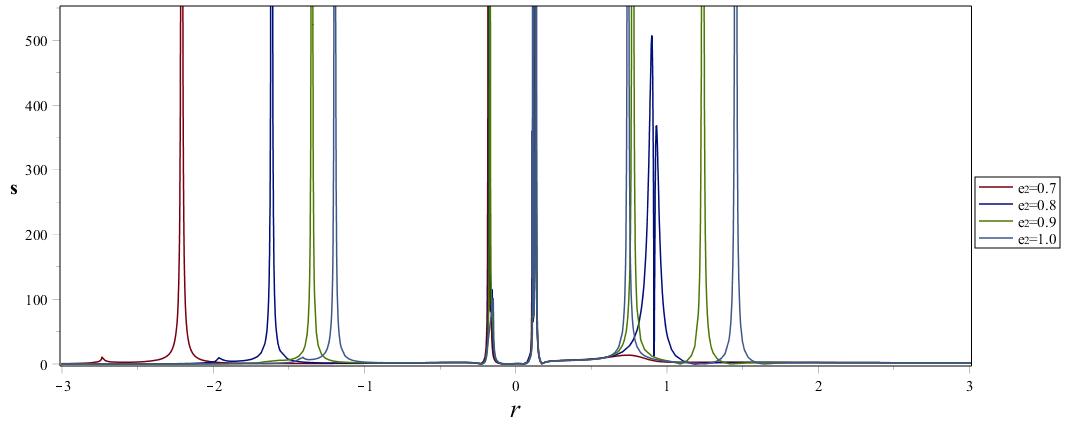}
        \label{fig:subfig4c}
        }
    \caption{Variation of GE density $ s $ of traversable NUT wormhole with $ r $ for different parameters. Here we have considered both the radial and angular contribution as in \eqref{sd2}. Here also the expression of GE density is being calculated using $ P_{1} $.}
   \label{NUT1rt}
\end{figure}





In FIG. \ref{NUT1r} and FIG. \ref{NUT1rt}, the gravitational entropy density for traversable NUT wormhole have been illustrated using the definition involving $ P_{1} $. To generate FIG. \ref{NUT1r}, we have considered only the radial component in the definition of gravitational entropy density, i.e., using \eqref{sd1}, whereas in FIG. \ref{NUT1rt} both the radial and angular contributions have been considered using \eqref{sd2}, as the metric has a nonzero $ g_{t\phi} $ component.

In FIG. \ref{NUT1r}(a), the GE density have been examined by varying the angular orientation $ \theta $. In this case we have taken $ m=0.2,\, n=0.3 $ and $ e_{2}=1 $.
In FIG. \ref{NUT1r}(b) the variation of the GE density with the parameter $ e_{2} $ (charge squared) is shown for the same values of the parameters, i.e. $ m=0.3,\, n=0.3 $ and with $ \theta=\frac{\pi}{4} $.

FIG. \ref{NUT1rt}(a) is drawn by assuming $ m=0.2,\, n=0.3 $ and $ e_{2}=1 $ to show the variation of the gravitational entropy density with angular orientation $ \theta $.
In the second case, i.e. in FIG. \ref{NUT1rt}(b), the effect of charge has been shown, where we have taken $ m=0.3,\, n=0.3 $ and $ \theta=\frac{\pi}{4} $ as our fixed parameters.
From both FIG. \ref{NUT1r} and FIG. \ref{NUT1rt}, which are drawn by considering the expression for $ P_{1} $, it is evident that there are too many discontinuities. This means that $ P_{1} $ does not serve as a good measure of gravitational entropy density in this case. Therefore we considered the alternative proposal with $ P_{2}=C_{abcd}C^{abcd} $ to calculate the GE. Here we will not quote the exact expressions for the GE density since these are extremely complicated and lengthy, but an idea regarding the nature of these expressions can be gathered if one refers to our work on accelerating black holes \cite{GC}.


\begin{figure}[ht]
    \centering
    \subfloat[]{\begin{minipage}[t][9cm][b]{.5\textwidth}
\centering
\includegraphics[width=1.0\textwidth, height=0.16\textheight]{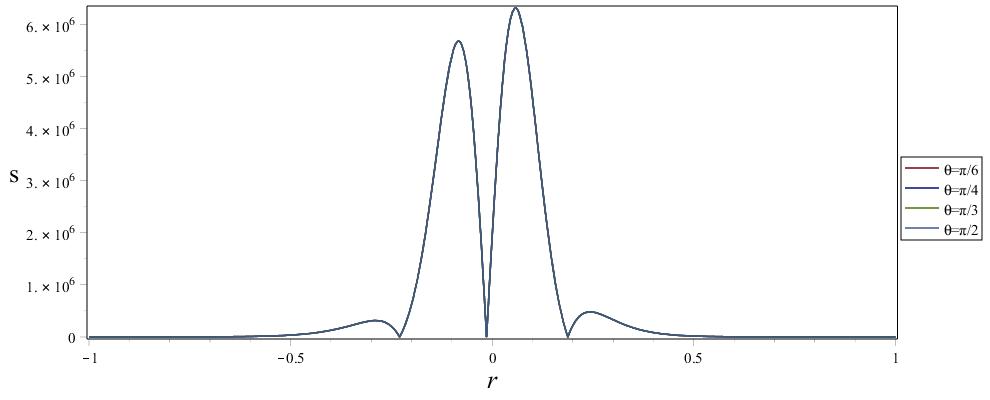}

\includegraphics[width=1.0\textwidth, height=0.16\textheight]{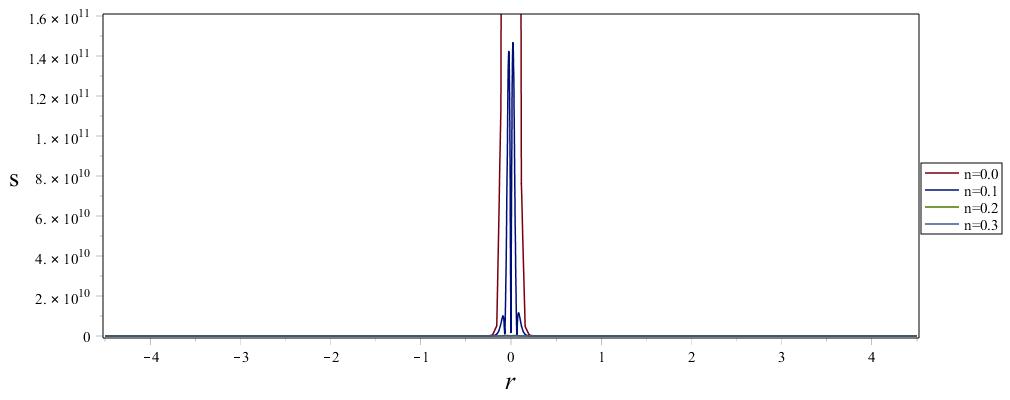}
\end{minipage}}%
       \subfloat[]{\begin{minipage}[t][9cm][b]{.5\textwidth}
\centering
\includegraphics[width=1.0\textwidth, height=0.16\textheight]{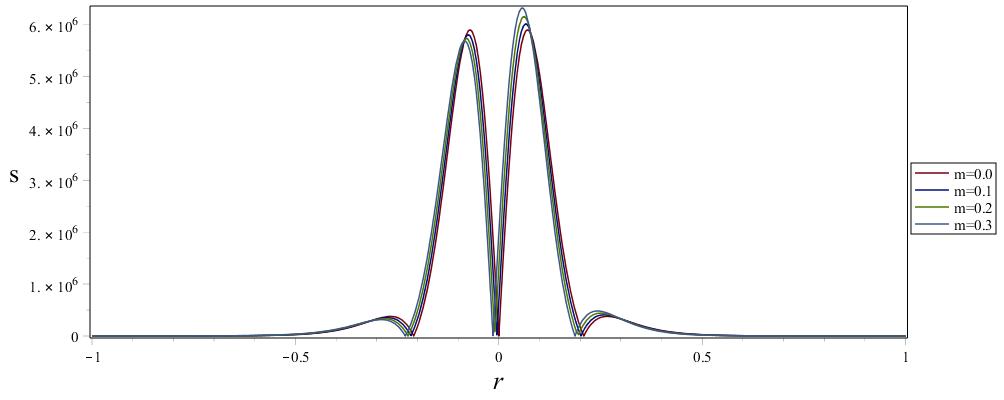}

\includegraphics[width=1.0\textwidth, height=0.16\textheight]{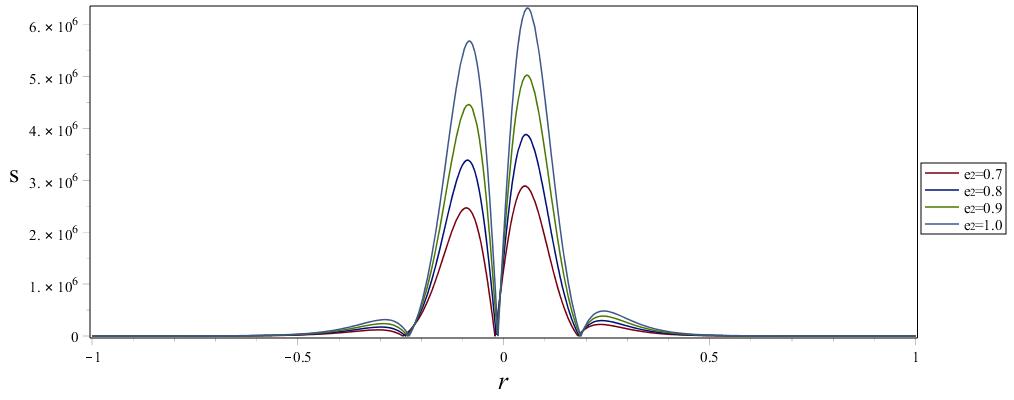}
\end{minipage}}%

    \caption{Variation of GE density $ s $ of traversable NUT wormhole with $ r $ for different parameters. Here we have only considered the radial contribution as in \eqref{sd1}. The definition of GE density is calculated using $ P_{2} $.}
   \label{NUT2r}
\end{figure}
\begin{figure}[ht]
    \centering
    \subfloat[]{\begin{minipage}[t][9cm][b]{.5\textwidth}
\centering
\includegraphics[width=1.0\textwidth, height=0.16\textheight]{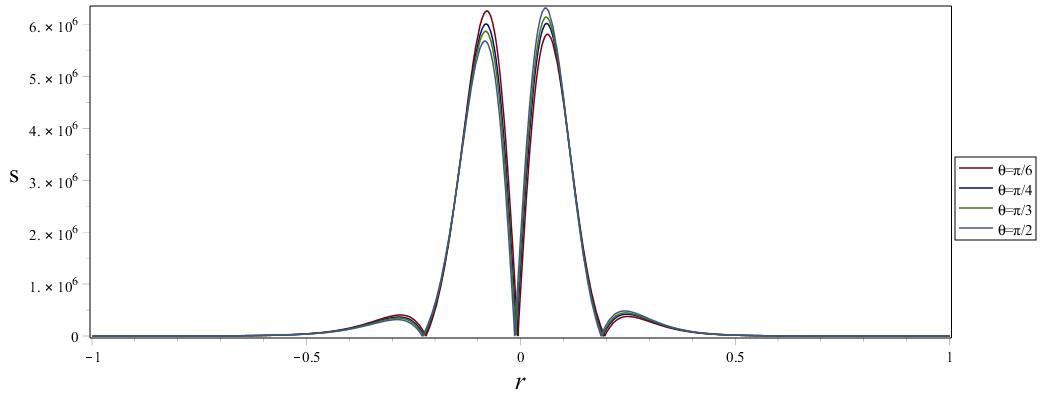}

\includegraphics[width=1.0\textwidth, height=0.16\textheight]{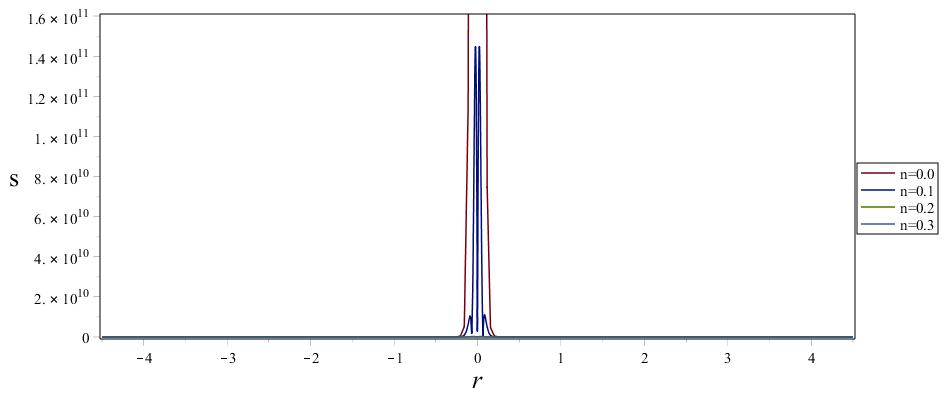}
\end{minipage}}%
       \subfloat[]{\begin{minipage}[t][9cm][b]{.5\textwidth}
\centering
\includegraphics[width=1.0\textwidth, height=0.16\textheight]{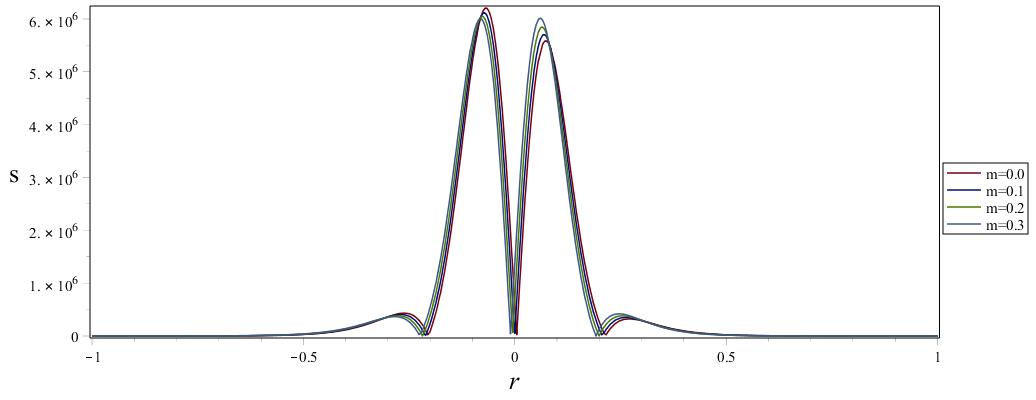}

\includegraphics[width=1.0\textwidth, height=0.16\textheight]{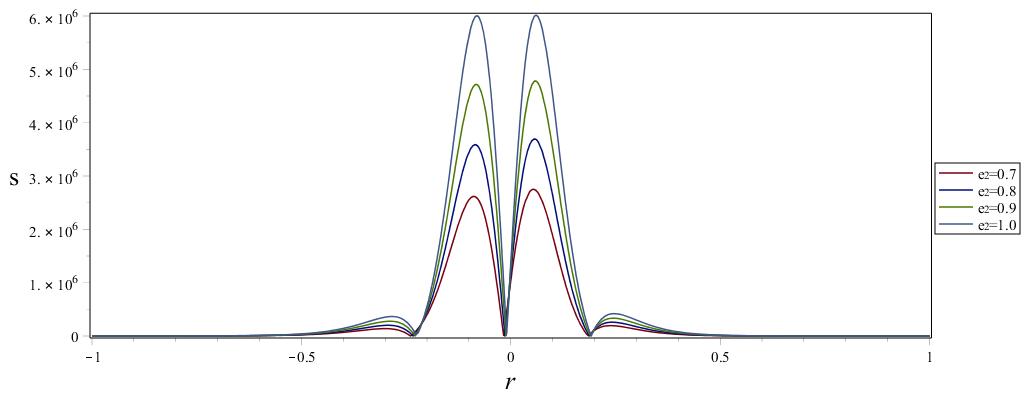}
\end{minipage}}%

    \caption{Variation of GE density $ s $ of traversable NUT wormhole with $ r $ for different parameters. Here we have considered both the radial and angular contribution as in \eqref{sd2}. The definition of GE density have been calculated using $ P_{2} $.}
   \label{NUT2rt}
\end{figure}

FIG. \ref{NUT2r} and FIG. \ref{NUT2rt} shows the investigation of the NUT wormhole using the expression \eqref{p2} for $ P_{2} $ in the definition of gravitational entropy density mentioned in \eqref{sd1} and \eqref{sd2} respectively.

FIG. \ref{NUT2r}(a) shows us the variation of GE density with angular orientation $ \theta $ (for fixed parameters $ m=0.2,\, n=0.3 $ and $ e_{2}=1 $) and the variation with NUT parameter while having fixed parameters as $ m=0.3,\, e_{2}=1 $ and $ \theta=\frac{\pi}{4} $. Similarly in FIG. \ref{NUT2r}(b), the top figure shows the variation of GE density with mass for $ n=0.3,\, e_{2}=1 $ and $ \theta=\frac{\pi}{4} $. The bottom figure of FIG. \ref{NUT2r}(b) gives us the effect of charge on the GE density for a traversable NUT wormhole with $ m=0.3,\, n=0.3 $ and $ \theta=\frac{\pi}{4} $. In FIG. \ref{NUT2r}, all the calculations are done by considering only the radial contribution in the definition of the GE density. In FIG. \ref{NUT2rt} all the figures are drawn by considering both the radial and angular components, which indicates small differences between the two cases. Parameters $ m=0.3,\, n=0.3 $ and $ e_{2}=1 $ are fixed in the top figure of FIG. \ref{NUT2rt}(a), which shows the variation with $ \theta $. In the bottom figure of the same column, the variation with NUT parameter is shown with parameters fixed as $ m=0.3,\, e_{2}=1 $ and $ \theta=\frac{\pi}{4} $. Finally in the column of FIG. \ref{NUT2rt}(b), the variation with mass and charge are shown for $ n=0.3,\, e_{2}=1 $, $ \, \theta=\frac{\pi}{4} $ and $ m=0.3,\, n=0.3 $, $\, \theta=\frac{\pi}{4} $ as the fixed parameters for these cases respectively.
\\
In all the above cases we note that the GE density rises near the throat region and vanishes in the asymptotic limit away from the wormhole. As the GE density function is a polynomial in $ r $ (in fact the curvature scalars themselves are such), for our chosen parameters it has at least two real roots. That is why we encounter two zeroes in the GE density function, meaning that the GE density is localized in two adjacent regions and the location of these regions depend entirely on the parameters of the wormhole, although maximum entropy occurs near the throat region only. From the above analysis we can clearly state that the Weyl scalar proposal involving $ P_{1} $ is not suitable for the analysis of this wormhole, but if we take $ P_{2} $ into consideration, then we can obtain a viable measure of GE density.
\item \textbf{CET proposal:} This spacetime is not strictly of Petrov type D, and hence the CET proposal cannot provide us a unique form of $ s_{grav} $ for this wormhole, as mentioned in the previous section when we provided an overview of the CET proposal.
\end{enumerate}


{$\bullet$} \textbf{NOTE:} In addition to the above mentioned traversable wormholes we have also analyzed and discussed two more types of traversable wormholes, namely, the AdS wormhole, and the very recently proposed wormhole ansatz by Juan Maldacena. Details of these two supplementary cases are presented in the Appendix section of this paper.

\section{Tolman law in wormhole spacetimes}
The Tolman law describes the spatial dependence of locally measured temperature distribution in a gravitational field \cite{SantiagoVisser}.
Of the four WHs that we considered till now, three are static spherically symmetric cases and they are in equilibrium. To expand our study on these systems we apply the famous Tolman law given by Richard Tolman in 1930 \cite{tolman}, which says that thermal equilibrium can exist within a temperature gradient provided that a gravitational field is present. Tolman assumed a sphere of fluid with the line element in the form
\begin{equation}\label{template}
ds^2=-e^{\mu}(dr^2+r^2 d\theta^2+r^2 sin^2\theta d\phi^2)+e^{\nu}dt^2 ,
\end{equation}
where $\mu$ and $\nu$ are functions of $r$. This represents the geometry of perfect fluid having spherical symmetry in full generality, and also the system is static, similar to the cases that we considered. The Tolman law, i.e., the relation between gravitational potential and equilibrium temperature measured by a local observer in proper coordinates, $ T_{0} $, is expressed as follows:
\begin{equation}\label{tol1}
\dfrac{d\, ln T_{0}}{dr}=-\frac{1}{2}\dfrac{d\nu}{dr}.
\end{equation}
In other words, the Tolman law shows that thermodynamic equilibrium in general relativistic spacetimes requires a temperature gradient, and a 4-acceleration to stop free fall. This law provides us with a way to compute temperature variation for a given stationary spacetime along a 4-velocity parallel to a timelike Killing vector field. The extension of this result was presented by R. C. Tolman and P. Ehrenfest in \cite{toleh}. They argued that
\begin{equation}\label{tol2}
T_{0}\sqrt{g_{00}}=\tilde{T}=\textsf{const.}
\end{equation}
i.e., the proper temperature measured by a local observer in thermal equilibrium depends on the position as given in the relation \eqref{tol2}, so that the product remains constant throughout the system. The quantity $ \tilde{T} $ remains constant for that system, and is called the ``Tolman temperature''. Let us now briefly examine the static WHs which we considered so far.
\\~~~~\\
\textbf{Ellis WH:}
Comparing the metric \eqref{ellism} of the Ellis WH with the general spherically symmetric spacetime metric \eqref{template}, and using the relations \eqref{tol1}, \eqref{tol2} we get $ T_{0} $, which is given below:
\begin{equation}
\dfrac{d\, ln T_{0}}{dr}=0 \Rightarrow T_{0}=\textsf{const.}=\tilde{T}.
\end{equation}
We find that the there is no gradient in local temperature. Here the 4-velocity of the static frame is a geodesic field which gives a constant temperature from Tolman's law. Hence the system is in Tolman temperature throughout.
\textbf{Darmour-Solodukhin WH:} Next we consider the DS wormhole. The temperature measured by the local observer is given by the expression \eqref{DST0}:
\begin{equation}\label{DST0}
\frac{1}{T_{0}}\dfrac{dT_{0}}{dr}=-\dfrac{m}{r^2(f(r)+\lambda^2)}\Rightarrow T_{0}=\dfrac{\mathbf{B}}{\sqrt{f(r)+\lambda^2}}.
\end{equation}
Here $\mathbf{B}$ is an integration constant. We can also validate the result \eqref{tol2} and express the local temperature in terms of Tolman temperature as given in \eqref{DST01}:
\begin{equation}\label{DST01}
T_{0}\sqrt{g_{00}}=\mathbf{B}=\tilde{T}\Rightarrow T_{0}=\dfrac{\tilde{T}}{\sqrt{f(r)+\lambda^2}}.
\end{equation}
Here we have the option to find the $ \tilde{T} $. As this metric reduces to the Schwarzschild metric for $ \lambda=0 $, the limit $ T_{0}(\infty)=\dfrac{\tilde{T}}{\sqrt{1+\lambda^2}} $, reduces to Hawking-Bekenstein temperature for $ \lambda=0 $. Therefore, $ \tilde{T}=T_{BH} $, and consequently $ T_{0}=\dfrac{T_{BH}}{\sqrt{f(r)+\lambda^2}} $ gives us the local temperature in terms of the Hawking-Bekenstein temperature $ (T_{BH}) $. Similarly at the throat, the temperature measured by the local observer is $ T_{0}(2m)=\frac{T_{BH}}{\lambda} $, which blows up at the event horizon for the Schwarzschild BH.
\\
\textbf{Exponential metric WH:} Finally we will discuss the exponential metric case. Like the previous cases we calculate the temperature measured by the local observer and find that
\begin{equation}
\frac{1}{T_{0}}\dfrac{dT_{0}}{dr}=-\frac{m}{r^2}\Rightarrow T_{0}=\mathbf{C} e^{\frac{m}{r}}.
\end{equation}
Here $ \mathbf{C} $ is the integration constant and it is indeed the Tolman temperature. In \eqref{EXPT01} we have expressed the local temperature in terms of the Tolman temperature:
\begin{equation}\label{EXPT01}
T_{0}\sqrt{g_{00}}=\mathbf{C}=\tilde{T}\Rightarrow
T_{0}=\tilde{T}e^{\frac{m}{r}}.
\end{equation}

In all the above three WH cases, we found that both $ \Theta=0 $ and $ \sigma_{ab}=0 $, resulting in a gravitational temperature $T_{grav} = \dfrac{1}{2\pi}\vert\dot{u_{a}}z^{a}\vert $, i.e., the gravitational temperature is proportional to the acceleration. We recall the Tolman law to write $ \dfrac{dT}{T} \sim -a_{l} dx^{l} $, and consequently infer about the relationship between $ T_{0} $ and $ T_{grav} $. For the Ellis WH, the relationship becomes trivial, but for the other two WHs the relationship is non-trivial. The relation between the two temperatures for the Darmour-Solodukhin WH becomes: $ T_{0}=T_{grav}\left(\dfrac{2\pi \tilde{T}}{m}\right)\dfrac{r^2}{\sqrt{r-2m}}$, and for the exponential metric WH it is $T_{0}=T_{grav} \left(\dfrac{2\pi\tilde{T}}{m}\right) r^2 e^{2m/r}$. It is clear that the two temperatures are somehow related to each other as the gravitational temperature is proportional to the acceleration, whereas in the Tolman case, the temperature gradient is proportional to the acceleration. For a fixed $ r $, if one temperature is known, then the another one can be computed from these relations.

\section{Summary}
In this paper we have analyzed the gravitational entropy of traversable wormholes from the perspective of two different proposals, and have tried to make a comparison between the two proposals in terms of their applicability to the different wormhole geometries. We considered a variety of traversable WHs in order to analyze thoroughly the behaviour of GE in such cases, and determine whether the GE proposals considered by us can provide us with a viable measure of GE. Indeed, we found that the GE proposals do give us a consistent measure of GE in most of them.

First we adopted a phenomenological approach of determining the gravitational entropy of traversable wormholes \cite{entropy1,entropy2}. Secondly, we have also checked the result of applying the CET proposal \cite{CET} on traversable wormhole geometries of Petrov type D (excepting one above and one in the Appendices). It is clear that although the CET proposal is very successful in various astrophysical and cosmological cases, the definition of temperature given by CET  fails in the case of Ellis wormhole. In a more cosmologically significant case, i.e., traversable wormholes that mimic black holes, namely the Darmour-Solodukin wormhole, we find that it possesses a viable gravitational entropy for both the Weyl scalar proposal and the CET proposal, with some differences between them.

Although both the Rudjord and the CET proposals were successful in the case of the exponential metric wormhole, their behavior differed which was expected as the former was a purely geometric proposal whereas the latter one is based on relativistic thermodynamics. The Rudjord proposal is directly inspired from the fact that black hole entropy is related to its geometry, but the CET proposal considers a much more local view of the system. In the case of NUT wormholes, both the WHs were checked thoroughly using the Rudjord method and additional changes were made according to \cite{entropy2} both in the magnitude of $ P_{i} $ and in the additional vector component contributions. For the case of the first NUT wormhole the definition of $ P_{1} $ is not suitable as it gives multiple discontinuities at different $ r $ for different combination of parameters, and it did not improve even after introducing the additional angular contribution of $ \theta $ in the definition of the gravitational entropy density. In the appendix, the case for the traversable AdS wormhole, which is also a NUT wormhole, the $ P_{1} $ definition is not so bad as in the previous one. The overall distribution of the gravitational entropy density changed significantly by tilting in a direction if we introduce additional angular contribution, but for $ P_{2} $ the behaviour is regular and it only changes shape because the angular component brings in additional entropy into the system. Overall, for wormholes where $ g_{t\phi} $ is nonzero, the definition using $ P_{2} $ and both the vector directional components gives us a more robust result. Below we are presenting our results in a systematic manner.

\begin{enumerate}
\item For the Ellis wormhole we got a well-behaved gravitational entropy density using the Weyl scalar (Rudjord) proposal. However for the CET proposal, though we have a non zero gravitational energy density, but the gravitational temperature becomes zero thereby making it impossible to calculate the GE. As the definition of the gravitational temperature is \emph{ad hoc} in the CET proposal, we have introduced a new gravitational temperature to obtain the GE, which we found to be well-behaved. The GE density in the Weyl scalar proposal is always vanishing at the throat of the WH, and the radial variation rate for the CET GE is always finite at the throat.

\item For the Darmour-Solodukhin wormhole, both the Weyl scalar proposal and the CET proposal gives us a finite viable measure of gravitational entropy. The rate of radial variation of GE and the local piecewise CET GE blows up at the throat of the WH.
\item For the exponential metric wormhole too, we obtain viable measures of GE for both the Weyl scalar and the CET proposals. In the Weyl scalar case, the GE density peaks near the throat on the other side, and the CET rate of radial variation of GE decreases and goes to zero, crossing the throat on the other side of the universe.
\item For the NUT wormhole we have computed the gravitational entropy density using both the functions $ P_{1} $ and $ P_{2} $, taking into account both the radial and angular components in the definition of $ s $. We have shown explicitly that for these wormholes, $ P_{2} $ gives us a viable measure of GE density. In this case, the CET proposal cannot provide us with an unique expression for gravitational entropy as the metric is not strictly of Petrov type D.
\end{enumerate}
Of the four WHs considered above, three of them are static spherically symmetric cases which are in equilibrium (i.e., the Ellis WH, the DS WH, and the exponential metric WH). Therefore, we applied the famous Tolman law on these spacetimes,  which says that thermal equilibrium can exist within a temperature gradient if a gravitational field is present there. We determined the Tolman temperature, and compared it to the temperature of the gravitational field.

In the Appendix section we have included the analysis for the NUT wormhole in the AdS spacetime as an additional case study. In this case, the Weyl proposal is applied for both the definitions of $ P_{1} $ and $ P_{2} $. In both these cases, the radial and the angular contributions together gives us a complete picture, but as these metrics have a nonzero $ g_{t\phi} $ component, we want to make a point that $ P_{2} $ (with both the radial and angular contributions) gives us a far more complete and viable measure of gravitational entropy density. Another traversable wormhole system proposed by Maldacena et al have been analyzed to examine the gravitational entropy behavior of a traversable wormhole connecting two black holes. In the Maldacena wormhole ansatz, both the Weyl proposal and the CET proposal gives us zero gravitational entropy density, making the system nonphysical from the thermodynamic perspective of gravitational entropy. For the sake of completeness we have analyzed the extremal magnetic BHs of this system and showed that the relevant functions, i.e., the gravitational energy density, the gravitational temperature, the ratio of curvature scalars and the gravitational energy density to gravitational temperature are continuous in the BH and WH junction. An important thing to note here is that the gravitational entropy of the extremal magnetically charged black holes on a horizon conforms with the Hawking-Bekenstein entropy of a black hole.

\section{Concluding remarks}

In this paper we have shown that the behavior of the gravitational entropy density function of a system depends on the definition employed to compute its value, and varies on a case by case basis. The two proposals which have been compared in this article are the Weyl scalar proposal and the CET proposal. In some cases the pure geometric method, i.e., the Weyl scalar proposal, provides us a good picture for the ratio of curvature scalars, $ P_{1} $. In the case of wormholes which have a nonzero $ g_{t\phi} $ term, the pure Weyl square $ P_{2} $ seems to work better. It is also important to consider both the radial and angular contributions in the definition of gravitational entropy. On the other hand, the CET proposal provides an unique gravitational entropy only for the Petrov type D and N spacetimes, but yields a far more nuanced result as it originates from the relativistic thermodynamic considerations. The definition of gravitational temperature proposed in the CET paper \cite{CET} does not give us an acceptable value in some of the cases. It has already been mentioned that the CET proposal is not dependent on the definition of temperature. Therefore a new definition of gravitational temperature can be used to suit the purpose. An important point which one must remember in this context is that, here we are considering only the wormhole geometry although the matter source may differ in each case. As we are interested in the gravitational entropy, the nature of matter source for the wormholes does not affect our analysis. Another important point to be noted is that the CET proposal have been applied to wormhole systems for the first time, and it is interesting to note its difference as compared to the Rudjord proposal.

We want to reiterate that by being frame-independent, the Weyl scalar proposals lost their connection
with the worldlines of physical fluids, and for this reason we cannot use these proposals to examine the GE for an important class of observers that traverse a WH, like the one we considered for the Ellis WH using the CET proposal. Since in a spherically symmetric static frame, the 4-velocity is aligned along the time axis, the only straightforward interpretation of the expressions in our calculations (obtained for the Weyl scalar proposals) is the variation of GE density along the radial direction, defined in this frame. These calculations demonstrate a GE that yields a different constant for each 2-sphere labelled by specific values of the coordinate $ r $. Other criticisms on the Weyl scalar proposals are available in \cite{CET}.

We argue that for any traversable wormhole to exist, it must have a viable gravitational entropy. So it is important to compare and study the proposals of gravitational entropy in the context of traversable wormholes. Obviously, the CET proposal of gravitational entropy is much more physically important, as it takes into account the restrictions imposed by relativistic thermodynamics. Not only is the interpretation of the gravitational energy density established on a firm footing, but it also provides the freedom to choose the gravitational temperature.
For spacetimes not belonging to Petrov types D and N,  several algebraic decompositions of the Bel-Robinson tensor can be computed, and the resulting expressions for CET gravitational entropy can also be studied. The CET proposal depends on frames, and this frame-dependence is a very important property that allows a link with thermodynamics. Though the Weyl scalar proposals does not contain the thermodynamic aspects in it, they can provide a good viable measure of GE for cases where the CET and other proposals may not. So we want to emphasize that a given proposal might work for some spacetimes and not for others. However, all spacetimes are not equally interesting or valid from a physical point of view. Therefore the GE proposals which work for the most physically meaningful spacetimes are the most interesting and relevant ones. From this view point, the connection to thermodynamics and dependence on frames makes the CET proposal more physically viable. Hence it appears that it will indeed be challenging to develop an universal proposal of gravitational entropy.

Another important point to note here is that we are not checking whether the entropy is increasing with time but whether these proposals can provide us with a viable expression of GE, as for the equilibrium states of the static cases the condition $ u^{a}\partial_{a}S_{grav}=0 $ holds. However, the thermodynamic study of equilibrium states for such self-gravitating systems are very interesting, like the study of Antonov instability in relativistic systems \cite{paddy,roupas} can important in these cases. Such studies of equilibrium states of these self-gravitating systems may provide us with new information.

In conclusion, from our analysis and the corresponding plots, it appears that for the traversable wormholes the GE density function will be well-defined if the definition of the vector field $ \mathbf{\Psi} $ is modified, either in the magnitude ($ P $), or in its direction (having additional angular components). A similar feature was observed in our earlier work \cite{GC} in the case of accelerating black holes. For the CET proposal, a new definition of gravitational temperature can be used in order to avoid the appearance of null gravitational temperature. All the static cases produced well defined unique GE in both the Weyl scalar and the CET GE proposals, with minor differences, indicating that the concept of gravitational entropy although new and contentious, does have some sense of theoretical robustness.

\section*{Data Availability}
Data sharing is not applicable to this article as no datasets were generated or analysed during the current study. All figures were plotted with Maple software using the theoretical equations.

\section*{Appendix: SOME ADDITIONAL NOTES}
In this section we will briefly analyse the case of two more traversable wormholes. These are additional case studies to support our conclusions, but less significant from the point of view of the observed features.

\subsection*{Appendix-I: Traversable AdS wormhole}

The following metric represents a stationary NUT wormhole with a negative cosmological constant with a nonlinear sigma model as source, which was  illustrated in \cite{WH2}:
\begin{equation}
ds^2=dz^2+\dfrac{\rho^2(z)}{4}\left[ -Q^2(d\tau+cos\theta d\phi)^2 +(d\theta^2+sin^2\theta d\phi^2) \right],
\end{equation}
where $ \rho(z)=\sqrt{\dfrac{3(K-8)}{4\vert\Lambda\vert}} \cosh\left(\dfrac{\vert\Lambda\vert^{1/2}}{\sqrt{3}}z\right), $ and $ Q^2= \dfrac{K}{4}.$
In addition to this, $ \Lambda<0 $ if the system has to satisfy the Einstein equations. Here $ Q $ is the NUT parameter and must be an even integer for the solution to be single-valued \cite{GP}. The asymptotic NUT-AdS regions with $ z\rightarrow \pm \infty $ are connected by this wormhole at the throat $ z=0 $. This spacetime has no curvature singularities and is locally regular, and is therefore an object of our interest in this paper.

The determinant of the metric for the traversable AdS wormhole is given by the following:
\\
\begin{equation}
g=-{\frac {27\, \left( K-8 \right) ^{3} \left( \cosh \left( 1/3\,\sqrt {
 \left| \Lambda \right| }\sqrt {3}z \right)  \right) ^{6}{K}^{2}
 \left( \sin \left( \theta \right)  \right) ^{2}}{65536\, \left(
 \left| \Lambda \right|  \right) ^{3}}}.
\end{equation}
Consequently the Weyl curvature scalar square is given by the following:
\begin{equation}
 W={\frac {4\, \left( {K}^{2}+16 \right) ^{2}{\Lambda}^{2}}{27\, \left(
\cosh \left( 1/3\,\sqrt { \left| \Lambda \right| }\sqrt {3}z \right)
 \right) ^{4} \left( K-8 \right) ^{2}}}.
\end{equation}
The Kretschmann curvature scalar ($ \tilde{K} $) for the AdS wormhole is given by the equation:
\begin{align}
\left. \tilde{K}\right.&= \frac {11\,{\Lambda}^{2}}{36\, \left( \cosh \left( 1/3\,\sqrt {
 \left| \Lambda \right| }\sqrt {3}z \right)  \right) ^{4} \left( K-8
 \right) ^{2}} \Bigg[ {\frac {96\, \left( \cosh \left( 1/3\,\sqrt {
 \left| \Lambda \right| }\sqrt {3}z \right)  \right) ^{4} \left( K-8
 \right) ^{2}}{11}}-\nonumber\\
 &{\frac { \left( 8\,K-64 \right)  \left( {K}^{2}+12
\,K-32 \right)  \left( \cosh \left( 1/3\,\sqrt { \left| \Lambda
 \right| }\sqrt {3}z \right)  \right) ^{2}}{11}}+{K}^{4}+ {\frac {8\,{K
}^{3}}{11}}+{\frac {368\,{K}^{2}}{11}}- \nonumber\\
&{\frac {256\,K}{11}}+{\frac {
3072}{11}} \Bigg].
\end{align}
\begin{figure}[ht]
    \centering
    \subfloat[Subfigure 1 list of figures text][]
        {
        \includegraphics[width=0.48\textwidth, height=0.155\textheight]{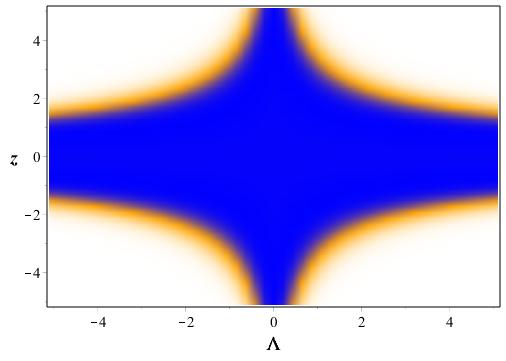}
        \label{fig:subfig3d}
        }
    \subfloat[Subfigure 2 list of figures text][]
        {
        \includegraphics[width=0.48\textwidth, height=0.157\textheight]{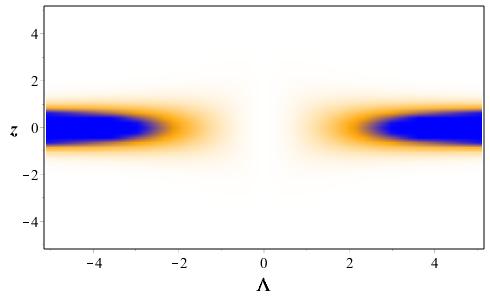}
        \label{fig:subfig4d}
        }
    \caption{(a)Variation of $ P_{1}^2 $ for traversable AdS wormhole with $ z $ and the parameter $\Lambda$. (b)Variation of $ P_{2} $ for traversable AdS wormhole with $ z $ and the parameter $\Lambda$. In both the cases the NUT parameter is $K = 16$. Here the blue region represents high positive values and it gradually decreases through the yellow region to the white colored region.}
   \label{AdSpp}
\end{figure}
\begin{enumerate}
\item \textbf{Weyl scalar proposal:} From the above expressions, the ratio of the two curvature scalars is obtained along straightforward calculations, and is given below:
\begin{align}\label{AdSp1}
\left. P_{1}\right.&= \frac {4\,\sqrt {33}}{33}\Bigg[ \left( {K}^{2}+16 \right) ^{2}
 \Bigg( {\frac {96\, \left( \cosh \left( 1/3\,\sqrt { \left| \Lambda
 \right| }\sqrt {3}z \right)  \right) ^{4} \left( K-8 \right) ^{2}}{11}}  \nonumber \\
  & - {\frac { \left( 8\,K-64 \right)  \left( {K}^{2}+12\,K-32 \right)
 \left( \cosh \left( 1/3\,\sqrt { \left| \Lambda \right| }\sqrt {3}z
 \right)  \right) ^{2}}{11}} + {K}^{4}+{\frac {8\,{K}^{3}}{11}}+ \nonumber \\
 &{\frac {368\,{K}^{2}}{11}}-{\frac {256\,K}{11}}+{\frac {3072}{11}} \Bigg) ^{-
1}\Bigg]^{\frac{1}{2}}.
\end{align}
The ratio of curvature scalars, i.e. $P_{1}^2$, is given in the expression \eqref{AdSp1}. For the sake of clarity, it is also illustrated graphically in FIG.\ref{AdSpp}(a).

\begin{figure}
\centering
\subfloat[]{\label{main:a1}\includegraphics[width=0.6\textwidth]{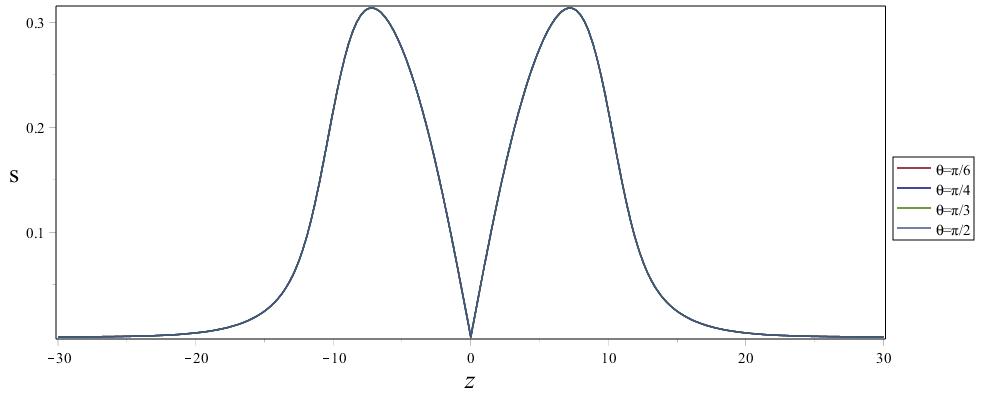}}
\par\medskip
\centering
\subfloat[]{\label{main:b1}\includegraphics[width=0.6\textwidth]{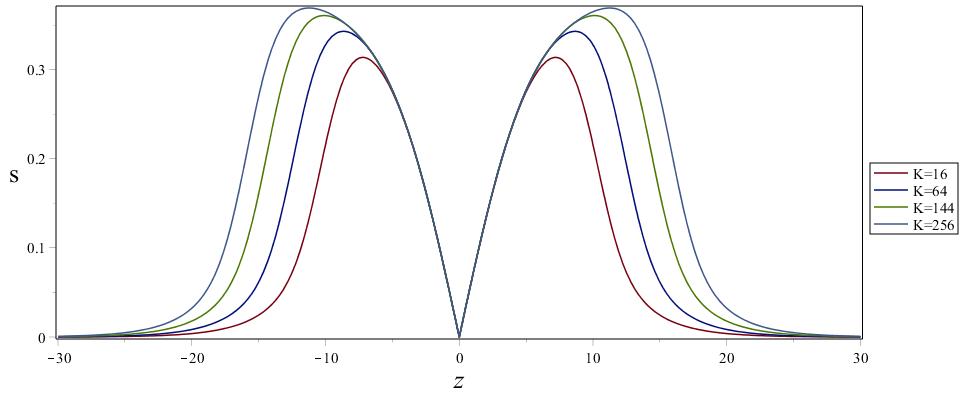}}
\par\medskip
\centering
\subfloat[]{\label{main:c1}\includegraphics[width=0.7\textwidth]{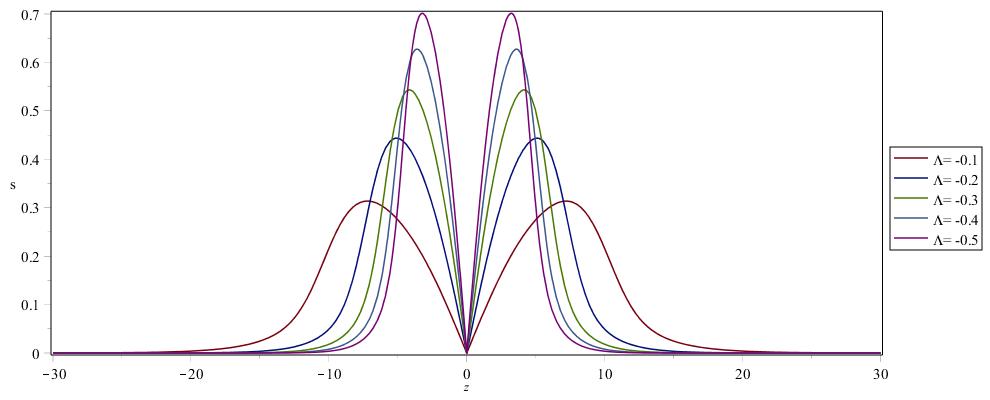}}

\caption{Variation of entropy density $ s $ of AdS traversable wormhole with $ z $ for different parameters. Here we have considered only the radial contribution as defined in \eqref{sd1}. Also the definition of gravitational entropy density is calculated using $ P_{1} $.}
\label{ads1z}
\end{figure}

\begin{figure}

\centering
\subfloat[]{\label{main:a2}\includegraphics[width=0.6\textwidth]{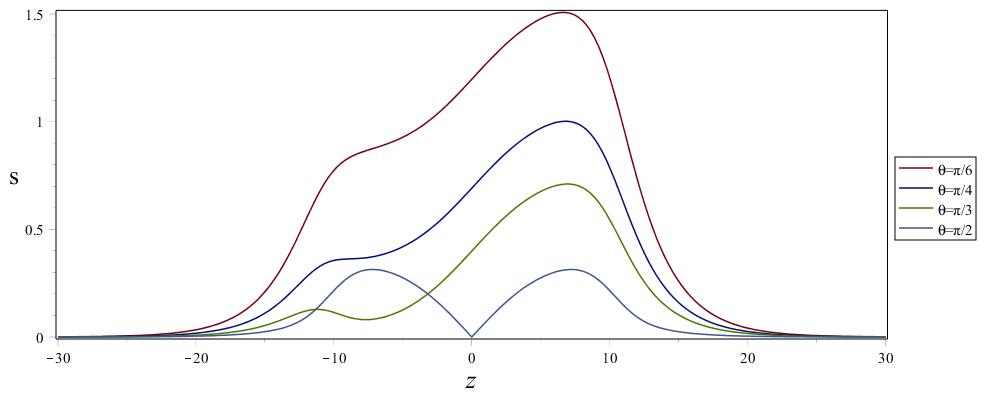}}
\par\medskip
\centering
\subfloat[]{\label{main:b2}\includegraphics[width=0.6\textwidth]{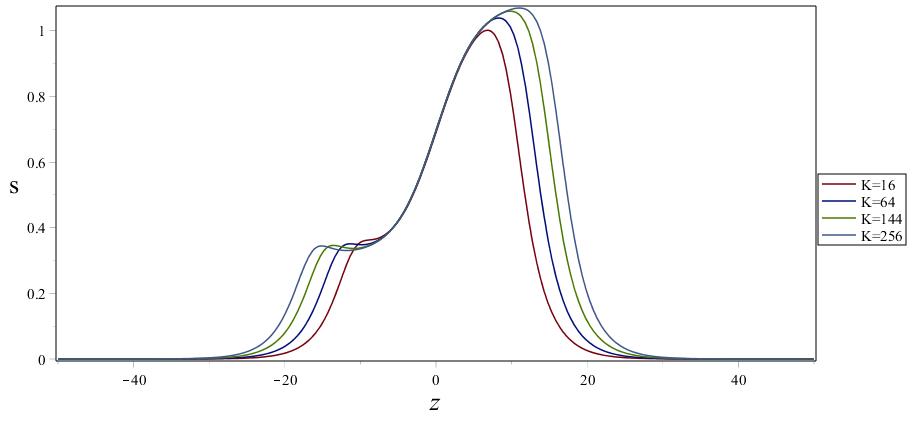}}
\par\medskip
\centering
\subfloat[]{\label{main:c2}\includegraphics[width=0.7\textwidth]{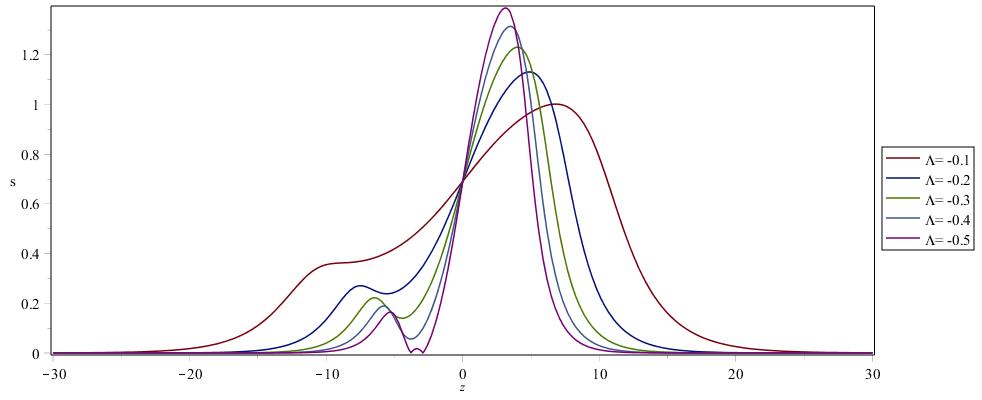}}

\caption{Variation of entropy density $ s $ of AdS traversable wormhole with $ z $ for  different parameters. We have considered both the radial and angular contribution \eqref{sd2} and the gravitational entropy density is calculated using $ P_{1} $.}
\label{ads1zt}
\end{figure}

In FIG.\ref{ads1z} and FIG. \ref{ads1zt} we have shown the variation of gravitational entropy density with different parameters. In both these cases, $ P_{1} $ is being used for the calculations. In FIG.\ref{ads1z} we have only taken the contribution of the radial component while determining the gravitational entropy density and in FIG. \ref{ads1zt} both the radial and angular contributions are taken into account.

In FIG.\ref{ads1z}(a) the variation of the gravitational entropy density of AdS wormhole is shown with angular orientation $\theta  $, where we have fixed the other parameters as $ \Lambda=-0.1, K=16 $. In this case we see no change as we did not include the angular contribution into our analysis. Next in FIG.\ref{ads1z}(b), the variation with the parameter $ K $ is being studied for $ \Lambda=-0.1, \theta=\dfrac{\pi}{4} $. As the value of $ K $ increases, so does the value of gravitational entropy density. In the last figure FIG.\ref{ads1z}(c), the variation of the gravitational entropy density of traversable AdS wormhole is shown with various negative values of cosmological constant $ \Lambda $ for the fixed parameter values as $ K=16, \theta=\frac{\pi}{4} $. Here we can clearly observe that with increasing negative value of the cosmological constant the peak value of the gravitational entropy density increases.
FIG. \ref{ads1zt} shows these variations with a higher sensitivity including both the radial and angular contributions in the entropy density. FIG. \ref{ads1zt}(a) shows the variation with the angular orientation while fixing the other parameters at $ \Lambda=-0.1, K=16 $. Here each angular orientation gives different gravitational entropy unlike the previous case. FIG. \ref{ads1zt}(b) gives us the nature of variation of entropy density with $ K $ when we fix $ \Lambda=-0.1, \theta=\dfrac{\pi}{4} $. In FIG. \ref{ads1zt}(c) the dependence of gravitational entropy density on the cosmological constant is being depicted, where we have chosen our free parameters as $ K=16, \theta=\frac{\pi}{4} $. Here the overall dependence remains the same but the extra contribution from the angular components in the gravitational entropy density makes it non zero at the throat region unlike in the previous case, which means that the gravitational entropy density is continuous through the wormhole throat.
 \begin{figure}
\centering
\subfloat[]{\label{main:a3}\includegraphics[width=0.6\textwidth]{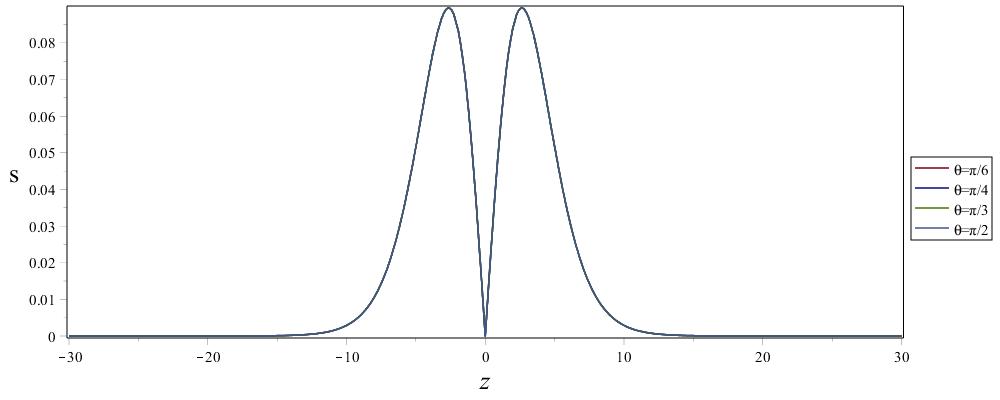}}
\par\medskip
\centering
\subfloat[]{\label{main:b3}\includegraphics[width=0.6\textwidth]{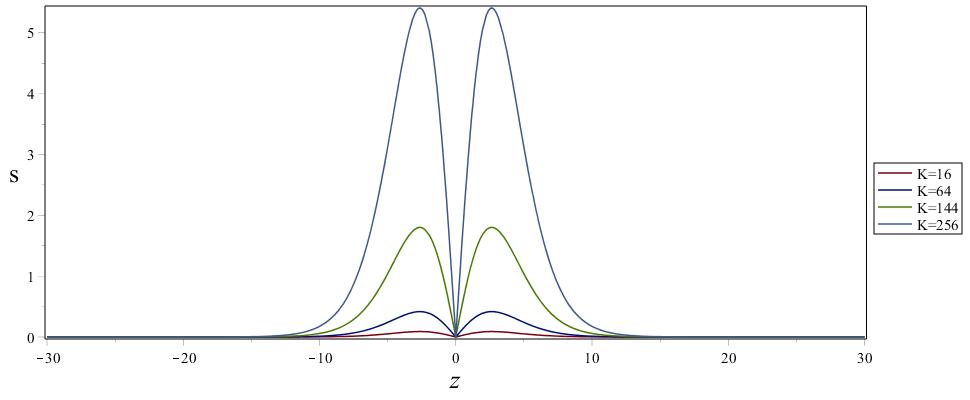}}
\par\medskip
\centering
\subfloat[]{\label{main:c3}\includegraphics[width=0.7\textwidth]{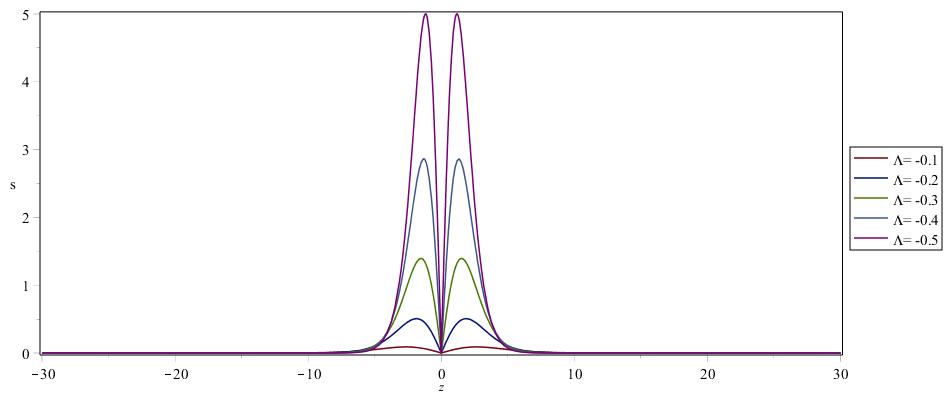}}

\caption{Variation of entropy density $ s $ of AdS traversable wormhole with $ z $ for  different parameters. Here we have considered only the radial contribution as in \eqref{sd1}. The gravitational entropy density is calculated using $ P_{2} $.}
\label{ads2z}
\end{figure}

\begin{figure}

\centering
\subfloat[]{\label{main:a4}\includegraphics[width=0.6\textwidth]{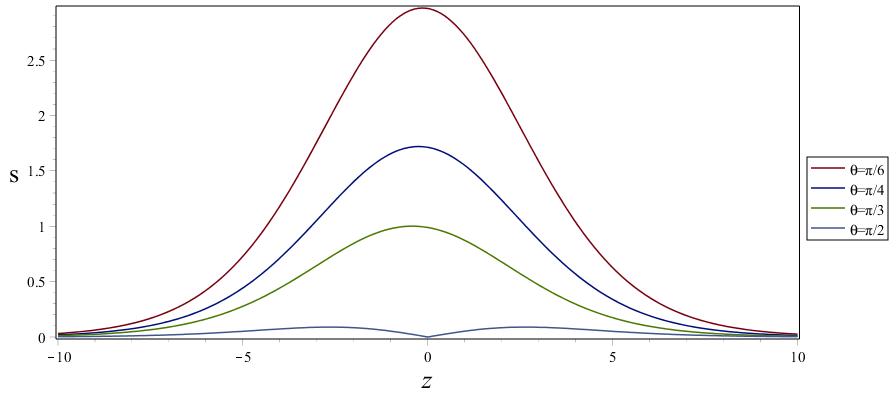}}
\par\medskip
\centering
\subfloat[]{\label{main:b4}\includegraphics[width=0.6\textwidth]{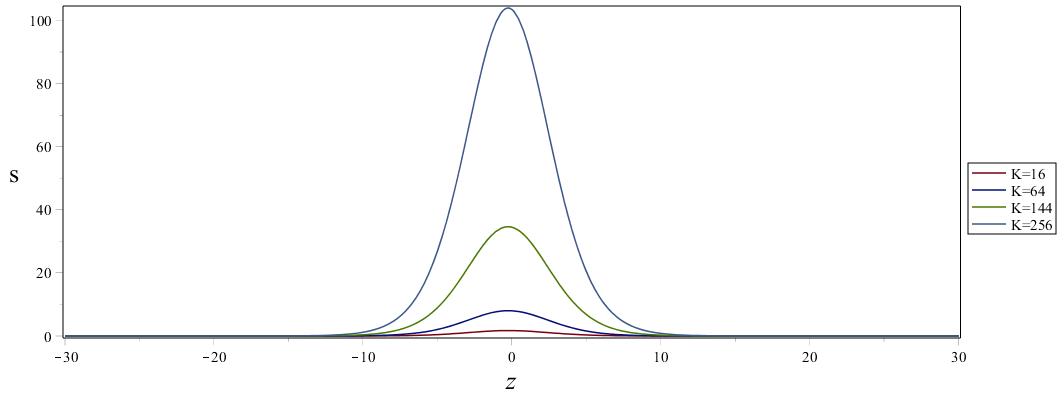}}
\par\medskip
\centering
\subfloat[]{\label{main:c4}\includegraphics[width=0.7\textwidth]{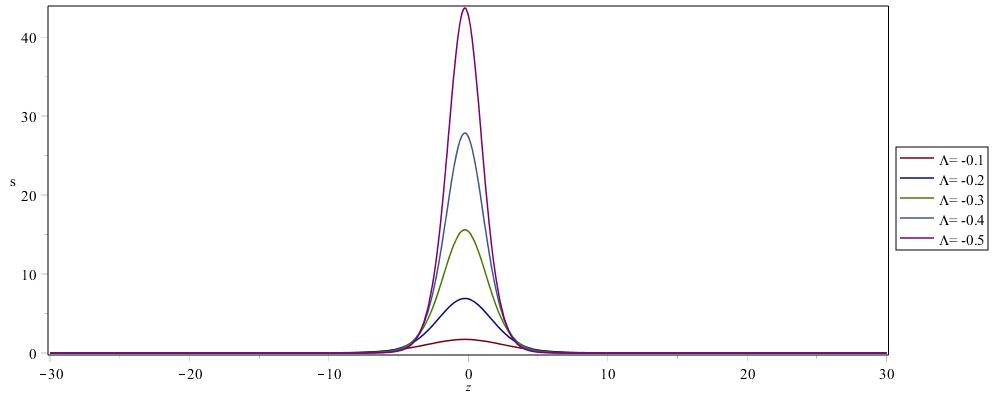}}

\caption{Variation of entropy density $ s $ of AdS traversable wormhole with $ z $ for  different parameters. Here we have considered both the radial and angular contribution \eqref{sd2}, and the gravitational entropy density is calculated using $ P_{2} $.}
\label{ads2zt}
\end{figure}

As we know that the $ g_{t\phi} $ component is also nonzero in the metric of AdS traversable wormhole, therefore the alternative definition of the gravitational entropy using $ P_{2} $ must be applied to see how the result differs from the former case. The graphical representation of $ P_{2} $ is shown in  FIG.\ref{AdSpp}(b).

Consequently in FIG. \ref{ads2z} and FIG. \ref{ads2zt} we have used $ P_{2} $ as the definition and used the radial contribution only to draw the graphs (in FIG. \ref{ads2z}), and in the later figures both the contributions of radial and angular components have been considered.

In FIG. \ref{ads2z}(a) we have chosen $ \Lambda=-0.1, K=16 $ to show the variation with the angular orientation while in FIG. \ref{ads2z}(b), the values $ \Lambda=-0.1, \theta=\dfrac{\pi}{4} $ are fixed to show the variation with the parameter $ K $. Similar to the case for $ P_{1} $, here too there are no changes in FIG. \ref{ads2z}(a) while the variations in FIG. \ref{ads2z}(b) are also similar to that of $ P_{1} $ except that the graphs are way more compact, i.e. the gravitational entropy density is localized in a much more smaller region when we consider $ P_{2} $. FIG. \ref{ads2z}(c) shows the variation of gravitational entropy density with the negative values of cosmological constant with the fixed parameters as $ K=16, \theta=\frac{\pi}{4} $ where we find that the magnitude of $ s $ increases with increasing negative values of the cosmological constant $ \Lambda $.

FIG. \ref{ads2zt}(a) shows the variation with $ \theta $ for $ \Lambda=-0.1, K=16 $. The introduction of the angular component changes the entropy density drastically for higher values of the angular orientation $ \theta $. If we fix $ \Lambda=-0.1, \theta=\dfrac{\pi}{4} $ as constants, then we obtain FIG. \ref{ads2zt}(b), which shows the variation with $ K $. Finally FIG. \ref{ads2zt}(c) shows the variation with $ \Lambda $ for the fixed parameter $ K=16, \theta=\frac{\pi}{4} $. The introduction of the angular component reduces the gravitational entropy density from a double peaked one to a single peaked function but the overall evenness with respect to the radius is not lost.

In general, the gravitational entropy density increases near the throat region when we consider both the radial and angular contribution for our analysis. Therefore we can say that the procedure involving the $ P_{2} $, which includes both the radial and angular contribution, is more suitable for the analysis of gravitational entropy in this case.
\end{enumerate}

\subsection*{Appendix-II: Maldacena wormhole ansatz}

Very recently Maldacena and Milekhin have discussed humanly traversable wormholes \cite{mal1}, where they have proposed a hypothetical connecting wormhole  between two oppositely charged magnetic blackholes. This is an interesting situation, worth analysing. The metric is given by
\begin{equation}\label{mwh}
ds^2=r_{e}^2[-(\rho^2+1)d\tau^2+\dfrac{d\rho^2}{(\rho^2+1)}+(d\theta^2+\sin^2\theta d\phi^2)],  \qquad \,\,\,\,\, -\rho_{c}\leq\rho\leq\rho_{c}
\end{equation}
\begin{figure}[ht]
    \centering
\includegraphics[width=0.5\textwidth]{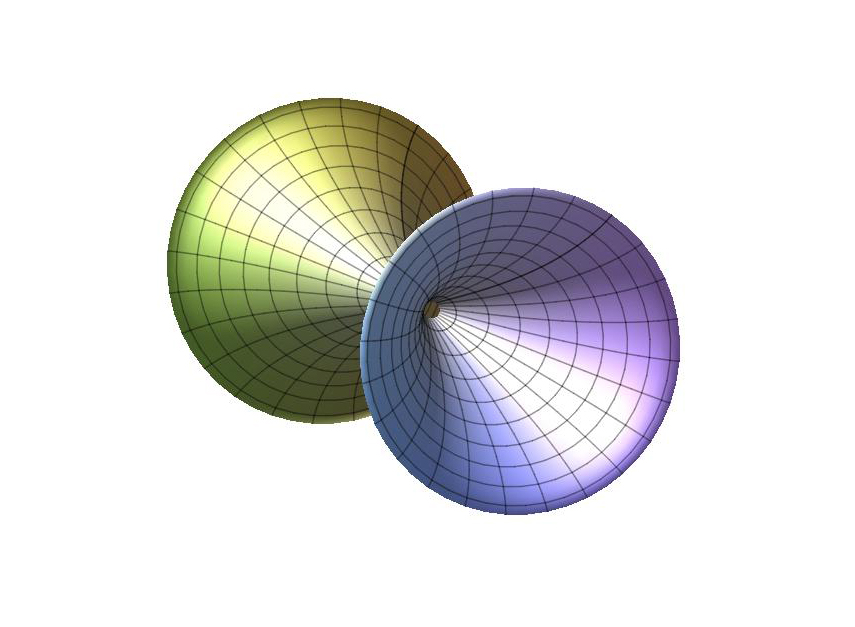}
\caption{Embedding diagram of Maldacena wormhole ansatz}
\label{Mal1}
\end{figure}
where $ \rho=\dfrac{l(r-r_{e})}{r_{e}^2} $ for $ \rho>>1 $, $ l=t/\tau $ and $ r-r_{e}<<r_{e} $ are the interrelations connecting the coordinates of \eqref{mwh} and \eqref{mbh}. Beyond the limit of $ \rho=\pm \rho_{c} $ of the wormhole region, the geometry is of two extremal magnetic blackholes. In Fig.\ref{Mal1} the wormhole is connecting two extremal magnetically charged black holes given by the following black hole metric \eqref{mbh}:

\begin{equation}\label{mbh}
ds^2=-fdt^2+\dfrac{dr^2}{f}+r^2(d\theta^2+sin^2\theta d\phi^2),
\end{equation}
where $ A=\dfrac{q}{2}cos\theta d\phi $; \,\,\,$\ell_{p}\equiv \sqrt{G_{4}}  $;\,\,\,$ r_{e}\equiv\dfrac{\sqrt{\pi}q\ell_{p}}{g_{4}} $; \,\,\,$ M_{e}=\dfrac{r_{e}}{G_{4}} $;\,\,\,$ f=\left(1-\frac{r_{e}}{r}\right)^2 $. Here $ q $ is the magnetic charge which is an integer and $M_{e}$ is the mass of the BH at extremality. In the near horizon region, $ r_{e} $ sets the radius of curvature and also the size of the 2-sphere. At the extremal limit as $ r\rightarrow r_{e} $, an infinite throat develops. Using the principal null tetrads we can easily get the following value of the Weyl scalar:

$$ \Psi_{2}=\dfrac{r_{e}(r_{e}-r)}{r^4} .$$
Using the CET proposal we obtain the gravitational energy density of the BH as:
\begin{equation}\label{mr}
\rho_{grav}=\dfrac{\alpha}{4\pi}\left\vert\dfrac{r_{e}(r_{e}-r)}{r^4}\right\vert.
\end{equation}
Similarly the gravitational temperature can be computed as
\begin{equation}\label{mt}
T_{grav}=\dfrac{1}{2\pi}\left\vert\dfrac{r_{e}}{r^2}\right\vert .
\end{equation}
Finally the ratio of gravitational energy density to gravitational temperature for this BH is given by
\begin{equation}\label{ms}
\dfrac{\rho_{grav}}{T_{grav}}=\frac{\alpha}{2}\left\vert\dfrac{(-r+r_{e})}{r^2}\right\vert .
\end{equation}
For the sake of completeness, and to check the validity of CET proposal, we can compute the gravitational entropy of this BH on a surface with radius $ R $ as:
\begin{equation}
S_{grav}=\int_{0}^{R}\int_{0}^{\pi}\int_{0}^{2\pi}\frac{\alpha}{2}\left\vert\dfrac{(-r+r_{e})}{r^2}\right\vert \dfrac{r^2 \sin\theta}{(1-\frac{r_{e}}{r})}dr d\theta d\phi=\alpha \pi R^2 .
\end{equation}
Thus the CET gravitational entropy of the magnetized extremal BH is directly proportional to the horizon area, conforming with the definition of Hawking-Bekenstein entropy.
Moreover, the ratio for the curvature scalars for the BH is given by the following expression:
\begin{equation}\label{mp}
P_{1}^2=\dfrac{6(-r+r_{e})^2}{7r_{e}^2-12r_{e}r+6r^2} .
\end{equation}

Next we calculate the Weyl scalar proposal gravitational entropy density for the connecting wormhole, which turns out to be:
$s=k_{s}\vert\nabla.\Psi\vert=0$.
Surprisingly, we find that the gravitational entropy of such wormholes vanishes in this proposal. This shows that either the proposal itself is not valid in this case, or the wormhole itself is nonphysical in nature.

Similarly for the CET proposal of gravitational entropy the gravitational energy density of the connecting wormhole is found to vanish, raising questions on the physical viability of such WHs. The relevant expressions are given below:
\begin{equation}
\rho_{grav}=0; \,\,\,\, T_{grav}=\dfrac{1}{2\pi}\left\vert  \dfrac{\rho}{r_{e}\sqrt{1+\rho^2}}   \right\vert;\,\,\,\,\, s_{grav}=0 \,\,(\rho\neq 0).
\end{equation}
As the gravitational entropy is zero even in the CET proposal, we can say that it conforms to our analysis of the extremal magnetic BHs. As $ r\rightarrow r_{e} $, the ratio of curvature scalars in \eqref{mp} reduces to zero, indicating a zero gravitational entropy density for the Weyl proposal. In this limit, for the CET proposal the gravitational energy density in \eqref{mr} also reduces to zero and the temperature in \eqref{mt} becomes $ \sim \left\vert\dfrac{1}{2\pi r_{e}} \right\vert$, which matches with the wormhole gravitational temperature in the limit $ \rho>>1 $. Consequently the ratio in \eqref{ms} becomes zero in this limit, thereby matching with the wormhole counterpart.
The interesting point to note here is that the gravitational temperature is nonzero, but at $ \rho=0 $. Thus at $ \rho=0 $ we may still have a finite entropy, as in that case the ratio of gravitational energy density to gravitational temperature assumes the form $ \frac{0}{0} $, but these proposals are not yet equipped to address such cases. ''




\part{Universal Thermodynamics}

\chapter{Thermodynamics of FRW Universe With Chaplygin Gas Models}




The contents of this chapter have been published in a journal, details of which are given below:\\

\textbf{JOURNAL REFERENCE:} General Relativity and Gravitation, 51:158 (2019)

\textbf{ARTICLE NAME:} Thermodynamics of FRW universe with Chaplygin gas models \\
DOI: 10.1007/s10714-019-2645-8 \\~~~\\

The paper is quoted below:\\

``

\section{Introduction}
Thermodynamics plays a very crucial role both in cosmological analyses as well as in the General Theory of Relativity.
The semi-classical description of black hole physics tells us that a black hole emits thermal radiation and behaves like a black body. This led to the successful description of a black hole as a thermodynamic system \cite{SWH1}. The introduction of the Bekenstein-Hawking entropy on the black hole event horizon yielded the complete development of the laws of black hole thermodynamics. Bekenstein had to assign an entropy function to a black hole in order to save the second law of thermodynamics (SLT) from becoming erroneous on the black hole horizon \cite{Bekenstein}. The temperature and the entropy of the black hole are proportional to the surface gravity on the horizon and the area of the horizon, respectively. Hence these parameters are related to the geometry of the black hole horizon. Moreover, the temperature, the entropy and the mass of the black hole were found to satisfy the first law of thermodynamics \cite{BCH}.

All these prompted physicists to search for a possible connection between black hole thermodynamics and the gravitational field equations. Jacobson \cite{Jacobson} was the first to derive the Einstein field equations from the proportionality of the black hole entropy and the horizon area together with the fundamental relation $\delta Q = TdS$, claiming that this relation is valid for all local Rindler causal horizons through each space time point,  with $\delta Q$ and $T$ as the energy flux and Unruh temperature seen by an accelerated observer just inside the horizon. Subsequently, Hayward \cite{Hayward} derived a unified first law of black-hole dynamics and relativistic thermodynamics in spherically symmetric general relativity. It was Padmanabhan \cite{Paddy} who formulated the first law of thermodynamics on ``any'' horizon for a general static spherically symmetric space time, starting from the Einstein equations. Thus the equivalence of the laws of thermodynamics with the analogous laws of black hole mechanics on one side and the Einstein equations of the classical theory of gravity on the other side, revealed a strong connection between quantum physics and gravity.\\
In the same way, on the cosmological scale, the SLT can be implemented by assuming that the universe is a closed system bounded by some horizon, preferably the cosmological apparent horizon. Applying the first law of thermodynamics to the apparent horizon of a FRW universe and considering the Bekenstein entropy on the apparent horizon, Cai and Kim \cite{caikim} derived the Friedmann equations for a universe with any spatial curvature. They used the entropy formulae for the static spherically symmetric black hole horizons in Gauss-Bonnet gravity and in Lovelock gravity, to obtain the Friedmann equations in these theories. Paranjpe et al \cite{PSP} showed that the field equations for the Lanczos-Lovelock action in a spherically symmetric spacetime can also be expressed in the form of the first law of thermodynamics. Akbar and Cai \cite{AC1} extended the work of Cai and Kim to the cases of scalar–tensor gravity and $f(R)$ gravity, and subsequently showed that \cite{AC2} the Friedmann equation of a FRW universe can be rewritten as the first law of thermodynamics on the apparent horizon of the universe and extended their procedure to the Gauss-Bonnet and Lovelock gravity. Cai and Cao \cite{CC} showed that the unified first law proposed by Hayward for the outer trapping horizon of a dynamical black hole could be applied to the apparent horizon of the FRW universe for the Einstein theory, Lovelock theory, and the scalar-tensor theories of gravity.

Although the cosmological event horizon does not exist in the big bang model of standard cosmology, but in a general accelerating universe dominated by dark energy, the cosmological event horizon separates out from the apparent horizon. Considering the physically relevant part of the Universe to be bounded by the dynamical apparent horizon, Wang et al \cite{WGA} showed that although both the first and the second laws
of thermodynamics are satisfied in such a case, but if the boundary of the Universe is assumed to be the cosmological event horizon, then both these laws break down at the event horizon, if the usual definition of temperature and entropy as applicable to the apparent horizon is extended to the event horizon. According to them, the first law may apply only to variations between nearby states of local thermodynamic equilibrium whereas the event horizon reflects the global properties of spacetime.

The conditions of validity of the generalized second law of gravitational thermodynamics in the phantom-dominated era of the flat FRW universe, was examined by Sadjadi \cite{Sadjadi}. Considering a homogeneous and isotropic universe, filled with perfect fluid having an arbitrary equation of state, Mazumder and Chakraborty \cite{MC1,MC2} have shown the validity of the GSLT of the universe with the event horizon as the boundary assuming the first law of thermodynamics, with some restrictions on the matter. Jamil et al \cite{JSS} investigated the validity of the GSLT in the cosmological scenario where dark energy interacts with both dark matter and radiation. They calculated separately the entropy variation for each fluid component and for the apparent horizon, and showed that the GSLT is always and generally valid, independently of the specific form of the equation of state (EOS) parameters of the fluids and of the background geometry. Tian and Booth \cite{TB} reexamined the thermodynamics of the Universe by requiring its compatibility with the holographic type gravitational equations which govern the dynamics of both the cosmological apparent horizon and the entire Universe. They proposed possible solutions to the existing problems regarding the temperature of apparent horizon and the evolution of cosmic entropy.

Yang et al \cite{YANG} showed that for a constant EOS of dark energy, the allowed interval of the EOS parameters for the validity of the GSLT has to be $ w_{D}\geq-1 $, in a universe enveloped by the apparent horizon and containing a Schwarzschild black hole. Xing et al \cite{XING} showed the validity of the thermodynamical properties of the universe in a new parametric model of dark energy with the equation of state $w = w_{0}+w_{1}.z(1+z)/(1+z^{2})$. In the spatially homogeneous and isotropic universe, assuming that the temperature and entropy in cosmology is as in a black hole, they examined the thermodynamical properties of the universe bounded by the apparent horizon and the event horizon respectively. They found that the first and the second laws of thermodynamics are valid inside the apparent horizon, while they break down inside the event horizon. Rani et al investigated the validity of the GSLT for a model of pilgrim dark energy interacting with cold dark matter in the frame-work of dynamical Chern-Simons modified gravity in a nonflat FRW universe \cite{RANI}. Sharif et al \cite{SZ} analyzed non-equilibrium aspects of thermodynamics on the apparent horizon of FRW universe in $f(R, T)$ gravity along with the validity of GSLT. Cardone et al studied \cite{CAR} two different dark energy models namely the Barboza-Alcaniz parameterization and the phenomenologically-motivated Hobbit model in the context of the GSLT. Iqbal et al \cite{AYE} investigated the validity of GSLT of the Ricci-Gauss-Bonnet dark energy and cold dark matter bounded by the apparent horizon and event horizon in flat FRW universe.
 Other authors also analyzed the GSLT in various theories and dark energy models like in \cite{MOR1,MOR2,LYM,SSS,JAW}.


Izquierdo and Pavon \cite{IP} explored the thermodynamics of dark energy by assuming the existence of the observer's event horizon in accelerated universes. They found that except for the initial stage of Chaplygin gas dominated expansion, the GSLT is valid in all such cases. The validity of the second law in an expanding Godel-type universe filled with generalized Chaplygin gas interacting with cold dark matter has also been examined \cite{godel}. Sharif and Saleem \cite{SS1} studied the validity of the GSLT in the presence of non-interacting magnetic field and new modified Chaplygin gas with FRW universe. Bamba et al. discussed the viability of the Generalized Chaplygin gas (GCG) as an alternative to $ \Lambda $CDM model to explain the origin of both Dark matter and Dark energy in a single fluid equation. They also discussed how the matter perturbation grows and how the sound speed limits the magnitude of free parameter $ \alpha $ in the EOS of GCG \cite{bamba}. Karami et al \cite{KAR} investigated the validity of the GSLT in a non-flat FRW universe in the presence of the interacting generalized Chaplygin gas with the baryonic matter, where the universe is assumed to be enclosed by the dynamical apparent horizon. Bandyopadhyay \cite{TB1} showed the validity of the GSLT in the braneworld scenario with induced gravity and curvature correction terms along with a dark energy component, namely, the Modified Chaplygin Gas on the 3-brane together with a perfect fluid as the dark matter.

It is therefore evident that although the apparent horizon is physically much more relevant to work with in a dynamical situation, but the event horizon has also its own importance. As we know that in a dynamically evolving universe or a black hole, both of these horizons are present, so it is justified to check for the validity of the generalized second law of thermodynamics (GSLT) on these horizons for a universe filled with various types of matter and/or energy, which in our case of study is a fluid like the Chaplygin gas. Chaplygin gas models are very versatile and useful cosmological models suitable for representing the different phases of evolution of the universe. In fact, the necessity of a model which can explain the evolutionary history of the universe successfully, led to the birth of the Chaplygin gas cosmology. Since the Chaplygin gas models can describe the accelerating expansion of the universe in the current epoch, hence they provide us a robust model for the mysterious Dark Energy. It is therefore quite prudent for one to compare the different Chaplygin gas models from a thermodynamic point of view, to identify the suitability of the different models in this group and hence comment on their merits. For this purpose we have examined the validity of the GSLT both on the cosmological apparent horizon and the cosmological event horizon for the different Chaplygin gas models. As each model in this group is distinct from the other, we obatin different cosmological consequences for the validity of the GSLT on both the horizons in these models.

For the analysis of the cosmological apparent horizon, we have considered the Kodama-Hayward temperature because the Kodama-Hayward surface gravity is more relevant for the description of dynamical horizons \cite{Faraoni}. In the case of the Variable modified Chaplygin gas, we have already determined the temperature of the FRW universe \cite{CGP} in another paper. This temperature is the bulk temperature. In this paper, after calculating the Kodama-Hayward temperature of the VMCG dominated FRW universe for the apparent horizon, we have compared these two types of temperatures to see how their behaviour affects the thermodynamics of the universe. To the best of our knowledge a comparison of this kind have not been done earlier. Further, we want to point out that our approach is much more general compared to other works as we did not assume any specific definition of surface gravity (i.e. temperature) for our analysis in the case of the cosmological event horizon. The analysis of generalized thermodynamics of FRW universe for models like the Variable modified Chaplygin gas (VMCG), New Variable modified Chaplygin gas (NVMCG), Generalized cosmic Chaplygin gas (GCCG), and Modified cosmic Chaplygin gas (MCCG) on both the cosmological horizons is also a completely new study. This will help further analysis on such models in future.

The plan of our work is as follows: in Section II we present a brief review of the various Chaplygin gas models which we have analyzed in this paper. This is followed by a general description of the theory of gravitational thermodynamics in Section III. In Sections IV and V, we analyze the criterion for which the GSLT will be valid on the cosmological apparent horizon and the cosmological event horizon, respectively, in the case of FRW universes filled with various types of Chaplygin gases. We follow up with some useful discussions in the penultimate section and end up with the conclusions in section VII.

\section{The Chaplygin gas models}
Two major problems of modern cosmology are those concerning Dark Energy and Dark Matter. Dark matter is the invisible mass in the universe or some invisible source of gravity which constitutes approximately $23$ percent of the composition of the observable universe. We also know that the universe is accelerating in its current state of expansion, an effect which is attributed to the presence of Dark Energy (constituting approximately $70$ percent of the observable universe). Although there is no clear understanding about the exact nature of this component, there are different models for explaining its effect. This led to the proposal of various types of exotic fluids as the matter content of the universe. An interesting type of such a fluid is the Chaplygin gas. It is found that some of the Chaplygin gas models can successfully describe all three phases of evolution of the universe, which therefore makes them very useful for cosmological studies. Hence it is necessary to examine the status of the GSLT on the cosmological horizons of FRW universes with matter content in the form of different variants of Chaplygin gas. Below we briefly present the Chaplygin gas models which we have analyzed in this work.

The equation of state for the Chaplygin gas \cite{CG1} is given by
\begin{equation}\label{cg_eos}
p=-B/\rho,
\end{equation}
where $ B $ is a positive constant, $ p $ is the pressure of the fluid, and $ \rho $ is the energy density. The Generalized Chaplygin gas (GCG) \cite{GCG}, is represented by the equation of state
\begin{equation}\label{gcg_eos}
p=-B/\rho^{\alpha},
\end{equation}
where $ \alpha $ is a positive constant lying within the range $0\leq \alpha \leq 1$. For the Modified Chaplygin gas (MCG) model \cite{MCG1}, the equation of state (EOS) is
\begin{equation}\label{mcg_eos}
p=A\rho-B/\rho^{\alpha},
\end{equation}
where $ A $ and $B$ are positive constants. The model which is further generalized is the Variable Modified Chaplygin gas (VMCG) \cite{VMCG1} with the EOS
\begin{equation}\label{vmcg_eos}
p=A\rho-B(a)/\rho^{\alpha},
\end{equation}
where $ B(a)=B_{0}a^{-n}=B_{0}V^{-n/3}$ is a function of the cosmological scale factor $ a $ of the FRW universe, $ B_{0} $ is a positive constant, $ n $ is any constant, and we have assumed $ V=a^{3} $ for the FRW universe.

Advancing further, we have the New Variable Modified Chaplygin gas (NVMCG) \cite{CD} with the EOS
\begin{equation}\label{nvmcg_eos}
p=A(a)\rho-B(a)/\rho^{\alpha},
\end{equation}
where $A(a)=A_{0}a^{-m}$, $ B(a)=B_{0}a^{-n} $ are functions of the cosmological scale factor $ a $, with $A_{0}, B_{0}, m $ as positive constants, $ n $ is any constant, and $ 0\leq \alpha \leq 1 $. The expression for energy density is \cite{CD}
\begin{align}
\rho&=a^{-3}e^{\frac{3A_{0}a^{-m}}{m}} \bigg[c_{0}+\frac{B_{0}}{A_{0}}\left( \frac{3A_{0}(1+\alpha)}{m} \right)^{\frac{3(1+\alpha)+m-n}{m}} \times \Gamma \left(\frac{n-3(1+\alpha)}{m},\frac{3A_{0}(1+\alpha)a^{-m}}{m}\right)\bigg]^{\frac{1}{1+\alpha}},
\end{align}
where $ \Gamma(x,y) $ is the upper incomplete gamma function and $ c_{0} $ is the integration constant.

Then comes the Generalized Cosmic Chaplygin gas (GCCG) \cite{GCCG} with the EOS
\begin{equation}\label{gccg}
p=-\rho^{-\alpha}[c+(\rho^{\alpha+1}-c)^{-w}],
\end{equation}
where $ c=\frac{E}{1+w} -1 $, and $ E $ can take both positive or negative constant values under the condition $ -L< w < 0 $, where $ L $ is a positive definite constant which can be larger than unity. The expression for the energy density in this case is
\begin{equation}
\rho=[c+(c_{1}NV^{-N}+1)^{\frac{1}{w+1}}]^{\frac{1}{\alpha+1}},
\end{equation}
where $c_{1}  $ is an arbitrary integration constant, and $ N=(1+\alpha)(1+w) $.

Finally, we have the Modified Cosmic Chaplygin gas (MCCG) with the EOS \cite{MCCG}
\begin{equation}\label{mccg_eos}
P=A\rho-\rho^{-\alpha}[(\rho^{\alpha+1}-C)^{-\gamma} + C],
\end{equation}
where $ 0<\alpha\leq 1 $, $ -b<\gamma<0 $ and $ b\neq 1  $. Here the parameter $ C=\frac{Z}{\gamma +1} -1 $, where $ Z $ is an arbitrary constant, and $ A $ is a positive constant. In the above EOS, if $ A\rightarrow 0 $, then we arrive at the EOS of GCCG. Using thermodynamic identity and binomial approximation, we obtain the approximate form of energy density of the MCCG as \cite{SA}:
\begin{equation}
\rho=\left[ \dfrac{C+(-C)^{-\gamma}+(\frac{\varepsilon}{V})^{M}}{A+1+\gamma(-C)^{-\gamma-1}}\right]^{\frac{1}{1+\alpha}},
\end{equation}
where $ \varepsilon=d(A+1)^{\frac{1}{M}} $, $ M=(1+\alpha)(1+A) $, and $ A+1+\gamma(-C)^{-\gamma-1}\neq 0 $. Here $ d $ is the constant of integration obtained during the calculation of energy density.

These expressions will be used in our subsequent calculations.

\section{Thermodynamic analysis}

We know that the Friedman equations can be written on a dynamical horizon in the form of the first law of thermodynamics \cite{caikim,AC1}
\begin{equation}
-dE_{H}=T_{H}dS_{H},
\end{equation}
where $ dE_{H} $ is the energy flowing across the horizon, $dS_{H}$ is the change of horizon entropy because of it, and $T_H$ is the horizon temperature.

Let us presume that the first law of gravitational thermodynamics holds on the cosmological horizons, and based on that premise the GSLT can be introduced in the form
\begin{equation}
\frac{dS_{T}}{dt}=\frac{dS_{b}}{dt}+\frac{dS_{H}}{dt}>0,
\end{equation}
where $ S_{T} $ is the total entropy, $ S_{H} $ is the horizon entropy and $ S_{b} $ is the bulk fluid entropy. Therefore, ultimately the GSLT is the direct extension of the SLT which says that the total entropy of the universe plus the cosmological horizon entropy should always increase.

\subsection{General formalism}

Let us assume that the universe bounded by the cosmological horizon is filled with a fluid {\color{blue}{of}} energy density $ \rho $ and pressure $ p $. The energy conservation relation is given by

\begin{equation}
\dot{\rho}+3H(p+\rho)=0,
\end{equation}
where $H$ is the Hubble parameter. The Einstein field equations for homogeneous, isotropic, flat FRW universe are
\begin{align}
3H^2 =\rho, \\
2 \dot{H}=-(p+\rho).
\end{align}

Assuming that the first law of thermodynamics holds on the cosmological horizon, we can write
\begin{equation}\label{firstlaw}
-dE_{H}=T_{H}dS_{H},
\end{equation}
where $ T_{H} $ is the temperature of the horizon, $dE_{H}$ is the amount of energy crossing the horizon in time $ dt $, and $dS_{H}$ is the amount of entropy change of the universe due to it. If $\dot{\rho}$ is the corresponding rate of change of the energy density of the universe, then we can write
\begin{align}
dE_{H}&=\frac{4\pi R_{H}^{3} \dot{\rho}dt}{3} \nonumber \\
&=-4\pi R_{H}^{3}H(p+\rho)dt.
\end{align}
Substituting the above expression in the equation (\ref{firstlaw}) depicting the first law of thermodynamics, we obtain the rate of change of horizon entropy as
\begin{equation}
\frac{dS_{H}}{dt} =\frac{4\pi R_{H}^{3} H(p+\rho)}{T_{H}}.
\end{equation}
We now use the Gibbs equation in the bulk to get the entropy of the fluid bounded by the horizon in the form
\begin{equation}
T_{H}dS_{b}=dE_{b} +pdV,
\end{equation}
where our underlying assumption is that the bulk temperature is equal to the horizon temperature ($ T_{H}=T_{b} $), i.e. the bulk and the horizon surface are in thermal equilibrium. Substituting $ V = 4\pi R_{H}^{3}/3  $ and $ E_{b} = 4 \pi R_{H}^{3}\rho/3 $ in the Gibbs relation, we have
\begin{equation}
T_{H}dS_{b} = (4\pi R_{H}^{2}\rho \dot{R}_{H} +\frac{4}{3}\pi R_{H}^{3}\dot{\rho}+4\pi p R_{H}^{2}\dot{R}_{H})dt.
\end{equation}
Therefore the rate of change of the fluid entropy is
\begin{equation}
\frac{dS_{b}}{dt}=\frac{4\pi R_{H}^{2} (p+\rho)(\dot{R_{H}}-H R_{H})}{T_{H}},
\end{equation}
and the rate of change of total entropy is
\begin{align}
\frac{dS_{T}}{dt} =\frac{d(S_{H}+S_{b})}{dt}   =\frac{4\pi R_{H}^{2}(p+\rho)\dot{R}_{H}}{T_{H}}.
\end{align}

\subsection{Cosmological apparent horizon}

It is known that for a spatially flat FRW universe, the apparent horizon and the Hubble horizon coincides \cite{AC2}. The area radius of the cosmological apparent horizon is given by

\begin{equation}\label{R_AH}
R_{AH}(t)=\frac{1}{H(t)}.
\end{equation}
Now for a flat FRW universe, the apparent horizon evolves according to the relation
\begin{align}
\dot{R}_{AH}(t) = -H \dot{H} R_{AH}^3 = \frac{(p+\rho)}{2H^2}.
\end{align}

Using the above relation in the expression for rate of change of total entropy, we obtain
\begin{equation}
\frac{dS_{T}}{dt}= \frac{4\pi (p+\rho)^2}{2 H^{4}T_{AH}},
\end{equation}
which is a positive quantity, and hence our usual expectation is that the GSLT will always be valid on the apparent horizon when the bulk fluid and the horizon are in thermal equilibrium.

\subsection{Cosmological event horizon}

The proper radius of the event horizon in the FRW universe is
\begin{equation}\label{R_EH}
R_{EH}(t)= a(t)\int^{+\infty}_{t} \frac{dt'}{a(t')},
\end{equation}
where $a(t)$ is the scale factor of the expanding universe. We know that if the above integral converges, then the universe will have an event horizon, and the equation according to which the cosmological event horizon evolves is given by
\begin{equation}
\dot{R}_{EH} =HR_{EH}-1 .
\end{equation}
Using the above relation in the expression for rate of change of  total entropy, we get
\begin{align}
\frac{dS_{T}}{dt}&=\frac{4\pi R_{EH}^{2}}{T_{EH}}(p+\rho)(HR_{EH}-1) \nonumber \\
&= \frac{4\pi R_{EH}^{2}H}{T_{EH}}(p+\rho)(R_{EH}-R_{AH}).
\end{align}

\section{Validity of GSLT on the Cosmological apparent horizon of various Chaplygin gas models}

Lets us now examine the status of the GSLT on the cosmological apparent horizon of FRW universes filled with the different variants of the Chaplygin gas listed in section II.

\subsection{VMCG}

\begin{figure}[ht]
\centering
 \includegraphics[width=0.42\textwidth]{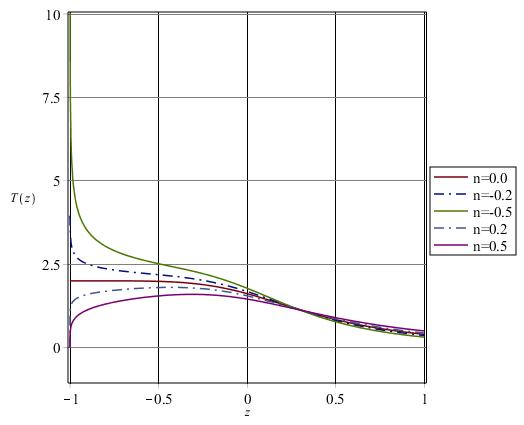}
 \caption{Variation of Kodama-Hayward temperature $T(z)$ on the AH for VMCG dominated FRW universe as a function of $ z $ for different values of $ n $.}
\label{khtemp}
\end{figure}

The Kodama-Hayward temperature (KHT) of the cosmological apparent horizon for a spatially flat FRW universe is given by \cite{BH}
\begin{equation}\label{Kodama}
k_B T = \left( \frac{\hbar G}{c} \right)\frac{(\rho - 3p)}{3H}.
\end{equation}
Here we can rewrite $H$ in terms of the redshift $z$. As (\ref{Kodama}) involves the pressure and energy density of the fluid, it captures the cosmological essence of the model. But this pressure and energy density are related by the EOS of the VMCG given by (\ref{vmcg_eos}) where $B(a)=B_{0}a^{-n}$. Thus we have studied the variation of the KHT on the Apparent Horizon (AH) for VMCG dominated FRW universe as a function of $z$ for different values of $n$. This is shown in FIG.\ref{khtemp}

\begin{figure}[ht]
\centering
\includegraphics[width=0.42\textwidth]{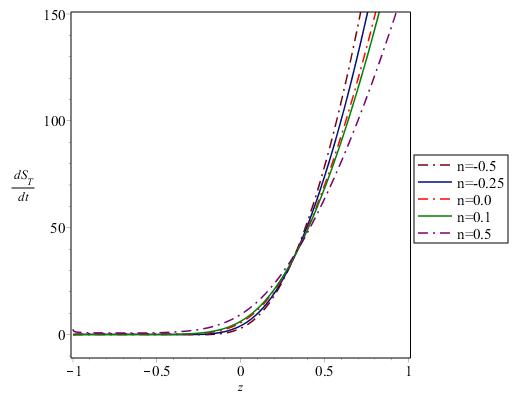}
\caption{Plot showing the variation of $ \dfrac{dS_{T}}{dt} $ with respect to the free parameter $n$ on the cosmological apparent horizon (using the Kodama temperature of the AH) for VMCG dominated FRW universe, validating the GSLT on the AH.}
\label{label-a}
\end{figure}

Using the expression of KHT on the cosmological apparent horizon, we have also studied the variation of $ \dfrac{dS_{T}}{dt} $ with respect to the free parameter $n$ for the VMCG dominated FRW universe, which is shown in FIG.\ref{label-a}. From this figure, it is evident that the total entropy on the AH always increases for different values of the parameter $ n $. Thus the GSLT is always valid on the apparent horizon of the VMCG dominated FRW universe when we consider the Kodama temperature of the horizon.

In a separate paper \cite{CGP} we have already determined the expression for the bulk temperature of the VMCG in a FRW universe.
It is therefore interesting to compare the validity of GSLT on the AH of the VMCG dominated FRW universe determined in terms of the KHT of the AH with the validity determined in terms of the temperature of the FRW universe filled with VMCG matter. For this purpose we have used the expression of the temperature of the VMCG dominated FRW universe which we have calculated in {\textbf{the}} paper \cite{CGP}. Substituting the EOS of the VMCG (given in equation \ref{vmcg_eos}) in the thermodynamic identity $\left(\frac{\partial U}{\partial V}\right)_s = -P $, we obtain the relation
\begin{equation}\label{07}
\left(\frac{\partial U}{\partial V}\right)_s = -A(U/V)+ B_0 V^{-n/3} (V/U)^{\alpha}.
\end{equation}
From equation (\ref{07}) the energy density is determined accurately up to an integration constant
\begin{equation}\label{08}
 \rho = \frac{1}{a^\frac{n}{1+\alpha}}\left[(1+\alpha)B_0/N + C / a^{3N}\right]^\frac{1}{1+\alpha},
\end{equation}
where $N=(A+1)(1+\alpha)-n/3$, and $C$ is the integration constant. The expression of energy density for the VMCG dominated FRW universe, as a function of scale factor, is obtained as
\begin{equation}\label{09}
\rho(a)=\dfrac{\rho_{0}}{a^{\frac{n}{1+\alpha}}}\left[\Omega_x + (a_0^{n}-\Omega_{x})(a_0/a)^{3N}\right]^\frac{1}{1+\alpha},
\end{equation}
where we defined the dimensionless parameter
\begin{equation}\label{10}
\Omega_x =\dfrac{(1+\alpha)B_0}{N \rho_{0}^{1+\alpha}}.
\end{equation}
Finally the temperature $ T(z) $ of this VMCG universe as a function of the redshift $z$ is obtained in the form \cite{CGP}
\begin{align} \label{T_z1}
\left. T(z)\right. =\frac{T_0(z+1)^{3N(1+\frac{\alpha}{1+\alpha}) +3A}(\frac{1}{\Omega_x})^{\frac{\alpha}{1+\alpha}}}{[1+(z+1)^{3N}(\frac{a_0^n}{\Omega_{x}}-1)]^\frac{\alpha}{1+\alpha}} \times\frac{[\frac{1}{\Omega_x}(1-\frac{n}{3N})-\frac{1}{a_0^n}]a_0^{n(1+\frac{\alpha}{1+\alpha})}}
 {[(1-\frac{n}{3N})(\frac{a_0^n}{\Omega_{x}}-1)(z+1)^{3N}-\frac{n}{3N}]}.
\end{align}

\begin{figure}[ht]
\centering
 \includegraphics[width=0.45\textwidth]{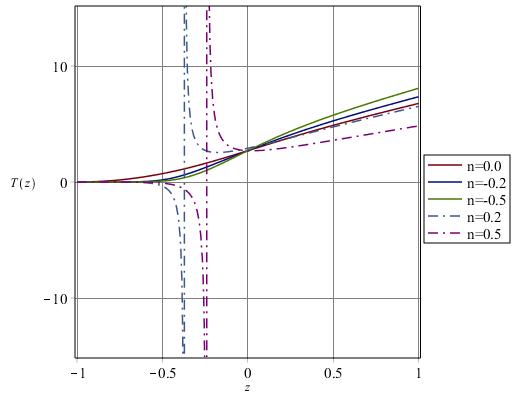}
 \caption{Variation of $T(z)$ as a function of $ z $ for different values of $ n $ for the VMCG dominated FRW universe.}
\label{vmcgtempfuture}
\end{figure}

The temperature of the VMCG dominated FRW universe is depicted in FIG.\ref{vmcgtempfuture} as a function of redshift. It can be seen that in the future epoch, the measure of temperature becomes negative and has an infinite discontinuity, which clearly indicates that the VMCG model is thermodynamically unstable for positive values of the parameter $ n $. This is consistent with the conclusions derived by Panigrahi and Chatterjee in \cite{vmcg}.

In the next two figures (shown in FIG.\ref{fig2}), we have used the temperature of VMCG dominated FRW universe in place of the horizon temperature for the near equilibrium scenario. In this case we observe that there is a clear difference from FIG.\ref{label-a}. For negative $ n $, the GSLT is always valid on the AH but for positive $ n $, the GSLT gets violated in the future epoch (i.e. for negative redshift).

As we have used the VMCG temperature in our entropy calculation, it is natural for us to come across a scenario in which the total entropy decreases for the range of positive values of $ n $. In actuality, this reflects the inherent thermodynamic behaviour of the model itself in the cosmological context. In FIG.\ref{fig2}(b), the negative rate of change of total entropy for $ n>0 $ is the result of the thermodynamic instability of VMCG for $ n>0 $ \cite{vmcg}. This is why we get to see different nature of the plots in FIG.\ref{label-a} and FIG.\ref{fig2}, respectively, when we use the two different temperatures (i.e. KHT and the temperature of VMCG, respectively).

\begin{figure}[ht]
    \centering
    \subfloat[Subfigure 1 list of figures text][]
        {
        \includegraphics[width=0.45\textwidth]{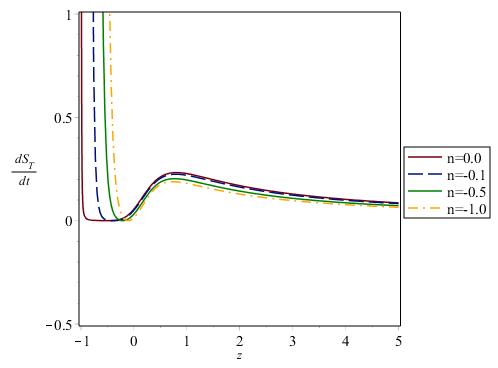}
        \label{fig:subfig1}
        }
    \subfloat[Subfigure 2 list of figures text][]
        {
        \includegraphics[width=0.43\textwidth]{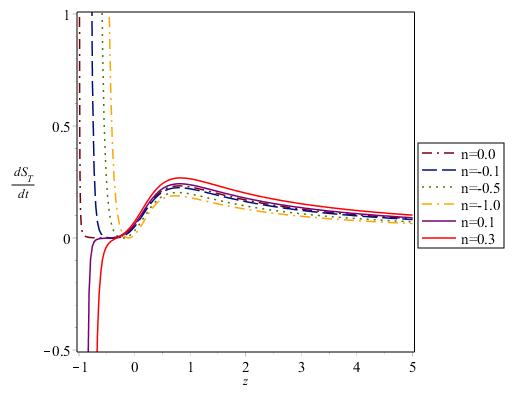}
        \label{fig:subfig2}
        }
    \caption{Variation of $ \dfrac{dS_{T}}{dt} $ wrt the free parameter $n$ preserving the validity of GSLT on the cosmological apparent horizon (using temperature of FRW universe dominated by VMCG).}
    \label{fig2}
\end{figure}

In this context it may be noted that Chen et al. \cite{CLX} have applied cosmological constraints on the VMCG model using the Markov chain Monte Carlo (MCMC) method. They have analyzed the validity of the generalized laws of thermodynamics on the horizons assuming the Hawking temperature of the horizons. However, in this paper we have used the KHT for our analysis. This approach is different from the method used by them. Thus the result of our analysis on the horizons, is more appropriate than those obtained by them. While inspecting the rate of variation of the entropy on the horizon as a function of redshift, they have extended their plot into the future epoch (i.e. to negative values of $z$). In this paper we have also extended our plots into the future epoch.

\subsection{MCG}

The case $ n=0 $ in the VMCG model represents the MCG dominated FRW universe. In FIG.\ref{fig2} we observe that the curve for $ n=0 $ always obeys the GSLT on the cosmological apparent horizon.

\subsection{GCCG}

In \cite{SS}, Sharif et al analyzed the thermodynamic stability of GCCG. Here we examine the validity of GSLT on the apparent horizon of the universe filled with GCCG. FIG.\ref{fig3} shows the variation of $ \dfrac{dS_{T}}{dt} $ with respect to the parameter $w$ in the EOS given in (\ref{gccg}), for the FRW universe filled with GCCG. We find that the total entropy always increases when $ \omega>-1 $, but for $ \omega<-1 $, the GSLT is violated in the future epoch.



\begin{figure}[ht]
    \centering
    \subfloat[Subfigure 1 list of figures text][]
        {
        \includegraphics[width=0.45\textwidth]{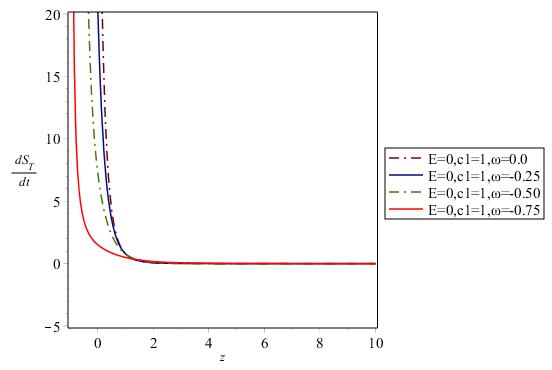}
        \label{fig:subfig3}
        }
    \subfloat[Subfigure 2 list of figures text][]
        {
        \includegraphics[width=0.48\textwidth]{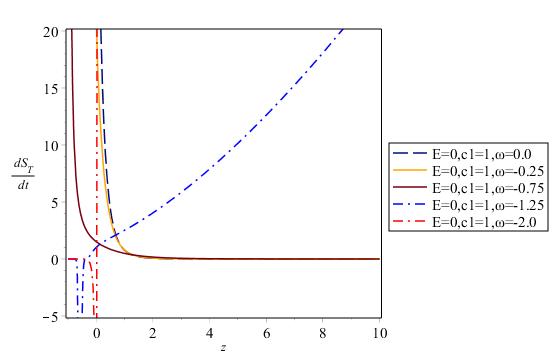}
        \label{fig:subfig4}
        }
    \caption{Variation of $ \dfrac{dS_{T}}{dt} $ wrt the free parameters $ \omega$ maintaining the validity of GSLT on the cosmological apparent horizon for GCCG.}
    \label{fig3}
\end{figure}


\begin{figure}[ht]
    \centering
    \subfloat[Subfigure 1 list of figures text][]
        {
        \includegraphics[width=0.45\textwidth]{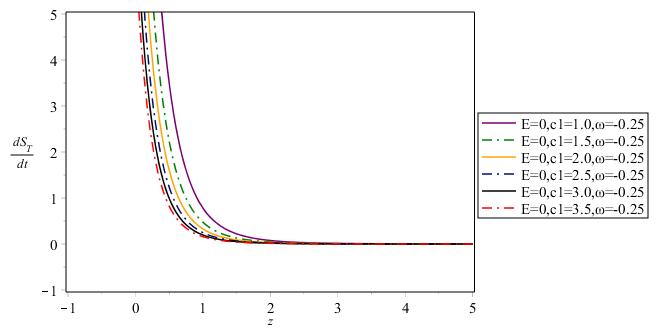}
        \label{fig:subfig5}
        }
    \subfloat[Subfigure 2 list of figures text][]
        {
        \includegraphics[width=0.47\textwidth]{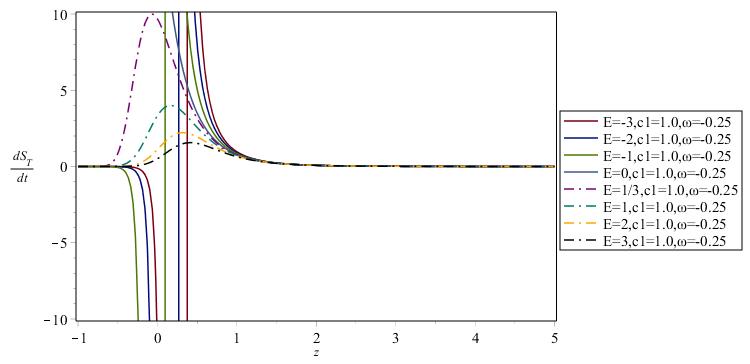}
        \label{fig:subfig6}
        }
    \caption{Variation of $ \dfrac{dS_{T}}{dt} $ wrt the free parameters $ \omega$ maintaining the validity of GSLT on the cosmological apparent horizon for GCCG.}
    \label{fig4}
\end{figure}

Next we study the variation of $ \dfrac{dS_{T}}{dt} $ with respect to the parameter $w$ of the GCCG dominated FRW universe for different values of the integration constant $ c_{1} $, and examine its variation in FIG.\ref{fig4}(a). From this figure it is clear that the GSLT is valid on the cosmological apparent horizon only for positive $ c_{1} $.

We also study the variation of $ \dfrac{dS_{T}}{dt} $ with respect to the parameter $w$ for a variation in $E$. This is shown in FIG.\ref{fig4}(b). It is evident that the GSLT is valid on the apparent horizon only for $ E>0 $. So we conclude that for the validity of GSLT on the cosmological apparent horizon, the GCCG model parameters must satisfy the conditions $ w>-1, E\geq 0$, and $ c_{1}>0 $.

\subsection{MCCG}
We now study the variation of $ \dfrac{dS_{T}}{dt} $ with respect to the arbitrary parameter $Z$ in the EOS of the MCCG dominated FRW universe and show the corresponding variation in the subsequent figures. In the FIG.\ref{fig5}(a), we find that the entropy of FRW universe filled with MCCG is well behaved and obeys the GSLT for negative $ Z $, and from FIG.\ref{fig5}(b) we find that the GSLT is valid for positive integration constant $ d $.

\begin{figure}[ht]
    \centering
    \subfloat[Subfigure 1 list of figures text][]
        {
        \includegraphics[width=0.50\textwidth]{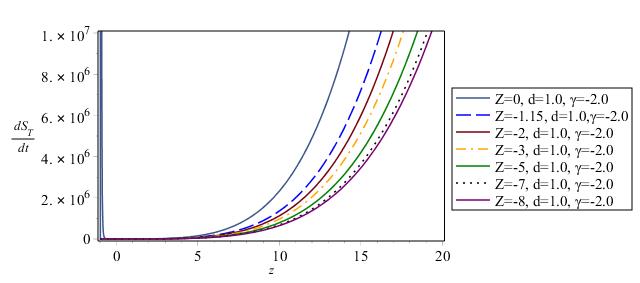}
        \label{fig:subfig7}
        }
    \subfloat[Subfigure 2 list of figures text][]
        {
        \includegraphics[width=0.45\textwidth]{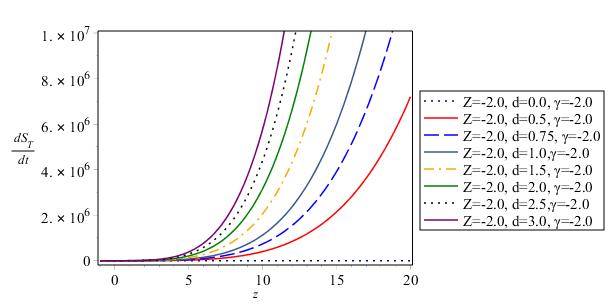}
        \label{fig:subfig8}
        }
    \caption{Variation of $ \dfrac{dS_{T}}{dt} $ wrt the free parameters $ Z,\gamma $ and the integration constant $d$, while preserving the validity of GSLT on the cosmological apparent horizon for MCCG, where we have chosen $ \alpha=0.1, \: \textrm{and} \: A=2 .$}
    \label{fig5}
\end{figure}

\begin{figure}[ht]
    \centering
    \subfloat[Subfigure 1 list of figures text][]
        {
        \includegraphics[width=0.46\textwidth]{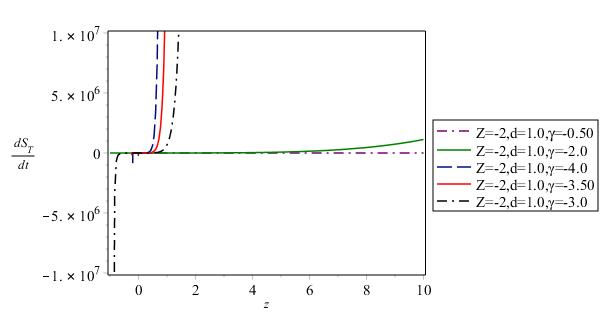}
        \label{fig:subfig9}
        }
    \subfloat[Subfigure 2 list of figures text][]
        {
        \includegraphics[width=0.48\textwidth]{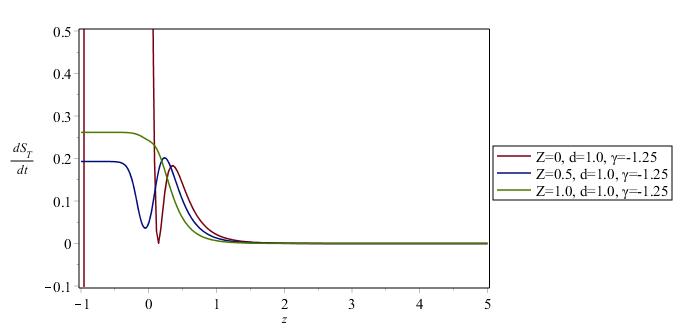}
        \label{fig:subfig10}
        }
    \caption{Variation of $ \dfrac{dS_{T}}{dt} $ wrt the free parameters $ Z,\gamma $ and the integration constant $d$, while preserving the validity of GSLT on the cosmological apparent horizon for MCCG.}
    \label{fig6}
\end{figure}

FIG.\ref{fig6}{\textbf{(a)}} depicts some more variations of the parameters and evidently the total entropy is well behaved only around $ Z=-2,\gamma=-2 $ and $ d=1 $. Here we have assumed the values of the parameters as  $\alpha=0.1$, and $ A=2$. In FIG.\ref{fig6}(b), we have explored the variations for different values of fixed parameters i.e. $ A=1$, and $ \alpha=1 $. We can see that the GSLT is valid in this case except for $ Z=0, d=1.0, \: \textrm{and} \: \gamma=-1.25 $. We also find that the GSLT is valid for positive $ Z $. Thus we conclude that the parameter $ d $ must be positive and accordingly there are distinct values of $ \gamma $ for the validity of GSLT on the cosmological apparent horizon in the MCCG model.

\subsection{NVMCG}

\begin{figure}[ht]
\centering
\includegraphics[width=0.63\textwidth]{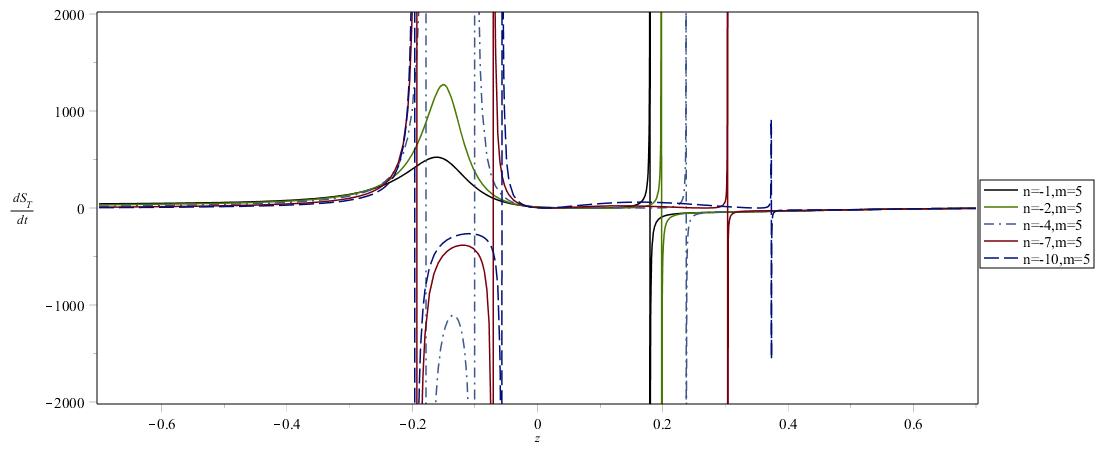}
\caption{Variation of $ \dfrac{dS_{T}}{dt} $ as a function of redshift on the cosmological apparent horizon for different values of  $ n $ with $ m=5 $ for NVMCG.}
\label{nvmcgnvary1}
\end{figure}

\begin{figure}[ht]
\centering
\includegraphics[width=0.63\textwidth]{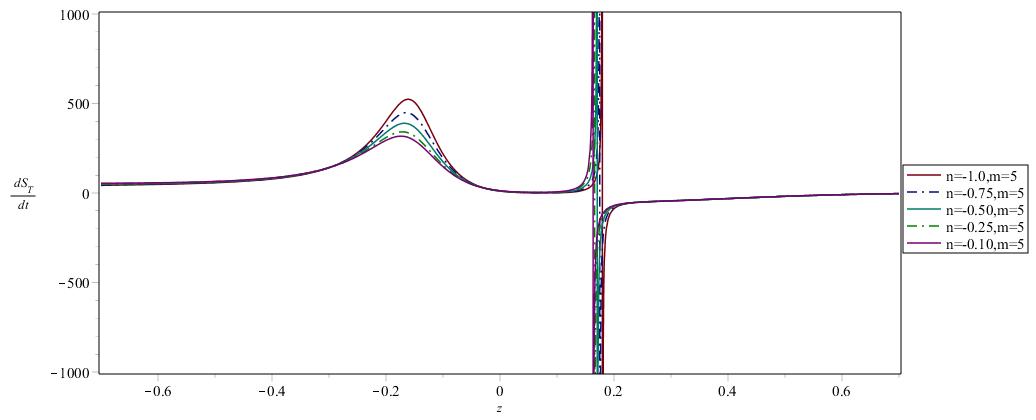}
\caption{Variation of $ \dfrac{dS_{T}}{dt} $ as a function of redshift on the cosmological apparent horizon for different values of  $ n $ with $ m=5 $ for NVMCG.}
\label{nvmcgnvary2}
\end{figure}

\begin{figure}[ht]
\centering
\includegraphics[width=0.60\textwidth]{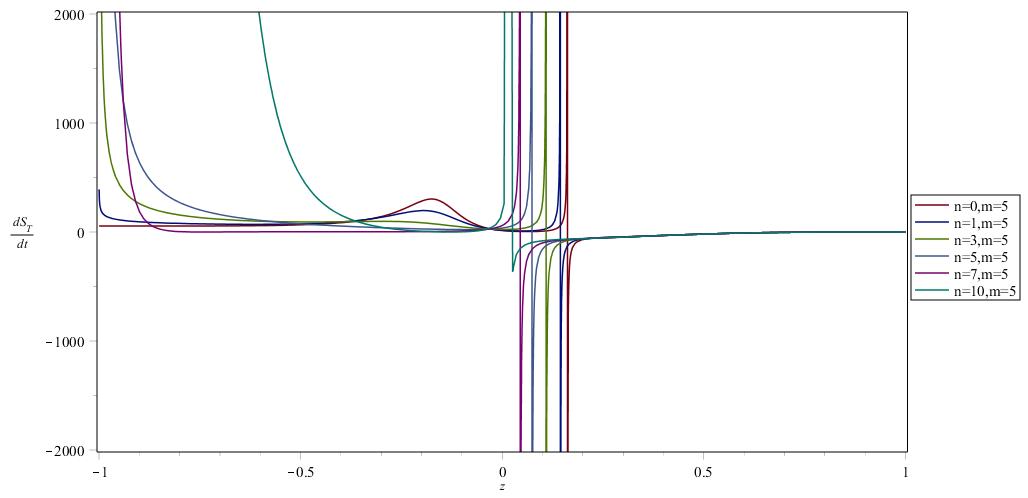}
\caption{Variation of $ \dfrac{dS_{T}}{dt} $ as a function of redshift on the cosmological apparent horizon for different values of  $ n $ with $ m=5 $ for NVMCG.}
\label{nvmcgnvary3}
\end{figure}

In this section we will do a similar analysis for the NVMCG model. In our calculations we have assumed the values $ A_{0}=1.0 $, $ B_{0}=10.0 $, $ \alpha=1.0 $, and $ c_{0}=1.0 $ for the parameters in the EOS of NVMCG. In FIG.\ref{nvmcgnvary1}, FIG.\ref{nvmcgnvary2} and  FIG.\ref{nvmcgnvary3}, we have fixed $ m=5 $ and varied $ n $. We can see that none of the plots are well-behaved and every curve indicates a conditional validity of the GSLT. Moreover all these plots show an abrupt variation in the rate of change of the total entropy during the evolution of the universe, which is difficult to explain.

\begin{figure}[ht]
\centering
\includegraphics[width=0.60\textwidth]{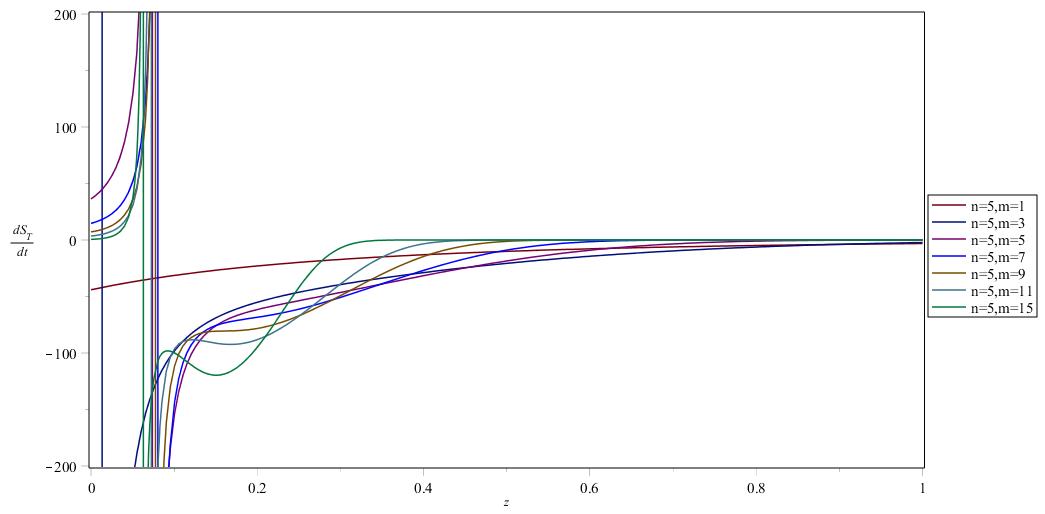}
\caption{Variation of $ \dfrac{dS_{T}}{dt} $ as a function of redshift on the cosmological apparent horizon for different values of $ m $ with $ n=5 $ for NVMCG.}
\label{nvmcgmvary}
\end{figure}

In FIG.\ref{nvmcgmvary} we have fixed the parameter $n$  and varied $ m $. Here also none of the curves satisfy the GSLT over their entire evolution. All the curves violate GSLT in the early phase of the universe and then suddenly the entropy increases in the recent epoch of the universe with vertical asymptotic discontinuities.

Thus from the above analysis we conclude that the NVMCG violates GSLT on the cosmological apparent horizon.

\section{Validity of GSLT on the Cosmological Event Horizon of various Chaplygin gas models}
We now proceed to examine the status of the GSLT on the cosmological event horizon (EH) of FRW universe dominated by the various Chaplygin gas fluids. We like to point out that our analysis on the event horizon is a general treatment without assuming any specific form of temperature.

As the event horizon is `teleological' in nature \cite{JN,HE}, we are only interested in analyzing the overall validity of the GSLT on it. In the calculations with regard to the cosmological EH, we have ignored the computation of $R_{EH}$ and the corresponding graphs, as the calculations become complicated due to increasing number of parameters involved in the EOS of these models. Though the radius of the cosmological event horizon can be calculated using equation (\ref{R_EH}), we do not require its explicit expression to understand the nature of validity of the GSLT on the cosmological EH.  Rather we have tried to examine the validity of GSLT using simple algebraic manipulations which clearly demonstrates the conditional nature of the validity.
However, we have examined the case of the VMCG dominated FRW universe in details as this model is cosmologically significant. Hence we have checked the validity of the GSLT on the cosmological event horizon in the VMCG case.

\subsection{VMCG}
In the case of the Variable Modified Chaplygin gas, the rate of change of total entropy on the cosmological event horizon is given by
\begin{equation}
\frac{dS_{T}}{dt}=\frac{4\pi R_{EH}^{2}H}{T_{EH}}\left(A \rho - \frac{B}{\rho^{\alpha}}+\rho \right)(R_{EH}-R_{AH}).
\end{equation}
Now for the validity of GSLT ($i.e. \dot{S}_{T}>0$), we need $ R_{EH}>R_{AH} $, which implies that we must have $ (A+1)\rho>\frac{B}{\rho^{\alpha}} $. This finally leads us to the condition
\begin{equation}\label{vmcg_cond}
\rho^{1+\alpha} >\frac{B_{0}a^{-n}}{(1+A)}.
\end{equation}

In the above condition (\ref{vmcg_cond}), we substitute the expression for energy density of the VMCG given by (\ref{09}), which is
\begin{equation}
\rho=\frac{\rho_{0}}{a^{\frac{n}{1+\alpha}}}\left[\Omega_{x}+(1-\Omega_{x})\left(\frac{1}{a}\right)^{3N}\right]^{\frac{1}{1+\alpha}}.
\end{equation}
This leads us to the following relation:
\begin{equation}
 \left[\Omega_{x} +(1-\Omega_{x})\frac{1}{a^{3N}}\right] > \frac{N\Omega_{x}}{(1+\alpha)(1+A)}.
\end{equation}

Replacing the scale factor $ a $ by the redshift $ z $ by using the substitution $ a=\frac{1}{z+1} $, we obtain
\begin{equation}\label{gslt_vmcg}
(z+1)^{3N}>\frac{-n\Omega_{x}}{3(1+A)(1+\alpha)(1-\Omega_{x})}.
\end{equation}

Considering the present day case we put $ z=0 $ in the above relation, and arrive at the limiting condition
\begin{equation}\label{limitvmcg}
n>3(1+A)(1+\alpha)\left(1-\frac{1}{\Omega_{x}}\right).
\end{equation}

\begin{figure}[ht]
    \centering
    \subfloat[Subfigure 1 list of figures text][]
        {
        \includegraphics[width=0.33\textwidth]{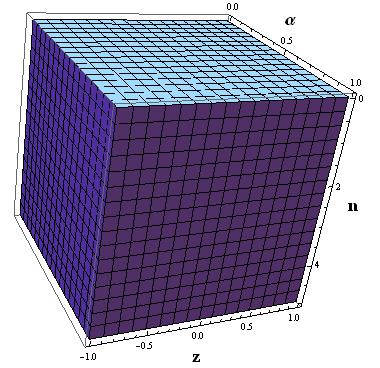}
        \label{fig:subfig11}
        }
     \hspace{0.75in}
    \subfloat[Subfigure 2 list of figures text][]
        {
        \includegraphics[width=0.30\textwidth]{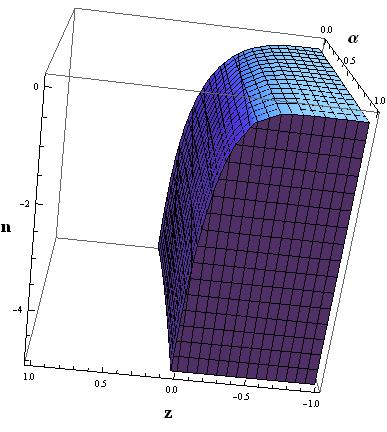}
        \label{fig:subfig12}
        }
    \caption{Variation of $ z $ wrt the free parameters $ n, \alpha $ while preserving the validity of GSLT on the cosmological event horizon for the condition of quintessence (fig. a) and phantom dominated universe (fig. b).}
    \label{fig_parameter}
\end{figure}

The above criterion depicts the condition for the validity of the GSLT on the event horizon of a VMCG dominated FRW universe. Let us assume $ A=1/3 $ , $ \Omega_{x}=0.7 $  and $ \alpha=0.25 $ in order to model a cosmologically viable evolution of the universe. This assumption leads us to the condition
\begin{equation}
(z+1)^{(5-n)}>-\left(\frac{n}{2.14}\right).
\end{equation}
From this relation we can immediately observe the explicit dependence of the redshift $z$ on the free parameter $ n $. We also see that when $ n $ is zero or positive, the relation becomes a trivial one, but when $ n $ becomes negative (signifying a phantom dominated universe), it prevents the redshift $ z $ from attaining the value of $ -1$ (indicating the future of the universe), as the left hand side becomes zero but the right hand side is still positive. This  indicates that for non-negative values of $ n $, $ (p+\rho)>0 $, and therefore, the above relation is only valid for $ n\geq 0 $.
Now putting $z=0  $ in the criterion (\ref{limitvmcg}) and setting  $ A=1/3 $, $ \alpha=0.25 $ and $ \Omega_{x}=0.7 $, we get
\begin{align}
n>3(1+A)(1+\alpha)\left(1-\frac{1}{\Omega_{x}}\right) \qquad \Rightarrow \qquad n>-2.14. \nonumber
\end{align}
Thus we arrive at the condition that $ n>-2.14 $ for the chosen values of the parameters for the validity of GSLT in the VMCG filled FRW universe bounded by the event horizon, provided we assume the \emph{a priori} condition that $ R_{EH}>R_{AH} $. From the cosmological analysis of VMCG in FRW universe by earlier workers, it has been found that for $ n\geq 0 $, the universe is dominated by quintessence \cite{VMCG1} or cosmological constant. Therefore the criterion (\ref{limitvmcg}) obtained above is consistent with it. From the above analysis we can conclude that for $n\geq 0$, GSLT is valid on the event horizon in a VMCG dominated FRW universe (Fig.~\ref{fig_parameter}(a)). However, we note here that the model VMCG itself is thermodynamically unstable in this range of $ n\geq 0 $ \cite{vmcg}.



Another condition for the validity of GSLT is $R_{EH}<R_{AH}$ with $ (p+\rho)<0 $. From the second condition quoted beside, we can say that
\begin{equation}\label{vmcg_another}
(z+1)^{3N}<\frac{-n\Omega_{x}}{3(1+A)(1+\alpha)(1-\Omega_{x})}.
\end{equation}
Now for the appropriate values of the free parameters $A=1/3, \Omega_{x}=0.7$ and $\alpha=0.25 $, equation (\ref{vmcg_another}) leads us to the relation
\begin{equation}\label{vmcg_particular}
(z+1)^{5-n} < - 0.47 n.
\end{equation}
This relation is not valid for $ n\geq0 $. It is known that the condition $ (p+\rho)<0 $ represents the case for $ n<0 $ (phantom dominated universe), which is consistent with the above relation (\ref{vmcg_particular}). It is easy to realize that for $ -1\leq z \leq 0 ,$ the relation certainly holds true, and for $ z=0 $ it yields us the condition $ n>-2.14 $. In fact, this condition is true up to some positive values of $ z $ depending on the corresponding value of $ n $. So we can certainly say that the GSLT holds for the FRW universe filled with the VMCG fluid during the current epoch and will also hold in the future.

However we can also see that for positive high values of $ z $ (in the early universe), the relation (\ref{vmcg_particular}) fails to hold. Consequently the GSLT is not valid in the early universe for a Phantom like VMCG (Fig.~\ref{fig_parameter}(b)).

Therefore we can safely claim that for $ n<0 $ (i.e. $R_{EH}<R_{AH}  $), the VMCG dominated FRW universe violates the GSLT on the event horizon in the early phase of the universe but it holds during the current epoch and will also hold in the future, and moreover in this range of $ n<0 $, the VMCG model itself is thermodynamically stable \cite{vmcg}.

At this juncture we like to note that in our other paper \cite{CGP}, we have checked the temperature variation of the FRW universe filled with VMCG, which is consistent with the thermodynamic dependence of the values of the parameter $n$. In the present paper also we find that the validity of the GSLT on the apparent horizon depends very much on the value of the parameter $n$, when we use the temperature of VMCG dominated FRW universe in place of the horizon temperature, for conditions approaching thermal equilibrium. Our analysis clearly shows that negative value of $n$ is thermodynamically stable (or favoured) in the sense that the GSLT is valid on the apparent horizon in such cases. In the same line of thinking we tried to investigate the validity of the GSLT on the event horizon of the VMCG universe and found that the parameter space of the GSLT validity condition for the phantom case is incomplete, but is complete for the quintessence case (Fig.~\ref{fig_parameter}).

\subsection{MCG}

Substituting $ n=0 $ in the equation of state for VMCG, we get the equation of state for MCG:

\begin{equation}
p=A\rho -\frac{B}{\rho^{\alpha}}.
\end{equation}
The condition of applicability of GSLT on the event horizon of a MCG dominated FRW universe is then
\begin{equation}\label{gslt_mcg}
(z+1)^{3(1+A)(1+\alpha)}>0
\end{equation}
which is a trivial relation given the equation of state of MCG. Now if we consider  $z=0$ for our present day, then by putting $ n=0 $ in the relation for $n $, we get
\begin{equation}
3(1+A)(1+\alpha)\left(1-\frac{1}{\Omega_{x}}\right)<0.
\end{equation}
For the chosen values of the parameters $A$, $\alpha$ and $\Omega_x$ mentioned in the earlier section, we find that the first two terms in the equation of state of MCG are positive, and we know that the third term must be negative. Therefore we can say that the GSLT is always valid on the event horizon of a MCG dominated universe iff $ R_{EH}>R_{AH} $.

\subsection{GCG}
The equation of state for GCG is given in (\ref{gcg_eos}) as
\begin{equation}
p=-\frac{B}{\rho^{\alpha}}.
\end{equation}
To consider the case of Generalized Chaplygin gas, we substitute $ n=0 $ and $ A=0 $ in the viability relation (\ref{gslt_vmcg}) and we obtain
\begin{equation}
 (z+1)^{3(1+\alpha)}>0,
\end{equation}
which is again a trivial condition. Moreover, if we put $ n=0 $ and $ A=0 $ in the relation (\ref{limitvmcg}) for $ n $, then for the present day value of $ z=0 $, we get
\begin{equation}\label{limitgcg}
 3(1+\alpha)\left(1-\frac{1}{\Omega_{x}}\right)<0.
\end{equation}
We know that the first term is always positive in the equation of state of GCG and the second term is always negative. Therefore the above condition (\ref{limitgcg}) is also trivial. Therefore we can say that the GSLT is always valid on the event horizon of a GCG dominated universe iff $ R_{EH}>R_{AH} $.

\subsection{GCCG}

The equation of state for the GCCG quoted in Section II is given by \cite{GCCG}
\begin{equation}
p=-\rho^{-\alpha}[c+(\rho^{\alpha+1}-c)^{-w}],
\end{equation}
and the expression for the energy density is
\begin{equation}
\rho=[c+(c_{1}NV^{-N}+1)^{\frac{1}{w+1}}]^{\frac{1}{\alpha+1}},
\end{equation}
where $c_{1}  $ is an arbitrary integration constant, and $ N=(1+\alpha)(1+w) $. The rate of change of total entropy in this model obeys GSLT on the cosmological event horizon if $ R_{EH}>R_{AH} $, and $ (p+\rho)>0 $. From the second condition beside, we get
\begin{align}
 \frac{c_{1}NV^{-N}}{(c_{1}NV^{-N}+1)^{\frac{w}{w+1}}}>0.
\end{align}
From the above expression we can say that the GSLT is valid in this case if $ c_{1}N>0 $. So there are two possibilities: either (i) $c_{1}>0  $ and $N>0  $, or (ii) $c_{1}<0  $ and $N<0  $.

But we know that $ (1+\alpha)>0 $. Therefore the condition $ N>0 $ means that $ -1<w<0 $, and when $ N<0 $, it means that $ w<-1 $. However the case $ N<0 $ is not possible because the equation of state parameter in this case should be greater than $ -1 $. Therefore when $ R_{EH}>R_{AH} $, the condition for the GSLT to be valid on the event horizon of a FRW universe filled with GCCG is $c_{1}>0  $, and $ -1<w<0 $. The initial conditions have to be chosen in such a manner that we have $c_{1}>0 $.

The other condition for the validity of the GSLT on the event horizon of this universe is $ R_{EH}<R_{AH} $ and $ (p+\rho)<0 $, which will then represent the phantom case. In the same way as before, we arrive at the relation
\begin{equation}
\frac{c_{1}NV^{-N}}{(c_{1}NV^{-N}+1)^{\frac{w}{w+1}}}<0,
\end{equation}
which is only possible when $c_{1}N<0  $. Now we know that for the phantom case, the equation of state parameter is $ w<-1 $, which suggests that $ N<0 $. So $ c_{1} $ must be positive.

Therefore when $ R_{EH}>R_{AH} $, the GSLT is valid when $c_{1}>0  $ and $ -1<w<0 $, and when $ R_{EH}<R_{AH} $,
the conditions are $c_{1}>0  $ and $ w<-1 $. So, if we choose the boundary conditions in such a way that $c_{1}>0  $, then the GCCG dominated FRW universe obeys GSLT on the event horizon.

\subsection{MCCG}

The equation of state for the Modified Cosmic Chaplygin gas is given by
\begin{equation}
P=A\rho-\rho^{-\alpha}[(\rho^{\alpha+1}-C)^{-\gamma} + C],
\end{equation}
where $ 0<\alpha\leq 1  $, $ -b<\gamma<0 $, $ b\neq 1  $, $ C=\frac{Z}{\gamma +1} -1 $, $ Z $ being an arbitrary constant, and $ A $ is a positive constant. The approximate form of energy density is
\begin{equation}
\rho=\left[ \dfrac{C+(-C)^{-\gamma}+(\frac{\varepsilon}{V})^{M}}{A+1+\gamma(-C)^{-\gamma-1}}\right]^{\frac{1}{1+\alpha}},
\end{equation}
where $ \varepsilon=d(A+1)^{\frac{1}{M}} $, $ M=(1+\alpha)(1+A) $, $ A+1+\gamma(-C)^{-\gamma-1}\neq 0 $, and $ d $ is the constant of integration. Now from the EOS of MCCG, we conclude that $ (\rho^{\alpha +1}-C)>0 $, which then leads us to the condition
\begin{equation}
\rho^{\alpha +1}>\dfrac{Z-(\gamma +1)}{(\gamma +1)}.
\end{equation}
For $ C>0 $, we have $ Z>(\gamma+1) $. In this case, if $ -1<\gamma<0 $, i.e. $ Z>0 $, then $ \rho^{\alpha +1} $ must be greater in magnitude than some positive constant, but if $ \gamma<-1 $, no specific conclusion can be drawn regarding the value of $ \rho $. Again when $ C<0 $, which means that $ Z<(\gamma +1) $, then $ \rho^{\alpha +1} $ is greater than some positive constant if $ \gamma<-1 $ (i.e. $ Z<0 $), but is greater than a negative value if $ -1<\gamma<0 $ (i.e. $ Z<1 $).

From the expression of energy density we can say that $ \gamma $ can only take integral values, and so we can discard the $ -1<\gamma<0 $ limit.
Clearly we can see that if $ C<0 $ and $ \gamma<-1 $ (i.e. $ Z<0 $), we can always have a lower positive bound of $ \rho $.

To check for the validity of the GSLT on the event horizon of MCCG dominated FRW universe, we note that the criterion for this validity is $ (P+\rho)>0 $ when $ R_{EH}>R_{AH} $, as in the previous cases. Using this inequality we obtain the following condition:
\begin{equation}
(A+1)\rho^{\alpha +1}-[(\rho^{\alpha+1}-C)^{-\gamma}+C]>0.
\end{equation}
Substituting the expression for energy density and using binomial expansion, we get
\begin{align}
 \left(\frac{\varepsilon}{V}\right)^{M}+(-C)^{-\gamma}-\rho^{\alpha+1}\gamma(-C)^{-\gamma-1}
  -[(-C)^{-\gamma}+(-\gamma)\rho^{\alpha+1}(-C)^{-\gamma-1}+...]>0,
\end{align}
and the end result is
\begin{equation}
\left(\frac{\varepsilon}{V}\right)^{M}>[\rho^{(\alpha+1)(-\gamma)}+(-\gamma)\rho^{(\alpha+1)(-\gamma-1)}(-C)+...].
\end{equation}
The right hand side is definitely a positive quantity if $ C<0 $. Denoting the entire term inside the square brackets by $ K $, we find that $ K>0 $ for $ C<0  $, but for $ C>0 $ nothing can be said regarding the sign of $K$. Also previous analyses of the equation of state of MCCG in cosmology showed that for negative $ C $ and $ Z<0 $, there is a positive lower bound for the energy density, and hence for the condition of validity of GSLT on the event horizon of MCCG we must have
\begin{equation}
d>V\left[ \frac{K}{(A+1)} \right]^{\frac{1}{M}}.
\end{equation}
This gives us a positive lower bound for the integration constant $d$. Therefore we can say that for $ C<0 $ and $ \gamma<-1 $ (or $ Z<0 $), the GSLT is valid on the event horizon of a MCCG dominated FRW universe when we choose our initial conditions in such a way that the integration constant stays above the lower bound $V\left[ \frac{K}{(A+1)} \right]^{\frac{1}{M}}$. In this limit of $ Z <0$, the model MCCG itself is also thermodynamically stable \cite{SA}.

\subsection{NVMCG}
In this section we analyze the the validity of GSLT for the NVMCG model. The equation of state for the New Variable Modified Chaplygin gas is already quoted earlier and is given by
\begin{equation}
p=A(a)\rho - \frac{B(a)}{\rho^{\alpha}},
\end{equation}
and the expression for energy density is
\begin{align}
\small \rho&=a^{-3}e^{\frac{3A_{0}a^{-m}}{m}} \bigg[c_{0}+\frac{B_{0}}{A_{0}}\left( \frac{3A_{0}(1+\alpha)}{m} \right)^{\frac{3(1+\alpha)+m-n}{m}} \times \Gamma \left(\frac{n-3(1+\alpha)}{m},\frac{3A_{0}(1+\alpha)a^{-m}}{m}\right)\bigg]^{\frac{1}{1+\alpha}},
\end{align}
where $ \Gamma(x,y) $ is the upper incomplete gamma function and $ c_{0} $ is the integration constant.
The rate of change of total entropy obeys GSLT on the cosmological event horizon if $ R_{EH}>R_{AH} $, and $(p+\rho)>0$. From the second condition we get the following relation
\begin{equation}
\rho^{\alpha +1}>\frac{B_{0}a^{-n}}{A_{0}a^{-m}+1}.
\end{equation}
It follows that if $ n>0 $ and $ m>0 $ in the limit of large $ a $, the above relation reduces to the form $\rho^{\alpha +1}>0  $,
which can correspond to the `quintessence' form of dark energy, whereas for $ n<0 $ and $ m>0 $, the relation becomes $\rho^{\alpha +1}>\infty  $, so that the energy density blows up, which corresponds to the phantom model of dark energy.

Now for the GSLT to be valid, the necessary condition is $ (p+\rho)>0 $, which means that the equation of state parameter is $ w>-1 $ (which corresponds to the quintessence model). Therefore we can say that in the NVMCG model, the GSLT is valid on the cosmological event horizon if $ n>0 $ and $ m>0 $ along with $ R_{EH}>R_{AH} $.

The other condition for the validity of GSLT on the cosmological event horizon is as usual $ (p+\rho)<0 $, when $ R_{EH}<R_{AH} $. The condition $ (p+\rho)<0 $  can be written as
\begin{equation}
\rho^{\alpha+1}<\frac{B_{0}a^{-n}}{A_{0}a^{-m}+1}.
\end{equation}
If we consider the case $ n>0 $ and $ m>0 $ in the large $ a $ limit, the above relation becomes $\rho^{\alpha+1}<0 $, which is not physically possible. So we can safely discard this case.

Next we consider the range $ n<0 $ and $ m>0 $, which yields the relation $\rho^{\alpha+1}<\infty  $, which is a perfectly acceptable criterion. Using the expression for energy density we arrive at the condition
\begin{align}
c_{0} & > \left[ \frac{B_{0}a^{-(n-3(1+\alpha))}}{(A_{0}a^{-m}+1)} \right]e^{\frac{-3A_{0}(1+\alpha)a^{-m}}{m}}
  -\frac{B_{0}}{A_{0}}\left(\frac{3A_{0}(1+\alpha)}{m}\right)^{\frac{m-(n-3(1+\alpha))}{m}}\times \nonumber\\
  & \Gamma\left(\frac{n-3(1+\alpha)}{m},\frac{3A_{0}(1+\alpha)a^{-m}}{m}\right).
\end{align}
This relation implies that $ m $ cannot be negative. When $ m\sim 0 $, the NVMCG model asymptotically approaches the VMCG model and for small positive values of $ m $, the above condition becomes
\begin{align}
c_{0} & > \left[ \frac{B_{0}a^{-(n-3(1+\alpha))}}{(A_{0}a^{-m}+1)} \right]e^{\frac{-3A_{0}(1+\alpha)a^{-m}}{m}} \nonumber \\
 & - \frac{B_{0}}{A_{0}}\left(\frac{3A_{0}(1+\alpha)}{m}\right)^{\frac{m-(n-3(1+\alpha))}{m}} \left[\left(\frac{3A_{0}(1+\alpha)a^{-m}}{m}\right)^{\frac{n-3(1+\alpha)}{m}-1} e^{-\frac{3A_{0}(1+\alpha)a^{-m}}{m}}\right],
\end{align}
that is
\begin{align}
c_{0}>\left[\frac{B(a)}{(A(a)+1)}-\frac{B(a)}{A(a)}\right]a^{3(1+\alpha)} e^{-\frac{3A_{0}(1+\alpha)a^{-m}}{m}}. \nonumber
\end{align}
For large $a$ and small positive $ m $ approaching zero, we find that the limit for $ c_{0} $ in the NVMCG model is
\begin{equation}
c_{0}>\left[\frac{B_{0}}{(A_{0}+a^{m})}-\frac{B_{0}}{A_{0}}\right]a^{3(1+\alpha)-n+m}.
\end{equation}
Thus the limit on $ c_{0} $ depends explicitly on the exponent $3(1+\alpha)-n+m  $. If $ n<3(1+\alpha)+m $, the value of the integration constant blows up (i.e. $ c_{0}>-\infty $) implying that the energy density blows up for large values of $ a $. But if $ n>3(1+\alpha)+m $, we must have $ c_{0}>0 $ in the large $ a $ limit. In the special case $ n=3(1+\alpha)+m $, the integration constant becomes $ c_{0}>\left[\frac{B_{0}}{(A_{0}+a^{m+3(1+\alpha)})}-\frac{B_{0}}{A_{0}}\right] $, i.e. it is greater than a negative constant value.
In the asymptotic limit of $ m\sim 0 $, we obtain the condition $ c_{0}>0 $ for large values of $ a $, for the validity of GSLT in the FRW universe filled with NVMCG and bounded by the event horizon.

\section{Discussions}

The entire analysis in this paper is based on the assumption that there is enough time for the fluid to attain thermal equilibrium with the cosmological horizons. If there is no thermal equilibrium, the validity of GSLT on the cosmological horizons becomes much more conditional.

In absence of thermal equilibrium between the fluid and the horizon \cite{TB} (i.e. $ T_{H}\neq T_{b} $), we may assume a near equilibrium situation and write an approximate relation for the rate of change of total entropy as
\begin{equation}
\frac{dS_{T}}{dt}=4\pi R_{H}^{2}(p+\rho)\left( \frac{H R_{H}}{T_{H}} + \frac{(\dot{R_{H}} - H R_{H})}{T_{b}} \right).
\end{equation}
We can always think of the bulk temperature to be very near to the horizon temperature i.e. $T_{b}=T_{H}+\delta T$, where we are assuming $ \dfrac{\delta T}{T_{H}}<<1 $. The rate of change of total entropy in near-equilibrium situation becomes
\begin{equation}
\dfrac{dS_{T}}{dt}\simeq \dfrac{dS^{0}_{T}}{dt} + \dfrac{\delta T}{T^{2}_{H}} 4\pi R^{2}_{H}(p+\rho)(HR_{H}-\dot{R_{H}}),
\end{equation}
where $ \dfrac{dS^{0}_{T}}{dt}\equiv \frac{4\pi R_{H}^{2}(p+\rho)\dot{R}_{H}}{T_{H}} $, is the rate change of total entropy when there is thermal equilibrium.

For the cosmological apparent horizon (with $ T_{AH}\neq T_{b} $), the rate of change of total entropy becomes
\begin{equation}
\frac{dS_{T}}{dt}= \frac{4\pi (p+\rho)^{2}}{2H^{4}T_{b}} + \left(\frac{4\pi}{H^{2}T_{AH}T_{b}} \right) (p+\rho)(T_{b}-T_{AH}).
\end{equation}
In that case, we can see that the GSLT is not always valid on the cosmological apparent horizon. The first term on the right hand side is always positive, and therefore either the second term is also positive (which suggests that $ (p+\rho)>0$ for $T_{b}>T_{AH}$, or $(p+\rho)<0,T_{b}<T_{AH}  $), or as a whole the right hand side of the above equation remains positive even if the second term becomes negative.

If $(p+\rho)>0  $, and $ T_{b}<T_{AH} $, we obtain the condition
\begin{equation}
\frac{(p+\rho)}{2H^{2}}>\left(1-\frac{T_{b}}{T_{AH}}\right),
\end{equation}
which represents the quintessence dominated universe. Another possibility is that if $(p+\rho)<0 $, and $ T_{b}>T_{AH} $, the condition for the validity of GSLT on the cosmological apparent horizon becomes
\begin{equation}
\frac{(p+\rho)}{2H^{2}}<\left(1-\frac{T_{b}}{T_{AH}}\right),
\end{equation}
which then represents the phantom dominated universe.

Now in the near-equilibrium scenario, the rate of change of total entropy on the cosmological apparent horizon can be written as
\begin{equation}
\dfrac{dS_{T}}{dt}\simeq\dfrac{dS^{0}_{T}}{dt} + \dfrac{\delta T}{T_{AH}^{2}}\left[\dfrac{4\pi(p+\rho)}{H^2} - \dfrac{4\pi (p+\rho)^{2}}{2H^{4}}\right] -\dfrac{4\pi(p+\rho)}{H^{2}}\dfrac{(\delta T)^{2}}{T^{3}_{AH}},
\end{equation}
which agrees with the general formula in the first order terms of $ \delta T $. Here $ \dfrac{dS^{0}_{T}}{dt} \equiv \frac{4\pi (p+\rho)^2}{2 H^{4}T_{AH}} $, is the rate of change of entropy on the cosmological apparent horizon when there is thermal equilibrium.

In the same way, if the temperature on the cosmological event horizon and the bulk fluid temperature are not in equilibrium, then we have the rate of change of total entropy as
\begin{equation}
\frac{dS_{T}}{dt}=4\pi R_{EH}^{2}H (p+\rho)\left(\frac{R_{EH}}{T_{EH}} - \frac{R_{AH}}{T_{b}}\right).
\end{equation}
In this case, for the GSLT to be valid on the cosmological event horizon, the conditions are either $ (p+\rho)>0,\frac{R_{EH}}{T_{EH}} > \frac{R_{AH}}{T_{b}}   $ (which represents the quintessence dominated universe) or the other possibility is that $ (p+\rho)<0,\frac{R_{EH}}{T_{EH}}< \frac{R_{AH}}{T_{b}}   $ (the phantom dominated universe).

In the near-equilibrium scenario, the rate of change of total entropy on the cosmological event horizon can be written as
\begin{equation}
\dfrac{dS_{T}}{dt}\simeq\dfrac{dS^{0}_{T}}{dt} + \dfrac{\delta T}{T^{2}_{EH}} 4\pi R^{2}_{EH}(p+\rho),
\end{equation}
which again agrees with the general formula given above. Once again, $ \dfrac{dS^{0}_{T}}{dt} \equiv \frac{4\pi R_{EH}^{2}}{T_{EH}}(p+\rho)(HR_{EH}-1)$ is the rate of change of entropy on the cosmological event horizon when there is thermal equilibrium.

Therefore in dynamical situations, if we assume that the difference in temperature is not very large, and do perturbative analysis around the horizon temperature, we can always get the equilibrium term in the leading order with some correction terms with higher orders in $ \delta T $. Hence, the study of the near equilibrium scenario becomes important from this point of view.

Thus we can see that in absence of thermal equilibrium, the validity of the GSLT on the cosmological horizons becomes far more conditional. Here also we can do the same analysis as done in our previous section to find out how the free parameters affect these conditions.

\section{Conclusions}

In this work we have examined the thermodynamic viability of some dark energy models which may be considered as alternatives to the $\Lambda$CDM model. The Chaplygin gas models, especially the VMCG model, is successful in explaining all three phases of evolution of the universe, namely, the radiation dominated phase, the matter dominated phase and the vacuum energy dominated phase. Thus it is more versatile than the $\Lambda$CDM model. Although strict constraints cannot be obtained from our analysis on the validity of GSLT for the different Chaplygin gas models, yet it provides us with clear ranges of parameters and in some cases these ranges conforms to the results obtained by other authors on those models, thereby establishing the appropriateness of our thermodynamic analysis of these models. Our analysis provides us with a clear picture on how to choose these models if the GSLT is to hold in a FRW universe filled with such fluids. It is to be noted that the universal thermodynamics is not the only factor determining the physics of these models. The validity of GSLT on cosmological horizons is a necessary condition for any cosmological model to be considered physically realistic.

In the case of the cosmological apparent horizon, from our consideration of the Kodama-Hayward temperature, we conclude that the MCG and VMCG models always obey the GSLT. For the case of the GCCG, the model parameters should lie in the range $ c_{1}>0, E\geq 0, -1<w<0 $, whereas for the MCCG model the parameter $ d $ must be positive. However, the NVMCG model does not obey the GSLT on the cosmological apparent horizon during the entire evolution of the cosmos. Either it has abrupt discontinuities, or the total entropy simply decreases in magnitude during certain phases of the evolution. Therefore the NVMCG violates the GSLT on the cosmological apparent horizon.

For the cosmological event horizon, we find that the validity of GSLT is always conditional, which we enlist below:
\begin{itemize}
  \item For the VMCG model we have shown that for $ n>0 $, the GSLT is valid on the cosmic event horizon with the automatic condition $ R_{EH}>R_{AH} $. For $ n<0 $, with the condition $ R_{EH}<R_{AH} $, the GSLT is violated on the cosmological event horizon (i.e. phantom dominated VMCG model violates GSLT on the event horizon). Therefore the validity of the GSLT on the event horizon favors the quintessence dominated ($n>0$) FRW universe in the VMCG model.
  \item In the MCG dominated FRW universe, the GSLT is always valid on the cosmological event horizon. It is also valid for GCG dominated FRW universe.
  \item In the case of GCCG dominated FRW universe, the GSLT is valid conditionally on the cosmological event horizon. In the case when $R_{EH}>R_{AH}  $,  the equation of state parameter has to be $ -1<w<0 $, and the integration constant $ c_{1} $ has to be chosen positive. In the case of $ R_{EH}<R_{AH} $, the parameter should be $ w<-1 $, and again the integration constant $ c_{1} $ has to be positive. So in both the cases, depending on the value of the parameter $ w $ and the initial conditions for $ c_{1} $ to be positive, the validity of GSLT on the cosmological event horizon for the GCCG model can be achieved in a FRW universe.
  \item In the NVMCG model, the GSLT is valid on the cosmological event horizon for two conditions. One is that if $R_{EH}>R_{AH}  $, then the GSLT is valid for the parameters $ n>0 $ and $ m>0 $. Therefore when $R_{EH}>R_{AH}  $, the validity of GSLT on the cosmological event horizon favors the quintessence dominated FRW universe for the NVMCG model. The second possible condition is obtained for $ R_{EH}<R_{AH} $. In this case we have shown that the only possibility is $ n<0 $ and $ m>0$ for the validity of the GSLT.
  \item We also want to point out that the limit on the value of $Z$ for the validity of GSLT on the event horizon of the universe filled with MCCG is consistent with the bound on $Z$ as obtained by Sharif \cite{SA} for the thermodynamic stability of MCCG. Both the analysis of \cite{SA} and that of ours yield the condition that $Z<0$.
\end{itemize}
Therefore as we always know that if we consider the universe to be bounded by the cosmological apparent horizon, then every fluid model satisfies the GSLT, but with the cosmological event horizon as the boundary surface, the situation changes, and different models demand different parameter ranges for the validity of the GSLT on the bounding surface. ''





\chapter{Evolution of FRW universe in variable modified Chaplygin gas model}



The contents of this chapter have been published in a journal, details of which are given below:\\


\textbf{ARTICLE NAME:} Evolution of FRW universe in variable modified Chaplygin gas model \\
arXiv:1906.12185v1  [gr-qc]  27 Jun 2019 \\~~~\\

The paper is quoted below:\\

``

\section{Introduction}

Einstein's General Theory of Relativity (GTR) revolutionized our understanding of gravity and the structure of space-time. It predicted many new things like the expansion of the universe, space-time singularity, and most recently, the discovery of the gravitational waves was another feather in the cap of GTR. But following the distance measurements of Type Ia supernova \cite{SN1a1,SN1a2,SN1a3,SN1a4,SN1a5}, astronomers came to the understanding that the universe's expansion is accelerated at the present time, an observation which could only be accounted for by the dynamics of a hitherto unknown form of energy, called ``Dark energy'' (DE). Coupled with this was the problem of explaining the observed rotation curve of the galaxies, which lead to the hypothesis of non-baryonic Cold Dark Matter (DM), constituted of particles which are yet to be detected directly. Since then, scientists have proposed a variety of theories to explain these observations. All these models can be broadly classified into two major groups: either one has to change the geometry part of the Einstein field equations to explain these observations, or change the matter-energy part. In an effort to modify the matter part, several researchers proposed the existence of various exotic fluids, a prominent one being the `quintessence', to explain the Dark energy. Dark Matter is gravitationally attractive, being responsible for the clustering of matter in the universe, whereas dark energy is repulsive and responsible for the accelerated expansion of the universe. At the same time, scientists were also looking for a model which could simultaneously explain the mechanism of both the DE and DM. These searches led to the development of the so-called \emph{Chaplygin Gas Cosmology}.

The \emph{Chaplygin gas} (described by the equation of state $P=-B/\rho$) \cite{CG1,CG2}, is an exotic perfect fluid. It explains both the aspects of DE and DM in a simple way and at the same time conforms to the observational data quite well. Several models of Chaplygin gas have been proposed in succession to explain the observational data more accurately. The simplest one is the generalized Chaplygin gas (GCG) \cite{GCG1,GCG2,GCG3,GCG4} with an equation of state
\begin{equation}\label{01}
P=-B/\rho^\alpha,
\end{equation}
where $ B $ is a positive constant and the parameter $\alpha$ takes on values such that $ 0<\alpha\leq 1 $.

The variable Chaplygin gas (VCG) was first proposed by Zhang and Guo \cite{VCG1,VCG2} with the equation of state
\begin{equation}\label{02}
P=-B(a)/ \rho,
\end{equation}
where the constant coefficient $B$ is replaced by a variable coefficient  $ B(a)=B_{0}a^{-n}$. Although it explains two important phases of the evolution of the universe: the dust phase and the present accelerated expansion phase, but it did not capture the earlier radiation-dominated phase of the universe. Hence came the next model: the modified Chaplygin gas (MCG) \cite{MCG1,MCG2} with the equation of state
\begin{equation}\label{03}
P= A\rho-B\rho^{-\alpha},
\end{equation}
where $A$ and $B$ are positive constants. This MCG model has the amazing capability to describe all three evolutionary phases of the universe, starting with the radiation phase (with $A=1/3$), then going through a pressureless phase (dust phase), and then transiting into the present negative pressure phase dominated by dark energy.

Subsequently, in order to explain the observational data even more accurately, researchers came up with more refined models in which the parameter $B$ was assumed to be a function of the scale factor $a(t)$ of the FRW universe. This led to two models, namely, the variable generalized Chaplygin gas (VGCG) and the variable modified Chaplygin gas (VMCG) \cite{VMCG1}. The VMCG equation of state is
\begin{equation}\label{04}
P= A\rho-B(a)\rho^{-\alpha},
\end{equation}
where $B(a)=B_{0}V^{-n/3}$, or $ B(a)=B_{0} a^{-n} $ (for FRW universe). Here the parameters $A$, $B_{0}$ are positive constants and $n$ is also a constant. This model can describe dark energy more accurately because of the extra free parameter $n$ appearing in the equation of state.

Once a cosmological model is proposed, it becomes necessary to examine the viability of such models from the point of view of the corresponding cosmological dynamics, as well as its thermodynamic stability. Several authors have already worked on these aspects (see for example \cite{VMCG2,VGCG3}). Here in this paper, we will deduce the temperature evolution of the FRW universe filled with VMCG as a function of red shift $ z $. We will also use observational data to determine the redshift at the epoch when the transition from deceleration to acceleration happened. We deduced the values of other relevant parameters like the Hubble parameter, the equation-of-state parameter and the speed of sound in terms of the redshift parameter and examined how these values differ from the results obtained from previous works on MCG and other Chaplygin gas models for the various values of $n$ permitted by thermodynamic stability. The temperature of decoupling is calculated with the value of decoupling redshift as $z\simeq 1100$.

\section{Thermodynamic analysis}
The metric corresponding to the flat FRW universe is given by

\begin{equation}\label{05}
ds^{2}=-dt^{2} + a^{2}(t)(dr^{2}+r^{2}d\theta^{2}+r^{2}sin^{2}\theta d\phi^{2}),
\end{equation}
where $ a(t) $ is the scale factor. For the sake of calculations, we have assumed $ V=a^{3} $ for the FRW universe.
The equation of state of the VMCG is
\begin{equation}\label{06}
P= A\rho-B\rho^{-\alpha},
\end{equation}
where $B=B_{0}V^{-n/3}$. Now we have the well known thermodynamic identity $\left(\frac{\partial U}{\partial V}\right)_s = -P $, in which we substitute (\ref{06}) to get
\begin{equation}\label{07}
\left(\frac{\partial U}{\partial V}\right)_s = -A(U/V)+ B_0 V^{-n/3} (V/U)^{\alpha}.
\end{equation}
From this equation (\ref{07}), the energy density is determined accurate up to the order of an integration constant in the form
\begin{equation}\label{08}
 \rho = \frac{1}{a^\frac{n}{1+\alpha}}\left[(1+\alpha)B_0/N + C / a^{3N}\right]^\frac{1}{1+\alpha},
\end{equation}
where $N=(A+1)(1+\alpha)-n/3$, and $C$ is the integration constant which can be an universal constant or a function of entropy $S$. Using the boundary condition i.e. the present day energy density $ \rho_0=\rho(a_0) $ in the above relation (\ref{08}), we can determine the integration constant in terms of $ \rho_0$ and $a_0 $. The resulting expression of energy density for the VMCG in FRW universe as a function of scale factor is
\begin{equation}\label{09}
\rho(a)=\dfrac{\rho_{0}}{a^{\frac{n}{1+\alpha}}}\left[\Omega_x + (a_0^{n}-\Omega_{x})(a_0/a)^{3N}\right]^\frac{1}{1+\alpha},
\end{equation}
where we have defined the dimensionless parameter
\begin{equation}\label{10}
\Omega_x =\dfrac{(1+\alpha)B_0}{N \rho_{0}^{1+\alpha}}.
\end{equation}
Introducing the parameter $ R=(1+A)(1+\alpha) $, and substituting $ n=0 $ in the above equation (\ref{09}), we get the energy density for MCG as
\begin{equation}\label{11}
\rho(a)=\rho_{0} \left[\Omega_x + (1-\Omega_{x})(a_0/a)^{3R}\right]^\frac{1}{1+\alpha}.
\end{equation}
The expression (\ref{09}) for the energy density can also be derived using the field equations for FRW cosmology. Here we have used purely thermodynamic approach and got the same expression. This in fact shows the close relation between GTR and thermodynamics.

We know that the first law of thermodynamics can be written in the form
\begin{equation}\label{12}
TdS=d(\rho/{m})+P d(1/{m}),
\end{equation}
where $ S $ is the entropy per particle, $ \rho $ is the total energy density, $ P $ is the pressure, $ T $ is the temperature in Kelvin, and $ m $ is the particle density in the system. Equation (\ref{12}) can be rewritten as
\begin{equation}\label{13}
dS=(1/Tm)d\rho-(P+\rho)/T m^2 dm.
\end{equation}
This leads us to two thermodynamic relations:
\begin{equation}\label{14}
\left(\dfrac{\partial S}{\partial \rho}\right)_m = 1/Tm,
\end{equation}
and
\begin{equation}\label{15}
\left(\dfrac{\partial S}{\partial m}\right)_\rho = -(p+\rho)/Tm^2.
\end{equation}

As $ T=T(\rho,m) $, the following identity becomes obvious:
\begin{equation}\label{16}
dT=\left(\dfrac{\partial T}{\partial m}\right)_\rho dm + \left(\dfrac{\partial T}{\partial \rho}\right)_m d\rho.
\end{equation}

Along with this we also have the integrability condition of the first law as
\begin{equation}\label{17}
T\left(\partial P /\partial \rho\right)_m = m \left(\partial T/ \partial m\right)_\rho + (p+\rho)\left(\partial T / \partial \rho\right)_m.
\end{equation}

The above two equations (\ref{16}) and (\ref{17}) can be solved for the unknowns $(\partial T/ \partial m)_\rho  $ and $ (\partial T / \partial \rho)_m $, and the condition for this is
\begin{equation}\label{18}
\dot{m}(p+\rho)-m\dot{\rho}=0.
\end{equation}

Now substituting this result back into the previous identities (\ref{16}) and (\ref{17}), we obtain the relation
\begin{equation}\label{19}
dT/T = (dm/m)(\partial P/\partial \rho)_m.
\end{equation}
If we now assume that the comoving particle number (proportional to $ma^3 $) of the fluid is conserved in the FRW universe, then we get the relation
\begin{equation}\label{20}
(\dot{m}/m)=-3 (\dot{a}/a),
\end{equation}
and substituting (\ref{20}) in the relation $$\dot{T}/T=(\dot{m}/m)\left(\dfrac{\partial p}{\partial \rho}\right)_m, $$ we obtain
\begin{equation}\label{21}
\dot{T}/T=-3 (\dot{a}/a)\left(\dfrac{\partial p}{\partial \rho}\right)_m.
\end{equation}
In order to determine $\left(\dfrac{\partial p}{\partial \rho}\right)_m  $, we use the equation of state of VMCG to arrive at the expression
\begin{equation}\label{22}
\left(\dfrac{\partial p}{\partial \rho}\right)_{m} =A +\frac{B_{0}\alpha a^{-n}}{\rho^{(\alpha + 1)}} + \frac{B_{0}n\rho^{-\alpha}}{a^{(n+1)}}\left(\dfrac{\partial a}{\partial \rho}\right)_{m}.
\end{equation}
Now using the conservation equation
\begin{equation}\label{23}
3\frac{\dot a}{a}(p+\rho)+ \dot \rho =0,
\end{equation}
and the equation of state for VMCG, we get
\begin{equation}\label{24}
\frac{B_{0}n\rho^{-\alpha}}{a^{(1+n)}}\left(\dfrac{\partial a}{\partial \rho}\right)_{m}=\frac{-nB_{0}a^{-n}}{3(1+A)\rho^{(1+\alpha)}-3B_{0}a^{-n}}.
\end{equation}
With the help of the equations (\ref{21}), (\ref{22}), (\ref{24}) and (\ref{09}) we finally obtain the following relation
\begin{align}
&\left(\dfrac{dT}{T}\right) =-3A\left(\dfrac{da}{a}\right)+\nonumber \\
&\frac{3nB_0 \left(\dfrac{da}{a}\right)}{3(1+A)\rho_0^{1+\alpha}[\Omega_{x}+(1-\Omega_{x}a_0^{-n})(a_0/a)^{3N} a_0^n]-3B_0}\nonumber \\
&- \frac{3B_0\alpha \left(\dfrac{da}{a}\right)}{\rho_0^{1+\alpha}[\Omega_{x}+(1-\Omega_{x}a_0^{-n})(a_0/a)^{3N} a_0^n]}. \label{25}
\end{align}
From (\ref{25}), we now calculate the temperature as a function of scale factor $a(t)$. This yields
\begin{align}
\left.T(a)\right. &= T_{0}\left(\frac{1}{a}\right)^{3A}\left(\dfrac{1+(\frac{1-\Omega_{x}}{\Omega_{x}})}{a^{3N} +(\frac{1-\Omega_{x}}{\Omega_{x}}) }\right) ^{\frac{\alpha}{1+\alpha}} \nonumber \\
&\times\left( \dfrac{(\frac{n}{n-3N})+{(\frac{1-\Omega_{x}}{\Omega_{x}})}}{a^{3N}(\frac{n}{n-3N})+(\frac{1-\Omega_{x}}{\Omega_{x}})}\right).\label{26}
\end{align}

To derive the temperature $ T(z) $ as a function of the redshift $z$, we substitute $(z+1)=\frac{a_0}{a}$ in $ T(a)$, and finally arrive at the expression

\begin{align} \label{T_z1}
\left. T(z)\right. &=\frac{T_0(z+1)^{3N(1+\frac{\alpha}{1+\alpha}) +3A}(\frac{1}{\Omega_x})^{\frac{\alpha}{1+\alpha}}}{[1+(z+1)^{3N}(\frac{a_0^n}{\Omega_{x}}-1)]^\frac{\alpha}{1+\alpha}} \nonumber \\
 &\times\frac{[\frac{1}{\Omega_x}(1-\frac{n}{3N})-\frac{1}{a_0^n}]a_0^{n(1+\frac{\alpha}{1+\alpha})}}
 {[(1-\frac{n}{3N})(\frac{a_0^n}{\Omega_{x}}-1)(z+1)^{3N}-\frac{n}{3N}]}.
\end{align}

\begin{figure}[ht]
 \centering
  \includegraphics[width=0.5\textwidth]{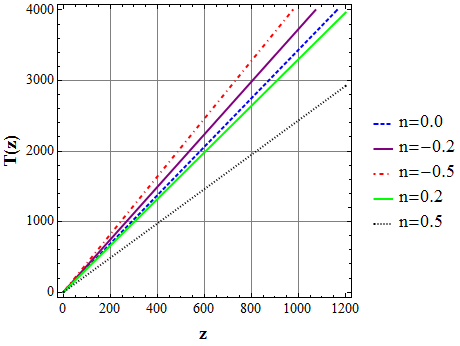}
 \caption{Variation of temperature  $ T(z) $ as a function of $z$ for different values of $ n $, where we have taken $ A=1/3 $, $ \Omega_x = 0.7 $, $ a_0=1 $, and $ \alpha=0.25 $ to see how the free parameter $ n $ affects the temperature $ T(z) $ in the VMCG model.}
\label{fig_1}
\end{figure}
This is the exact expression of temperature in terms of the redshift $ z $, for VMCG. In Fig.~\ref{fig_1}, we have plotted the temperature $ T(z) $ for different values of the free parameter $ n $. We can see that for large $ z $, the temperature decreases linearly with decreasing $ z $, but for small $ z $ it falls to zero in a gradual nonlinear fashion as $ z $ goes to negative values, indicating the possible future evolution of temperature of the universe. In this paper, wherever possible, we have extended the plots up to $z=-1$ in order to take into account the future evolution of the model.

\begin{figure}[ht]
\centering
\includegraphics[width=0.55\textwidth]{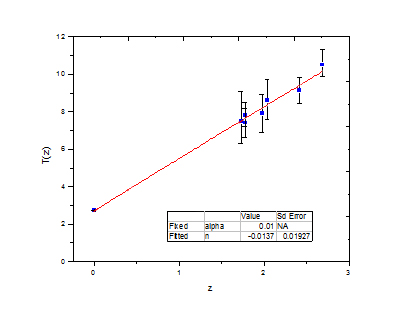}
\caption{A fit of the temperature $T(z)$ using some observational points available in literature (listed in TABLE I), where we have taken $ A=1/3 $, $ \Omega_x = 0.7 $, $ a_0=1 $, and $ \alpha=0.25 $ in the VMCG model. Here the value of $ \alpha $ was fixed to $ 0.01 $ and the fitted best value of $ n $ is $ -0.0137 $.}
\label{fig_1_b}
\end{figure}

The  cosmic microwave background radiation (CMBR) is a fundamental consequence of the hot Big-Bang. It is the radiation leftover after the decoupling from matter in the early evolutionary phases of the Universe. This radiation excites the rotational levels of some interstellar molecules, including carbon monoxide (CO), which can serve as a measuring device for the astronomers. Indirect measurement of $T(z)$ is one of the most powerful cosmological tests available.

Assuming that the CMB is the only source of excitation, Songaila et al. (1994) \cite{Songaila7.4} determined its temperature to be $T_{CMB} = 7.4 \pm 0.8 K$ from neutral carbon atoms at $z = 1.776$ in a cloud towards the quasar Q$1331+170$. Subsequent improvements have placed the estimate for the present CMB black-body temperature at the value of $T_{CMB} = 2.725\pm 0.002 K $ (Mather et al. 1999 \cite{mathercobe}), which was measured locally (at redshift z = 0).

Lima et al. \cite{lima} argued that the CMB temperature at high $z$ may be smaller than the predicted standard values, which opened the scope of alternative models for the big bang. Using new elements in the form of decaying vacuum energy density and gravitational `adiabatic' photon creation along with the late inflationary models driven by a scalar field, they deduced a new temperature law and compared its predictions with the standard cosmological results.

Srianand et al. (2008) \cite{sri9.15}, also assumed the CMB to be the only source of excitation and imposed stringent upper-limits on $T_{CMB}$ for a large sample of C {\small I} fine structure absorption lines detected in high signal to noise, high resolution spectra. They detected carbon monoxide in a damped Lyman$-\alpha$ system at $z_{abs} = 2.41837$ in the SDSS database towards SDSS J$143912.04+111740.5$, and from the CO excitation temperatures they determined $T_{CMBR} = 9.15 \pm 0.72 K$.

In their paper, J. Ge et al. \cite{BECHTOLD7.9} presented the detection of absorption lines from the ground state and excited states of C {\small I} in the $z=1.9731$ damped Lyman$\alpha$ system of the QSO $0013-004$ and estimated other contributions to the excitation of the C {\small I} fine-structure levels. They used the population ratio of the excited state to the ground state and estimated the CMBR temperature of $T =7.9 \pm 1.0 K$ at $0.61$ mm and $z=1.9731$, which matched with the predictions of standard cosmology at that time.

Noterdaeme et al. (2010) \cite{Noterdaeme10.5}, in  their paper, presented the analysis of a sub damped Lyman-$\alpha$ system with neutral hydrogen column density at $z_{abs}=2.69$ toward SDSS J$123714.60 + 064759.5$. The excitation of CO was found to be dominated by radiative interaction with the CMBR and they derived $T_{ex}(CO) = 10.5 K$ corresponding to the expected value of $T_{CMBR}(z=2.69)=10.05 K $.

Using three new and two previously reported CO absorption line systems detected in quasar spectra during a
systematic survey carried out using VLT$/$UVES, P. Noterdaeme et al. \cite{ Noterdaeme} constrained the
evolution of $T_{CMB}$ to $z\sim3$. Combining their measurements with previous constraints, they obtained $T_{CMB}(z)=(2.725 \pm 0.002)\times(1 + z)^{1-\beta} K$ with $\beta = -0.007\pm 0.027$.

All these have motivated us to derive exact expression for the temperature of the FRW universe dominated by VMCG matter as a function of redshift $z$, in order to check for the viability of this cosmological model by using this observational constraint.

\begin{table}[htbp]
  \centering
  \caption{ T(z) table for different values of redshift as obtained from different references mentioned in the column}
    \begin{tabular}{|rrr|}
    \hline
    z \,&\, T(z) \,&\, Reference \\
     \hline
      \hline
    1.776 \,&\, $7.4_{-0.8}^{+0.8}$ \,&\, \cite{Songaila7.4}  \\
    1.7293 \,&\, $7.5_{-1.2}^{+1.6}$ \,&\, \cite{Noterdaeme}\\
    1.7738 \,&\, $7.8_{-0.6}^{+0.7}$ \,&\, \cite{Noterdaeme}\\
    2.6896 \,&\, $10.5_{-0.6}^{+0.8}$ \,&\, \cite{Noterdaeme10.5,Noterdaeme}\\
    2.4184 \,&\, $9.15_{-0.7}^{+0.7}$ \,&\, \cite{sri9.15,Noterdaeme}\\
    2.0377 \,&\, $8.6_{-1.0}^{+1.1}$ \,&\, \cite{Noterdaeme}\\
    1.9731 \,&\, $7.9_{-1.0}^{+1.0}$ \,&\, \cite{BECHTOLD7.9}\\
    0     \,&\, $2.725_{-0.002}^{+0.002}$ \,&\, \cite{mathercobe}\\
    \hline
    \end{tabular}%
  \label{tab:addlabel}%
\end{table}%

We have used some of the observational temperature data points for different redshifts in the Fig.\ref{fig_1_b}  from the Table \ref{tab:addlabel} and used our theoretical curve to show the overall agreement with the cosmological observations.

If we substitute the parameter values $\alpha=0.25$, $A=1/3$ and $ \Omega_{x}=0.7$ in (\ref{T_z1}), then we get the expression of temperature as a function of the free parameter $n$:
\begin{align}
 T(z) &=2.9(z+1)\left[(z+1)^{n-5}+\frac{3}{7}\right]^{-1/5} \nonumber \\
 & \times \left(\dfrac{(13n-15)}{7n(z+1)^{n-5}+6n-15}\right).\label{28}
\end{align}
We should be able to get the corresponding expression for MCG if we put $n=0$ in equation (\ref{28}). After substituting $n=0$ in (\ref{T_z1}), we obtain
\begin{align}
T(z) =& T_0(z+1)^{3(R-1)}\nonumber \\
 & \times [\Omega_x + (1-\Omega_x)(z+1)^{3R}]^{-\alpha/1+\alpha}. \label{29}
\end{align}
This result matches exactly with the corresponding expression for the MCG as calculated by Bedran et al \cite{MCG3}. We now have a working formula for the temperature. We can use the boundary conditions as $ T_0=2.7K $ and substitute the values of other variables like $ A=1/3 $ (for the radiation phase) and $ \Omega_x = 0.7 $ (most commonly used and accepted dark energy parameter) in equation (\ref{T_z1}) to get the expression of temperature in terms of $ \alpha $ and $n$. If we substitute $ n=0 $ (the MCG case) and $ \alpha=\frac{1}{4} $ in the resulting expression, then we obtain
\begin{equation}
T(z)=2.90K (z+1)[(z+1)^{-5}+3/7]^{-\frac{1}{5}},\label{30}
\end{equation}
which is consistent with the expression of temperature in the paper \cite{MCG3}. Unfortunately for VMCG we don't know the constraints on $ \alpha $ and $ n $. From the consideration of thermodynamic stability in the case of VMCG, one of the authors \cite{VMCG2} have shown that the condition for stability is $ n\leq 0 $.

\begin{figure*}[ht]
 \centering
  \includegraphics[width=0.9\textwidth]{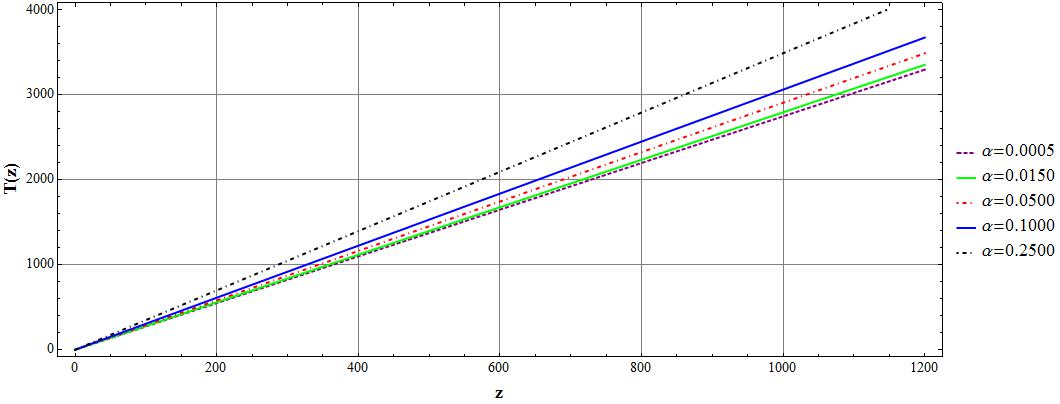}
 \caption{Variation of temperature  $ T(z) $ as a function of $z$ for different values of $ \alpha $, where we have taken $ A=1/3 $, $ \Omega_x = 0.7 $, and $ n=-0.01 $ to see how the free parameter $ \alpha $ affects the temperature $ T(z) $ in the VMCG model.}
\label{fig_1a}
\end{figure*}
In Fig.~\ref{fig_1a} we have shown the temperature evolution of the universe as a function of redshift for different values of $ \alpha $ where we have chosen a fixed value of the parameter $ n=-0.5 $. We can clearly see that as the value of $ \alpha $ increases, the temperature increases for a particular value of $n $. For the MCG model, from the above equation (\ref{30}), using the decoupling redshift as $ z\approx1100 $, the temperature of decoupling ($T_{d}$) is found to be $ T_{d}\approx 3800K $. For VMCG the decoupling temperature is complicated, and depends on the parameters $ n$ and $\alpha $. Putting the values $ A=1/3 $, $z=1100$ and $\Omega_{x}=0.7$ in equation (\ref{T_z1}), we arrive at the following relation
\begin{align} \label{T_d}
\left. T_{d}(n,\alpha)\right. &=\frac{(2.7)(1101)^{3N(1+\frac{\alpha}{1+\alpha}) +1}(1.43)^{\frac{\alpha}{1+\alpha}}}{[1+(1101)^{3N}(0.43)]^\frac{\alpha}{1+\alpha}} \nonumber \\
 &\times\frac{[(1.43)(1-\frac{n}{3N})-1]}
 {\left[(1-\frac{n}{3N})(0.43)(1101)^{3N}-\frac{n}{3N}\right]}.
\end{align}
where $3N=4(1+\alpha)-n$.

\begin{figure}[ht]
 \centering
  \includegraphics[width=0.50\textwidth]{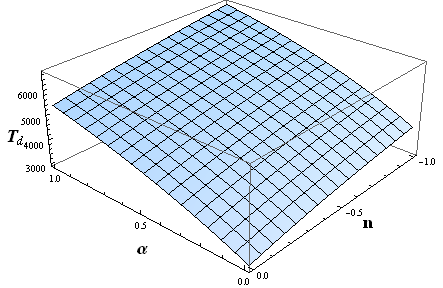}
 \caption{Variation of decoupling temperature $T_{d}$, as a function of $n$ and $\alpha$, where we have taken $ A=1/3 $, $ \Omega_x = 0.7 $, $ a_0=1 $ and $ z=1100 $ to see how the two free parameters affect the decoupling temperature.}
\label{fig_2}
 \end{figure}

In Fig.~\ref{fig_2} we have shown the dependence of decoupling temperature on the parameters $ n $ and $ \alpha $ for the VMCG model using the expression for decoupling temperature in equation (\ref{T_d}). We have calculated the values of decoupling temperatures for the VMCG model using $ \alpha=1/4 $, and $ n=-0.1 , -1.0 , -2.0 $, and the corresponding values are $ T_{d}\approx $ 3941 K, 5033 K, and 5764 K respectively. Thus we can see that the decoupling temperature increases with higher negative values of $ n $ for a fixed value of $ \alpha $ in the VMCG model.

The expression for energy density of VMCG in FRW universe can be used to determine the Hubble parameter $ H=\frac{\dot a}{a} $ as a function of redshift $ z $. We know that for the flat FRW universe we have
\begin{equation}
3({\dot a}/{a})^2 = \rho,
\end{equation}
which gives $ H^2=\rho/3 $, and so we have
\begin{equation}
H^2=H_{0}^2 (\rho/\rho_{0}).
\end{equation}
Now using the expression for $ \rho(a) $ and changing our variable to the redshift $ z $, we get
\begin{align}
H(z)= & H_{0}(z+1)^{n/2(1+\alpha)}\nonumber \\
 & \times [\Omega_{x} + (a_{0}^n - \Omega_{x})(1+z)^{3N}]^{1/2(1+\alpha)}.
\end{align}

\begin{figure}[ht]
\centering
  \includegraphics[width=0.5\textwidth]{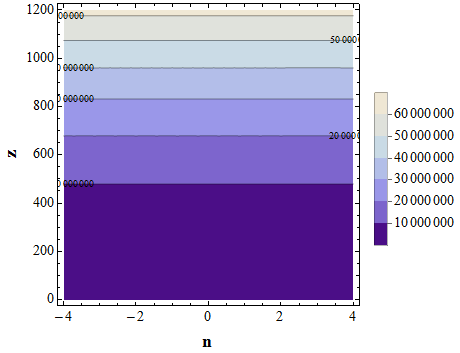}
\caption{Contour plot of Hubble parameter $H(z)$ as a function of $ z $ and $n$, where we have taken $ H_0=70 $, $ A=1/3 $, $ \Omega_x = 0.7 $, $ a_0=1 $, and $ \alpha=0.25 $ to see how the free parameter $ n $ affects $ H(z) $ in the VMCG model.}
\label{fig_3}
\end{figure}

In the contour plot of Fig.~\ref{fig_3} we have shown how the Hubble parameter varies with $ z $ for different values of $ n $. We can see that for high values of $ z $ there is no significant shift in the magnitude of the Hubble parameter for different values of $ n $.

\begin{figure}[ht]
 \centering
  \includegraphics[width=0.5\textwidth]{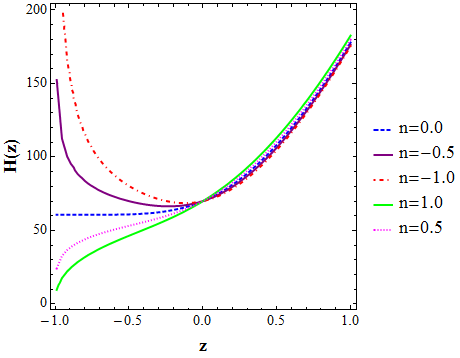}
 \caption{Plot of the Hubble parameter $H(z)$ as a function of $ z $ and $n$, where we have taken $ H_0=70 $, $ A=1/3 $, $ \Omega_x = 0.7 $, $ a_0=1 $, and $ \alpha=0.25 $ to see how the free parameter $ n $ affects $ H(z) $ in the VMCG model.}
 \label{fig_4}
 \end{figure}

We can also see in Fig.~\ref{fig_4} that for positive $ n $ the value of $ H $ decreases and approaches zero, whereas for negative $ n $ (i.e. phantom dominated universe) it increases rapidly as $ z $ approaches $ -1 $, indicating the Big rip that will occur in future. As the square of the Hubble parameter is proportional to the energy density $ \rho(z) $, it is clear that for negative $ n $ (thermodynamically stable condition) the energy density increases rapidly to infinity, as it should be in a phantom dominated universe as $ z $ approaches $ -1 $.\\
 In Fig.~\ref{fig_4} and in the subsequent figures, we have included the range $ -1<z<0 $ (which indicates blueshift with respect to the present epoch) to show how the different parameters of this VMCG model, like $H(z),\, z,\, W(z),\, q(z)\, \textrm{and}\, v^2(z)$, will vary in the future, and how they will differ from each other depending on the values of the other parameters in this model.

\begin{figure}[ht]
 \centering
  \includegraphics[width=0.5\textwidth]{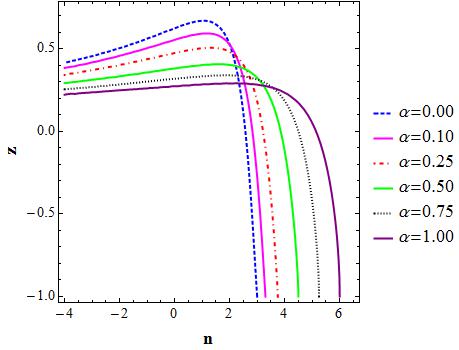}
 \caption{Variation of the redshift $z$ (for $ P=0 $) as a function of $n$, where we have taken $ A=1/3 $, $ \Omega_x = 0.7 $, $ a_0=1 $, $ \alpha=0.25 $ to see how the free parameter $ n $ affects the redshift of the dust phase in the VMCG model.}
 \label{fig_5}
\end{figure}

We can also find the redshift $ z $ when the pressure passes through the zero value ($ P=0 $). We substitute $ P=0 $ in the equation of state and use the expression for $ \rho(a) $ and substitute $ B_0 $ in terms of $ \Omega_x $ to arrive at the following expression:\\
\begin{equation}
[\Omega_x + (1-\Omega_x)(z+1)^{4(1+\alpha)-n}]= \left[\dfrac{4(1+\alpha)-n}{(1+\alpha)}\right]\Omega_{x}.
\end{equation}
\\
With $ A=1/3 $, $ \Omega_x = 0.7$ and setting $ n=0 $ (for the MCG) and $ \alpha = 1/4 $, we obtain the redshift as $ z=0.48 $. For nonzero values of $ n $, we get (for $ P=0 $) the dependence of $ z $ on $ n $ for different values of $ \alpha $ as shown in the Fig.~\ref{fig_5}.

It is evident that as we move towards more negative values of $ n $, the value of redshift (for $ P=0 $) decreases very slowly. As we vary $ \alpha $, we can see that as it increases for a fixed negative value of $ n $, the value of $ z $ decreases.\\ Therefore, for a phantom dominated universe in the VMCG model, the redshift for dust phase must be $ z<0.48 $ for the chosen values of the parameter $ \alpha = 0.25 $, whereas for positive $ n $ there is no such bound.

\begin{figure}[ht]
 \centering
  \includegraphics[width=0.5\textwidth]{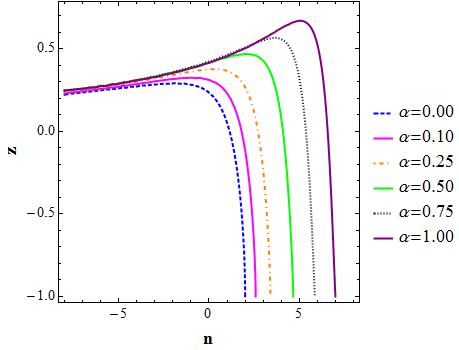}
 \caption{Variation of redshift (for $ \ddot{a}=0 $) as a function of $n$, where we have taken $ A=1/3 $, $ \Omega_x = 0.7 $, $ a_0=1 $ and $ \alpha=0.25 $ to see how the free parameter $ n $ affects the redshift at the time of flip in acceleration in the VMCG model.}
 \label{fig_6}
 \end{figure}

We can also calculate the redshift of transition of the expansion of the universe from deceleration to acceleration. Using the equation of state and setting the condition $ \ddot{a}=0 $, i.e. $ 3P+\rho=0 $, we get the following relation\\
\begin{equation}
[\Omega_x + (1-\Omega_x)(z+1)^{4(1+\alpha)-n}]= \left[\dfrac{4(1+\alpha)-n}{2(1+\alpha)}\right]\Omega_{x},
\end{equation}
\\
where we have assumed that $ A=1/3 $. Now if we substitute $ \Omega_x = 0.7$, $ \alpha=0.25 $ and $ n=0 $ (representing the MCG), we get $ z=0.18 $ (for MCG). The variation of $ z $ (for transition from deceleration to acceleration phase) as a function of $ n $ for different values of $ \alpha $ can also be seen in Fig.~\ref{fig_6}.\\
As we vary $ \alpha $, we can see that as it increases for a fixed negative value of $ n $, the value of $ z $ increases. From the plot it is clear that if $ n>0 $ (i.e. the Big rip is avoided), the value of the redshift for the flip in acceleration in the VMCG model must be $ z<0.18 $ for the chosen values of the parameter $ \alpha=0.25 $, but for negative value of $ n $, such a conclusion cannot be drawn.

\newpage
\section{Discussions}

\begin{figure}[ht]
 \centering
  \includegraphics[width=0.5\textwidth]{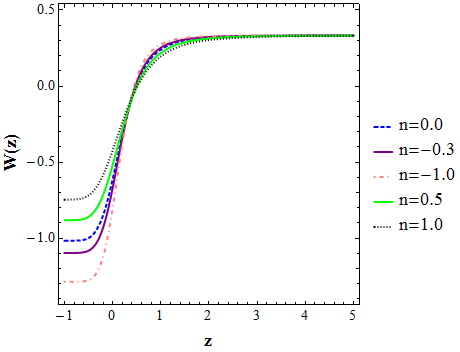}
 \caption{Variation of EOS parameter $\textit{W(z)}$ for different values of $n$, where we have taken $ A=1/3 $, $ \Omega_x = 0.7 $, $ a_0=1 $ and $ \alpha=0.25 $ to see how the free parameter $ n $ affects $\textit{W(z)} $ in the VMCG model.}
 \label{fig_7}
 \end{figure}

In this section we discuss the variation of some useful cosmological parameters of the VMCG in terms of the redshift $ z $, which is a parameter which can be easily measured through observations. First we consider the equation of state (EOS) for the VMCG and use the expression of $ \rho(a) $ to derive the equation of state parameter $ \textit{W(z)}=P/\rho $ in the form
\begin{equation}\label{W_z}
 \textit{W(z)}=A - \frac{N}{(1+\alpha)}\frac{1}{[1+(\frac{1}{\Omega_{x}}-1)(z+1)^{3N}]}.
\end{equation}
Analysing (\ref{W_z}) we find that for high $ z $, the EOS parameter approaches $\textit{W(z)}\simeq A$, and as it should correspond to the `Radiation-dominated phase' of the universe, we can safely say that $ A $ must have the value $ 1/3 $. For small $ z $, the EOS parameter approaches the value $ \textit{W(z)}\simeq -1+\frac{n}{3(1+\alpha)} $. As this expression explicitly depends on $ n $, it means that if $ n $ is negative, then $\textit{W(z)}<-1 $, which corresponds to the phantom-dominated universe and Big rip is unavoidable, whereas for $ n\geq 0 $, the EOS parameter becomes $ \textit{W(z)}\geq -1 $, so that Big rip is avoided. From the plot in Fig.~\ref{fig_7} we find that for different values of $ n $, as $ z $ increases, the value of the EOS parameter approaches $1/3$, and further the plot also shows the position where the pressure becomes zero and then negative, approaching different negative values for different values of $ n $. The value of the redshift for the `dust phase' ($ P=0 $) is very close to the value of $ z\simeq 0.48 $ depending on the value of $ n $, which agrees with our analysis in the previous section.

\begin{figure}[ht]
 \centering
  \includegraphics[width=0.5\textwidth]{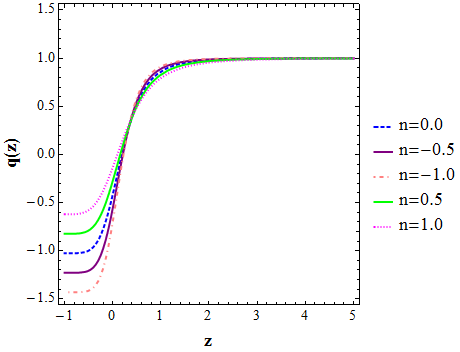}
 \caption{Variation of deceleration parameter ($q(z) $) for different values of $n$, where we have taken $ A=1/3 $, $ \Omega_x = 0.7 $, $ a_0=1 $ and $ \alpha=0.25 $ to see how the free parameter $ n $ affects $q(z) $ in the VMCG model.}
 \label{fig_8}
 \end{figure}

It is known that the deceleration parameter $ q(z) $ is related to $\textit{W(z)}$ by the relation \cite{VMCG2}
\begin{equation*}
 q(z)=1/2 + 3\textit{W(z)}/2.
\end{equation*}
Thus the expression for the deceleration parameter $ q(z)  $ as a function of redshift $ z $ is
\begin{align}
q(z)= & 1/2 + (3/2) \nonumber \\
 & \times \left[A - \frac{N}{(1+\alpha)}\frac{1}{[1+(\frac{1}{\Omega_{x}}-1)(z+1)^{3N}]}\right].
\end{align}
Fig.~\ref{fig_8} clearly indicates the variation of $q(z)$ with $ z $ for different values of $ n $. For large $ z $, the expression becomes  $ q(z)\simeq 1/2 + 3A/2 $, which is constant, and for small $ z $ the deceleration parameter takes the form $ q(z)\simeq -1 + \frac{n}{2(1+\alpha)} $, which again explicitly depends on $ n $. \\
This means that $q(z)$ was constant in the radiation phase, then gradually decelerated while passing through the dust phase and then entered the current accelerating dark energy dominated phase. Depending on the value of $ n $, $q(z)$ approaches different values for small $ z $. For positive values of $ n $, $ q(z)>-1 $ and for $ n \leq 0 $, $ q(z) \leq -1 $. From the figure one can easily see that $q(z)$ crosses zero (i.e. the moment of flip from deceleration to acceleration) near $ z\simeq 0.18 $, depending on the value of $ n $, which conforms to our calculations in the previous section.

Similarly we can calculate the velocity of sound $ v^2_{s}=(\partial P/\partial \rho)_{s} $. Using the equation of state for VMCG we obtain the relation
\begin{equation}\label{v_s}
v^2_{s} = A +\frac{B\alpha}{\rho^{(1+\alpha)}}-\frac{\rho^{-\alpha}B_{0}(-n)a^{-(n+1)}}{\partial \rho/\partial a}.
\end{equation}
If we calculate $(\partial \rho/\partial a) $ from (\ref{09}) and substitute it in (\ref{v_s}), then we get the expression for the velocity of sound as a function of $ z $:
\begin{align}
\left.v^2_{s}(z)\right. &= A + \frac{(N\alpha/(1+\alpha))}{[1+(\frac{1-\Omega_{x}}{\Omega_{x}})(z+1)^{3N}]} \nonumber \\
&-\frac{Nn}{n +(\frac{1-\Omega_x}{\Omega_x})(1+z)^{3N}(n+3N)}
\end{align}

\begin{figure}[ht]
 \centering
  \includegraphics[width=0.5\textwidth]{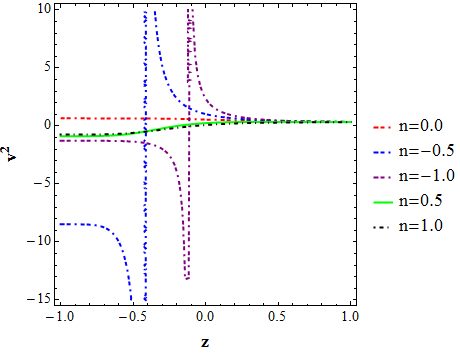}
 \caption{Variation of velocity of sound $ v^2_{s}(z) $ for different values of $n$, where we have taken $ A=1/3 $, $ \Omega_x = 0.7 $, $ a_0=1 $ and $ \alpha=0.25 $ to see how the free parameter $ n $ affects $v^2_{s}(z) $ in the VMCG model.}
 \label{fig_9}
\end{figure}
From Fig.~\ref{fig_9} we find that the velocity has a magnitude lying below unity and the nature is consistent at large $ z $, because for large value of redshift i.e., in the early phase of the universe, the velocity was  $v^2_{s}(z)\simeq A =1/3  $, and then as the redshift $ z $ became smaller, the velocity of sound increased rapidly for $ n<0 $ (phantom dominated universe). \\
After that the velocity becomes imaginary. \\
Whereas for $ n>0 $, as the redshift decreases, it slowly decreases and becomes negative. For small values of $ z $, the velocity of sound is given by \\
\begin{equation}
v^2_{s}(z)\simeq -1+ \frac{n}{3(1+\alpha)}, \nonumber
\end{equation}
\\
 which for negative $ n $ is always negative (signifying imaginary speed), and for the speed to be real, we must have $ n>3(1+\alpha) $, which is thermodynamically unstable for positive $ \alpha $. \\
Therefore this scenario for $ n<0 $ indicates a perturbative cosmology and favours structure formation in the universe \cite{sound}. Whereas for $ n=0 $, the velocity approaches a constant positive value for small values of $ z $.

\newpage
\section{Conclusions}

In this paper, we have determined the energy density of VMCG matter in a FRW universe using a thermodynamic approach, and derived the exact expression for the temperature $ T(z) $ and the Hubble parameter $ H(z) $ of the corresponding universe as a function of redshift $ z $. We have derived the redshift for the `dust phase' ($ P=0 $) and for  the epoch of transition from deceleration to acceleration ($ \ddot{a}=0 $) of the FRW universe. We have also determined the dependence of the redshift during the different phases of expansion of the universe on the free parameter $ n $. Subsequently we have shown the dependence of the equation of state parameter $ \textit{W(z)} $, deceleration parameter $ q(z) $ and the velocity of sound $ v^2(z) $ on the free parameter $ n $ as a function of redshift $z$. We find that the VMCG model perfectly represents the three different phases of the universe namely the `Radiation phase' ($ P=\rho/3 $), the `dust phase' ($ P=0 $), and later the negative pressure epoch dominated by the so called `Dark energy'. We also find that the VMCG model with $ n<0 $ (for thermodynamic stability) and other accepted values of the parameters, explains the value of decoupling temperature very well.

Therefore from the above analysis we can conclude that a FRW universe filled with thermodynamically stable variable modified Chaplygin gas not only represents the three phases of evolution of the universe very well along with the change of expansion rate from deceleration to acceleration, but also it shows a consistent temperature evolution of universe.

\newpage

In this context we like to mention that two of the authors have also examined the validity of the generalized second law of thermodynamics (GSLT) on the cosmological apparent horizon (AH) and the event horizon (EH) of FRW universe dominated by various types of Chaplygin gas fluids, one of them being the VMCG \cite{ChakGuha}. The GSLT is always valid on the apparent horizon of the VMCG dominated FRW universe. But for $ n<0 $ (i.e. $R_{EH}<R_{AH}  $), the VMCG dominated FRW universe violates the GSLT on the event horizon in the early phase of the universe but it holds in our current epoch and will also hold in the future, where we know that in this range of $ n<0 $, the VMCG model itself is thermodynamically stable \cite{VMCG2}. ''

\part{Epilogue}
Finally in this section we will conclude our thesis and discuss an overall outlook about the work we have done. 

\bigskip

In the introductory portion of this thesis we introduced the basic concepts regarding the thermodynamic studies in gravitational physics. We also highlighted the importance of such studies as it has the potential to uncover the deeper theoretical connections between gravity and other fundamental forces. No doubt this is a very vast and important field encompassing every fundamental aspect of gravity from spacetime structure to black holes physics. Throughout the literature we found that thermodynamical studies of gravity point towards the similarities between thermodynamical laws and gravitational physics. In addition, it also indicates the close connection between the laws of black hole mechanics and the generalized second law of thermodynamics (GSLT). In fact, in a sense, these two things cannot be considered to be separate as the preservation of the second law of thermodynamics required the introduction of the BH horizon entropy. 

Interestingly, even from the perspective of information theory, the concept of entropy is intertwined with the idea of a horizon. Speaking in simple terms, if some surface hides some information from an observer then that observer can attach the notion of an entropy to that surface. This idea is also applicable to BHs as no information can escape from inside the BH, and hence, intuitively, it makes perfect sense to attach the entropy on the BH horizon, as it is the causal horizon separating the events. Also from purview of the equivalence principle, a similar treatment can be thought of for any Rindler horizon, effectively subverting the uniqueness of the BH horizon and making the entropy observer dependent. This line of investigation again puts the GSLT in a broader context, as one can attach entropy to cosmological horizons and argue that in order to get the total entropy of the universe one must also count the cosmological horizon entropy, effectively generalizing the second law. 

In our work we argued that the GSLT should be taken into account when considering the models of various cosmological fluid. We have concentrated our studies on various Chaplygin gas models which have the potential to model the dark energy and in some cases both the Dark and Baryonic matter. We have shown conclusively that many of such models when tested against the GSLT falls short or gets restricted in their parameter space, potentially introducing a new method to check such theoretical models' validity. This is very important because this can rule out various models from the thermodynamic standpoint and if in future any conclusive model emerges from the astronomical studies then that model can be tested in this manner to check the validity of the GSLT itself, acting as a litmus test of both the models and the theory. This work will also help us to model the class of Chaplygin gases or similar fluids in a far more theoretically viable manner. To show the usefulness of the thermodynamic approach, We have also considered one of the important variant of the Chaplygin gas, namely, the VMCG and considered it as a cosmic fluid model for the universe. Though in this study we did not explicitly invoke the idea of horizon but we considered the universe as a bounded system with a volume, and then used different thermodynamical laws and identities to arrive at the temperature profile of the universe, which showed remarkable overall similarity with the actual temperature profile of the cosmos. This is a remarkable result, as the model is a crude approximation of the overall mass profile of the universe with no nuances, but if there also we see such temperature profiles, then we can conclude that the laws of thermodynamics are actually applicable on an universal scale and the Chaplygin gas models are actually very good approximations for the study of the behaviour of the universe. This temperature profile is an exact theoretical result, which can reproduce other Chaplygin gas temperature profiles in appropriate parameter limits and can also be used to determine different temperatures at different phases of the cosmos, like the temperature of decoupling. In fact we have used this result to check the entropy profile of the cosmological horizon, which differed from the result when we used the Kodama temperature, indicating that such fluid temperature profiles cannot be used for the horizon temperature. The primary reason for such a consequence might be that there are unresolved issues with thermal equilibrium when we use such approximated large scale fluid models, which means that one needs to look at the cosmological horizon more  minutely and be mindful of the new physics involved there. Another reason might be that such near equilibrium thermodynamic laws are not sufficient for the dynamical nature of the cosmological horizon. Overall, the two studies on Chaplygin gas models showed us that we should not assume thermal equilibrium between the horizon and the bulk. It indicates that such approximations may only be possible when we include the horizon dynamics when calculating bulk temperature. In such scenarios the horizon thermodynamics should be included with the bulk thermodynamics. One important observation that we must mention here is that during the determination of the temperature profile we neglected the cosmological horizon. Possibly for this reason this mismatch may have occurred. A more fundamental issue arises from this: Is there any relation between the bulk temperature and the horizon temperature? From the BH laws it is evident that the horizon temperature is related to the surface gravity whereas the bulk temperature arises out of the internal energy dynamics of the matter. Therefore if there are universal thermodynamic laws, then how should we relate these two temperatures and how does the thermal equilibrium happen? We were confronted with these questions and will take up the investigations in the near future.

After this we took up the issue of gravitational entropy (GE). Although there were studies regarding GE, we found much of the areas of investigation are still unexplored. For that reason we wanted to investigate the consequences that we are led to when we try to generalize the study of GE to more general systems like accelerating BHs and we found that the concept of GE is more or less applicable to these systems and in such cases the GE also has the form of the Hawking-Bekenstein entropy. The most remarkable finding is the appearance of the acceleration parameter in the expression of GE, indicating a deep connection between acceleration and temperature(like the Unruh effect). This means that, if a BH is accelerating, then its GE is different from that of a non-accelerating BH. In other words, an accelerated observer will measure a different GE from that of a non accelerated one. This is a very important finding because it shows the generality of GE (BH entropy is a special case of GE) even in extreme systems like accelerating BHs. We did similar studies on different cosmological spacetimes (representative of different epochs of evolution of the universe) also and found in each case that the GE increases monotonically with time and tends to zero near the cosmological singularity, which validated the Weyl curvature hypothesis. In this study we also showed conclusively that the GE is directly correlated with the structure formation of the universe. The entire investigation validated the basic ideas of GE proposed by Clifton, Ellis and Tavakol (CET). We then tested the different GE proposals on traversable wormhole systems and found that there also more or less the idea of GE is viable. The idea behind this study was to check the idea that if these systems are traversable then it must have a viable GE. We found no detailed work in this area that is why this work is the first one on a large class of WHs. We not only studied the WH GE in great detail but also made a comparison between two important proposals which we called the Weyl scalar proposal and the CET proposal. The relation between gravitational temperature and the Tolman temperature of WHs were also studied. We hope that this work will pave the way for new investigations in this field. These studies were new, extensive and detailed, which showed many flaws and strengths of the GE proposals. We consider these studies to be very important in this field of research as it not only validated many ideas but also gave us new issues to ponder upon.

One might get tempted to establish a possible relation between the thermodynamic entropy of horizons ($ S_{H} $) and the gravitational entropy ($ S_{\textrm{grav}} $) of cosmological horizons, but we must be careful before asserting such a claim. Let us take the simplest example of the FLRW universe as a test case and try to find out these possible connections. On the cosmological horizon we can employ the previously discussed Weyl scalar proposal and find the gravitational entropy on that surface.
Such considerations quickly reveal that the GE on the surface of the cosmological horizon is zero as the FLRW spacetime is conformally flat, therefore the rate of change of the GE is also zero amounting to a net zero contribution of the free gravitational field to the total entropy change of that universe. The same is true for the CET proposal, as the FLRW universe is conformally flat, resulting in a zero gravitational energy density.

We can clearly see from this simple example that the gravitational entropy of the cosmological horizon, irrespective of the nature of proposal, is different from the usual horizon entropy, which is due to the energy flux crossing from the bulk under the conditions near thermal equilibrium, because the gravitational entropy has nothing to do with matter and purely comes from the free gravitational field.

Similarly, another interesting comparison can be made regarding the horizon temperature. Although the temperature of the horizon cannot be deduced from the Weyl scalar proposal, but for the FLRW case the Weyl scalar GE of the cosmological horizon is zero anyway, and therefore the temperature is unspecified. Moving on to the CET proposal, we find that there we obtained the gravitational temperature as: $ T_{grav}=H/2\pi$, for the FLRW spacetime, which is distinct from the Kodama-Hayward temperature of the horizon. Hence, one must not confuse between the two types of entropies studied in this thesis. The GE of the cosmological horizon arises from the free gravitational field, whereas the usual thermodynamic horizon entropy used in the later section of this work is a consequence of the validity of the first law of thermodynamics on the horizon.

Let us also comment on the state of second law in the light of this topic. If we consider the BHs, then the total entropy is $S_{total}= S_{BH}+ S_{Univ}$, i.e., we must consider the BH entropy in order to have a complete picture of the generalized second law. Similarly when we again think of the cosmological horizons, then the total entropy must be taken as : $S_{total}= S_{Hor}+ S_{Univ}$. These two generalizations are somewhat similar as in both the cases it is the horizon entropy that is being included within the usual thermodynamic entropy. This is also very interesting to see that the proposal of GE also suggests that the free gravitational field also carry entropy and it must be included in the total entropy of the universe, moreover the studies on GE also indicates that the BH entropy is a special case of GE when applied on BHs. Therefore it seems to suggest that if we take $ S_{total}= S_{grav}+ S_{Univ} $, then the inclusion of GE automatically covers the previous two generalizations to some extent. This statement is true only if we are to apply the definition of GE (e.g. Weyl scalar proposal) on cosmological horizons also like BH horizon. Hence from our studies and the other viability studies of the GE, it appears that the GSLT should be considered by taking into account the contribution of the GE, i.e. total entropy$=$ thermodynamic entropy $+$ gravitational entropy, although, more rigorous and exhaustive studies are needed on several issues before arriving at such a conclusion definitively.


\chapter{Appendix : List of Figures}
\bgroup
\hypersetup{linkcolor = black}
\listoffigures
\egroup



\end{document}